\newcommand{\fonttypesizestring}{12pt}

\documentclass[a4paper,\fonttypesizestring,twoside,onecolumn,openright,final]{memoir}

\newcommand{\othersfigdir}{0.graphics/figs.others}

\usepackage{amsmath,amssymb}
\usepackage{amsfonts}
\usepackage{dsfont}
\usepackage{fixmath}
\usepackage{enumerate}
\usepackage{mathrsfs}
  \usepackage{enumitem}
\usepackage{tocloft}

\cftsetindents{part}{0in}{0.3in}
\cftsetindents{chapter}{0in}{0.3in}
\cftsetindents{section}{0.3in}{0.5in}
\cftsetindents{subsection}{0.8in}{0.5in}

\usepackage{ifthen}

\usepackage[iso-8859-7]{inputenc}

\usepackage[greek,english]{babel}
\usepackage{teubner}
\usepackage{amsmath, bm}

\usepackage{amssymb}

\usepackage{amsthm}

\usepackage{psfrag}

\usepackage{numprint}

\usepackage{calc}

\usepackage[printonlyused,smaller]{acronym}

\usepackage[intoc,noprefix]{nomencl}

\usepackage[numbers,sort&compress]{natbib}

\makeatletter\let\c@lofdepth\relax\let\c@lotdepth\relax\makeatother
\usepackage[small,bf,tight]{subfigure}

\usepackage{dsfont}

\usepackage{dtklogos}

\usepackage{bigstrut}

\usepackage{multirow}

\usepackage{blkarray}
\makeatletter\let\BA@quicktrue\BA@quickfalse\makeatother

\usepackage{textcomp}

\usepackage{mathrsfs}

\usepackage{balance}
 \usepackage{algpseudocode}
  \usepackage[chapter]{algorithm}
  
  \usepackage{rotating}
\usepackage{array,multirow}
\usepackage{diagbox}

\setstocksize{297mm}{210mm}

\settrimmedsize{1.0\stockheight}{1.0\stockwidth}{*}

\settrims{0pt}{0pt}

\setlist[itemize]{noitemsep, topsep=0pt}
\setlist[enumerate]{noitemsep, topsep=0pt}

\settypeblocksize{24cm}{15.9cm}{*} %JUST FOR THE DRAFT VERSION

\setlrmargins{*}{*}{1} %JUST FOR THE DRAFT VERSION

\setulmargins{*}{*}{1} %JUST FOR THE DRAFT VERSION

\checkandfixthelayout[classic]

\newcommand{\beforechapskipstring}{0pt} %default: 50pt

\makechapterstyle{vkapstylemain}{%
\chapterstyle{default} %other interesting styles: bianchi, lyhne
\setlength{\beforechapskip}{\beforechapskipstring}
}

\makechapterstyle{vkapstyleothers}{%
\chapterstyle{default}
\setlength{\beforechapskip}{\beforechapskipstring}
}

\makeevenhead{headings}{\thepage}{}{\footnotesize\leftmark}

\makeoddhead{headings}{\footnotesize\rightmark}{}{\thepage}

\nouppercaseheads

\makeheadrule{headings}{\textwidth}{\normalrulethickness} %(this is 0.4pt)



\makeatletter
\renewcommand{\fnum@table}[1]{\textbf{\tablename~\thetable} }
\renewcommand{\fnum@figure}[1]{\textbf{\figurename~\thefigure} }
\makeatother

\newlength{\halflinewidth}
\newlength{\quarterlinewidth}

\makeatletter\renewcommand{\@biblabel}[1]{[#1]}\makeatother

\newcommand{\flag}[2]{\newboolean{#1}\setboolean{#1}{#2}}

\makeatletter\newcommand*{\lptnum}[1]{\npdecimalsign{.}\numprint{\strip@pt#1}}\makeatother

\newcommand{\pttoin}{0.013837000138370001383700013837} %1pt in inches
\newcommand{\linnum}[1]{\setlength{\auxlength}{\pttoin#1}\lptnum{\auxlength}}

\newcommand{\pttocm}{0.03514598035145980351459803514598} %1pt in centimeters
\newcommand{\lcmnum}[1]{\setlength{\auxlength}{\pttocm#1}\lptnum{\auxlength}}

\newcommand{\filenamenlo}{\jobname.nlo}
\newcommand{\filenamenls}{\jobname.nls}

\makeatletter\newcommand{\nomenclaturesortline}[3]{\nomenclature[#1\two@digits{\inputlineno}]{#2}{#3}}\makeatother

\newcommand{\nomgroupname}[1]%
{\ifthenelse{\equal{#1}{C}}{Complex Numbers}%
{\ifthenelse{\equal{#1}{D}}{Distributions}%
{\ifthenelse{\equal{#1}{F}}{Functions}%
{\ifthenelse{\equal{#1}{H}}{Set Theory}%
{\ifthenelse{\equal{#1}{M}}{Matrices \& Vector Spaces}%
{\ifthenelse{\equal{#1}{O}}{Operators}%
{\ifthenelse{\equal{#1}{R}}{Random Variables}%
{\ifthenelse{\equal{#1}{S}}{Stochastic Processes}%
{\ifthenelse{\equal{#1}{T}}{Transformations \& Mappings}%
{\ifthenelse{\equal{#1}{Z}}{Other Notations}%
}}}}}}}}}}

\renewcommand{\nomgroup}[1]%
{\ifthenelse{\equal{#1}{C}}{\medskip\item[\textbf{\nomgroupname{#1}}]\rule{\linewidth}{\normalrulethickness}}%
{\ifthenelse{\equal{#1}{D}}{\medskip\item[\textbf{\nomgroupname{#1}}]\rule{\linewidth}{\normalrulethickness}}%
{\ifthenelse{\equal{#1}{F}}{\medskip\item[\textbf{\nomgroupname{#1}}]\rule{\linewidth}{\normalrulethickness}}%
{\ifthenelse{\equal{#1}{H}}{\medskip\item[\textbf{\nomgroupname{#1}}]\rule{\linewidth}{\normalrulethickness}}%
{\ifthenelse{\equal{#1}{M}}{\medskip\item[\textbf{\nomgroupname{#1}}]\rule{\linewidth}{\normalrulethickness}}%
{\ifthenelse{\equal{#1}{O}}{\medskip\item[\textbf{\nomgroupname{#1}}]\rule{\linewidth}{\normalrulethickness}}%
{\ifthenelse{\equal{#1}{R}}{\medskip\item[\textbf{\nomgroupname{#1}}]\rule{\linewidth}{\normalrulethickness}}%
{\ifthenelse{\equal{#1}{S}}{\medskip\item[\textbf{\nomgroupname{#1}}]\rule{\linewidth}{\normalrulethickness}}%
{\ifthenelse{\equal{#1}{T}}{\medskip\item[\textbf{\nomgroupname{#1}}]\rule{\linewidth}{\normalrulethickness}}%
{\ifthenelse{\equal{#1}{Z}}{\medskip\item[\textbf{\nomgroupname{#1}}]\rule{\linewidth}{\normalrulethickness}}%
}}}}}}}}}}

\makeatletter
\newcommand*{\defeq}{\mathrel{\rlap{\raisebox{0.3ex}{$\m@th\cdot$}}\raisebox{-0.3ex}{$\m@th\cdot$}}=}
\makeatother

\makeatletter
\newcommand*{\defeqr}{=\mathrel{\rlap{\raisebox{0.3ex}{$\m@th\cdot$}}\raisebox{-0.3ex}{$\m@th\cdot$}}}
\makeatother

\theoremstyle{plain} %Default
\newtheorem{proposition}{Proposition}[chapter]
\newtheorem*{proposition*}{Proposition}

\newtheorem*{lemma*}{Lemma}
\newtheorem{theorem}{Theorem}[chapter]
\newtheorem*{theorem*}{Theorem}

\newtheorem*{corollary*}{Corollary}

\theoremstyle{definition}

\newtheorem*{definition*}{Definition}

\newtheorem*{example*}{Example}

\theoremstyle{remark}
\newtheorem{remark}{Remark}[chapter]
\newtheorem*{remark*}{Remark}

\newtheorem*{conclusion*}{Conclusion}

\def\execute{\begingroup\catcode`\%=12\catcode`\\=12\executeaux}
\def\executeaux#1{\immediate\write18{#1}\endgroup}

\execute{makeindex mythesis.en.nlo -s nomencl.ist -o mythesis.en.nls}
\makenomenclature

\newlength{\auxlength}

\newlength{\fonttypesize}

\theoremstyle{plain}

\setsecnumdepth{subsubsection}

\settocdepth{subsection}

\setbeforesubsecskip{-\onelineskip}
\setaftersubsecskip{\onelineskip}

\usepackage{titlesec}

\titleformat{\paragraph}
{\normalfont\normalsize\bfseries}{\theparagraph}{1em}{}
\titlespacing*{\paragraph}
{0pt}{3.25ex plus 1ex minus .2ex}{1.5ex plus .2ex}

\flag{coverpageflag}{true}
\flag{titlepageflag}{true}
\flag{copyrightpageflag}{true}
\flag{dedicatepageflag}{true}
\flag{abstractflag}{true}
\flag{perilhpshflag}{false}
\flag{acknowledgementsflag}{true}
\flag{tocflag}{true}
\flag{listfigures}{true}
\flag{listtables}{true}
\flag{listalgorithms}{true}
\flag{abbreviationsflag}{true}
\flag{nomenclatureflag}{true}
\flag{bibliographyflag}{true}
\flag{publicationsflag}{false}
\flag{colophonpageflag}{false}
\flag{arenachapterflag}{false}
\flag{chapteroneflag}{true}
\flag{chaptertwoflag}{true}
\flag{chapterthreeflag}{true}
\flag{chapterfourflag}{true}
\flag{chapterfiveflag}{true}
\flag{chaptersixflag}{true}
\flag{chaptersevenflag}{true}
\flag{chaptereightflag}{true}
\flag{chapternineflag}{true}
\flag{chaptertenflag}{true}
\flag{appendicesflag}{true}

\begin{document}

\ifcoverpageflag
%COVER PAGE%%%%%%%%%%%%%%%%%%%%%%%%%%%%%%%%%%%%%%%%%%%%%%%%%%%%%%%%%%%%%%%%%%%%%
\begin{titlingpage}
\centering%
\Large{\textsc{Aristotle University of Thessaloniki\\
School of Electrical and Computer Engineering\\
Telecommunications Department}}\\[0.5cm]

\begin{table}[h]
\centerfloat%
\includegraphics[width=3.0cm]{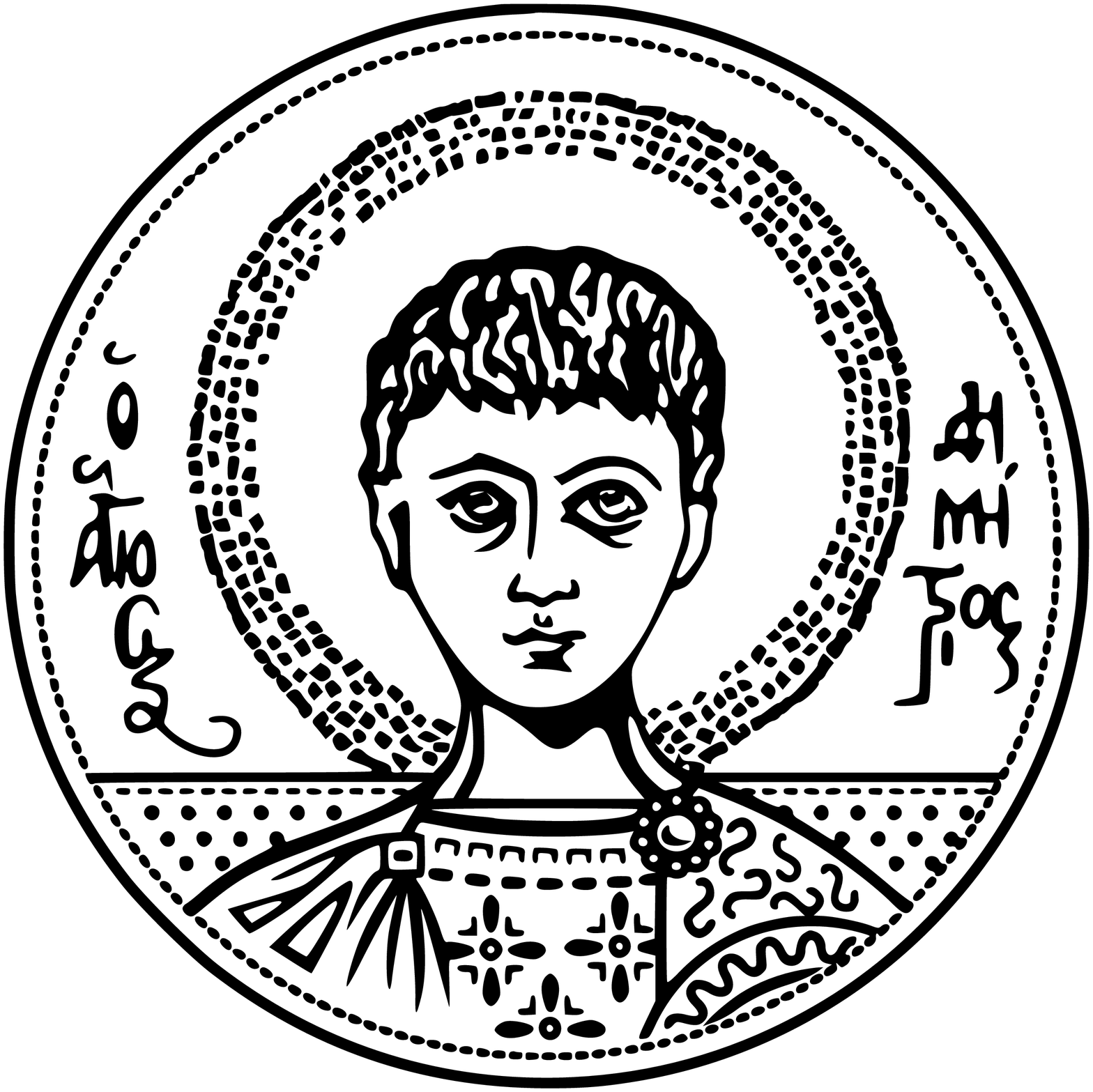}
\end{table}
\vspace{2cm}

\LARGE{\textbf{Panagiotis D. Diamantoulakis}}\\
\Large{\textit{Diploma in Electrical and Computer Engineering}}\\[1cm]

\Large{RESOURCE ALLOCATION IN WIRELESS NETWORKS WITH ENERGY CONSTRAINTS}\\[1cm]

\LARGE{\textbf{Ph.D Thesis}}
\null\vfill
%\Large{\textsc{Thessaloniki\\September 2017}}
\end{titlingpage}

%COVER PAGE%%%%%%%%%%%%%%%%%%%%%%%%%%%%%%%%%%%%%%%%%%%%%%%%%%%%%%%%%%%%%%%%%%%%%
\else\fi

\newpage

\iftitlepageflag
%TITLE PAGE%%%%%%%%%%%%%%%%%%%%%%%%%%%%%%%%%%%%%%%%%%%%%%%%%%%%%%%%%%%%%%%%%%%%%
%The headers and footers are empty (no page numbering as well)
\thispagestyle{empty}

\begin{centering}
\Large{RESOURCE ALLOCATION IN WIRELESS NETWORKS WITH ENERGY CONSTRAINTS.}\\[1cm]

\large{\textbf{by}}\\[1cm]

\large{\textbf{PANAGIOTIS D. DIAMANTOULAKIS}}\\[2cm]
A Dissertation submitted to the\\
\textbf{Aristotle University of Thessaloniki}\\
in partial fulfillment of the requirements\\
for the Degree of Doctor of Philosophy\\[1cm]

%September 2017\\[2cm]
\end{centering}

\textbf{Approved by the Thesis Committee:}

\begin{itemize}
\item Dr.~George~K.~Karagiannidis, Supervisor\\
Professor, Aristotle University of Thessaloniki, Greece
\item Dr.~Dimitris~P.~Labridis, Advisory Committee Member\\
Professor, Aristotle University of Thessaloniki, Greece
\item Dr.~Fotini-Niovi~Pavlidou, Advisory Committee Member\\
Professor, Technological Educational Institute of Central, Greece
\item Dr.~Leonidas~Georgiadis, Examination Committee Member\\
Professor, Aristotle University of Thessaloniki, Greece
\item Dr.~Traianos Yioultsis, Examination Committee Member\\
Associate Professor, Aristotle University of Thessaloniki, Greece
\item Dr.~Sami~Muhaidat, Examination Committee Member\\
Associate Professor, Khalifa University, UAE
\item Dr.~Harilaos~G.~Sandalidis, Examination Committee Member\\
Assistant Professor, University of Thessaly, Greece
\end{itemize}

%TITLE PAGE%%%%%%%%%%%%%%%%%%%%%%%%%%%%%%%%%%%%%%%%%%%%%%%%%%%%%%%%%%%%%%%%%%%%%
\else\fi

\newpage

\ifcopyrightpageflag
%COPYRIGHT PAGE%%%%%%%%%%%%%%%%%%%%%%%%%%%%%%%%%%%%%%%%%%%%%%%%%%%%%%%%%%%%%%%%%
\thispagestyle{empty}
\noindent Copyright \copyright \ 2017 Panagiotis D. Diamantoulakis. All Rights Reserved.\\[0.5cm]
\noindent Copyright \copyright \ 2017 Aristotle University of Thessaloniki. All Rights Reserved.\\[0.5cm]
Resource Allocation in Wireless Networks with Energy Constraints.
\null\vfill
\noindent\textquotedblleft The approval of the present Doctor of Philosophy Thesis from the School of Electrical and
Computer Engineering of Aristotle University of Thessaloniki does not indicate the acceptance of the author's
opinions\textquotedblright
\begin{flushright}
(L.~5343/1932, article~202, par.~2)
\end{flushright}

%COPYRIGHT PAGE%%%%%%%%%%%%%%%%%%%%%%%%%%%%%%%%%%%%%%%%%%%%%%%%%%%%%%%%%%%%%%%%%
\else\fi

\newpage

\ifdedicatepageflag
%DEDICATION PAGE%%%%%%%%%%%%%%%%%%%%%%%%%%%%%%%%%%%%%%%%%%%%%%%%%%%%%%%%%%%%%%%%
%These calculations serve the correct shifting of the dedication text
\newlength{\dedichshift}\setlength{\dedichshift}{\textwidth}
\settowidth{\auxlength}{\large{sti}}
\addtolength{\dedichshift}{-\auxlength}
\greektext
%Flag for dedication alignment
%true: Centers dedication
%false: Shifts dedication to the right
\flag{dedicationcenterflag}{true}

\ifdedicationcenterflag
\setlength{\dedichshift}{0.5\dedichshift}
\else\fi

\thispagestyle{empty}

\textit{\\[5cm]
\begin{adjustwidth}{\dedichshift}{0cm}
%\gr{Sth magiopo'ula mou, th Margar'ita}
\latintext
To my family \\
\end{adjustwidth}}
\latintext
%DEDICATION PAGE%%%%%%%%%%%%%%%%%%%%%%%%%%%%%%%%%%%%%%%%%%%%%%%%%%%%%%%%%%%%%%%%
\else\fi

\frontmatter

\ifabstractflag
%ABSTRACT%%%%%%%%%%%%%%%%%%%%%%%%%%%%%%%%%%%%%%%%%%%%%%%%%%%%%%%%%%%%%%%%%%%%%%%
\chapter{Abstract}\label{ch:abstract}

The opportunities arising from the recent advances in multimedia, along with the emerging future internet-of-things (IoT) applications, such as smart cities, health monitoring devices, and driverless cars, are limited by the operational expenses (OPEX) of the base stations (BSs), as well as the finite battery capacity of the involved wireless communication devices. This calls for conscious utilization of energy and other resources, especially taking into account the trade-off among the utilized energy, bandwidth, and time. An alternative/complementary approach to reduce the OPEX and increase the networks lifetime is energy harvesting (EH), which refers to harnessing energy from the environment. Interestingly, except for harvesting energy from ambient light sources, which are uncontrollable, wireless power transfer can also be used, in order to remotely charge wireless devices with low energy requirements in a wide area. However, EH creates several new challenges, with the main focus of this dissertation being on the development of novel scheduling and resource allocation schemes, that take into account and regulate the energy constraints imposed by the levels of harvested energy.

To this direction, the first chapter of this thesis investigates the optimal energy, time, and bandwidth allocation problem for the downlink of energy harvesting base stations (EHBSs), with the main focus being on autonomous EHBS. The presented analysis takes into account the impact of the energy constraint on users preferences and the BS's revenue. In order to model the competitive nature of the problem, game theory is used, and more specifically the framework of a generalized Stackelberg game. Also, an efficient iterative method is proposed to facilitate all the players to reach the variational equilibrium, i.e., the optimal solution of the game.

The next two chapters focus on wireless powered networks (WPNs) and simultaneous wireless information and power transfer (SWIPT) using radio frequency (RF) technology, with emphasis being given on the created trade-offs. One of the main contributions of these chapters is the introduction of both uplink and downlink non-orthogonal multiple access (NOMA) for WPNs.
Moreover, the individual data rates and fairness are improved, while the formulated problems are optimally and efficiently solved. It is shown that, compared to orthogonal multiple access, NOMA offers a considerable improvement in throughput, fairness, and energy efficiency. Rather than this, proportional fairness is maximized and uplink/downlink of WPNs are jointly optimized, in which cases, except for NOMA, time divison multiple access is also investigated. Also, the role of interference is considered, which has been recognized as one of the main reasons of the asymmetric overall degradation of the users' performance, due to different path-loss values, called from now on as \textit{cascaded near-far problem}. Moreover, SWIPT is investigated and efficiently optimized in the context of multicarrier cooperative communication networks.

Finally, simultaneous lightwave information and power transfer (SLIPT) for indoor IoT applications is introduced, as an alternative technology to RF-based SWIPT. It is noted that light WPT is fundamentally different to RF, due to the divergent channels characteristics, transmission/reception equipment, and EH model among others. For this reason, novel and fundamental SLIPT strategies are proposed, which can be implemented through Visible Light or Infrared communication systems, equipped with a simple solar panel-based receiver. In order to balance the trade-off between the harvested energy and communication performance two optimization problems are formulated and optimally solved. Moreover, simulation results are presented, in order to illustrate the performance of the proposed methods.
%\include{\greekabstractpath}
%ABSTRACT%%%%%%%%%%%%%%%%%%%%%%%%%%%%%%%%%%%%%%%%%%%%%%%%%%%%%%%%%%%%%%%%%%%%%%%
\else\fi

\ifacknowledgementsflag
%ACKNOWLEDGEMENTS%%%%%%%%%%%%%%%%%%%%%%%%%%%%%%%%%%%%%%%%%%%%%%%%%%%%%%%%%%%%%%%
\chapter{Acknowledgements}\label{ch:Acknowledgements}
\vspace{1.8cm}
Optimization theory, which used in this thesis to enhance the performance of communication networks with energy constraints, can be generally used in numerous applications, too many to cite here, such as economics, politics, etc. However, optimization, apart from a mathematical tool, is a way of thinking, acting, and living. When applied in daily routine, it can substantially improve the quality of life. To achieve so, critical thinking, careful strategy, and collaboration with others are necessary elements. This has been one of the most important lessons I have learned during the years of pursuing my Ph.D.

At this point, I would like first to thank my advisory and mentor, Dr. George Karagiannidis, Professor at the Department of Electrical and Computer Engineering of Aristotle University of Thessaloniki and leader of the Wireless Communications Systems Group (WCSG), who believed in me and gave me the opportunity to participate in the group, conduct research, and seek a Ph.D. in Electrical Engineering. His enthusiasm, guidance, patience, insight, and his ability to actively participate in all of his students works helped me in every moment of researching and writing of this thesis. I am grateful to have been influenced by personality, as well as his way of thinking and problem solving, which has greatly aided me, in every aspect of my life.  Also, my special thanks to my colleague, friend, and member of WCSG Dr. Koralia Pappi, an exemplary scientist and human being, who offered to me irreplaceable support. Her guidance was a catalyst for the completion of this dissertation. Also, she taught me by her living example that it's not the struggle that defines you but rather how you respond.

I would like to thank the members of my advisory committee, Dr. Dimitris P. Labridis and Dr. Fotini - Niovi Pavlidou, Professors at the Department of Electrical and Computer Engineering of Aristotle University of Thessaloniki, for their continuous support and our valuable collaboration that was necessary for the preparation of this thesis.

My special thanks to Dr. Sami Muhaidat, Associate Professor at the Department of Electrical and Computer Engineering of Khalifa University, and Dr. Paschalis Sofotasios, Assistant Professor at the Department of Electrical and Computer Engineering of   Khalifa University, for their valuable guidance and support during my visit to Khalifa University.
 
I would also like to thank Dr. Vincent Poor, Professor at Department of Electrical and Computer Engineering of Princeton University,  Dr. Robert Schober, Professor at the Institute for Digital Communications of Friedrich-Alexander University Erlangen-N\"{u}rnberg, Dr. Arumugam Nallanathan, Professor at the School of Electronic Engineering and Computer Science at Queen Mary University of London, Dr. Zhiguo Ding, Professor at School of Computing and Communications of Lancaster University, and Dr. Michail Matthaiou, Reader at Queen's University of Belfast, UK, for their assistance, our constructive conversations, and valuable collaboration.

My special thanks to Dr. Diomidis Michalopoulos, Dr. Athanasios Lioumpas, Dr. Vasileios Kapinas, Dr. Nestor Chatzidiamantis, and Dr. Ioanna Diamantoulaki, researchers whom I greatly admire and respect, and have provided valuable guidance and assistance during various stages of my studies. 

I would also like to thank  my colleagues and friends, Dr. Alexandros-Apostolos Boulogeorgos, Dr. Dimitrios Karas,  Ms. Georgia Ntouni, Mr. Sotirios Mihos, and Ms. Efi Manopoulou, for our excellent collaboration, their valuable company, and support.

I would also like to thank the new members of WCSG, Mr. Stelios Trevlakis, Mr. George Katsikas, for the fruitful cooperation and for making me feel confident about the future of the group.

This thesis is dedicated to my beloved parents, Dimitrios and Aikaterini, for their unconditional support and love.

Also, I would like to express my deepest gratitude to my family, Ioanna, Xristina, Michalis, Maria-Eirini, and Rafael, who stood me throughout my studies, encouraged me in every step toward my goals, and gave me hope for a better future.

Last but not least, I owe my gratitude to Aristotle University of Thessaloniki that has been a second home for me, all these years. 

%\begin{flushright}
%\rule{3.6cm}{0.02cm}\\
%Panagiotis D. Diamantoulakis\\
%September 27, 2017
%\end{flushright}

%ACKNOWLEDGEMENTS%%%%%%%%%%%%%%%%%%%%%%%%%%%%%%%%%%%%%%%%%%%%%%%%%%%%%%%%%%%%%%%
\else\fi

\iftocflag
%TOC%%%%%%%%%%%%%%%%%%%%%%%%%%%%%%%%%%%%%%%%%%%%%%%%%%%%%%%%%%%%%%%%%%%%%%%%%%%%
\tableofcontents %Put asterisk for not including it in the TOC
%TOC%%%%%%%%%%%%%%%%%%%%%%%%%%%%%%%%%%%%%%%%%%%%%%%%%%%%%%%%%%%%%%%%%%%%%%%%%%%%
\else\fi

\iflistfigures
%TOC%%%%%%%%%%%%%%%%%%%%%%%%%%%%%%%%%%%%%%%%%%%%%%%%%%%%%%%%%%%%%%%%%%%%%%%%%%%%
\newpage
\listoffigures %Put asterisk for not including it in the TOC
%TOC%%%%%%%%%%%%%%%%%%%%%%%%%%%%%%%%%%%%%%%%%%%%%%%%%%%%%%%%%%%%%%%%%%%%%%%%%%%%
\else\fi

\iflisttables
%TOC%%%%%%%%%%%%%%%%%%%%%%%%%%%%%%%%%%%%%%%%%%%%%%%%%%%%%%%%%%%%%%%%%%%%%%%%%%%%
\newpage
\listoftables %Put asterisk for not including it in the TOC
%TOC%%%%%%%%%%%%%%%%%%%%%%%%%%%%%%%%%%%%%%%%%%%%%%%%%%%%%%%%%%%%%%%%%%%%%%%%%%%%
\else\fi

%\iflistalgorithms
%TOC%%%%%%%%%%%%%%%%%%%%%%%%%%%%%%%%%%%%%%%%%%%%%%%%%%%%%%%%%%%%%%%%%%%%%%%%%%%%
%\newpage
%\listofalgorithms
%\addcontentsline{toc}{chapter}{List of Algorithms}
%TOC%%%%%%%%%%%%%%%%%%%%%%%%%%%%%%%%%%%%%%%%%%%%%%%%%%%%%%%%%%%%%%%%%%%%%%%%%%%%
%\else\fi

\chapterstyle{vkapstylemain}

\ifabbreviationsflag
%ABBREVIATIONS%%%%%%%%%%%%%%%%%%%%%%%%%%%%%%%%%%%%%%%%%%%%%%%%%%%%%%%%%%%%%%%%%%
\chapter{Abbreviations}\label{ch:abbreviations}

%Note: In the option of the acronym environment put the longest acronym

\begin{table}[!h]\renewcommand{\arraystretch}{1.2}
\begin{tabular}{m{3.5cm}l}
AC & Alternating Current\\
AEHBSs & Autonomous Energy Harvesting Base Stations\\
AF & Amplify-and-Forward\\
AP & Access Point\\
AS & Antenna Switching\\
ATSC & Advanced Television Systems Commitee\\
AWGN & Additive White Gaussian Noise\\
BS & Base Station\\ 
C & Constraint\\
CAPEX & Capital Expenditures\\
CnfP & Cascaded Near-Far Problem\\
CRAN & Cloud Radio Access Network\\
D& Destination\\
DC& Direct Current\\
DF & Decode-and-Forward\\
DoF & Degrees of Freedom\\
ECO & Economy\\
EH & Energy Harvesting\\
%EE & Energy efficiency\\
EHBS & Energy Harvesting Base Station\\
EM & Electromagnetic\\
FoV & Field-of-View\\
FSO & Free Space Optical\\
GNE & Generalized Nash Equilibrium \\
GNEP & Generalized Nash Equilibrium problem \\
GSE & Generalized Stackelberg Equilibrium \\
HAP   & Hybrid Access Point\\
IEEE & Institute of Electrical and Electronics Engineers\\
IoT& Internet-of-Things \\
IR & Infrared\\
IS & Interfering Source\\
KKT & Karush-Kuhn-Tucker \\
LED & Light Emitting Diode\\
\end{tabular}
\end{table}

\begin{table}[!h]\renewcommand{\arraystretch}{1.2}
\begin{tabular}{m{3.5cm}l}
LM & Lagrange Multiplier\\
LoS & Line of Sight\\
LPF & Low-Pass Filter\\
LTE & Long Term Evolution\\
LTE-A & Long Term Evolution-Advanced\\
MIMO & Multiple-input multiple-output\\
MUST & Multi-user superposition transmission \\
NOMA & Non-Orthogonal Multiple Access\\
NOMA-TS & Non-Orthogonal Multiple Access with Time-Sharing\\
OFDMA & Orthogonal Frequency Division Multiple Access\\
OPEX & Operational Expenses\\
OWC& Optical Wireless Communication\\
OWPT & Optical Wireless Power Transfer\\
PB & Power Beacon  \\
PD & Photodetector\\
PV & Photovoltaic \\
PIN & Positive-Intrinsic-Negative\\
PS & Power Splitting\\
QoS & Quality-of-Service\\
R & Relay\\
RFID  &Radio Frequency Identification \\
RF & Radio Frequency \\
Rectenna & Rectifying Antenna\\
SE & Stackelberg Equilibrium \\
SIC & Successive interference cancellation\\
SINR & Signal-to-Interference-plus-Noise Ratio \\
SNR & Signal-to-Noise Ratio\\
SLIPT & Simultaneous Lightwave Information and Power Transfer \\
SS & Spatial Switching\\
SVD & Singular Value Decomposition\\
SWIPT & Simultaneous Wireless Information and Power Transfer \\
s.t. & Subject to\\
TS & Time-Sharing\\
TSw & Time-Switching\\
TSp & Time-Splitting\\
TDMA & Time Division Multiple Access\\
UAV & Unmanned Aerial Vehicle \\
UV & Ultraviolet\\
\end{tabular}
\end{table}

\begin{table}[!h]\renewcommand{\arraystretch}{1.2}
\begin{tabular}{m{3.5cm}l}
VE & Variational Equilibrium\\
VI & Variational Inequality\\
VL & Visible Light\\
VLC & Visible Light Communication\\
WD  & Wireless Device\\
WEE & Weighted Energy Efficiency\\
Wi-Fi & Wireless Fidelity\\
WPN & Wireless Powered Network\\
WPT & Wireless Power Transfer  \\
w.r.t. & with respect to\\
3GPP & 3rd Generation Partnership Project\\
5G & Fifth Generation\\
\end{tabular}
\end{table}
%ABBREVIATIONS%%%%%%%%%%%%%%%%%%%%%%%%%%%%%%%%%%%%%%%%%%%%%%%%%%%%%%%%%%%%%%%%%%
\else\fi

\ifnomenclatureflag
%ABBREVIATIONS%%%%%%%%%%%%%%%%%%%%%%%%%%%%%%%%%%%%%%%%%%%%%%%%%%%%%%%%%%%%%%%%%%
\chapter{List of Symbols}\label{ch:nomenclature}

\begin{table}[!h]\renewcommand{\arraystretch}{1.2}
\begin{tabular}{m{3.5cm}l}
$\alpha$& Weight that corresponds to the downlink rate\\
$\beta$ & Weight that corresponds to the uplink rate\\
$\gamma_0$ & Received signal-to-noise plus interference ratio\\
$\gamma_n$ & $\gamma_n=|h_n|^2\mathcal{G}_{0}\mathcal{G}_{n}$\\
$\Gamma(\kappa,\zeta)$ & Gamma distribution with shape parameters $\kappa,\zeta>0$\\
$\Gamma(\cdot)$ & Gamma function\\
$\tilde{\gamma}_n$ & $\tilde{\gamma}_n=|\tilde{h}_n|^2\tilde{\mathcal{G}}_{0}\mathcal{G}_{n}$\\
$\gamma_{\mathrm{s},i}$ & Signal-to-noise ration of the source-relay link\\
$\gamma_{\mathrm{r},i}$ & Signal-to-noise ration of the source-relay link\\
$\gamma_{\mathrm{tot},i}$ & End-to-end signal-to-noise ration\\
$\tilde{\gamma}_{\mathrm{tot},i}$ & Approximation of end-to-end signal-to-noise ratio\\
$\gamma_\mathrm{th}$ & Signal-to-noise plus interference ratio threshold\\
$\Delta_i$ & $i$-th region of a graph\\
$\epsilon$ & $\epsilon\rightarrow 0^+$\\
$\varepsilon$ & $\varepsilon\rightarrow 0^+$\\
$\eta_1$ & Energy harvesting efficiency\\
$\eta_2$ & Efficiency of the amplifier\\
$\eta$ & $\eta=\eta_1\eta_2$\\
$\eta_0$ & Photo-detector responsivity\\
$\theta_n$ & Power fraction used for information by the $n$-th user\\
$\tilde{\theta}_n$ & $\tilde{\theta}_n=\ln(\theta(n))$\\
$\boldsymbol{\theta}$ & Set of values  $\theta_1,...,\theta_n,\theta_N$\\
$\tilde{\boldsymbol{\theta}}$ & Set of values  $\ln(\theta_1),...,\ln(\theta_n),\ln(\theta_N)$\\
${\boldsymbol{\lambda}}$ & LM set with elements $\lambda_1,...,\lambda_n,...,\lambda_N$\\
$\lambda_i$ & Lagrange multiplier\\
$\mu$ & Lagrange multiplier\\
$\hat{\lambda}_i$ & Positive step size\\
$\nu_n$ & Additive noise at the $n$-th user\\
$\nu_0$ & Additive noise at the BS\\
$\nu_{r,i}$ & Additive noise  at the relay at the $i$-th channel\\
$\nu_{d,i}$ & Additive added at the destination at the $i$-th channel\\
\end{tabular}
\end{table}

\begin{table}[!h]\renewcommand{\arraystretch}{1.2}
\begin{tabular}{m{3.5cm}l}
$\xi$ & Path loss exponent\\
$\xi_0$ & $\xi_0=-\frac{1}{\log_2\cos(\Phi_{1/2})}$\\
$\rho$ & refractive index\\
$\rho_0$ & $\rho_0=\frac{P_0}{N_0W}$\\
$\Phi_{1/2}$ & Ssemi-angle at half luminance\\
$\phi_i$ & The $i$-th eigenvalue of the Hessian matrix\\
$\varphi$ & Irradiance angle\\
$\tau_m$ & Time-sharing variable, i.e., portion of time that corresponds to \\
& the $m$-th permutation\\
$\mathbold{\tau}$ & Set of values of $\tau_1,...,\tau_m,...,\tau_M$\\
$\psi$ & Incidence angle\\
$\Psi(\cdot,\cdot)$ & Digamma function\\
$\Psi_\mathrm{fov}$ & Field-of-view\\
$\Psi_\mathrm{fov,1}$ & Field-of-view during phase 1\\
$\sigma^2$ & Noise power\\
$\Omega$ & Strategic form of game\\
$A$& Peak amplitude of electrical signal\\
$A_1$ & Peak amplitude of electrical signal during phase 1\\
$A_2$ & Peak amplitude of electrical signal during phase 2\\
$A_n'$ & Peak amplitude of electrical signal transmitted by\\
& the $n$-th neighboring LED\\
$\tilde{A}_n$ & $\tilde{A}_n=\frac{\eta \rho_0 g_nT}{t_n}+1$\\
$\mathbf{A}$ & Matrix with elements $\mathbf{A}(m,j_{m,n})$ \\
$\mathbf{A}(m,j_{m,n})$& Index of $n$-th user with decoding order $j_{m,n}$ \\
&during the $m$-th permutation that correspond to\\
$a_n$  & Weight that couples the $n$-th user's utility\\
$\tilde{a}_n$ & $\tilde{a}_n=\eta\rho_0 g_n$\\
$B$ & Direct current bias\\
$B_1$ & Direct current bias during phase 1\\
$B'_n$ & Direct current bias of $n$-th neighboring LED\\
$B_2$ & Direct current bias during phase 2\\
$\boldsymbol{b}$ & Rate profile set with elements  $b_1,...,b_n,..,b_N$\\
$b_n$ & Rate profile parameter of the $n$-th user\\
$\tilde{b}_n$ & $\tilde{b}_n=\eta\rho_0\sum_{j=n+1}^Ng_j$\\
$\tilde{B}_i$ & $\tilde{B}_i=1+\frac{\eta\rho_0 \sum_{n=N+1-i}^N{g}_iT_i}{1-T_i}$\\
$c_1$ & Price per utilized energy unit\\
$c_2$ & Price per unit of resources (energy/time)\\
$\mathcal{CN}(\mu,s^2)$ & Complex Gaussian distribution with mean $\mu$ and variance $s^2$\\
\end{tabular}
\end{table}

\begin{table}[!h]\renewcommand{\arraystretch}{1.2}
\begin{tabular}{m{3.5cm}l}
$\tilde{c}_n$ &$\tilde{c}_n=\eta \rho\sum_{i=n}^{N}g_i$ \\
$\cos(x)$ & Cosine of x\\
$\tilde{d}_n$ & $\tilde{d}_n=N+1-n$\\
$d_n$ & Distance between the $n$-th user and the BS/AP\\
$\frac{df(x)}{dx}$ & Derivative of function $f$ with respect to x\\
$D_0$ & Distance between the interfering source (IS) and the BS\\
$D_i$ & Distance of the $i$-th group of users from the BS\\
$d_\mathrm{sr}$ & Distance between the source and the relay\\
$d_\mathrm{rd}$ & Distance between the relay and the destination\\
$\mathcal{E}$ & Energy efficiency\\
$g_n$ & $g_n=\gamma_n^2$\\
$\mathbb{E}[\cdot]$ & Statistical expectation\\
$e$ & Base of the natural logarithm\\
$\exp(x)$ & Exponential of x\\
$\mathbf{E}$ & Set of values with elements $E_{\mathrm{BS},1},...,E_{\mathrm{BS},n},...,E_{\mathrm{BS},N}$\\
$\mathbf{E}_{-n}$ & Set  $\mathbf{E}$ excluding $E_{\mathrm{BS},n}$ \\
$E_{\mathrm{BS}}$ & Amount of harvested energy by the BS\\
$E_{\mathrm{BS},n}$ & The fraction of the energy dedicated for the transmission to\\
& the $n$-th user\\
$E_n$ & Harvested energy by the $n$-th user\\
$\tilde{E}_n$ & Energy harvesting rate of the $n$-th user\\
$E_{TS}$ & Harvested energy when time-splitting is used\\
$\tilde{E}_\mathrm{TSBO}$ & Harvested energy when time-splitting with DC bias\\
& optimization is used\\
$\tilde{E}_\mathrm{TSBO}$ & Reformulation of $\tilde{E}_\mathrm{TSBO}$\\
$\boldsymbol{F}$ & Vector defined as $(\nabla_{E_{\mathrm{BS},1}} U_1, \nabla_{q_1} U_1,...,\nabla_{E_{\mathrm{BS},N}} U_N, \nabla_{q_N} U_N)$\\
$f$ & Scalar function\\
$F_0$ & Fill factor\\
$\tilde{\mathcal{G}}_{0}$ & Directional antenna gain of the PB\\
$\mathcal{G}_{0}$ & Directional antenna gain of the BS\\
$\mathcal{G}_{n}$ & Directional antenna gains of the $n$-th user\\
$G_n$ & Normalized channel power gain, given by $G_n=\frac{L_n|H_n|^2}{N_0}$\\
$G_{\mathrm{r},i}$ & Gain of the relay over the $i$-th channel\\
$g_\mathrm{f}$ & Gain of the optical filter\\
$g_\mathrm{c}(\psi)$ & Gain of the optical concentrator\\
$H_n(t)$ & Small scale fading coefficient of the channel between the BS and\\
& the $n$-th user\\
$\tilde{h}_n$ & Coefficient of the channel between the PB and the $n$-th user\\
\end{tabular}
\end{table}

\begin{table}[!h]\renewcommand{\arraystretch}{1.2}
\begin{tabular}{m{3.5cm}l}
$h_n$ & Coefficient of the channel between the BS/AP and the $n$-th user\\
$h'_n$ & Coefficient of the channel between the $n$-th neighboring LED\\
& and the user's receiver\\
$\tilde{H}_n$ & Small scale coefficient of the channel between the PB and the\\
& $n$-th user\\
$h_{\mathrm{r},i}$ & Coefficient of the channel between the relay and the destination\\
& $h_{\mathrm{r},i}\sim \mathcal{CN}(0,\sigma_{r,i}^2)$\\
$h_{s,i}$ & Coefficient of the channel between the source and the relay\\
& $h_{\mathrm{s},i}\sim\mathcal{CN}(0,\sigma_{s,i}^2)$\\
$i_\mathrm{AC}$ & AC component of electrical current\\
$i_{\mathrm{AC},1}$ & Part of $i_\mathrm{AC}$ provoked by the dedicated LED\\
$i_{\mathrm{AC},2}$& Part of $i_\mathrm{AC}$ provoked by other interfering \\
& sources (not the dedicated LED)\\
$i_0$ & Output current of Schottky diode\\ 
$I_0$ & Dark saturation current\\
$i_r(t)$ & Electrical current at the output of the photodetector\\
$I_\mathrm{DC}(t)$ & Direct current (DC) signal \\
$I_{\mathrm{DC},1}$ & Part of $i_\mathrm{DC}$ provoked by the dedicated LED\\
$I_{\mathrm{DC},2}$ & Part of $i_\mathrm{DC}$ provoked by other interfering \\
&sources the dedicated (not the dedicated LED)\\
$I_n$ & Interfering signal received by the $n$-th user\\
$I$ & Interfering signal received by the BS\\
$I_\mathrm{L}$ & Minimum input bias current\\
$I_\mathrm{H}$ & Maximum input bias current\\
%$I_\mathrm{sc}$ & Short circui current\\
$\mathcal{J}$ & Jain's fairness index\\
$j_{m,n}$& Decoding order of $n$-th user during $m$-th permutation\\
$K$ & Maximum number of iterations\\
$K_1$ & $K_1=\frac{R_\mathrm{th}}{\log_2\left(1+\frac{e\left(\eta_0 h_{1}P_\mathrm{LED}(I_H-I_L)\right)^2}{8\pi\left(P_I(\Psi^*_\mathrm{fov,1})+\sigma^2\right)}\right)}$\\
$K_2$ & $K_2=\min\left(\frac{R_\mathrm{th}}{\log_2\left(1+\frac{e\gamma_\mathrm{th}}{2\pi}\right)},1\right)$\\
$l_i$ & The sum of the LM, $\lambda_i$ and the price $c_i$\\
$L_n$ & Path loss/shadowing coefficient of the channel between the  \\
 & BS and the $n$-th user\\
$\tilde{L}_n$ & Path loss/shadowing coefficient of the channel between the  \\
 & PB and the $n$-th user\\
$L_r$ & Physical area of the photo-detector\\
$\log_2(x)$ & Base-2 logarithm of $x$\\
$\ln(x)$ & Natural logarithm of $x$\\
\end{tabular}
\end{table}

\begin{table}[!h]\renewcommand{\arraystretch}{1.2}
\begin{tabular}{m{3.5cm}l}
$\mathcal{L}$ & Lagrangian function\\
$\tilde{\mathcal{L}}$ & Simplified Lagrangian function\\
$M$ & Number of permutations\\
$\max (\cdot)$ & Maximum value of a set of elements\\
$\min (\cdot)$ & Minimum among of a set of elements\\
$\mathcal{M}_k$ & Subset of users, $\mathcal{M}_i\subseteq\mathcal{N}$\\
$m_\mathrm{s}(t)$ & Modulated electrical signal\\
$\mathcal{N}$ & Set of users\\
$N$ & Number of users\\
$N_\mathrm{c}$ & Number of carriers\\
$N_i$ & Number of users of $i>1$-th group\\
$N_0$ & Noise power spectral density\\
$P_0$ & Maximum transmission power of the BS\\
$P^\mathrm{d}_n$ & Transmit power from the BS to the $n$-th during downlink\\
$p^\mathrm{d}_n$ &$p^\mathrm{d}_n=\frac{P^\mathrm{d}_n}{N_0W}$ \\
$P^\mathrm{u}_n$ & Transmit power  of the $n$-th user during uplink\\
$P_{\mathrm{I},j}$ & Power of received interference by the $j$-th user\\
$P_{\mathrm{I},0}$ &  power of the interference received by the BS\\
$p_{\mathrm{I},j}$ & $p_{I,j}=\frac{P_{I,j}}{N_0W}$\\
$P_\mathrm{IS}$ & Transmit power of the interfering source\\
$p_{IS}$ & $p_\mathrm{IS}=\frac{P_\mathrm{IS}}{N_0W}$\\
$p_{\mathrm{I},0}$ & $p_{\mathrm{I},0}=\frac{P_{\mathrm{I},0}}{N_0W}$\\
$P_\mathrm{ro}$ & Received optical signal from other sources\\
$\tilde{p}_n$ & $\tilde{p}_n=\ln(p^\mathrm{d}_n)$\\
$\boldsymbol{p}$ & The set of values $p^\mathrm{d}_1,...,p^\mathrm{d}_n,,...p^\mathrm{d}_N$\\
$\tilde{\boldsymbol{p}}$ & The set of values $\ln(p^\mathrm{d}_1),...,\ln(p^\mathrm{d}_1),...,\ln(p^\mathrm{d}_N)$\\
$P_\mathrm{rt}$ & The total harvested power at the relay\\
$P_{\mathrm{r}0}$ & Available power from other sources (apart from SWIPT) \\
& and/or other fixed power needs of the relay\\
$P_{\mathrm{s},i}$ & Transmitted power by the source on the $i$-th channel\\
$P_{\mathrm{r},i}$ &Power transmitted by the relay over the $i$-th channel\\
$P_\mathrm{rm}$ & Maximum allowable transmit power by the relay\\
$P_\mathrm{sm}$ & Maximum allowable transmit power by the source\\
$\mathbf{P_\mathrm{s}}$ & Set of values $P_{\mathrm{r},1},...,P_{\mathrm{r},i},...,P_{\mathrm{r},N}$\\
$\mathbf{P_\mathrm{s}}$ & Set of values $P_{\mathrm{s},1},...,P_{\mathrm{s},i},...,P_{\mathrm{s},N}$\\
$P_\mathrm{LED}$ & LED power\\
$P_\mathrm{t}$ & Transmitted optical signal\\
$\mathbf{q}$ & Set of values with elements $q_1,...,q_n,....,q_N$\\
\end{tabular}
\end{table}

\begin{table}[!h]\renewcommand{\arraystretch}{1.2}
\begin{tabular}{m{3.5cm}l}
$\mathbf{q}_{-n}$ & Set $\mathbf{q}$ excluding $q_n$\\
$q_n$ & Product of time and bandwidth allocated to the $n$-th user\\
$Q$& Product of total available time and bandwidth\\
$\mathcal{R}$ & Auxiliary variable\\
$\tilde{\mathcal{R}}$ & $\tilde{\mathcal{R}}=\ln(\mathcal{R})$\\
$R^\mathrm{d}_n$ & Achievable downlink rate of the $n$-th user\\ 
$R^\mathrm{d}_{\mathrm{t},n}$ & Targeted downlink rate of the $n$-th user\\
$R^\mathrm{u}_n$ & Achievable uplink rate of the $n$-th user\\ 
$\tilde{R}^\mathrm{u}_n$ & Achievable uplink rate of the $n$-th user\\
& when NOMA-TS is used\\
$\mathcal{R}_\mathrm{eq}$ & Equal rate among users\\
$\mathcal{R}_{\mathrm{min}}$ & Minimum achievable throughput among users\\ 
$\mathcal{R}_\mathrm{tot}$ & System throughput\\
$\tilde{\mathcal{R}}_\mathrm{tot}$ & Simplified system throughput after the end-to-end\\
& signal-to-noise ratio approximation\\
$\mathcal{R}_\mathrm{sum}$ & Sum of individual data rates\\
$R_0$ & Lambertian radiant intensity\\
$R^\mathrm{d}_{n\rightarrow j}$ & Rate at which the $j$-th user detects the message intended \\
& for the $n$-th user\\
$R_{\mathrm{th}}$ & Information rate threshold\\
$r_{c1}$ & Radius of inner circle\\
$r_{c2}$ & Radius of outer circle\\
$s^\mathrm{d}_0$ & Baseband signal transmitted by the BS\\
$s^\mathrm{d}_n$ & Baseband signal transmitted by the BS to the $n$-th user\\
$s^u_n$ & Baseband signal transmitted by the $n$-th user\\
$S_n(t)$ & Time allocation  parameter, $S_n(t)\in\{0,1\}$\\
$T$& Duration of the first phase in two-phase protocols\\
$T_i$ & Auxiliary variable\\
${\boldsymbol{t}}$ & Set with elements $t_1,...,t_n,...t_N$\\
$\tilde{{\boldsymbol{t}}}$ & Set with elements $\ln({t}_1),...,\ln({t}_n),...\ln({t}_N)$\\
$t_\mathrm{in}$ & Time instant\\
$t_n$ & Time allocated to the $n$-th user\\
$\tilde{t}_n$ & $\tilde{t}=\ln(t_n)$\\
$T_\mathrm{tot}$ & Total duration of a time frame\\
$U_\mathrm{BS}$ & Utility function of the base station\\
$U_\mathrm{n}$ & Utility function of the $n$-th user\\
$V_\mathrm{oc}$ & Open circuit voltage\\
$V_\mathrm{t}$ & thermal voltage\\
\end{tabular}
\end{table}

\begin{table}[!h]\renewcommand{\arraystretch}{1.2}
\begin{tabular}{m{3.5cm}l}
$\mathcal{U}_i$ & Projection operator\\
$W$ & Total available bandwidth\\
$W_n$ & Bandwidth of the $n$-th user or the $n$-th channel\\
$y_n$ & Baseband signal received by the $n$-th user\\
$y_{\mathrm{r},i}$ & Baseband received signal by the relay over the $i$-th channel\\
$y_{\mathrm{d},i}$ & Baseband received signal by the destination over the $i$-th\\
& channel\\
$\mathcal{W}(x)$ & Lambert W function\\
$\mathcal{W}_0(x)$ & Principal branch of Lambert W function\\
$X$ & $X=\frac{\eta_2 \tilde{E}_n \gamma}{N_0W}$\\
$y_\mathrm{RF}$ & Received RF band signal \\
$z^*$ & Solution of $z\ln(z)-z-\sum_{n=1}^Ng_n+1=0$\\
$z^*_n$& Solution of $\ln(1+z_n)-\frac{z_n}{1+z_n}=\frac{b_n}{\lambda_n}\sum_{n=1}^N\frac{\lambda_ng_n}{b_n(1+z_n)}$\\
$Z_n$ &  $Z_n=1+\mathcal{W}_0 \left(\frac{-l_1+l_2K_n}{e l_1}\right)$\\
${\tilde{z}_1}$ & ${\tilde{z}_1}=\frac{\alpha \exp(\tilde{\mathcal{R}})}{ T}$\\
${\tilde{z}_2}$ & ${\tilde{z}_2}=\alpha \exp(\tilde{\mathcal{R}}-\tilde{t}_n)$\\
$\nabla$ & Gradient\\
$|\cdot|$ & Absolute value\\
$||\cdot||$ & Cardinalilty of a set\\
$(\cdot)^*$ & Optimality or solution of an equation\\
$\frac{\partial f(x_1,...,x_i,...,x_m)}{\partial x_i}$ & Partial derivative of function $f$ with respect to $x_i$\\
$[.]_{\xi}$ &  $\max (.,\xi)$\\
$[\cdot]^+$ & $[\cdot]^+=\min(\cdot,0)$\\
$(\cdot)^\mathrm{d}$ & Value for the downlink phase\\
$(\cdot)^\mathrm{u}$ & Value for the uplink phase\\
$\bar{(\cdot)}$ & Conjugate of $(\cdot)$\\
$(\cdot)^{[i]}$& The value of $(\cdot)$ during phase $i$\\
$\subseteq$ & Subset\\
$x!$& Factorial of $x$\\
$\sum(\cdot)$ & Sum of elements\\
$\{\cdot\}$& A set of terms\\
$\sim$& Follows distribution\\
$\approx$ & Approximately equal to\\
$f|_{(\cdot)}$ & Simplified representation of function $f$ considering $(\cdot)$\\
$\infty$ & Infinity\\
$\rightarrow 0$ & Tends to zero\\
$\forall$ & For each\\
\end{tabular}
\end{table}
%ABBREVIATIONS%%%%%%%%%%%%%%%%%%%%%%%%%%%%%%%%%%%%%%%%%%%%%%%%%%%%%%%%%%%%%%%%%%
\else\fi

%\ifnomenclatureflag
%%NOMENCLATURE%%%%%%%%%%%%%%%%%%%%%%%%%%%%%%%%%%%%%%%%%%%%%%%%%%%%%%%%%%%%%%%%%%%
%%Add some text in the preamble and postamble of the nomenclature
%\renewcommand{\nompreamble}{\textit{- Items within each group are sorted alphabetically with respect to their
%description (right column).\\- Quantities defined in the context of this thesis are framed for the ease of the
%reader.}}
%\renewcommand{\nompostamble}{\textit{Note: Obviously, properties involving complex matrices apply also to complex
%scalars and real matrices}.}
%%This fixes the problem (if any) of mistaken headings
%\clearpage\markboth{\nomname}{\nomname}
%%Prints the Nomenclature just after the 'List of abbreviations'
%%The optional argument defines the \nomlabelwidth (default: 1cm)
%\printnomenclature[2.4in]
%%File with the list of symbols (the items can also be spread all over the chapters)
%\input{\nomenclaturepath}
%%Executes MakeIndex
%%\execute{makeindex \filenamenlo -s nomencl.ist -o \filenamenls}
%%Opens the nomenclature file (*.nlo) and writes in itzd
%\makenomenclature
%%NOMENCLATURE%%%%%%%%%%%%%%%%%%%%%%%%%%%%%%%%%%%%%%%%%%%%%%%%%%%%%%%%%%%%%%%%%%%
%\else\fi

\mainmatter
%
%\ifarenachapterflag
%%Chapter for testing
%\include{\arenachapterpath}
%\else\fi

%\part{Introduction}\label{Part:Introduction}

%\ifchapteroneflag
%CHAPTER 1%%%%%%%%%%%%%%%%%%%%%%%%%%%%%%%%%%%%%%%%%%%%%%%%%%%%%%%%%%%%%%%%%%%%%%
\chapter[Introduction and Thesis Layout][Introduction and Thesis Layout]{Introduction and Thesis Layout}\label{ch:chapter1}

Energy is a major concern for current wireless technology implementation, as there is always a trade-off between the associated costs, communication bandwidth, and battery lifetime. A feasible solution that can considerably reduce the operational expenses (OPEX) and increase the networks lifetime is energy harvesting (EH). The EH poses several new challenges and calls for an interdisciplinary approach, integrating the design of wireless networks with advances
from circuits and devices that harvest and transfer energy. Also, it requires the development of novel scheduling and resource allocation schemes that take into and regulate the energy constraints imposed by the levels of harvested energy. Due to these tight constraints, EH can have a negative impact on the quality-of-service (QoS) achieved by the wireless networks, which can be counterbalanced by the exploitation of transmission schemes and communications protocols with higher energy and spectral efficiency.

\section{Energy Harvesting}
The opportunities arising from the recent adnavces in multimedia, along with the emerging future internet-of-things (IoT) applications, such as smart cities, health monitoring devices, and driverless cars, are limited by the OPEX of the base stations (BSs) \cite{Roy}, as well as the finite battery capacity of the involved wireless communication devices \cite{Sude, Magazine}. In this context, EH, which refers to harnessing energy from the environment, such as solar, wind, and geothermal heat, or other energy sources, such as finger motion, vibrations, breathing, and blood pressure, and converting to electrical energy, is regarded as a promising solution for energy-sustainability of wireless nodes in communication networks and reduction of the OPEX \cite{Sude}. Also, EH is regarded as a disruptive technological paradigm to prolong the lifetime of energy-constrained wireless networks. 

\subsection{Applications and Advantages of EH in Wireless Communication Networks}

Recent years have seen a surge in research on the power consumption aspect of wireless and cellular networks, due to
the increasing concern on rising global energy demand and decreasing the industry's overall carbon footprint. The power
consumption of cellular networks also constitutes a large portion of OPEX for service providers \cite{Vereecken, Roy}. Regarding a cellular network, the power needed to run a BS, as well as the corresponding cooling facilities,
forms the major share of energy consumption, which motivates the utilization of EH for the power supply of the BSs. EH can be used as a alternative/complementary technology to energy saving, such as the \textit{economy (ECO) mode}, e.g., turning the transceivers on/off during low traffic conditions, and \textit{adjustment of the coverage area}. Also, apart from increasing the energy efficiency in traditional communication networks with grid connected BSs, EH is useful when the connection of the BS to the power grid is technically and economically challenging, such as in developing countries. Although these BSs can be possibly powered by diesel generators \cite{5783984}, this solution might not be feasible, due to the inefficiency of diesel fuel power generators and high transportation costs of diesel fuel to BSs located in remote areas \cite{5700169, Kwan}. As a result, wireless networks with energy harvesting base stations (EHBSs) are not only envisioned to be cost-efficient, but also self-sustained.

Moreover, in today's widely used devices and IoT applications, such as smartphones, tablets, wearables, and sensor networks, which require connection to the power cord, EH can be seen as the final challenge to true mobility. Also, especially in low power devices, such as those used in sensor networks, EH can be used to minimize, if not to eliminate the use of battery power, replacing the traditional batteries with super capacitors \cite{krik3}, and leading to even lower energy consumption levels. This is extremely important especially when replacing or recharging the batteries is inconvenient, costly, or dangerous, such as in remote areas, harsh industrial environments, e.g., rotating and moving platforms, human bodies, or vacuum equipment. Furthermore, EH can counterbalance the tremendous increase of the number of batteries needed for such off grid devices, which will bring difficulties for material supplying, recycling and will cause high environmental impact.

%Energy-Efficient Resource Allocation in OFDMA Systems with Hybrid Energy Harvesting Base Station: 
%"In practice, BSs may not be connected to the power grid, especially in developing countries. Thus, the assumption
%of a continuous energy supply made in [7]-[13] is overly optimistic in this case. Although these BSs can be possibly
%powered by diesel generators [14], the inefficiency of diesel fuel power generators and high transportation costs of diesel
%fuel are obstacles for the provision of wireless services in remote areas [15]. In such situations, energy harvesting is
%particularly appealing since BSs can harvest energy from natural renewable energy sources such as solar, wind, and
%geothermal heat, thereby reducing substantially the operating costs of the service providers. As a result, wireless networks
%with energy harvesting BSs are not only envisioned to be energy-efficient in providing ubiquitous service coverage, but
%also to be self-sustained.

\subsection{Wireless Power Transfer}

The main disadvantage of traditional EH methods is that they rely solely on ambient energy sources, such as solar, wind energy, and vibrations, which are uncontrollable and in some cases unpredictable. For this reason, harvesting energy from sources that intentionally generate an electromagnetic (EM) field seems to be an interesting alternative. This concept, termed as wireless power transfer (WPT), was initiated by Nikola Tesla in 1910's. The utilized technology to enable WPT depends on the distance between the source and the terminal \cite{book, krik3, 6951347}. On near-field WPT, non-radiative techniques are used, i.e., capacitive or inductive coupling, the performance of which can be significantly improved by using a resonant scheme. On the other hand, far-field energy transfer is based on the transmission/reception of propagating EM waves. More specifically, the rectifying antenna (rectenna) circuit, invented by  William C. Brown in 1964, can be used to convert the received radio frequency (RF) signal, into direct current (DC), while lasers of light emitting diodes (LEDs), photocells, and lenses are used in order to transfer energy with light waves. Starting from 1970, WPT has been explored in the context of solar power satellites, which harvest energy by sunlight and transfer this to earth,  using microwave power transmission or laser beam, e.g., by NASA \cite{niyato2016book}. 

Recently, research and development in WPT have received tremendous momentum \cite{niyato2016book}. Inductive coupling is the most widely used wireless technology. Its applications include charging handheld devices like phones and electric toothbrushes, radio frequency identification (RFID) tags, and chargers for implantable medical devices like artificial cardiac pacemakers, or electric vehicles. Note that near-field WPT is out of the scope of the technical analysis provided in this dissertation, which focuses on far-field WPT with radio and light waves.

\begin{figure}
\centering
\includegraphics[width=\linewidth]{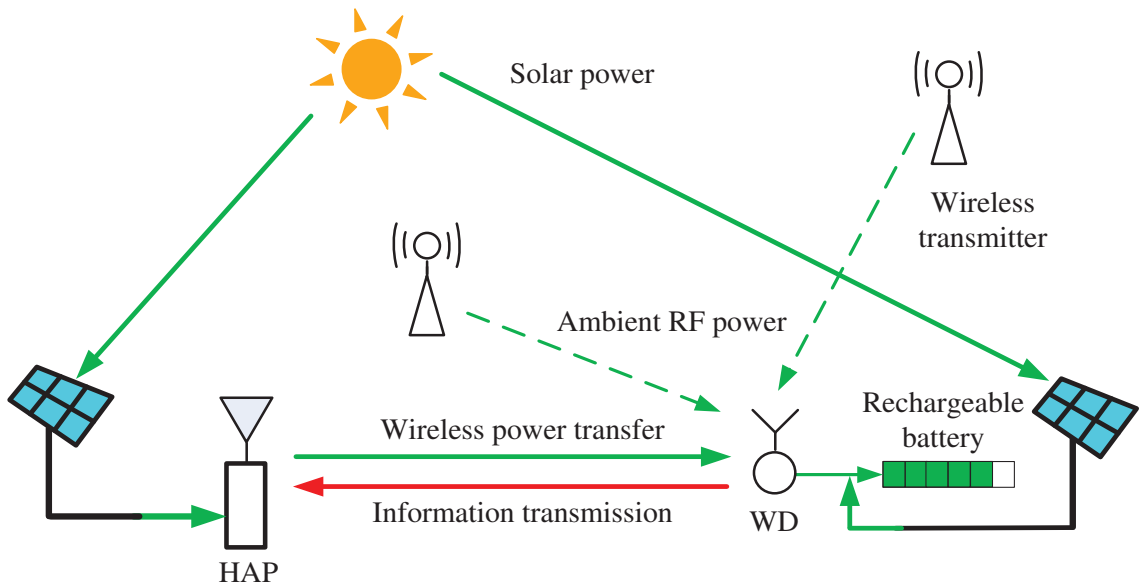}
\caption{A wireless powered communication network with hybrid energy sources \cite{7462480}.}
\label{green}
\end{figure}

Far-field based WPT is particularly important in communication networks, as it allows wireless devices (WDs) with low energy requirements to be remotely powered in a wide area, therefore, it provides a feasible solution for applications where EH and remote energy supply is the only powering option \cite{Ju}. Interestingly, it has been reported that 3.5 mW and 1 $\mu$W of wireless power can be harvested from RF signals, at distances of 0.6 and 11 m, respectively, with an energy harvester operating at 915MHz \cite{zungeru2012radio}. Furthermore, recent advances in electronics and, specifically, in rectifying antennas designing, will further increase the efficiency of EH from RF signals in the near future \cite{vullers2009micropower, 7462480}. In a communication system employing WPT, energy can be harvested both from dedicated sources in a fully controlled manner and opportunistically from ambient signals, which increases the overall energy efficiency and redefines the role of interference. The dedicated sources are usually assumed to be more powerful nodes, termed as power beacons (PBs), which can coincide with the access points (APs). Compared to other EH technologies, which are uncontrollable,  WPT enlarges the potential to reduce the utilization of batteries, as well as their frequent replacement. Thus, WPT is fairly regarded as an eco-friendly technology. Moreover, as illustrated in Fig. \ref{green}, hybrid access points (HAPs) can exploit conventional forms of renewable energy, strengthening the green character of WPT technology \cite{7462480}. Also, as illustrated in the same figure, WDs can still harvest energy from other sources rather than WPT, with the latter being used only when necessary.  

WPT creates unique challenges in the design of communication systems, since, in some cases, it conflicts with the information trasnmission, especially when the same resources are used, e.g., transmitted power, time, spectrum, and antennas, and a compromise is required. This introduces a new area of investigation, in which the vision of Nikola Tesla with WPT meets the basic brinciples of information theory, as formulated by Claude Shannon \cite{Grover, Varshney}. More specificially, in WPT, the nodes use the power of the received signal to charge their batteries \cite{KwanWPT}, or to transmit the information to other nodes, e.g., to the access point (AP) \cite{Ju, QWu}, as illustrated in Fig. \ref{green}. But, in practice, nodes cannot  harvest energy and receive/transmit information simultnaeously \cite{Trung1, Suraweera, Ju, krik3, Zoran}, which complicates the optimization of communication networks with WPT and brings trade-offs on their design. Also, an important issue associated with all wireless power systems is limiting the exposure of people and other living things to potentially injurious EM fields, as well as the unwanted interference and EM pollution, which calls for conscious utilization of the radiated energy. This can be achieved by unifying the information and energy transmission, which is the basis of \textit{simultaneous wireless information and power transfer (SWIPT)}. For example, sensor networks can be controlled and charged with the same signal \cite{krik3}. SWIPT, when properly optimized, can result in significant gains in terms of spectral efficiency, time delay, energy consumption, and interference management by superposing information and power transfer. However, this approach, calls for the redesign of existing wireless networks.

\subsection{Industrial Interest}
The application of EH and WPT on wireless devices/sensors have attracted the interest not only from academia, but also from industry.  Some of the involved companies are:
\begin{itemize}
\item The idea of EHBSs has already been adopted by many mobile network operators, worldwide \cite{LeonGarcia}. 
%Also, a program called “Green Power for Mobile” to
%use renewable energy resources for BSs has been started
%by 25 leading telecoms including MTN Uganda and Zain,
%united under the Global Systems for Mobile communications
%Association (GSMA) [39]. [Green Cellular Networks: A Survey, Some Research
%Issues and Challenges]
\item \textit{LaserMotive} works on WPT for long transmission spans, using laser lights for unmanned aerial vehicles (UAVs) \cite{lasermotive}, which in turn can be used for communication purposes.
\item \textit{Wi-charge} manufactured a novel system for WPT using infrared (IR) \cite{wicharge}. 
\item \textit{EnOcean Alliance} works on the EH wireless sensor technology for collecting energy out of the air, e.g., kinetic motion, pressure, light, differences in temperature, and converting them into the energy for wireless communications \cite{enocean}.
\item \textit{ABB} manufactured an EH wireless temperature transmitter \cite{abb}.
\item \textit{PowerbyProxi}, {Humavox}, and {Powercast} work on wireless power sensors, mainly focusing on the power transfer over a short range using the resonant inductive technology \cite{powerbyproxi}, \cite{powercastco}, \cite{humavox}. 
\item \textit{Sunpartner Technologies} has created Wysips Crystal, which is a device combining optical and photovoltaic (PV) technologies, and can produce electricity from an natural or artificial light source, using invisible PV cells \cite{bialicwysips}.
\item \textit{Philips} is also involved in exploring applications of WPT, considering rotational system with high power WPT, remote sensing applications, etc \cite{philips}.
\end{itemize}

\section{Resource Allocation in Wireless Networks}

The integration of new heterogeneous services and applications, e.g. IoT ones, in the design of future communication networks, e.g., of the upcoming 5G networks, creates diverse QoS requirements \cite{wong2017key}. The general goal is to improve the QoS, as specified by each user or the network, while using the available resources, e.g., bandwidth, power, and time as efficiently as possible. This leads to the formulation of constrained optimization problems, which aim to the maximization of specific metrics that take into account the service criteria. When the aim is to maximize spectral efficiency and the users utilize different channels, the user's utility is usually defined as a logarithmic, concave function of the user's signal-to-interference-plus-noise ratio (SINR) \cite{Poor}.  Different utilities functions can be used when error performance \cite{GKK1, GKK2, GK6, GK8}, users' prioritization, fairness, power consumption, or the cost of service are the main issues, or when the users share the same communication channels. Also, in some cases, resource allocation problems can also be modeled using game-theoretic approaches. Although there is an overlap between the game-theoretic and optimization-theoretic approaches, game theory tends to focus on the multiuser competitive nature of the problem and on the users' interaction. Moreover, in most practical scenarios, distributed algorithms are preferred over centralized ones, with their main advantages including the reduction of complexity and scalability \cite{Poor}.

It is highlighted that efficient resource allocation in EH communication networks involves new challenges, since it involves the constraints imposed by the levels of energy harvested by each node. Since a node is energy-limited only
till the next harvesting opportunity (recharge cycle), it can optimize its energy usage to maximize performance during
that interval \cite{Sude}. Moreover, nodes powered solely by EH may not be able to maintain a stable operation and to guarantee a fixed QoS \cite{Kwan}. 

\section{Thesis Layout}
The rest of this dissertation is organized as follows. Chapter 2 examines optimal resource allocation for the downlink of autonomous energy harvesting base station (AEHBS) from a game theoretic point of view, by considering measures, such as the users' utilities and the base stations revenues. Chapter 3 focuses on wireless powered communication networks, and specifically on the optimization of individual data rates and fairness. Chapter 4 is dedicates on simultaneous wireless information and power transfer. Chapter 5 investigates simultaneous lightwave information and power transfer (SLIPT). Finally, chapter 5 concludes the dissertation and presents some possible future extensions. The structure of these chapters is described below.

\subsection{Energy Harvesting Base Stations}

Section 2.1 serves as introduction to resource allocation problem in EHBSs.

In Section 2.2, optimal energy and resource allocation for the downlink of an AEHBS
is investigated. In particular, the joint maximization of the users' utilities and the base station's revenue is considered, while these hierarchical decision problems are matched to the framework of a generalized Stackelberg game. Also, an efficient iterative method is proposed to facilitate all the players to reach the variational equilibrium, i.e. the optimal solution of the game. Simulation results validate the effectiveness of the proposed method.

\subsection{Wireless Powered Networks}
Section 3.1 serves as an introduction to the basic principles behind wireless powered networks (WPNs). Also, the created trade-offs are described, taking into account throughput maximization, energy efficiency, and fairness and we present and discuss the solution of several optimization problems, considering different scenarios for the network consistence, the adopted protocol, and the energy arrival knowledge.

In Section 3.2, the concept of non-orthogonal multiple access (NOMA) is proposed, as a mean to increase the performance of wireless-powered uplink communication systems, consisting of one BS and multiple EH users. More specifically, the main focus is on the individual data rate optimization and fairness improvement and it is shown that the formulated problems can be optimally and efficiently solved by either linear programming or convex optimization. In the provided analysis, two types of decoding order strategies are considered, namely \textit{fixed decoding order} and \textit{time-sharing}. Furthermore, an efficient greedy algorithm is proposed, which is suitable for the practical implementation of the time-sharing strategy. Simulation results illustrate that the proposed scheme outperforms the baseline orthogonal multiple access scheme. More specifically, it is shown that NOMA offers a considerable improvement in throughput, fairness, and energy efficiency. Also, the dependence among system throughput, minimum individual data rate, and harvested energy is revealed, as well as an interesting trade-off between rates and energy efficiency. Finally, the convergence speed of the proposed greedy algorithm is evaluated, and it is shown that the required number of iterations is linear with respect to the
number of users.

In section 3.3 the analysis of Section 3.2 is extended to the case of proportional fairness maximization, in order to balance  the fundamental trade-off between sum rate and fairness of a wireless-powered uplink communication system. Two well known communication protocols are considered, namely time division multiple access (TDMA) and NOMA with time-sharing (NOMA-TS). It is shown that NOMA-TS outperforms the considered benchmark scheme, which is the NOMA with fixed decoding order and adaptive power allocation, while TDMA proves to be an appropriate choice, when all the users are located in similar distances from the BS.

\subsection{Simultaneous Wireless Information and Power  Transfer (SWIPT)}
Section 4.1 serves as an introduction to the basic principles of simultaneous wireless information and power transfer.

Section 4.2 is a non-trivial extension of Chapter 3, such as that except for energy, information is also transmitted during downlink, using SWIPT, while also taking into account co-channel interference. More specifically, it investigates the downlink/uplink of WPNs, which are exposed to the effect of the cascaded near-far problem (CnfP), i.e., the asymmetric overall degradation of the users' performance, due to different path-loss values. More specifically, assuming that the users are able to harvest energy both from interference and desired signals, higher path-loss reduces the downlink rate of the far user, while it also negatively affects its uplink rate, since less energy can be harvested during downlink. Furthermore, if the far user is located at the cell-edge, its performance is more severely impaired by interference, despite the potential gain due to EH from interference signals. To this end, the downlink/uplink users' rates are fairly maximized, by utilizing corresponding priority weights. Two communication protocols are taken into account for the downlink, namely TDMA and NOMA, while NOMA-TS is considered for the uplink. The formulated multidimensional non-convex optimization problems are transformed into the equivalent convex ones and can be solved with low complexity. Simulations results illustrate that: i) a relatively high downlink rate can be achieved, while the required energy is simultaneously harvested by the users for the uplink, ii) dowlink NOMA is a more appropriate option with respect to the network topology, especially when a high downlink rate is desired.

Section 4.3 investigates optimal SWIPT in a multicarrier two-hop link with a wireless powered relay. First, the corresponding optimization problem is formulated, which consists of the joint optimization -in terms of achievable rate- of, i) the dynamic power allocation among multiple channels and, ii) the selection of the power splitting
ratio between information processing and EH at the relay, when amplify-and-forward is applied. This is a nonconvex
optimization problem which is mapped to a convex one and optimally solved using one-dimensional search and dual
decomposition, while a suboptimal efficient iterative method is also proposed. Simulations reveal a significant increase in the
throughput, and verify the effectiveness of the fast-converging iterative solution.

\subsection{Simultaneous Lightwave Information and Power Transfer (SLIPT)}
Section 5.1 serves as an introduction to optical wireless communications (OWCs), which emphasis on indoor IoT applications.

In Section 5.2, the concept of SLIPT for indoor optical wireless systems is presented. Specifically,  novel and fundamental SLIPT strategies are proposed, which can be implemented through Visible Light (VL) or IR communication systems, equipped with  a simple solar panel-based receiver. These strategies are performed at the transmitter or at the receiver, or at both sides, named \textit{Adjusting transmission}, \textit{Adjusting reception} and \textit{Coordinated adjustment of transmission and reception}, correspondingly. Furthermore, the fundamental trade-off between harvested energy and  QoS is compromised,  by maximizing the harvested energy, while achieving the required user's QoS. To this end, two optimization problems are formulated  and optimally solved. Computer simulations validate the optimum solutions and reveal that the proposed strategies considerably increase  the harvested energy, compared to SLIPT with fixed policies.

\subsection{Conclusions and Future Work}
Chapter 5 concludes the thesis, where the main conclusions of the presented research are
drawn, while possible future extensions of this work are also proposed.

%CHAPTER 1%%%%%%%%%%%%%%%%%%%%%%%%%%%%%%%%%%%%%%%%%%%%%%%%%%%%%%%%%%%%%%%%%%%%%%
%\else\fi

%\part{Information-Theoretic Physical Layer Security}\label{Part:PhyLayerSecurity}

%\ifchaptertenflag
%CHAPTER 5 (CONCLUSIONS) %%%%%%%%%%%%%%%%%%%%%%%%%%%%%%%%%%%%%%%%%%%%%%%%%%%%%%%
\chapter[Energy Harvesting Base Stations][]{Energy Harvesting Base Stations}\label{ch:chapter2}
The introduction of energy harvesting capabilities for BSs poses many new challenges for resource allocation algorithm design, due to the time varying availability of the harvested energy. This chapter focuses on dynamic resource allocation in EHBSs, with emphasis on AEHBSs, which are based solely to renewable sources of energy and suffer to guarantee a certain QoS.

\section{Introduction to Resource Allocation in Energy Harvesting Base Stations}\label{ch:chapter2_1}

\begin{figure}
\centering
\includegraphics[width=\columnwidth]{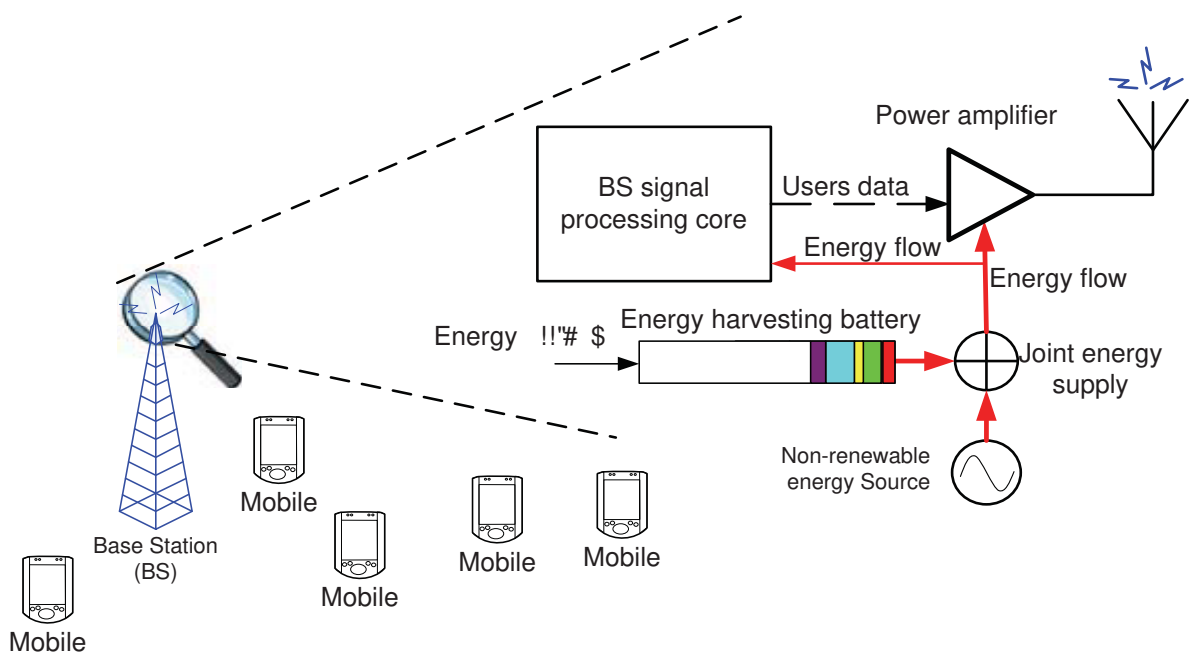}
\caption{A system with an hybrid energy harvesting base station\cite{Kwan}.}
\label{Fig0chapter2}
\end{figure}

EHBSs use the harvested power from the wind, solar radiation, etc. to satisfy at least a part of their energy demand \cite{Wang, Diamantoulakis} and reduce the related costs \cite{PD6}. In the general case, the BS utilizes multiple energy harvesting and non-renewable energy sources, being termed as hybrid energy harvesting base station (HEHBS). A typical paradigm of a BS using two energy sources, i.e., a renewable energy harvesting source and a non-renewable energy one, is depicted in Fig. \ref{Fig0chapter2}. 

Although the amount of renewable energy is potentially unlimited, the availability of renewable energy at the BS is a random event. This calls for the redesign of the corresponding resource allocation schemes, since new criteria and constraints need to be considered, such as the cost efficiency and the availability of renewable energy. For example, a practical and well-investigated problem is the minimization of the \textit{completion time} with constraints on the consumed energy, either assuming a point-to-point communication system \cite{16,19}, or broadcasting \cite{20,21,22}.  An alternative objective is the maximization of \textit{throughput} with optimal power control time sequences, which has been considered in \cite{17,18,19}. It is also noted that all works \cite{16,17,18,19,20,21,22} focus on systems with a single energy storage, while \cite{16,19,21,22} investigate the case of energy harvesting transmitters with both energy and data queues. 

On the other hand, a decentralized \textit{delay} minimization approach, which is applicable to HEHBSs, has been presented in \cite{6151187}. More specifically, a decentralized online algorithm is proposed using the \textit{Lagrangian theory}. Also, a different approach is presented in \cite{Kwan}, which proposes a novel resource allocation scheme that maximizes the \textit{weighted energy efficiency (WEE)} in multiuser systems with hybrid EHBSs (HEHBSs), using both the electricity grid and EH as power sources. The WEE is defined as the total number of bits successfully delivered to the mobile users, considered over a time frame per joule of consumed energy. In turn, the part of the consumption that corresponds to the utilization of the harvested energy is coupled by a positive constant, with value lower than 1, in order to discourage the BS to consume non-renewable energy. Also, in the problem formulation, minimum system data rate requirements are considered for each user, which are imposed for fairness. The corresponding optimization problem is optimally solved using \textit{Lagrange dual decomposition} and convex optimization tools. 

In general, such multi-variable and multi-constraint nonlinear resource allocation problems are quite complicated and, unfortunately, cannot be solved in closed form. Also, the design of an iterative solution method with acceptable complexity is often challenging. To this direction, convex optimization is quite prelevant, since if a problem is formulated as a convex one, then it can be solved very reliably and efficiently, by using interior-point or other methods for convex optimization, e.g., based on \textit{Lagrange dual decomposition}. The main advantage of Lagrange dual decomposition is that it can lead to smaller subproblems, that can be solved in parallel. This approach facilitates the design of an efficient distributed method for solving the initial optimization problem \cite{Boyd1}. Also, the associated dual problem often has an interesting interpretation, in terms of the original problem. 

% However, optimality of the derived solution is still an issue and needs to be verified by checking specific conditions, such as the convexity of the initial optimization problem and constraint qualification. 

%\newpage
\section{Autonomous Energy Harvesting Base Stations with Minimum Storage Requirements}\label{Autonomous Energy Harvesting Base Stations}
This section investigates optimal energy, time, and bandwidth allocation problem for the downlink of an AEHBS with very low storage requirements, taking into account both the users preferences and the BS's revenue. The corresponding hierarchical decision problems are formulated using game-theoretic tools. Also, an efficient iterative and distributed method is proposed in order to enable the users and the BS to choose the unique optimal solution of the game, i.e., to reach the equilibrium.

The results of the research presented in this section are included in \cite{harvesting}.

\subsection{Related Work and Motivation}
With the advances in the speed of wireless backhaul, that eliminate the need for wired connection, AEHBSs are of particular interest, especially for small cells \cite{Dhillon} or for use as relay stations \cite{Song}. In HEHBSs, user fairness can be achieved by setting a minimum rate requirement for each user as a minimum level of satisfaction \cite{Kwan}. However, in an AEHBS that is solely powered by EH, this is not always possible, since the BS may not have enough energy \cite{Dhillon}. The probability of meeting this requirement can be increased by employing larger harvesting and storage equipment, however, this also substantially increases the capital expenditures (CAPEX).

\subsection{Contribution}
This section investigates user fairness in AEHBSs with minimum energy storage requirements. A novel scheme is proposed, in which the users pay per amount of utilized resources and energy used for transmission. A utility function for each user is defined, which reflects the user experience and satisfaction, as an alternative fairness requirement. In this framework, the users' objective is to maximize their levels of satisfaction, while the BS aims to maximize its revenue, accommodating all the users over the finite available spectrum. In order to find the optimal solution of this hierarchical decision problem, game theoretic tools are utilized, and specifically the Stackelberg games \cite{Basar}. Also, in order to minimize the storage requirements, it is considered that in a specific time frame, the BS can only use the energy harvested in the previous time frame. Results show that previously proposed solutions, such as maximizing the total capacity for a specific amount of energy, are not fair solutions and do not optimize the users' experience in an AEHBS.

\subsection{System model}\label{S:Systemchapter2}
A wireless network consisting of $N$ users and one AEHBS is assumed. The communication is divided into time frames of duration $T_\mathrm{tot}$, while the total available bandwidth is denoted by $W$. The communication with each user occupies a distinct fraction of time and frequency, so that the signals of different users do not interfere. At the time instant $t_\mathrm{in}$ within a frame $k$, the BS transmits to the $n$-th user with power $P^\mathrm{d}_n(t_\mathrm{in})$, using the bandwidth fraction $W_n(t_\mathrm{in})$. For the link between the BS and the $n$-th user, $L_n(t_\mathrm{in})$ and $H_n(t_\mathrm{in})$ denote the path loss/shadowing coefficient and the small scale fading coefficient, respectively. The time fraction allocated to the $n$-th user is denoted by $t_n$.

%Thus, during the time allocated for the $n$-th user, considering unitary antenna gains, the $n$-th user receives
%\begin{equation}
%y_n(t_\mathrm{in})=h_n(t_\mathrm{in})\sqrt{P^\mathrm{d}_n(t_\mathrm{in})}s_n^{\mathrm{d}}(t_\mathrm{in})+\nu_n(t_\mathrm{in}),
%\end{equation}
%where $s_n^{\mathrm{d}}(t_\mathrm{in})(t_\mathrm{in})$, with $\mathbb{E}[|s_n^{\mathrm{d}}(t_\mathrm{in})|^2]=1$, is the message for the $n$-th user, with $\mathbb{E}[\cdot]$ and $|\cdot|$ denoting the statistical expectation and the absolute value, respectively. Also, $v_n$ denotes the additive noise at the $n$-th user and $h_n=H_n(t_\mathrm{in})\sqrt{L_n(t_\mathrm{in})}$.

It is considered that the energy consumed by the BS is limited by the amount of harvested energy, i.e., in time frame $k$, the BS can only use the energy that was harvested during the time frame $k-1$, a strategy that creates very low storage requirements. For notational simplicity, it is assumed that the BS consumes energy only for transmission, which is denoted by $E_{\mathrm{BS}}$. If $E_{\mathrm{BS},n}$ is the fraction of the energy dedicated for the $n$-th user, it holds that
$\sum_{n=1}^NE_{\mathrm{BS},n}\leq E_{\mathrm{BS}}$.

The channel capacity between the BS and the $n$-th user in a frame $k$ is given by the effective rate (bits per time frame) that each user can achieve during the frame $T_\mathrm{tot}$, which, assuming additive white Gaussian noise (AWGN), is given by \cite{Kwan}
\begin{equation}
R^\mathrm{d}_n=\int_0^{T_\mathrm{tot}} S_n(t)W_n(t_\mathrm{in})\log_2\left(1+\frac{P^\mathrm{d}_n(t_\mathrm{in})G_n(t_\mathrm{in})}{W_n(t_\mathrm{in})}\right)\mathrm{d}t_\mathrm{in}.
\label{capacity}
\end{equation}
In (\ref{capacity}), $S_n(t_\mathrm{in})\in\{0,1\}$ denotes the time allocation with $\int_0^{T_\mathrm{tot}}S_n(t_\mathrm{in})\mathrm{d}t_\mathrm{in}=t_n$, and $G_n(t_\mathrm{in})=\frac{L_n(t_\mathrm{in})|H_n(t_\mathrm{in})|^2}{N_0}$, where $N_0$ is the noise power spectral density. In order to simplify theoretical studies it is assumed that the coherence time and bandwidth of the channel are larger than $T_\mathrm{tot}$ and $W$ respectively (see \cite{Chao} and references therein). It is also assumed that the power transmitted towards each user is constant, and thus $P^\mathrm{d}_n=\frac{E_{\mathrm{BS},n}}{t_n}$. Therefore all values are frequency flat and constant within each frame $k$, that is, independent of $t_\mathrm{in}$. The analysis focuses on a specific time frame, and $t_\mathrm{in}$ is dropped hereafter. If the time and bandwidth resources for each user are denoted by $q_n$, where $q_n=t_nW_n$, then the channel capacity can be simplified to
\begin{equation}
R^\mathrm{d}_n=q_n\log_2\left(1+\frac{E_{\mathrm{BS},n}}{q_n}G_n\right).
\end{equation}
Note that, in this section, the term ``resource'' herein refers to bandwidth/time and not energy.
The sum of the available resources must satisfy the condition
$\sum_{n=1}^Nq_n\leq Q$,
where $Q=T_\mathrm{tot}W$. Hereafter, $\mathbf{E}=\{E_{\mathrm{BS},1},\ldots,E_{\mathrm{BS},1},\ldots,E_{\mathrm{BS},N}\}$ and $\mathbf{q}=\{q_1,\ldots,q_n,\ldots,q_N\}$ denote the sets of values of the energy and resource allocation among users within a frame $k$, respectively.

\subsection{Problem Formulation}\label{S:Problemchapter2}
It is assumed that the BS charges the users according to the resources and the transmitted energy they use during a time frame $k$.  Both the users and the BS want to maximize their utility functions, i.e., their levels of satisfaction and their revenue, respectively.

\subsubsection{Objective of the BS}
The BS can control the price $c_1$ per utilized energy unit (e.g. \$/J) and $c_2$ per unit of resources (e.g. \$) in order to maximize its utility function. The utility function $U_\mathrm{BS}$ of the BS captures the revenue that the operator can receive by selling its resources and energy, and it is given by
\begin{equation}
U_\mathrm{BS}=c_1\sum_{n=1}^NE_{\mathrm{BS},n} +c_2\sum_{n=1}^Nq_n.
\end{equation}

\subsubsection{Objective of the Users}
The users respond to the prices by demanding a certain amount of resources and energy, in order to maximize their utilities, $U_n$. For each user ${n\in\mathcal{N}}$, where $\mathcal{N}$ is the set of all the users, the utility function corresponds to a level of satisfaction that a user obtains \cite{Poor}. The following assumptions are made for the utility function of the users:
\begin{enumerate}[label=\roman*.]
\item It is affected by a satisfaction parameter $a_n$ of each user. This parameter is different according to the appliance of the user or the service it requires, i.e., a smart phone or a tablet that is usually online will have higher demands for rate than a simple cell phone or a smart meter. Generally, a higher $a_n$ implies a higher level of satisfaction. Thus, $U_n$ must be an increasing function with respect to (w.r.t.) $a_n$. 
\item It is a decreasing function w.r.t. the prices $c_1$ and $c_2$, since higher prices lead to a lower level of satisfaction. \item It reflects the users' desire for higher capacity as well as the saturation of their satisfaction levels as higher capacity is achieved \cite{Gajic}. This is ensured when, apart from the payment ($c_1E_{\mathrm{BS},n}+c_2q_n$), $U_n$ includes a term that is an increasing and concave function of $R^\mathrm{d}_n$.
\end{enumerate}

To this end, the following utility function for the users is considered:
\begin{equation}
U_n=a_n\log_2(1+R^\mathrm{d}_n)-c_1E_{\mathrm{BS},n}-c_2q_n, \quad a_n,c_1,c_2>0.
\end{equation}
In the above, the weight $a_n$, multiplied by the term concerning the rate, captures the satisfaction acquired by the quality of service, while the subtraction - or equivalently, a negative unitary weight multiplied by the term concerning the payment - captures the dissatisfaction induced by the users' charges. Therefore, $U_n$ is measured in units of satisfaction.
Thus, for a fixed set of prices $c_1$ and $c_2$,  the objective of any user $n$ is
%===========================================================
\begin{equation}
\begin{array}{ll}
\underset{q_n,E_{\mathrm{BS},n}}{\text{\textbf{max}}}& U_n(c_1, c_2, {E_{\mathrm{BS},n}},\mathbf{E}_{-n}, q_n,\mathbf{q}_{-n}) \\
\,\,\,\,\,\text{\textbf{s.t.}}&\mathrm{C}_1: \sum\limits_{n=1}^NE_{\mathrm{BS},n}\leq E_\mathrm{BS},\,\mathrm{C}_2: \sum\limits_{n=1}^Nq_n\leq Q,   \\
\end{array}
\label{optchapter2}
\end{equation}
%=============================================
where $\mathbf{E}_{-n}$  and $\mathbf{q}_{-n}$ denote the sets $\mathbf{E}$ and $\mathbf{q}$, excluding $E_{\mathrm{BS},n}$ and $q_n$, respectively.

\subsection{Stackelberg Game}
The hierarchical decision problem presented in Section \ref{S:Problemchapter2} can be effectively solved using a Stackelberg game. In this hierarchical game, the followers decide in response to the decision taken by the leader. In this context, at each decision point the users (the followers) know the prices set by the BS (the leader), while the BS has knowledge of the users' strategies, i.e., the demanded energies and resources. Consequently, the $N$ users form a lower level game.  When this lower level game achieves an equilibrium, the equilibrium together with the operator (BS) form the higher level game. Then, the equilibrium of the higher level is called the Stackelberg Equilibrium (SE) and it is the solution of the game \cite{Basar}.
This game is defined in its strategic form as follows:
\begin{equation}
\Omega=\left\{\left(\mathcal{N}\cup\textit{BS}\right),\{E_{\mathrm{BS},n},q_n\}_{n\in\mathcal{N}},\{U_n\}_{n\in\mathcal{N}},U_\mathrm{BS},c_1,c_2\right\}.
\end{equation}

\subsubsection{Generalized Stackelberg Equilibrium}
In (\ref{optchapter2}), the amount of resources and energy requested by each user depend not only on its own strategy (resources and energy demand) but also on the strategies of the other users, since they share the same constraints. Thus, the noncooperative game among the users represents a generalized Nash Equilibrium problem (GNEP), the solution of which is the generalized Nash Equilibrium (GNE) \cite{GNE, Tushar}. Moreover, since the objective of the BS is to maximize its utility, $U_\mathrm{BS}(c_1,c_2,\mathbf{E^*},\mathbf{q^*})$, taking into account the GNE of the users' game, the formulated game $\Omega$ represents a generalized Stackelberg game, the solution of which is the generalized Stackelberg Equilibrium (GSE) \cite{Leitmann}. The set of strategies ($c_1^*,c_2^*,\mathbf{E^*},\mathbf{q^*}$) that constitutes the GSE is given by solving the following inequalities, where $(\cdot)^*$ denotes a solution value:
\begin{equation}
U_n(c^*_1,c^*_2,E_{\mathrm{BS},n}^*,{\mathbf{E}^*_{-n}},q_n^*,{ \mathbf{q}^*_{-n}})\geq U_n(c^*_1,c^*_2,E_{\mathrm{BS},n},{\mathbf{E}^*_{-n}},q_n,{\mathbf{q}^*_{-n}}),\,\forall{n\in \mathcal{N}},
\end{equation}
\begin{equation}
U_\mathrm{BS}(c^*_1,c^*_2,\mathbf{q^*},\mathbf{E^*})\geq U_\mathrm{BS}(c_1,c_2,\mathbf{q^*},\mathbf{E^*}), \mathrm{C}_1, \mathrm{C}_2.
\end{equation}
\subsubsection{Existence and Uniqueness of the GNE}
A GNEP can be solved by using a \emph{variational inequality (VI)} problem
reformulation. The GNEP might have multiple or even infinitely many solutions; however each solution of the GNEP is not always a solution of the VI \cite{scutari}. A solution of the GNEP which is also a solution of the VI is called a variational equilibrium (VE) and is considered to be the most socially stable GNE \cite{Tushar}.

Since ${\boldsymbol{F}}=(\nabla_{E_{\mathrm{BS},1}} U_1, \nabla_{q_1} U_1,...,\nabla_{E_{\mathrm{BS},N}} U_N, \nabla_{q_N} U_N)$ is the gradient of a scalar function, i.e., ${\boldsymbol{F}}=\nabla f$ with
$f=\sum_{n=1}^NU_n$, the variational solutions are the solutions of the following optimization problem \cite{scutari}:
%===========================================================
\begin{equation}
\begin{array}{ll}
\underset{\mathbf{q},\mathbf{E}}{\text{\textbf{max}}}&f(\mathbf{q},\mathbf{E}) \\
\,\,\,\text{\textbf{s.t.}}&\mathrm{C}_1, \mathrm{C}_2.\\
\end{array}
\label{opt1chapter2}
\end{equation}
%=============================================

\begin{theorem}\label{existence1}
For fixed prices $c_1$ and $c_2$, a social VE exists, which is unique.
\end{theorem}
\begin{proof}
In order to accommodate the proof of the existence and uniqueness of VE, it is assumed that $q_n \in [\epsilon,Q]$ and $E_{\mathrm{BS},n} \in [\varepsilon,E_\mathrm{BS}]$, where $\epsilon,\varepsilon \rightarrow 0^+$. With the aid of the above assumption, it can be shown that the Hessian matrix of $U_n$
has negative eigenvalues, so $U_n$ is jointly strictly concave with respect to the optimization variables $q_n$ and $E_{\mathrm{BS},n}$. Also, the sum over $n$ preserves the concavity of the objective function, $f$, in (\ref{opt1chapter2}), while the constraints $\mathrm{C}_1$ and $\mathrm{C}_2$ are linear. Thus, (\ref{opt1chapter2}) is strictly concave, which means that it has a unique solution and so does the equivalent VI.
\end{proof}

\subsubsection{Solution of the Game}

\paragraph{Solving the Variational Inequality}
Using the Lagrange multiplier method, the Lagrangian for the optimization problem (\ref{opt1chapter2}) is
\begin{equation}
\mathcal{L}=f(\mathbf{q},\mathbf{E})-\lambda_1\left(\sum_{n=1}^NE_{\mathrm{BS},n}-E_\mathrm{BS}\right)-\lambda_2\left(\sum_{n=1}^Nq_n-Q\right),
\end{equation}
where $\lambda_1,\lambda_2\geq 0$ are the Lagrange multipliers (LMs) related to $\mathrm{C}_1,\mathrm{C}_2$, respectively. The values $\mathbf{q^*},\mathbf{E^*}, \lambda_1^*$ and $\lambda_2^*$ that satisfy the Karush-Kuhn-Tucker (KKT) conditions can be obtained iteratively as follows: in \textit{Layer 1} the users simultaneously choose their demands for a fixed set of $\lambda_1$ and $\lambda_2$, the values of which are updated in \textit{Layer 2} by the BS.

\textit{Layer 1 (Solved by the users)}: From the KKT conditions it must hold that $\frac{\partial \mathcal{L}}{\partial E_{\mathrm{BS},n}}=0$ and $\frac{\partial \mathcal{L}}{\partial q_n}=0$. So, with direct calculations the optimal values of $E_{\mathrm{BS},n}$ and $q_n$ are obtained in a distributed way, since they can be calculated in parallel by each user according to
\begin{equation}
E_{\mathrm{BS},n}^* = \left[\frac{\left(\exp({Z_n})-1\right)\left(\frac{a_nG_n\exp({-Z_n})}{l_1} - (\ln(2))^2\right)}{Z_nG_n\ln(2)}\right]_{\varepsilon},
\end{equation}
\begin{equation}
q_n^* = \left[\frac{\frac{a_nG_n\exp({-Z_n})}{l_1} - (\ln(2))^2}{Z_n\ln(2)}\right]_{\epsilon},
\end{equation}
where $[.]_{\xi}=\max (.,\xi)$, $\exp(x)$ is the exponential of $x$, and $Z_n$ is given by
\begin{equation}
Z_n=1+\mathcal{W}_0 \left(\frac{-l_1+l_2G_n}{e l_1}\right),
\label{lambert}
\end{equation}
where $e$ is the base of the natural logarithm, $l_1=\lambda_1+c_1$ and $l_2=\lambda_2+c_2$. Moreover, $\mathcal{W}_0(x)$  returns the principal branch of the Lambert W function, $\mathcal{W}(x)$, also called omega function or product logarithm, defined as $x=W(x)\exp({\mathcal{W}(x)})$ \cite{lambert}. Note that $\mathcal{W}(x)$ is a built-in function in most well-known mathematical software packages, such as Matlab, Mathematica, etc.

\textit{Layer 2 (Solved by the BS)}: Using the dual-domain and subgradient methods, $\lambda_1$ and $\lambda_2$ can be obtained by
\begin{equation}
\lambda_1(j+1)=\left[\lambda_1(j)-\hat{\lambda}_1(j)\left(E_\mathrm{BS}-\sum_{n=1}^nE_{\mathrm{BS},n}\right) \right]_0,
\end{equation}
\begin{equation}
\lambda_2(j+1)=\left[\lambda_2(j)-\hat{\lambda}_2(j)\left(Q-\sum_{n=1}^nq_n\right)\right]_0,
\end{equation}
where $\hat{\lambda}_i(j)\,, i \in \{1, 2\}$ are positive step sizes, chosen
in order to satisfy the \textit{diminishing step size rules} and $j>0$ is the iteration index.
Since the optimization problem (\ref{opt1chapter2}) is concave, it is guaranteed that the iteration between the two layers converges to the optimal solution of (\ref{opt1chapter2}) \cite{Boyd1, Boyd2}.

It is remarkable that by adopting the proposed method, each user can separately calculate its own optimized demand in each iteration, and thus the users can reach the VE with no communication among themselves. In contrast, they are only required to communicate with the BS, a fact that substantially reduces the overhead.  These benefits fully justify the selection of the proposed distributed solution method.

\paragraph{Optimizing the prices}
From the KKT conditions it holds that
\begin{equation}\frac{\partial U_n}{\partial E_{\mathrm{BS},n}}-\lambda^*_1=0,\quad\frac{\partial U_n}{\partial q_n}-\lambda^*_2=0.
\end{equation}
Since, $\lambda^*_1\geq 0$ and $\lambda^*_2 \geq 0$, the prices $c_1$ and $c_2$ must satisfy the following inequalities:
\begin{equation}
c_1\leq \frac{a_nG_nq_n^*}{\left(G_nE_{\mathrm{BS},n}^*+q_n^*\right)\left(\ln(2)+q_n^*\ln\left(1+\frac{G_nE_{\mathrm{BS},n}^*}{q_n^*}\right)\right)\ln(2)}
\label{price1}
\end{equation}
\begin{equation}
c_2\leq\frac{a_n\left(-G_nE_{\mathrm{BS},n}^*+\left(G_nE_{\mathrm{BS},n}^*+q_n^*\right)\ln\left(1+\frac{G_nE_{\mathrm{BS},n}^*}{q_n^*}\right)\right)}{\left(G_nE_{\mathrm{BS},n}^*+q_n^*\right)\left(\ln\left(2\right)+q_n^*\ln\left(1+\frac{G_nE_{\mathrm{BS},n}^*}{q_n^*}\right)\right)\ln(2)}.
\label{price2}
\end{equation}
The revenue maximizing prices $c_1^*$ and $c_2^*$ are given by (\ref{price1}) and (\ref{price2}), respectively, when these hold with equality.

\begin{figure}
\centering
\includegraphics[width=0.8\columnwidth]{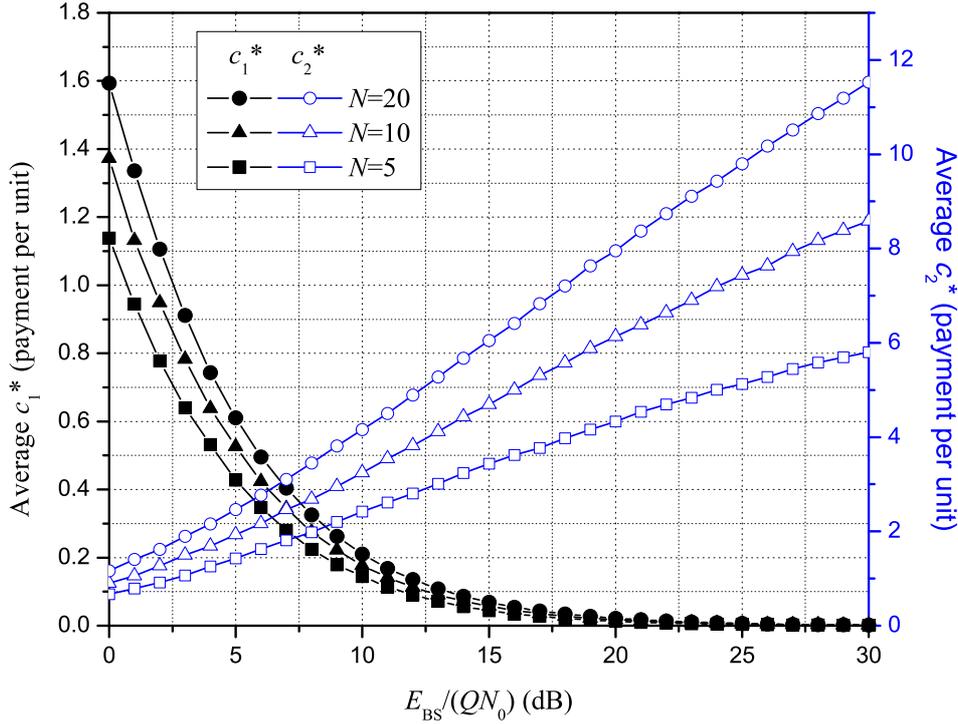}
\caption{Average prices $c_1^*$ and $c_2^*$ per unit of energy and resources, respectively.}
\label{Fig1chapter2}
\end{figure}

\subsection{Simulations and Discussion}
The simulation results consider the case $L_n=1$, $Q=1$ and $a_n$ chosen randomly and uniformly in the range of $[1,2]$. All statistical results are averaged over random values of $H_n$ and $a_n$, where $H_n\sim\mathcal{CN}(0,1)$.
All the provided results take into account the effects of the number of users and the different amount of harvested energy. To this end, the desired metric is plotted vesrus the harvested energy, $\frac{E}{QN_0}$, for three cases for the number of users, i.e., $N=5,10,20$.

In Fig. \ref{Fig1chapter2} the average prices $c_1^*$ and $c_2^*$ are illustrated. It is seen that the optimal prices increase with the number of users. This implies that the demand is increasing, while the available resources and energy are limited. On the other hand, when the harvested energy increases, the price per unit of energy reduces (e.g. $c_1^*=0.00109$ for $N=5$ and $\frac{E}{QN_0}=30$ dB), since the BS has more energy to cover the demand. In contrast to $c_1^*$, increasing the harvested energy leads to an increase in the optimum price $c_2^*$. This is a notable observation, because it indicates that the more the transmitted power increases, the smaller the increase of the users' utilities is, and thus, instead of demanding energy, the demand for the other resources increases.

\begin{figure}
\centering
\includegraphics[width=0.8\columnwidth]{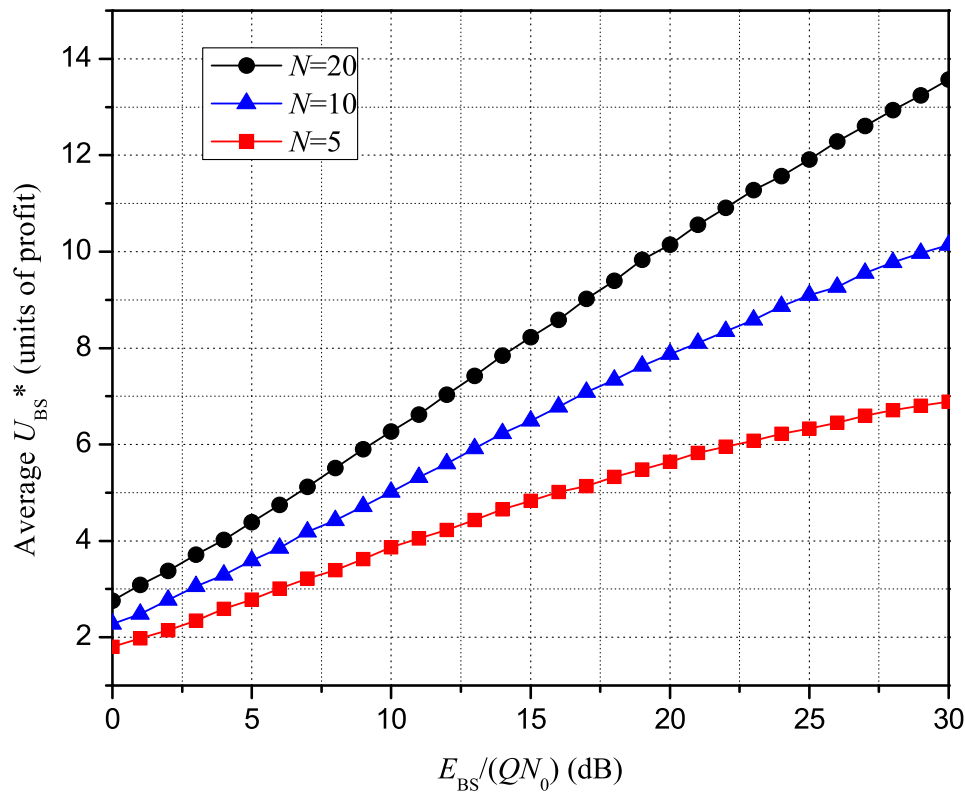}
\caption{Average revenue $U_\mathrm{BS}^*$ of the BS.}
\label{Fig2chapter2}
\end{figure}

\begin{figure}
\centering
\includegraphics[width=0.8\columnwidth]{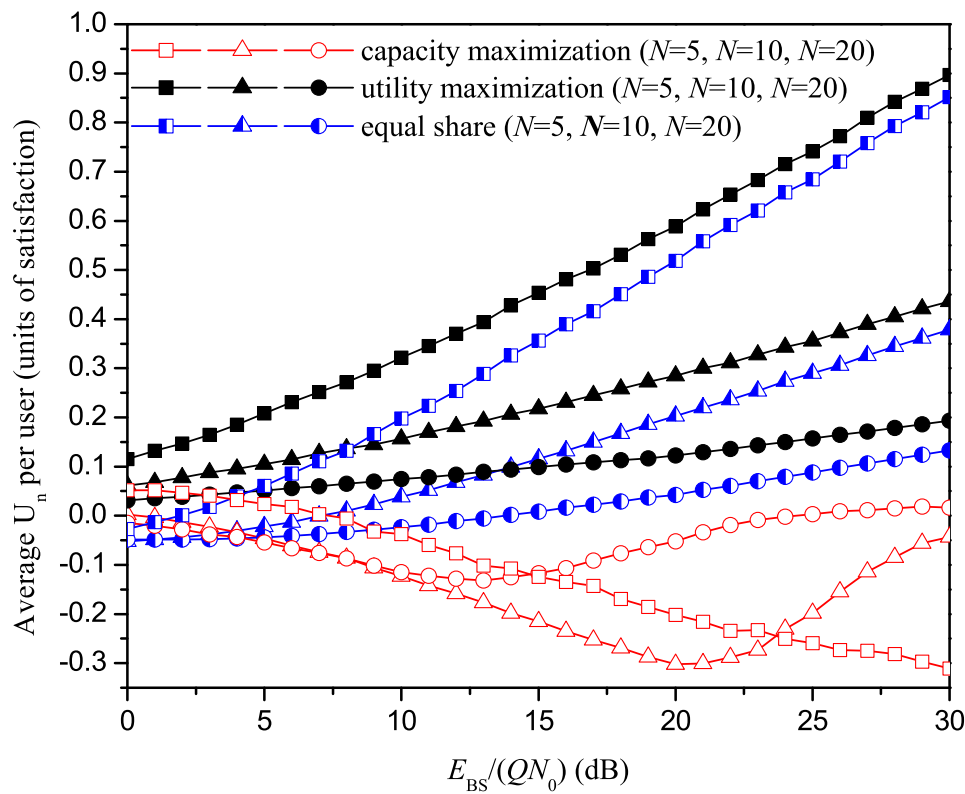}
\caption{Comparison of the average utility per user.}
\label{Fig3chapter2}
\end{figure}

Figs. \ref{Fig2chapter2} and \ref{Fig3chapter2} show the average BS revenue and the average utility per user, respectively. It can be observed that when the proposed method (utility maximization) is applied, by increasing the harvested energy, both $U_\mathrm{BS}^*$ and $U_n^*$ also increase, i.e. when the BS harvests more energy it increases its revenue, because of the increase of prices in this case, while the users are more satisfied, since their spectral efficiency increases. In Fig. \ref{Fig3chapter2}, in conjunction with the utility maximization, two other energy and resource allocation strategies are illustrated: maximization of the total capacity of the network (capacity maximization), and equal power and resource sharing among the users. One can observe that the utility maximization method outperforms both schemes, reflecting the fairness that is achieved among users and underscoring the value of the game-theoretic approach followed in this work.

%\section{Conclusions}
%In this paper, we have used a \textit{generalized Stackelberg game} to investigate the optimal energy and resource allocation problem of an autonomous energy harvesting base station. We have proved existence and uniqueness of the variational equilibrium. We have also proposed an efficient distributed algorithm, which can be adopted by the players in order to maximize their utility functions. The simulation results have revealed that by applying the proposed method, the users maximize their level of satisfaction while the BS maximizes its revenue for all values of harvested energy and for an arbitrary number of users.
%CHAPTER 5 (CONCLUSIONS) %%%%%%%%%%%%%%%%%%%%%%%%%%%%%%%%%%%%%%%%%%%%%%%%%%%%%%%
%\else\fi

%%\ifchaptersixflag
%%CHAPTER 5 (CONCLUSIONS) %%%%%%%%%%%%%%%%%%%%%%%%%%%%%%%%%%%%%%%%%%%%%%%%%%%%%%%
\chapter[Wireless Powered Networks][]{Wireless Powered Networks}\label{ch:chapter3}
This chapter focuses on wireless powered communications, where energy harvested from RF signals is used to power the wireless terminals and enable the information transmission during the uplink. More specifically, it investigates the joint design of downlink energy transfer and uplink information transmission in multiuser communications systems, with the aim to maximize specific metrics, such as throughput and fairness. First, the main trade-offs in the existing literature are presented. Next, uplink NOMA is proposed in order to improve performance and utilize more efficiently the harvested energy. To this end, either the achievable system throughput or equal individual data rate is maximized. Also, as a means to balance these two metrics, the proportional fairness is considered and maximized for both TDMA and NOMA. Simulation results illustrate the effectiveness of the proposed methods, which outperform the existing baseline ones.

\section{Introduction to Trade-offs in Wireless Powered Networks}\label{Introduction to Trade-offs in Wireless Powered Networks}
This section provides useful insights into the main trade-offs that have been explored in the literature in the context of wireless powered communications, that employ the \textit{harvest-then-transmit} protocol \cite{Ju}, where the users first harvest energy, and then they transmit their independent messages to the BS by using the harvested energy. It is noted that this type of networks has initially investigated by \cite{harsource1} and \cite{Ju}, assuming a single user and multiple users that employ TDMA, respectively. The presented analysis provides insights on the dependence among throughput, fairness, and energy efficiency. Besides, extended simulations illustrate that
\begin{itemize}
\item The increase of the energy arrival rate reduces the portion of time that is allocated to EH.
\item Stochastic knowledge of the energy rate arrival only slightly reduces the throughput compared to the deterministic case.
\item Sophisticated optimization methods can be used to improve fairness, at the expense of sum-throughput.
\item The maximization of the sum-throughput reduces considerably the achieved energy efficiency.
\end{itemize}

%

%Let's assume that the received RF band signal is $\bar{y}(t)=\sqrt{2}\Re{\tilde{y}(t)e^{\hat{j}2\pi f_ct}}$, where  $f_c$
%\begin{equation}
%\tilde{y}(t)=x(t)+\tilde{n}_A(t)
%\end{equation}, 
%in which
%$x(t)=\sqrt{P_r}A(t)e^{j\varphi(t)}$ is a complex baseband signal with amplitude $P_rA(t)$ and phase $\varphi(t)$, $P_r=\mathbb{E}[|x(t)|^2]=1$ with $\mathbb{E}[\cdot]$ and $|\cdot|$ denoting the statistical expectation and the absolute value, respectively, $\tilde{n}_A(t)=n_{I}(t)+n_Q(t)$ is the antenna noise with $n_{I}(t)$ and $n_{I}(t)$ denoting the in-phase, and quadrature noise components.
%
%A typical energy receiver that converts RF energy directly via a rectenna architecture is ilustrated in Fig. [XXX]. In
%the rectenna, the received RF band signal $\bar{y}(t)$ is converted to a direct current (DC) signal iDC(t) by a rectifier, which consists of a Schottky diode and a passive low-pass filter (LPF). The DC signal $i_{DC}(t)$ is then used to charge the battery to store the energy. With an input voltage proportional to $y(t)$, the output current $i(t)$ of a Schottky diode is given by [14]:
%\begin{equation}
%i(t)=I_s(e^{\zeta\bar{y}(t)}-1)=a_1
%\end{equation}

\subsection{System model}
\label{System_model_WPN}
The uplink in a wireless network is considered, consisting of $N$ EH users, one BS, and one PB. It is assumed that the PB supplies wirelessly energy to the users and does not participate in the information transmission. However, it is assumed that the PB is empowered with the functionality of a communication entity, and is capable of performing tasks such as channel estimation \cite{Caijun1, OWPR}.  It is also assumed that all nodes are equipped with a single antenna, while they share the same frequency band.

The communication is divided into timeslots of unitary duration. Hereinafter, the notation $(\cdot)_n$ will be used to denote the value of the variable $(\cdot)$ for the $n$-th user. Besides, $\mathcal{N}$ will denote the set of all users, while $(\cdot)^*$ will be utilized to denote optimality. Note that when the provided analysis focuses to a single user $n$, 
$(\cdot)_n$ will be omitted.

\subsubsection{Energy receiver}
\begin{figure}[t!]
\centering
\includegraphics[width=0.9\linewidth]{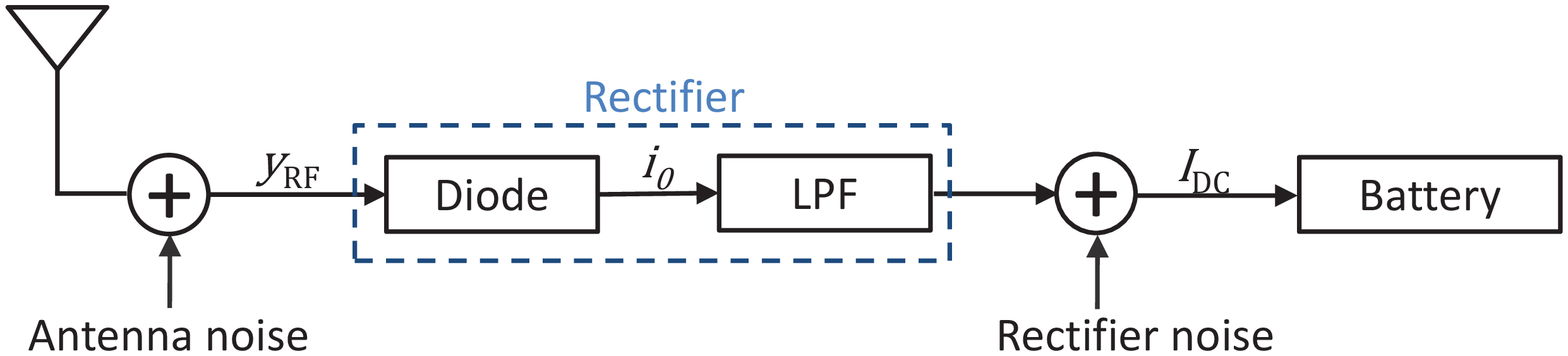}
\caption{Energy receiver.}
\label{energy receiver}
\end{figure}
A typical energy receiver that converts RF energy directly via a rectenna architecture is ilustrated in Fig. \ref{energy receiver}. In
the rectenna, the received RF band signal $y_\mathrm{RF}$ is converted to a DC signal $I_\mathrm{DC}(t)$ by a rectifier, which consists of a Schottky diode, the output current of which is $i_0$, and a passive low-pass filter (LPF). The DC signal $I_\mathrm{DC}$ is then used to charge the battery to store the energy. Also, it is assumed that the converted energy to be stored is linearly proportional to $I_\mathrm{DC}$. Thus, the EH rate, denoted by $\tilde{E}_n$ is given by
\begin{equation}
\tilde{E}_n=\eta_1\mathbb{E}[I_\mathrm{DC}],
\end{equation}
where  $\eta_1$ is the EH effciency and $\mathbb{E}[\cdot]$ denotes the statistical expectiation. Note that the
harvested energy due to the noise (including both the antenna
noise and the rectifier noise) is a small constant and thus
ignored. After this assumption, $\mathbb{E}[I_\mathrm{DC}]$ is  proportional to the average received power of the RF band signal \cite{architecture}.

\subsubsection{Harvest-then-transmit protocol}
It is assumed that users cannot receive and transmit simultaneously. For this purpose, the harvest-then-transmit protocol is employed and there are the following two distinct phases during a timeslot \cite{Ju}:
\begin{itemize}
\item \textit{Phase 1}: The users harvest energy in order to charge their batteries. The duration of this phase is denoted by $T$.
\item \textit{Phase 2}: The remaining amount of time, i.e. $1-T$ is assigned to the users, in order to transmit their messages.
\end{itemize}
%More information about the two phases are given in the following subsections.

\paragraph{Phase 1}

The harvested energy, denoted by $E_n$, is
\begin{equation}
E_n=\tilde{E}_nT.
\end{equation}
Assuming that users harvest energy from the signals transmitted by a dedicated PB, in order to properly express the EH rate, the signal received by each user is defined. To this end, it is assumed that the PB transmits a baseband signal with power $P_0$, that is, $s^\mathrm{d}_0P_0$, where $s^\mathrm{d}_0$ is an arbitrary complex signal and  $\mathbb{E}[|s^\mathrm{d}_0|^2]=1$, with $\mathbb{E}[\cdot]$ and $|\cdot|$ denoting the statistical expectation and the absolute value, respectively. Also, $(\cdot)^\mathrm{d}$ denotes a value for the downlink phase. Note that $s^\mathrm{d}_0$ can also be used to send downlink information at the same time, however, this usage is out of the scope of this chapter, and will be separately investigated in the next chapter.  The baseband signal received by the $n$-th user, $y_n$, can be expressed as
\begin{equation}
y^\mathrm{d}_n=\tilde{h}_n\sqrt{\tilde{\mathcal{G}}_{0}\mathcal{G}_{n}P_0}s^\mathrm{d}_0+\nu_n,
\end{equation}
where $\nu_n$ denotes the AWGN and $\tilde{h}_n=\tilde{H}_n \sqrt{\tilde{L}_n}$, $\tilde{H}_n$, and $\tilde{L}_n$  denote the channel coefficient, the small scale fading coefficient, and the path loss factor from the PB to the $n$-th user. Also, $\tilde{\mathcal{G}}_{0}$ and $\mathcal{G}_{n}$ are the directional antenna gains of the PB and the $n$-th user, respectively. Assuming that energy that can be harvested due to the receiver noise is negligible, the EH rate is given by 
\begin{equation}
\tilde{E}_n=\eta_1\tilde{\gamma}_nP_0,
\end{equation}
where $\tilde{\gamma}_n=|\tilde{h}_n|^2\tilde{\mathcal{G}}_{0}\mathcal{G}_{n}$.

At the special case that the users harvest energy solely from the signals transmitted by the BS, i.e. when the BS has the power beacon (PB) functionality during the $1$-st phase, $\tilde{\gamma}_n$, $\tilde{h}_n$, $\tilde{H}_n$, $\tilde{L}_n$, and $\tilde{\mathcal{G}}_{0}$ can be replaced by $\gamma_n$, ${h}_n$, $H_n$, ${L}_n$, and $\mathcal{G}_0$, where $h_n$, $H_n$ and $L_n$ denote the channel coefficient, the small scale fading, and the path loss factor from the BS to the $n$-th user, respectively, and $\mathcal{G}_{0}$ denotes the antenna gain of the BS. It is assumed that the channel conditions remain constant during a timeslot, and their exact values are known by the coordinator of the network, e.g., the BS.

\paragraph{Phase 2}
Assuming channel reciprocity, the channel coefficient and the path-loss factor from the $n$-th user to the BS are given by $\bar{h}_n$, and $L_n$ respectively, where $\bar{(\cdot)}$ denotes the conjugate of $(\cdot)$, while  $N_0$ denotes the white power spectral density of the AWGN and $W$ is the channel bandwidth. 
Each user transmits its message, $s^\mathrm{u}_n$, where $\mathbb{E}[|s^\mathrm{u}_n|^2]=1$, with transmit power $P^\mathrm{u}_n$, where the superscript $(\cdot)^\mathrm{u}$ denotes a value for the uplink phase. Thus, when there is a sole active transmitter, the observation at the BS is given by
\begin{equation}
y^\mathrm{u}_n=\bar{h}_n\sqrt{\mathcal{G}_{0}\mathcal{G}_{n}P^\mathrm{u}_n}s^\mathrm{u}_n+\nu_0,
\end{equation}
where  $\nu_0$ denotes the AWGN at the BS.

Moreover, let $R_n^\mathrm{u}$ denote the achievable throughput of the $n$-th user and $t_n$ the time that is allocated to the $n$-th user, in order to transmit its information. Since it has been assumed that users cannot receive and transmit simultaneously, it must hold that
\begin{equation}
t_n\leq 1-T.
\end{equation}
Finally, it is considered that WPT is the sole energy source and, unless otherwise stated, all nodes consume energy only for transmission.

\subsection{Main Trade-offs}
The harvest-then-transmit transmit protocol infers several interesting trade-offs that have considerably attracted the research interest \cite{architecture, Ju, Zoran, harsource1, HTC,atanasovski2015future}. First, there is a nontrivial trade-off between the time dedicated to EH and that for information transmission \cite{LLiu, Ju}. A solution to this trade-off depends on both the available channel state information and the knowledge of the energy arrival rate. To this end, two cases have been considered in the literature \cite{harsource1}:
\begin{itemize}
\item The deterministic case, where the EH rate is known in advance.
\item The stochastic case, where the EH rate is unknown and only its statistical properties are available.
\end{itemize}

Moreover, there is an interesting trade-off between performance and fairness. In general, when the sum-throughput is maximized, fairness is considerably reduced, due to the ``\textit{doubly near-far}'' problem. This phenomenon appears when a user far from the BS receives a smaller amount of wireless energy than a nearer user, while it needs to transmit with more power \cite{Ju}. In this case, in order to achieve fairness the following three schemes can be used:
\begin{itemize}
\item The \textit{weighted sum-throughput maximization}, which aims to maximize the scaled sum of the users throughputs.
\item The \textit{rate profile}, which aims to maximize the sum-throughput under the constraint that each user's throughput is proportional to the sum-throughput. In this method, a predetermined proportionality parameter is utilized, which ensures a minimum level of fairness.
\item The \textit{common throughput maximization}, which guarantees equal throughput allocations to all users and simultaneously maximizes their sum-throughput.
\end{itemize}
Note that the level of fairness of each scheme has a direct impact on the achieved sum-throughput.

Finally, although most of the research focuses on throughput maximization and fairness improvement, there is also an interesting trade-off between throughput and energy efficiency, which was studied in \cite{QWu, EE_uplink_downlink}. Because of the rapidly rising energy costs and the tremendous carbon footprints of existing systems, energy efficiency, is gradually accepted as an important design criterion for future communication systems. Moreover, in wireless powered communication networks, significant amount of energy may be consumed during EH, in order to combat the channel attenuation, which makes the consideration of energy efficiency even more interesting \cite{QWu}.

\subsubsection{Trade-off between Information Transmission and Energy Transfer}
This section focuses on the trade-off between the time that is allocated to energy transfer and information transmission. More specifically, it is shown that the achieved throughput is strongly affected by the time that is allocated to each phase, while for simplicity, a single user is considered.
The achievable throughput in bits/second/Hz in each timeslot is given by \cite{harsource1}
\begin{equation}
R^\mathrm{u}=t\log_2\left(1+\frac{P^\mathrm{u}\gamma}{N_0W}\right).
\label{throughput1}
\end{equation}
Since given $T$, $R^\mathrm{u}$ is an increasing function of $t$, it can be written as
\begin{equation}
R^\mathrm{u}=(1-T)\log_2\left(1+\frac{P^\mathrm{u}\gamma}{N_0W}\right).
\label{throughput2}
\end{equation}

\begin{figure}[t!]
\centering
\includegraphics[width=0.8\linewidth]{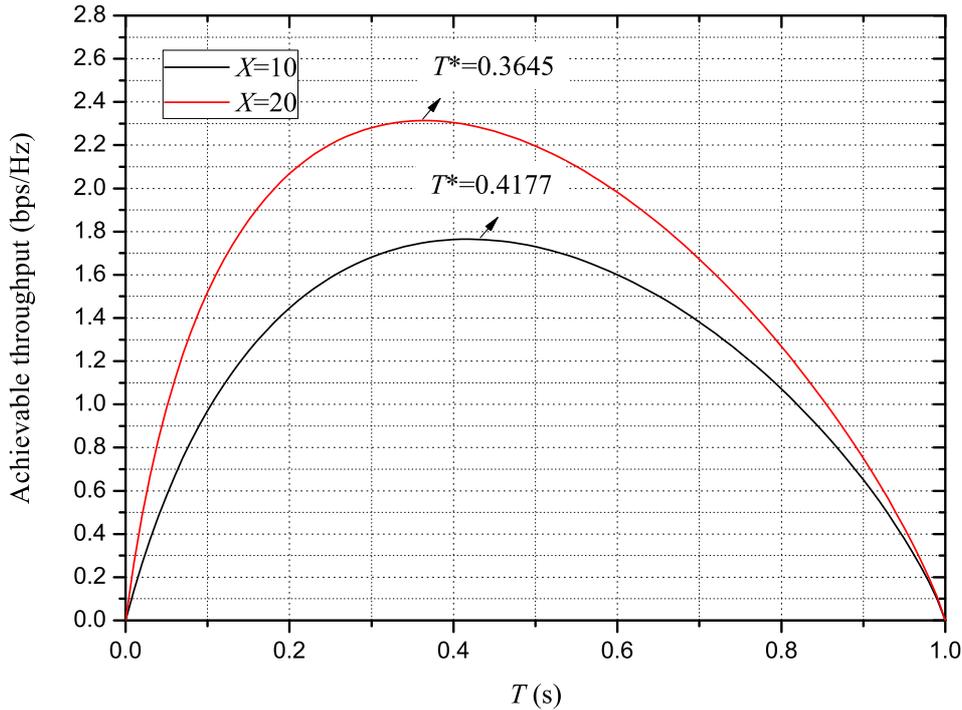}
\caption{Throughput versus the time allocated to EH.}
\label{stochastic1}
\end{figure}

Taking into account that $R^\mathrm{u}$ is an increasing function with respect to $P^\mathrm{u}$, it can be replaced by 
\begin{equation} 
P^\mathrm{u}=\frac{\eta_2E}{1-T},
\end{equation}
with $\eta_2$ being the efficiency of the user's amplifier.
Thus, the achievable throughput in each timeslot is given by \cite{harsource1}
\begin{equation}
R^\mathrm{u}=(1-T)\log_2\left(1+\frac{XT}{(1-T)}\right),
\label{throughput}
\end{equation}
where $X=\frac{\eta_2 \tilde{E} \gamma}{N_0W}$, with $\eta_2$ being the efficiency of the user's amplifier,

Fig. \ref{stochastic1} shows the throughput given in (\ref{throughput}) for $X=10$ and $X=20$ versus the time allocated to EH. It is observed that the throughput is zero when $T=0$, as well as when $T=1$. Also, it is illustrated that the achievable throughput has one global maximum, i.e., it first increases when $T<T^*$, while it decreases when $T>T^*$.  This can
be explained as follows. With small $T$, the amount of energy
harvested by the user is small. As the user harvests more energy with increasing $T$, i.e., more energy
is available for the information transmission, the
throughput increases with $T$ \cite{Ju}. However, as $T$ becomes larger
than $T^*$ the throughput is decreased due
to the reduction in the time allocated to information transmission. Also, it is observed that when $X=10$, $T^*=0.4177$, while when $X=20$, $T^*=0.3645$. Thus, it can be concluded that when $X$ increases the optimal time allocated to EH decreases. This is because when the user has enough energy to transmit, its sensitivity to the resource of time dedicated to information transmission increases.

\paragraph{Deterministic Energy Arrival} Assuming that $X$ can be perfectly estimated, the achievable throughput maximization problem can be written as \cite{harsource1}
\begin{equation}
\begin{array}{ll}
\underset{T}{\text{\textbf{max}}}& R^\mathrm{u}\\
\,\,\,\text{\textbf{s.t.}}&\mathrm{C}:0<T<1.
\end{array}
\label{maxthroughput}
\end{equation}

It can easily be proved that $R^\mathrm{u}$ is strictly concave with respect to $T$ in $(0,1)$, $\forall X$, since it holds that \cite{harsource1}
\begin{equation}
\frac{d^2R^\mathrm{u}}{dT^2}=-\frac{X^2}{\ln(2)(1-T)((X-1)T+1)^2}<0.
\label{concavity1}
\end{equation}

Thus, the optimal value for $T$ in $(0,1)$ that maximizes $R^\mathrm{u}$ is unique and can be obtained through
\begin{equation}
\frac{dR_{sum}}{dT}=0.
\end{equation}
After some mathetmatical manipulations, the optimal value can be expressed as \cite{harsource1}
\begin{equation}
T^*=\frac{X-1-\mathcal{W}_0\left(\frac{X-1}{e}\right)}{(X-1)(\mathcal{W}_0\left(\frac{X-1}{e}+1\right))},
\label{optimal T}
\end{equation}
where $(\cdot)^*$ denotes a solution value and $\mathcal{W}_0(x)$ returns the principal branch of the Lambert W function.

%, also called omega function or product logarithm. This function is defined as the set of solutions of the equation $x=\mathcal{W}_0(x)\exp({\mathcal{W}_0(x)})$ \cite{lambert}. Note that $\mathcal{W}_0(x)$ can be easily evaluated since it is a built-in function in most of the well-known mathematical software packages as Matlab, Mathematica, etc. \cite{harvesting}.

\paragraph{Stochastic Energy Arrival} When the EH rate is non-deterministic, the appropriate metric to optimize is the long term expectation of the achievable throughput, which is denoted by $\mathbb{E}[R^\mathrm{u}]$. Next, it is assumed that $X$ follows a Gamma distribution, i.e. $X\sim\Gamma(\kappa,\zeta)$, where $\kappa,\zeta>0$ refer to the shape parameter, because Gamma distribution can model accurately the EH rate \cite{savethentransmit}. Note that the channel condition between the user and the BS, described by $\gamma$, is assumed to be invariable within each timeslot, thus $\gamma$ is considered as constant coefficient of EH rate. The corresponding maximization problem of the expected achievable throughput is given by
\begin{equation}
\begin{array}{ll}
\underset{T}{\text{\textbf{max}}}& \mathbb{E}[R^\mathrm{u}]\\
\,\,\,\text{\textbf{s.t.}}&\mathrm{C}:0<T<1.
\end{array}
\label{maxthroughputstochastic}
\end{equation}

Considering the signal-to-noise ratio (SNR), there are two asymptotic cases:

\begin{itemize}
\item \textit{Low SNR Approximation}: Under a low SNR condition, i.e. when $\frac{XT}{1-T}\ll 1$, $\mathbb{E}[R^\mathrm{u}]$ is given by \cite{harsource1}
\begin{equation}
\mathbb{E}[R^\mathrm{u}]=\frac{T}{\ln(2)}\mathbb{E}[X].
\end{equation}
Apparently, in this case, $\mathbb{E}[R^\mathrm{u}]$ is maximized when $T\rightarrow 1$.
\item \textit{High SNR Approximation}: Assuming a high SNR condition, $\mathbb{E}[R^\mathrm{u}]$ can be expressed as \cite{harsource1}
\begin{equation}
\mathbb{E}[R^\mathrm{u}]=\frac{1-T}{\ln(2)}\left(\Psi(\kappa)+\ln\left(\frac{T}{1-T}\zeta\right)\right),
\end{equation}
where $\Psi(\cdot)$ denotes the digamma function, defined by
\begin{equation}
\Psi(x)=\frac{d}{dx}\ln\left(\Gamma(x)\right),
\end{equation}
with $\Gamma(\cdot)$ being the Gamma function.
The second derivative of $\mathbb{E}[R^\mathrm{u}]$ can be expressed as
\begin{equation}
\frac{d^2\mathbb{E}[R^\mathrm{u}]}{dT^2}=-\frac{1}{\ln(2)}\left(\frac{1}{T(1-T)+\frac{1}{T^2}}\right).
\end{equation}
Apparently, it holds that $\frac{d^2\mathbb{E}[R^\mathrm{u}]}{dT^2}<0$, for $0<T<1$ and, thus, the optimization problem in (\ref{maxthroughputstochastic}) is concave and has a unique solution.
Finally, by setting $\frac{d\mathbb{E}[R^\mathrm{u}]}{dT}=0$ and after some mathematic manipulation, the value of $T$ which corresponds to the solution of (\ref{maxthroughputstochastic}), for the high SNR approximation,  is given by \cite{harsource1}
\begin{equation}
T^*=\frac{1}{\mathcal{W}_0\left(\zeta \exp({\Psi(k)-1})\right)+1}.
\end{equation}
\end{itemize}
Note that the derived solution is suboptimal, i.e. it does not necessarily maximize the throughput in (\ref{throughput}).

\paragraph{Comparison between the Deterministic and the Stochastic Case} 

Next, the stochastic and the deterministic case are compared in terms of time dedicated to EH and average throughout versus the parameter $\kappa$, assuming that $\kappa=\zeta$. Note that lower value of $\kappa$ implies lower average $X$. For reference, the throughput when $T=0.5$, i.e. the \textit{half-half} case, is also depicted.
\begin{figure}[h!]
\centering
\includegraphics[width=0.8\columnwidth]{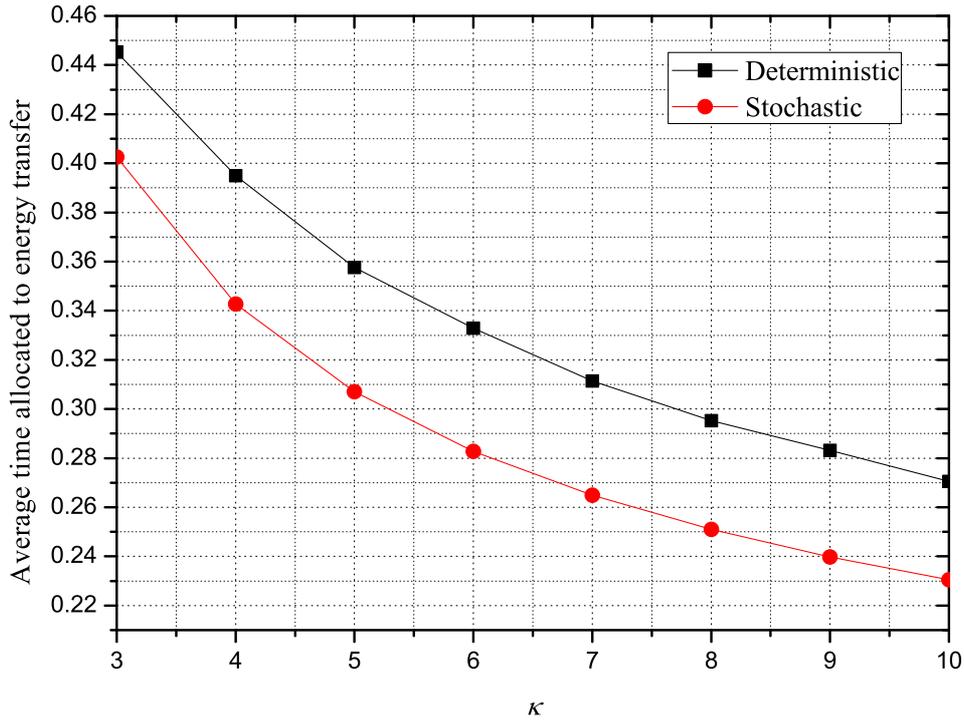}
\caption{Comparison of the deterministic and stochastic case in terms of time allocated to EH.}
\label{time}
\end{figure}

\begin{figure}[h!]
\centering
\includegraphics[width=0.8\linewidth]{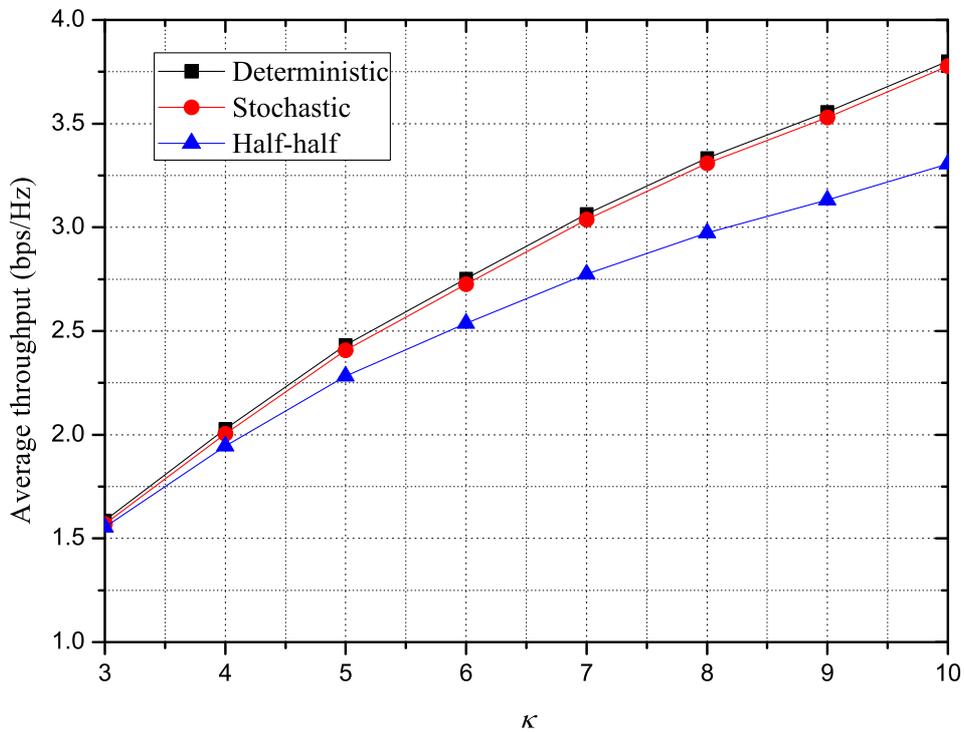}
\caption{Comparison of the deterministic and stochastic case in terms of throughput.}
\label{comparison}
\end{figure}

As it is shown in Fig. \ref{time} the solution of the optimization problem in (\ref{maxthroughputstochastic}), which corresponds to the stochastic case, usually leads to the selection of lower value of $T$ than the optimal one that is selected when the optimization problem in (\ref{maxthroughput}) is solved. Therefore, as it can be observed in Fig. \ref{comparison}, the deterministic case slightly outperforms the stochastic case for all values of $\kappa$. This is reasonable, since in the former one the optimal save-ratio can be exactly derived, in contrast to the second one, where only the expected optimal choice can be made.

On the other hand, both the deterministic and stochastic cases outperform the half-half, which validates that both schemes can efficiently use the level of knowledge of the EH rate in order to maximize the throughput \cite{harsource1}.

\subsubsection{Trade-off between Fairness and Performance}

This subsection focused on methods available in the existing literature, that can be used to increase the sum-throughput and/or fairness among users when TDMA is used. Also, it is assumed that the users harvest energy solely from the signals transmitted by the BS, i.e., during the first phase the BS has the PB funcionality. According to TDMA, considering the amount of time that is allocated to each user, it is limited by \cite{Ju}
\begin{equation}
\sum_{n=1}^Nt_n\leq 1-T.
\label{time constraint}
\end{equation}
Taking into account (\ref{throughput}), the achievable throughput in bits/second/Hz  of the $n$-th user can be written as \cite{Ju}
\begin{equation}
R^\mathrm{u}_n=t_n\log_2(1+\frac{\gamma_nP^\mathrm{u}_n}{N_0W})=t_n\log_2\left(1+\frac{\eta\rho_0 Tg_n}{t_n}\right),
\label{userthroughputtdma}
\end{equation}
where $\eta=\eta_1\eta_2$, $\rho_0=\frac{P_0}{N_0W}$ and $g_n=\gamma_n^2$.

\paragraph{Sum-Throughput Maximization}
The sum-throughput of all users is given by \cite{Ju}
\begin{equation}
\mathcal{R}_\mathrm{sum}=\sum_{n=1}^NR^\mathrm{u}_n,
\end{equation}
and the corresponding sum-throughput maximization problem can be formulated as \cite{Ju}
\begin{equation}
\begin{array}{ll}
\underset{T,\boldsymbol{\tau}}{\text{\textbf{max}}}& \mathcal{R}_\mathrm{sum},\\
\,\,\,\text{\textbf{s.t.}}&\mathrm{C}_1: T+\sum_{n=1}^Nt_n\leq 1,\\
&\mathrm{C}_2: T>0,\\
&\mathrm{C}_n: t_n\geq 0, \forall n\in\mathcal{N}.
\end{array}
\label{system throughput tdma}
\end{equation}
The eigenvalues of the Hessian matrix of $R_n$ denoted by $\phi_i,i\in\{1,2\}$ are given by $\phi_1=0$ and
\begin{equation}
\phi_2=-\frac{g_n^2\eta^2\rho_0^2(t_n^2+T^2)}{t_n(t_n+g_n\eta\rho_0 T)^2\ln(2)}<0,
\end{equation}
i.e. they are both non-positive. Thus, $R_n$ is jointly concave with respect to $T$ and $t_n$, since its Hessian matrix has non-positive eigenvalues and, thus, it is negative semi-definite. Therefore, since $R_\mathrm{sum}$ is the summation of concave functions, i.e. $R_n$, it is also concave \cite{Boyd1}. Besides, all the constraints are linear. Consequently, the optimization problem in (\ref{system throughput tdma}) can be also solved by using convex optimization techniques. Next, this problem is solved using dual decomposition. For this reason, the Lagrangian of (\ref{system throughput tdma}) is needed which is given by
\begin{equation}
\mathcal{L}(\lambda,{\boldsymbol{t}},T)=\mathcal{R}_\mathrm{sum}-\lambda(T+\sum_{i=1}^Nt_n-1),
\end{equation}
where $\lambda$ is the LM that corresponds to the constraint $\mathrm{C}_1$ and ${\boldsymbol{t}}$ the set with elements $t_n$.

Using the KKT conditions, the optimal time allocated to EH and information transmission by the $n$-th is given by \cite{Ju}
\begin{equation}
T^*=\frac{z^*-1}{\eta\rho_0\sum_{n=1}^Ng_n+z^*-1}
\label{optimalT}
\end{equation}
and
\begin{equation}
t_n^*=\frac{\eta \rho_n g_n}{\eta\rho_0\sum_{n=1}^Ng_n+z^*-1},
\end{equation}
where $z^*$ is the unique solution of the following equation:
\begin{equation}
z\ln(z)-z-\eta\rho_0\sum_{n=1}^Ng_n+1=0.
\end{equation}

\paragraph{Weighted Sum-Throughput}
In order to consider fairness, the weighted sum-throughput can be used used and maximized, which is given by
\begin{equation}
\mathcal{R}_\mathrm{wsum}=\sum_{n=1}^Na_n R^\mathrm{u}_n.
\end{equation}
In the above, higher $d_n$ implies higher $b_n$, however the exact values of $a_n$ depend by the specific application. The formulation and the solution of the corresponding optimization problem is similar to the one in (\ref{system throughput tdma}). An alternative to the maximization of the weighted sum-throughput is the rate-profile method, which will be further discussed in the following subsection.

\paragraph{Rate-Profile Method}
In order to realize the rate-profile method, the rate profile vector needs to be introduced, which is denoted by $\boldsymbol{b}=\{b_1,..,b_n,...,b_N\}$ and defined as
\begin{equation}
b_n=\frac{R^\mathrm{u}_n}{\mathcal{R}_\mathrm{sum}},
\end{equation}
where $R^\mathrm{u}_n$ denotes the required throughput of the $n$-th user and is given by (\ref{userthroughputtdma}).  Note that the vector $\boldsymbol{b}$ is specified by the specific application. For example, if $b_n>b_m$ it means that the $n$-th user requires higher rate than the $m$-th user.

The corresponding fairness aware optimization problem can be written as
\begin{equation}
\begin{array}{ll}
\underset{T}{\text{\textbf{max}}}& \text{{min}}(\frac{R^\mathrm{u}_1}{b_1},\frac{R^\mathrm{u}_2}{b_2},...,\frac{R^\mathrm{u}_N}{b_N}) \\
\,\,\,\text{\textbf{s.t.}}&\mathrm{C}_1:T+\sum_{n=1}^Nt_n\leq 1,\\
&\mathrm{C}_2:T>0,\\
&\mathrm{C}_3:t_n\geq 0, \forall n\in\mathcal{N}.
\end{array}
\label{rate profile problem 1}
\end{equation}
%=============================================

The objective function of the optimization problem in (\ref{rate profile problem 1}) is not a purely analytical expression. For this purpose, (\ref{rate profile problem 1}) is transformed into its epigraph form by utilizing the auxiliary variable $\mathcal{R}$. Therefore, it can now be written as \cite{Ju}

\begin{equation}
\begin{array}{ll}
\underset{T,\boldsymbol{\tau}}{\text{\textbf{max}}}& \mathcal{R} \\
\,\,\,\text{\textbf{s.t.}}&\mathrm{C}_1:T+\sum_{n=1}^Nt_n\leq 1,\\
&\mathrm{C}_2:T>0,\\
&\mathrm{C}_3:t_n\geq 0, \forall n\in\mathcal{N},\\
&\mathrm{C}_4: \frac{R^\mathrm{u}_n}{b_n}\geq\mathcal{R}.
\end{array}
\label{rate profile problem 2}
\end{equation}
%=============================================

Note that the optimization problem in (\ref{rate profile problem 2}) is concave and, thus, it can also be solved by dual decomposition. Its Lagrangian is given by
\begin{equation}
\mathcal{L}(\mu,{\boldsymbol\lambda},{\boldsymbol{t}},T,\mathcal{R})=\mathcal{R}-\mu\left(T+\sum_{n=1}^Nt_n-1\right)+\sum_{n=1}^N\lambda_n\left(\frac{R^\mathrm{u}_n}{b_n}-\mathcal{R}\right),
\label{lagrangiansection1}
\end{equation}
where $\lambda_n$ and $\mu$ are the LMs that correspond to the constraints $\mathrm{C}_4$ and $\mathrm{C}_1$, respectively, and ${\boldsymbol{\lambda}}$ is the LM set with elements $\lambda_n$.

The dual problem is given by
\begin{equation}
\underset{\mu,{\boldsymbol{\lambda}}}{\text{\text{min}}}\,\,\,\,\underset{T,R_\mathrm{eq}}{\text{\text{max}}}\,\,L(\mu,{\boldsymbol\lambda},{\boldsymbol{t}},T,\mathcal{R}).
\label{dual problemsection1}
\end{equation}
Considering the parts of the Lagrangian related to $\mathcal{R}$, it holds that
\begin{equation}
\underset{\mathcal{R}}{\text{\text{max}}}\left(1-\sum_{n=1}^N \lambda_n\right)\mathcal{R}=\begin{cases} 0 &\textit{if}\,\,\sum_{n=1}^N\lambda_n=1,\\ \infty  &\mathrm{otherwise}. \end{cases}\\
\end{equation}
Thus, the dual problem in (\ref{dual problemsection1}) is bounded if and only if $\sum_{n=1}^N\lambda_n=1$.
By setting
\begin{equation}
\lambda_N=1-\sum_{n=1}^{N-1}\lambda_n
\label{simplification}
\end{equation}
in (\ref{lagrangiansection1}), the variable $\mathcal{R}$ vanishes and the dual problem in (\ref{dual problemsection1}) is simplified to
\begin{equation}
\underset{\boldsymbol\lambda}{\text{\text{min}}}\,\,\,\,\underset{T,{\boldsymbol{t}}}{\text{\text{max}}}\,\,\tilde{\mathcal{L}}(\mu,{\boldsymbol{\lambda,t}},T).
\label{simplified dual problem}
\end{equation}
where $\tilde{\mathcal{L}}(\mu,{\boldsymbol{\lambda}},{\boldsymbol{t}},T)=\mathcal{L}(\mu,{\boldsymbol\lambda},{\boldsymbol{t}},T,\bar{R})|_{(\ref{simplification})}$.

According to the KKT conditions, given $\lambda^*$,  the optimal $\boldsymbol{t}$ and $T$ are given by
\begin{equation}
T^*=\frac{1}{1+\sum_{n=1}^N\frac{\eta\rho_0 g_n}{z_n^*}},
\label{optimalT2section1}
\end{equation}
\begin{equation}
t_n^*=\frac{\frac{\eta\rho_0 g_n}{z_n^*}}{1+\sum_{n=1}^N\frac{\eta\rho_0 g_n}{z_n^*}},
\label{optimalboldt}
\end{equation}
where $z_n^*,\forall n\in \mathcal{N}$ is the solution of the following set of equations:
\begin{equation}
\ln(1+z_n)-\frac{z_n}{1+z_n}=\frac{b_n}{\lambda_n}\sum_{n=1}^N\frac{\lambda_ng_n}{b_n(1+z_n)}
\end{equation}

Since the dual function in (\ref{simplified dual problem}) is differentiable, it can be solved iteratively. In each iteration, $T$ and ${\boldsymbol{t}}$ are calculated for a fixed LM vector, using (\ref{optimalT2section1}) and (\ref{optimalboldt}), while ${\boldsymbol{\lambda}}$ is then updated using the gradient method as follows \cite{Boyd1,Boyd2}
\begin{equation}
\begin{split}
\lambda_n[j+1]=&\left[\lambda_n[j]-\hat{\lambda}_n[j]\left(\frac{t_n}{b_n}\log_2(1+\frac{\eta\rho_0g_nT}{t_n})-\frac{t_N}{b_N}\log_2(1+\frac{\eta\rho_0g_NT}{t_N})\right)\right]^+_{\mathcal{U}_n},\\
&\forall n \in \{1,..,N-1\},
\end{split}
\end{equation}
where $j$ is the iteration index, $\hat{\lambda}_n,\,n\in\{1,...,N-1\}$ are positive step sizes, $[\cdot]^+=\max(\cdot,0)$, and $\mathcal{U}_n$ denotes the projection operator on the feasible set
\begin{equation}
\mathcal{U}_n=\{\lambda_n|\sum_{n=1}^N \lambda_n=1\}.
\end{equation}
The projection can be simply implemented by a clipping function $\big[\lambda_n[j+1]\big]_0^{1-\sum_{i=1}^{n-1}\lambda_i}$ and $\lambda_N$ can be obtained from (\ref{simplification}). Since Problem 1 is concave, it is guaranteed that the iterations between the two layers converge to the optimal solution if the size of the chosen step satisfies the infinite travel condition \cite{Boyd2}
\begin{equation}
\sum_{j=1}^{\infty}\hat\lambda_n[j]=\infty,\,n\in\{1,...,N-1\}.
\end{equation}

Finally, the optimal $\mathcal{R}$ can be evaluated by
\begin{equation}
\mathcal{R}^*=\underset{n\in\mathcal{N}}{\text{min}}\left(\frac{t_n^*}{b_n}\log_2\left(1+\frac{\eta\rho_0 g_n T^*}{t_n^*}\right)\right),
\end{equation}
where $T^*$ is given by (\ref{optimalT2section1}). This is because $\mathcal{R}^*$ is actually limited by the most stringent constraint.

\paragraph{Common Throughput Maximization}
The common throughput approach corresponds to the rate-profile method with parameters $b_n=\frac{1}{N},\forall{n\in\mathcal{N}}$ \cite{Ju}. This approach guarantees equal throughput allocations to all users, while it maximizes the sum-throughput, which now is defined as $\mathcal{R}_\mathrm{sum}=N\mathcal{R}$. Notice that the sum-throughput maximization in (\ref{system throughput tdma}) and the common-throughput maximization deal with two extreme cases of throughput allocation to the users in a wireless-powered communication network where the fairness is completely ignored and a strict equal fairness
is imposed, respectively.
\begin{figure}
\centering
\includegraphics[width=0.8\linewidth]{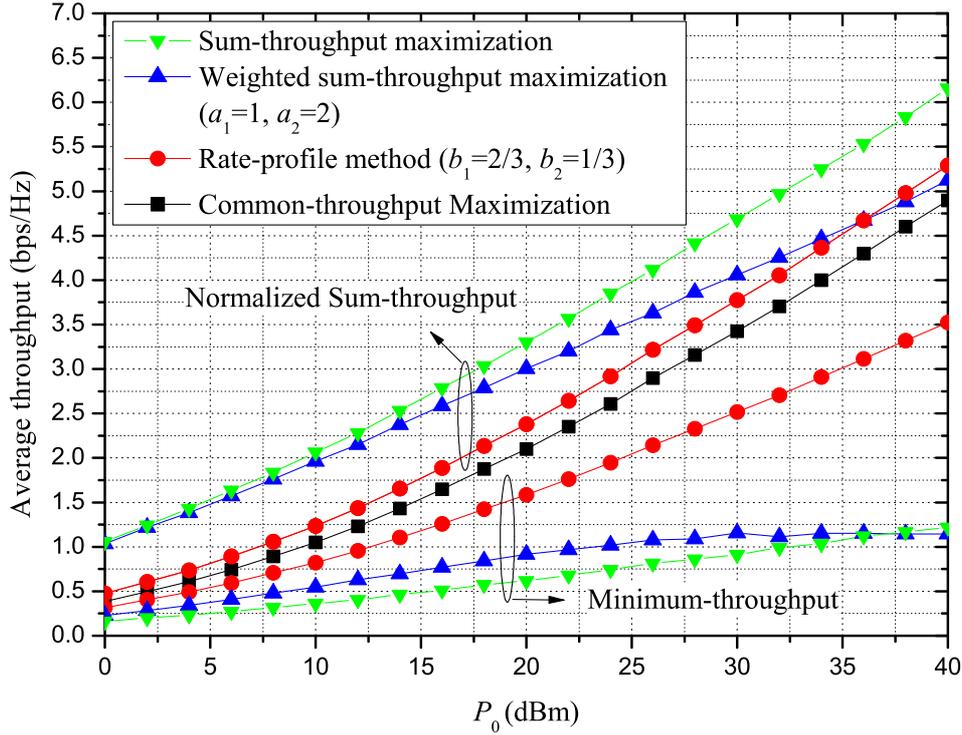}
\caption{Comparison among fairness aware schemes.}
\label{rates}
\end{figure}

\paragraph{Comparison}
\label{Comparison of Fairness Aware Schemes}
Next, the weighted sum-throughput maximization, the weighted sum-throughput maximization, the rate-profile method, and the common-throughput maximization are compared, in terms of performance and fairness. To this end, in Fig. \ref{rates} the sum and the minimum-throughput are illustrated when $T$ and $\boldsymbol{t}$ are set according to each one of the aforementioned methods. It is assumed that the path loss is given by $L_n=10^{-3}d_n^{-2}$ \cite{Ju}, where $d_n$ is the distance between the $n$-th user and the BS. It is further assumed that $d_1=5$ m, $d_2=10$ m, $N_0W=-114$ dBm, and $\mathcal{G}_0=\mathcal{G}_n=0$ dB. Besides, it is assumed that the small scale fading coefficient is given by the complex random variable $H_{n}\sim\mathcal{CN}(0,1)$. All statistical results are averaged over $10^5$ random realizations. Besides, it is assumed that $a_1=1,a_2=2$ and $b_1=2/3, b_2=1/3$ for the rate-profile method and the weighted sum-throughput maximization, respectively. This means that for the weighted sum-throughput maximization the user with $d_n=10$ m has double weight than the user with $d_n=5$ m, while for the rate-profile method the user with the longer distance from the BS must achieve at least half the throughput that the other user achieves. As one can observe, there is a trade-off between the sum-throughput and the minimum-throughput. The rate profile method achieves a good balance between the sum and the minimum-throughput, while for the high region of $P_0$, it outperforms the weighted sum-throughput maximization both in terms of throughput and fairness. On the other hand, the common-throughput maximization achieves the highest minimum-throughput and the lowest sum-throughput. Consequently, this method should be selected when it is important for the users to transmit with equal rate, such as when only symmetrical rates are permitted.

\begin{figure}
\centering
\includegraphics[width=0.8\linewidth]{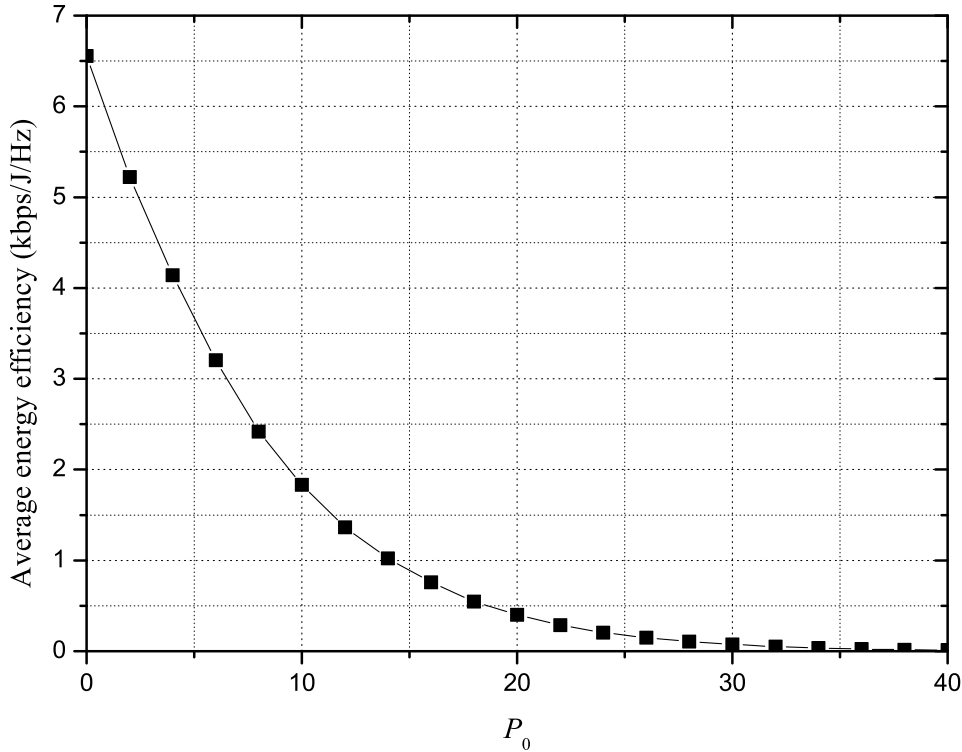}
\caption{Average energy efficiency when the sum-throughput is maximized.}
\label{efficiency}
\end{figure}
\subsubsection{Trade-off between Energy-Efficiency and Throughput}
Next, the energy efficiency optimization is investigated, while it is considered that the processing cost is negligible. Thus, assuming that energy is consumed only for transmission, the efficiency of the energy transmitted by the BS, denoted by $\mathcal{E}$, and defined as the ratio of the sum-throughput and the consumed energy
\begin{equation}
\mathcal{E}=\frac{\mathcal{R}_{sum}}{P_0T}.
\end{equation}
It can easily be shown that $\mathcal{E}$ is maximized when $T\rightarrow 0$.

In Fig. \ref{efficiency} the energy efficiency versus $P_0$ is depicted when $T$ and ${\boldsymbol{t}}$ are chosen in order to maximize the sum-throughput, i.e. by solving (\ref{system throughput tdma}). The simulation parameters are set according to \ref{Comparison of Fairness Aware Schemes}. As it can be observed, the energy efficiency is decreased considerably when the value of $P_0$ is increased. On the other hand, as it has already been illustrated in Fig. \ref{rates}, the sum-throughput increases as $P_0$ increases. Consequently, there is a clear trade-off between achievable throughput and energy efficiency. More details about this trade-off have been presented in \cite{QWu}, where a detailed power consumption model has been considered, taking into account the circuit power consumed for hardware processing.

\section{Wireless Powered Communications with Non-Orthogonal Multiple Access (NOMA)}\label{Wireless Powered Communications with Non-Orthogonal Multiple Access} \label{chapter1section2}

In this section, the utilization of NOMA for the uplink of multiuser wireless powered communications is proposed. The presented analysis focuses on the joint design of downlink energy transfer and uplink information transmission. More specifically, the related variables are optimized, taking into account two different criteria: the sum-throughput and the equal rate maximization. The formulated optimization problems are optimally and efficiently solved by either linear programming methods
or convex optimization. Simulation results illustrate that the proposed scheme outperforms the baseline orthogonal multiple access scheme, while they reveal the dependence among sum-throughput, minimum data rate, and harvested energy. The results of the research presented in this section are included in \cite{NOMA_WPT_joutnal, conference_NOMA, chapterinbook}.

\subsection{Related Work and Motivation}
As it has already been shown in the previous section, the dependence of the wireless powered nodes on EH can have a negative impact on the individual data rates that can be achieved. Consequently, existing methods, which increase power-bandwidth efficiency, should be carefully explored \cite{Matthaiou, Xiaoming, QWu, OFDMA}. Toward this direction, the utilization of orthogonal multiple access schemes, such as TDMA \cite{Ju}, might not be the most appropriate choice.

On the other hand, NOMA was proved to increase spectral efficiency \cite{Saito, Tafazoli, SurveyNOMA}.  For this reason, the two-user downlink special case of NOMA, termed as multi-user superposition transmission (MUST), has been included in the 3rd Generation Partnership Project (3GPP) Long Term Evolution Advanced (LTE-A) \cite{3gpp}. Also, it has been recognized as a promising multiple access technique for fifth generation (5G) networks \cite{DOCOMO, Ding3, VLC, Tafazoli, Linglong}. In addition to its applications in cellular networks, NOMA has also been applied to other types of wireless networks. For example, a variation of NOMA, termed Layer Division Multiplexing (LDM), has been proposed to the next general digital TV standard Advanced Television Systems Commitee (ATSC) 3.0 \cite{LDMTV}.

NOMA is substantially different from orthogonal multiple access schemes, since its basic principle is that the users can achieve multiple access by exploiting the power domain \cite{SurveyNOMA}. In non-orthogonal techniques, all the users can utilize resources simultaneously, which lead to inter-user interference \cite{PWang, Xu, verdu, Zhao}. Consequently, multi-user detection techniques are required to retrieve the users' signals at the receiver, such as joint decoding or successive interference cancellation (SIC), with the later being the most famous one. The implementation of downlink NOMA is based on superposition coding at the BS \cite{Ding3}, in contrast to uplink NOMA, where electromagnetic waves are naturally superimposed with received power \cite{Tafazoli}.  The performance of a downlink NOMA scheme with randomly deployed users has been investigated in \cite{Ding3, 7361990}, while the application of NOMA for the downlink of cooperative communication networks was proposed in \cite{Ding2}, among others. The application of multiple-input multiple-output (MIMO) and cloud radio access networks (CRANs) techniques to NOMA has been considered in \cite{DingNOMA,7460209,7277111, 7555306} and \cite{7962295}, respectively. 

In \cite{Tafazoli}, the authors investigate NOMA for the uplink of a communication network, consisting of traditional nodes with fixed energy supplies. Uplink NOMA is able to achieve the system upper bound, while it can be used as a means to improve fairness among the users \cite{DNCTse}. Although the decoding order does not affect the system throughput, it does affect the individual rates the users can achieve, since previously decoded messages are subtracted from the observation when decoding subsequent messages. Therefore, the decoding order optimization is a crucial issue in fairness aware uplink NOMA systems \cite{Tafazoli}. The decoding order can be either fixed during a communication frame (fixed-decoding order) \cite{Tafazoli}, or a set of different decoding orders can be used for corresponding fractions of time, by using \textit{time-sharing (TS)} \cite{BRimoldi, Zhao, Jaramillo, honig}. However, when uplink NOMA is combined with wireless powered communications, the capacity region is significantly affected by the amount of the harvested energy, a fact that has not been fully investigated yet. More specifically, the main difference between NOMA with SWIPT and conventional NOMA is the fact that the users are unequal in terms of transmission power capabilities, since the transmission power of self-sustained nodes which perform EH is upper-limited by the harvested energy. Consequently, in this case, rate maximization and user fairness are still open problems.

%\vspace{-0.15in}
\subsection{Contribution}

Unlike recent literature, this section investigates the application of NOMA for a wireless-powered uplink communication system, which consists of one BS and multiple EH users, in order to increase the individual data rates and the user fairness. As a joint processing technique at the BS, SIC is implemented, which affects user fairness through the choice of decoding order of the users' messages. In order to explore the possibilities to increase user fairness, two different decoding order strategies are proposed, namely fixed decoding order and TS. These decoding strategies are combined with two optimization objectives; i) maximizing the system throughput while improving the minimum individual rate (asymmetric rates case), ii) maximizing the minimum individual data rate (symmetric rates case).

Based on the above, four different decoding schemes with corresponding optimization problems are formulated, solved, and evaluated. More specifically, the analysis offers the following:
\begin{itemize}
\item For each of the cases above, the time used for EH is optimized.
\item It is shown that all formulated problems can be optimally solved by either linear programming  or convex optimization tools, which is important for the practical implementation of the proposed schemes. 
\item The use of TS is proposed, in order to increase fairness, so that a user whose message suffers from strong interference for one decoding order can experience a better reception reliability for another decoding order during the implementation of SIC. In this case, a tractable reformulation of the the initial problems is provided, while a greedy, fast-converging algorithm is also proposed for the calculation of the TS variables. This algorithm aims at choosing only a subset of possible decoding orders of the users' messages, so that i) the optimization problem dimensions are reduced and ii) the decoding order changes only a few times, avoiding synchronization issues.
\end{itemize}

The evaluation of the proposed strategies through extensive simulations reveals that NOMA can offer substantial improvement in user fairness compared to TDMA, by improving the individual rates of the weak users. The results are further improved when TS is also applied, and the proposed algorithm can offer a simple calculation of the TS parameters. It must be noted that the implementation of NOMA in the uplink is not a burden for the users, i.e., the encoding complexity at the users' side is not affected, since joint processing through SIC is only applied at the BS. Finally, the user synchronization is usually simpler than the case of TDMA. 
\subsection{System Model}
\begin{figure}
\centering
\includegraphics[width=\linewidth]{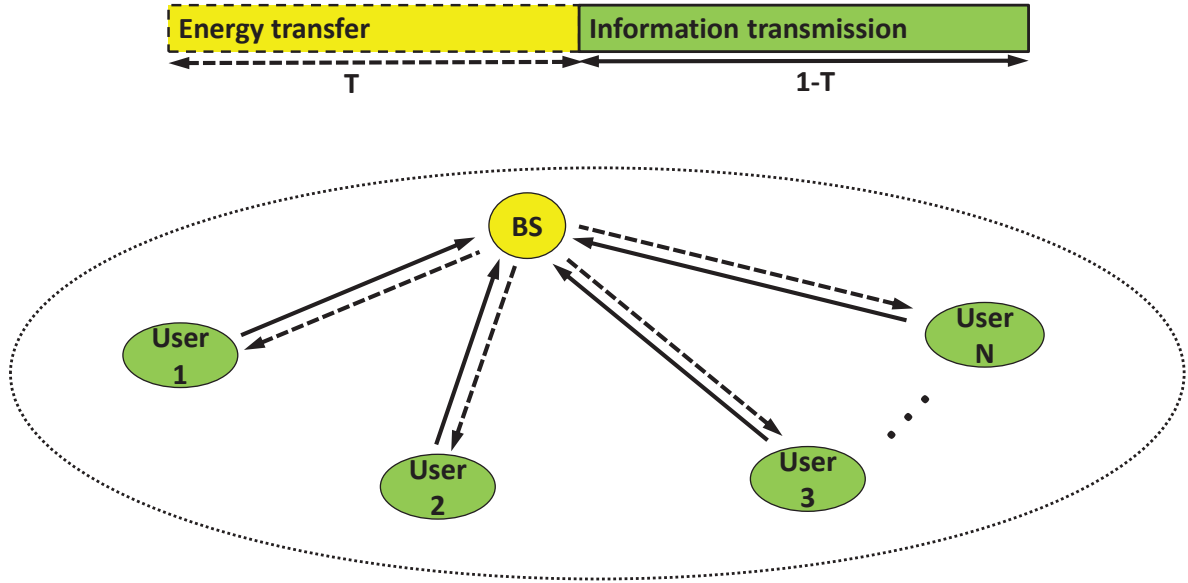}
%\vspace{-0.2in}
\caption{Sequential energy transfer and information transmission in NOMA communication networks.}
\label{fig1}
\end{figure}

%\vspace{-0.05 in}
A wireless network consisting of multiple users and one BS is considered, where all nodes are equipped with a single antenna.  It is assumed that the channel state remains constant during a time frame, and can be perfectly estimated by the BS. The considered system model is presented in Fig.\ref{fig1}.
Similarly to \ref{System_model_WPN}, it is considered that  the network adopts a harvest-then-transmit protocol, i.e. at first, the amount of time $T$ is assigned to the BS to broadcast wireless power, $P_0$, to all users \cite{Ju}. In other words, during the first phase the BS has the PB funcionality. The remaining time, $1-T$, is assigned to users, which, in contrast to the analysis presented in the previous section, simultaneously transmit their independent information to the BS by using the energy harvested from the first phase. In order to detect the users' signals, the BS implements a joint processing technique \cite{Ding3, honig}, and for this purpose, it employs NOMA \cite{Tafazoli}, where the BS decodes the users' messages utilizing SIC. It is assumed that the energy transmitted by each user is limited by the amount of harvested energy, i.e. during time portion $1-T$, each user can only use the energy that was harvested during $1-T$. The energy harvested by the $n$-th user is
\begin{equation}
E_n=\eta_1\gamma_nP_0T,
\end{equation}
where $\gamma_n$ includes the channel power gain, the path loss factor, and the antennas gains and $\eta_1$ is the EH efficiency. Note that the definition $E_n$ has already been provided in \ref{System_model_WPN} and it is only reincluded here for the sake of convenience. The transmit power of the $n$-th user is given by 
\begin{equation}
P^\mathrm{u}_n=\frac{\eta_2E_n}{1-T},
\end{equation}
with $\eta_2$ being the efficiency of the user's amplifier.
Also, the observation at the BS, during Phase 2, is given by
\begin{equation}
y^\mathrm{u}=\sum_{n=1}^N\bar{h}_n\sqrt{\mathcal{G}_{0}\mathcal{G}_{n}P^\mathrm{u}_n}s^\mathrm{u}_n+\nu_0,
\end{equation}
where $\bar{h}_n$ is the channel coefficient between the $n$-th user and the BS, with $h_n$ being its reciprocal, $\nu_0$ denotes the AWGN at the BS and $\mathcal{G}_{0}$ and $\mathcal{G}_{n}$ are the directional antenna gains of the BS and the $n$-th user, respectively.

%\vspace{-0.15in}
\subsubsection{Optimization Objectives}
In this section, two schemes are discerned, namely the symmetric rates and asymmetric rates cases. When users transmit their messages with equal individual data rates (symmetric rates), the goal is to maximize the common user rate, denoted by $\mathcal{R}_{\mathrm{eq}}$, which coincides with the maximization of the minimum individual data rate (the rate of the weakest user). However, in the case when the users are allowed to transmit with different data rates (non-symmetric rates), the system performance optimization usually focuses on the maximization of the system total throughput, denoted by $\mathcal{R}_{\mathrm{tot}}$. Note that the above objectives are not equivalent, since the achievable system throughput might be maximized at the expense of the minimum individual data rate and vice versa. Additionally, the aim is to improve user fairness for both cases. To this end, two distinct objectives are considered, for optimizing the provided QoS, as described below.

\emph{Maximization of the achievable system throughput with minimum rate improvement:} In the case of non-symmetric rates,  the achievable rate region of the network is optimized, so that it contains the points which correspond to the maximum system throughput. In order to increase user fairness, the individual data rate of the weakest user is also maximized, through optimization of the decoding order of the users' messages. 

\emph{Maximization of the equal individual data rates:} When the users transmit with equal rate, this corresponds to the minimum individual data rate among users. When this objective is set, the achievable rate region of the network is optimized, so that it contains those points that maximize the achievable throughput of the weakest user, without necessarily seeking to maximize the achievable system throughput. 

In the above cases, the sum-rate of the network, denoted by $\mathcal{R}_{\mathrm{sum}}$, is the sum of the individual data rates of the users. When maximizing the system throughput, users are allowed to transmit with asymmetric rates, thus $\mathcal{R}_{\mathrm{sum}}=\mathcal{R}_{\mathrm{tot}}$. When maximizing the common rate, all users transmit with symmetric rates, that is, with rate $\mathcal{R}_{\mathrm{eq}}$. In that case, $\mathcal{R}_{\mathrm{sum}}=N\mathcal{R}_{\mathrm{eq}}$.

In the NOMA scheme, the decoding order of the users' messages at the BS affects the individual rate of each user, without affecting the total throughput, as it will be discussed in the following subsections. Therefore it can be used to improve user fairness without affecting the total throughput. To this end, two different approaches are explored, concerning the decoding order of the users' messages, in order of ascending complexity: i) fixed decoding order and ii) TS. TS assumes multiple decoding orders of the users' messages, each of which is applied for a fraction of decoding time $1-T$. Thus, it improves the QoS at the expense of higher computational complexity, which depends on the number of utilized user permutations. Therefore, a balance between optimality and complexity can be achieved when selecting the number of distinct decoding order permutations. The resulting optimization problems can be classified into the following four schemes, which will be referred to as (a)-(d) hereafter:
\begin{enumerate}[label=(\alph*)]
\item Achievable system throughput maximization and minimum individual data rate optimization with fixed decoding order.
\item Achievable system throughput maximization and minimum individual data rate optimization with TS.
\item Equal individual data rate maximization with fixed decoding order.
\item Equal individual data rate maximization with TS.
\end{enumerate}
The above schemes are also summarized in Table \ref{Table:schemes}, where $(\cdot)^*$ denotes the optimization solution value, and $\mathcal{R}_{\mathrm{eq,c}}^*\leq\mathcal{R}_{\mathrm{eq,d}}^*$ due to the use of TS in (d).

\begin{table*}[t!]
\centering 
%\vspace{-0.1in}
\caption{Comparison of proposed optimization problems}\label{Table:schemes}
{\begin{tabular}{llllll}
\hline &\textbf{Objective}&\textbf{Decoding Order} &\textbf{Complexity}&\textbf{Sum-rate}&\textbf{Fairness}\\
\hline
\textbf{(a)} &$\mathcal{R}_{\mathrm{tot}}$&fixed &low&$\mathcal{R}_{\mathrm{tot}}^*$, maximized&medium\\
\textbf{(b)} &$\mathcal{R}_{\mathrm{tot}}$&TS &medium&$\mathcal{R}_{\mathrm{tot}}^*$, maximized&high\\
\textbf{(c)} &$\mathcal{R}_{\mathrm{eq}}$&fixed &medium&$N\mathcal{R}_{\mathrm{eq,c}}^*$, medium&maximized\\
\textbf{(d)} &$\mathcal{R}_{\mathrm{eq}}$&TS &high&$N\mathcal{R}_{\mathrm{eq,d}}^*$, high&maximized\\
\hline
\end{tabular}}
\end{table*}

%\vspace{-0.15in}
\subsubsection{User and System Throughput Evaluation}
In this subsection, the achievable user and system throughput are calculated, depending on the decoding order strategy, and it is proven that the system throughput is independent of the decoding order. The achievable throughput of the $n$-th user with fixed decoding order is denoted by $R^\mathrm{u}_n$, while the the achievable throughput of the $n$-th with TS, is denoted by $\tilde{R}^\mathrm{u}_n$.

\paragraph{Achievable user throughput with fixed decoding order}

Let us first consider a fixed decoding order of the users' messages at the BS, according to their index, $n$. Then, for decoding the first user's message ($n=1$), interference is created due to all other users $n=2,...,N$, while on the second user's message, interference is created due to users $n=3,...,N$, and so on. Then, the achievable throughput in bits/second/Hz of the $n$-th user, $1\leq n \leq (N-1)$, denoted by $R^\mathrm{u}_n$ in the case of fixed decoding order, is given by \cite{Tafazoli}
\begin{equation}
\begin{split}
R^\mathrm{u}_n&=(1-T)\log_2\left(1+\frac{P^\mathrm{u}_n\gamma_n}{\sum_{j=n+1}^N (P^\mathrm{u}_j\gamma_j)+N_0W}\right)\\
&=(1-T)\log_2\left(1+\frac{\frac{\eta\rho_0 T g_n}{1-T}}{\frac{\eta\rho_0 T\sum_{j=n+1}^Ng_j}{1-T}+1}\right),
\end{split}
\label{user rate}
\end{equation}
while the achievable throughput of the $N$-th user is 
\begin{equation}
R^\mathrm{u}_N=(1-T)\log_2\left(1+\frac{\eta\rho_0 T g_N}{1-T}\right).
\label{N user rate}
\end{equation}
In (\ref{user rate}) and (\ref{N user rate}), $\rho_0=\frac{P_0}{N_0W}$, $\eta=\eta_1\eta_2$, where $N_0$ is the power spectral density of the AWGN. Also, assuming channel reciprocity, $g_n$ is given by $g_n=\gamma_n^2$.

\paragraph{Achievable user throughput with TS} The basic principle of TS is that the order of decoding for the users can change for specific fractions of the duration $1-T$ \cite{honig}. In general, there are $N!$ configurations with different decoding order, as many as the different permutations of the $N$ users.  Let $\tau_m$, with $\sum_{m}\tau_m=1$, denote the portion of time $1-T$ for which the BS decodes the users' messages, according to the $m$-th permutation, $m=1,\ldots,M$, where $M\leq N!$ is the number of used permutations. Hereinafter, $\mathbold{\tau}$ denotes the set of values of $\tau_m\forall m$.

In order to express the achievable user throughput $\tilde{R}^\mathrm{u}_n$, let $\mathbf{A}$ be the matrix, which represents the set of specific $M$ permutations, with elements $\mathbf{A}(m,j_{m,n})$, corresponding to the indices of the users, i.e. $\mathbf{A}(m,j_{m,n})=n$. The decoding order of the users during the $m$-th permutation is determined by the indices of the columns, $j_{m,n},\forall n$, for the $m$-th row of matrix $\mathbf{A}$, i.e. if $j_{m,n}<j_{m,z}$, the message of the $n$-th user will be decoded before the message of the $z$-th. More specifically, the value of a matrix  element is the index of a user. The index of the row denotes a specific  permutation, and the index of the column denotes the decoding order of the user in that permutation. For example, if $\mathbf{A}(2,4)=3$, it means that, when the 2-nd permutation is applied, the message of the $3$-rd user will be decoded in the $4$-th order. 

Thus, taking the TS configuration into account, the achievable throughput of the $n$-th user, denoted by $\tilde{R}_n$ in the case of TS, can be written as
\begin{equation}
\tilde{R}^\mathrm{u}_n(T)\!=\!\sum_{m=1}^{M}\tau_m(1-T)\log_2\left(1+\frac{\frac{\eta\rho_0T g_n}{1-T}}{\frac{\eta\rho_0T\sum_{i>j_{m,n}}g_{\mathbf{A}(m,i)}}{1-T}+1}\right)\!\!.
\label{with sharing}
\end{equation}

\paragraph{Achievable system throughput} 

In order to calculate the achievable system throughput, the TS case is first considered. It is
\begin{equation}\label{sum_trhoughput_proof}
\begin{split}
{\mathcal{R}}_{\mathrm{tot}}&=\sum_{n=1}^N\tilde{R}^\mathrm{u}_n=\sum_{n=1}^N\sum_{m=1}^{M}\tau_m(1-T)\log_2\left(1+\frac{\frac{\eta\rho_0T g_n}{1-T}}{\frac{\eta\rho_0T\sum_{i>j_{m,n}}g_{\mathbf{A}(m,i)}}{1-T}+1}\right)\\
&=(1-T)\sum_{m=1}^M\tau_m\sum_{n=1}^N\log_2\left(1+\frac{\frac{\eta\rho_0T g_n}{1-T}}{\frac{\eta\rho_0T\sum_{i>j_{m,n}}g_{\mathbf{A}(m,i)}}{1-T}+1}\right)
\end{split}
\end{equation}

\begin{theorem}
The achievable system throughput, when TS is used, is given by
\begin{equation}
{\mathcal{R}}_{\mathrm{tot}}=(1-T)\log_2\left(1+\frac{\eta \rho_0\sum_{n=1}^{N}g_n}{\frac{1-T}{T}}\right).
\end{equation}
\end{theorem}

\begin{proof}
From the definition of the matrix $\mathbf{A}$, it holds that $n={\mathbf{A}(m,j_{m,n})}$, and thus $g_n=g_{\mathbf{A}(m,j_{m,n})}$. Furthermore, for a specific permutation $m$, the inner sum of \eqref{sum_trhoughput_proof} over all users is equivalent to the sum of all columns $j_{m,n}$ for fixed $m$ (a row in $\mathbf{A}$), since there is one-to-one mapping. Thus, \eqref{sum_trhoughput_proof} can be written as
\begin{equation}\label{sum_throughput_proof2}
\begin{split}
{\mathcal{R}}_{\mathrm{tot}}&=(1-T)\sum_{m=1}^M\tau_m\sum_{j_{m,n}=1}^N\log_2\left(1+\frac{\frac{\eta\rho_0T g_{\mathbf{A}(m,j_{m,n})}}{1-T}}{\frac{\eta\rho_0T\sum_{i>j_{m,n}}g_{\mathbf{A}(m,i)}}{1-T}+1}\right)\\
&=(1-T)\sum_{m=1}^M\tau_m\left(\sum_{j_{m,n}=1}^{N-1}\left(\log_2\left(\eta \rho_0\sum_{i=j_{m,n}}^{N}g_{\mathbf{A}(m,i)}+\frac{1-T}{T}\right)\right.\right.\\
&-\left.\left.\log_2\left(\eta \rho_0\sum_{i=j_{m,n}+1}^{N}g_{\mathbf{A}(m,i)}+\frac{1-T}{T}\right)\right)\right.\\
&+\left.\left(\log_2\left(\eta \rho_0 g_{\mathbf{A}(m,N)}+\frac{1-T}{T}\right)-\log_2\left(\frac{1-T}{T}\right)\right)\right)\\
&=(1-T)\sum_{m=1}^M\tau_m\log_2\left(1+\frac{\eta \rho_0\sum_{j_{m,n}=1}^{N}g_{\mathbf{A}(m,j_{m,n})}}{\frac{1-T}{T}}\right)\\
&=(1-T)\sum_{m=1}^M\tau_m\log_2\left(1+\frac{\eta \rho_0\sum_{n=1}^{N}g_n}{\frac{1-T}{T}}\right)\\&=(1-T)\log_2\left(1+\frac{\eta \rho_0\sum_{n=1}^{N}g_n}{\frac{1-T}{T}}\right).
\end{split}
\end{equation}
\end{proof}

It is evident from the above derived expression that the achievable system throughput is independent of the TS scheme, that is, of the permutations which are used and their time portion. Similarly, when fixed decoding order is considered, the achievable system throughput, denoted by $\mathcal{R}_\mathrm{tot}$, is again given by \eqref{sum_throughput_proof2} as a special case with the same result, in which $M=1$, that is there is only one permutation ($m=1$), while $\tau_1=1$.

\subsection{System Throughput Maximization with Minimum Throughput Improvement}
In this section, the achievable system throughput maximization problem is formulated and solved. Then, elaborating on this solution,  the minimum individual data rate is improved, considering descending decoding order and TS (schemes (a) and (b), respectively). Thereafter, in order to reduce the complexity of scheme (b),  a greedy algorithm, which efficiently optimizes the TS configuration, is proposed. 

It can be easily observed that, when $T=0$ or $T=1$, no energy or no time, respectively, is available to the users in order to transmit, and thus the system throughput is zero. The optimization problem, which aims at maximizing the system throughput, can be written as
\begin{equation}
\begin{array}{ll}
\underset{T}{\text{\textbf{max}}}& \mathcal{R}_{\mathrm{tot}} \\
\,\,\,\text{\textbf{s.t.}}&\mathrm{C}:0<T<1.
\end{array}
\label{system throughput}
\end{equation}
%=============================================
In (\ref{system throughput}), $\mathcal{R}_{\mathrm{tot}}$ is strictly concave with respect to $T$ in $(0,1)$, since it holds that
\begin{equation}
\frac{d^2\mathcal{R}_{\mathrm{tot}}}{dT^2}=-\frac{(\eta \rho_0 \sum_{n=1}^{N}g_n)^2}{\ln(2)(1-T)(1-T+\eta \rho_0 \sum_{n=1}^{N}g_nT)^2}<0.
\label{cancavity}
\end{equation}
Thus, the optimal value for $T$ in $(0,1)$ that maximizes $\mathcal{R}_{\mathrm{tot}}$ is unique and can be obtained through
\begin{equation}
\frac{d\mathcal{R}_{\mathrm{tot}}}{dT}=0.
\end{equation}
After some mathetmatical manipulations, the optimal value can be expressed as 
\begin{equation}
T^*=1-\frac{\eta \rho_0 \sum_{n=1}^{N}g_n}{\eta \rho_0 \sum_{n=1}^{N}g_n+\frac{\eta \rho_0 \sum_{n=1}^{N}g_n-1}{W_0(\frac{\eta \rho_0 \sum_{n=1}^{N}g_n-1}{e})}-1},
\label{optimal T2}
\end{equation}
where $(\cdot)^*$ denotes a solution value and $W_0(x)$ returns the principal branch of the Lambert W function. Furthermore, it is $0<T^*<1$, because from $x=\mathcal{W}_0(x)\exp({\mathcal{W}_0(x)})$, taking into account $\mathcal{W}_0(x)\geq -1, \forall x$, it is
\begin{equation}
\frac{\eta \rho_0 \sum_{n=1}^{N}g_n-1}{W_0(\frac{\eta \rho_0 \sum_{n=1}^{N}g_n-1}{e})}=\exp\left({W_0(\frac{\eta \rho_0 \sum_{n=1}^{N}g_n-1}{e})+1}\right)\geq 1.
\end{equation}

\subsubsection{Minimum Achievable Throughput Improvement with Descending Decoding Order}\label{maxthrdes}
Having optimized the achievable system throughput using (\ref{system throughput}), the next step is the selection of the decoding order of the users' messages. The simplest case is to adopt a fixed decoding order among users, that is, according to their indices. For fairness, the users' indices are assigned in a way that the values $g_n\forall N$  are sorted in descending order, i.e., $g_1\geq...\geq g_N$, since this allows decoding the weakest user's message without interference.  Therefore, this scheme (scheme (a)) increases both fairness and minimum achievable throughput, $\mathcal{R}_{\mathrm{min}}$, compared to schemes with other decoding order, e.g., compared to ascending decoding order.
 %\vspace{-0.15in}
\subsubsection{Minimum Achievable Throughput Improvement with TS} 
Next, the TS technique is utilized and optimized in order to improve the minimum achievable throughput among users, while the system throughput is kept at its maximum by setting $T=T^*$, where $T^*$ is given by (\ref{optimal T2}). In contrast to fixed decoding order, the TS technique has the benefit that, by proper selection of $\mathbold{\tau}$, any point of the capacity region can be achieved, and, thus, it can be exploited in order to improve fairness among the users. Also, as it has already been proven, the achievable system throughput is independent of the decoding order of the messages and, thus, the corresponding optimization scheme does not degrade the achievable system throughput. The resulting optimization problem (scheme (b)) is 
\begin{equation}
\begin{array}{ll}
\underset{\mathbold{\tau}, \mathcal{R}_{\mathrm{min}}}{\text{\textbf{max}}}& \mathcal{R}_{\mathrm{min}} \\
\,\,\,\,\text{\textbf{s.t.}}&\mathrm{C}_n: \tilde{R}^\mathrm{u}_n(T^*)\geq  \mathcal{R}_{\mathrm{min}},\,\forall{n\in\mathcal{N}},\\
&\mathrm{C}_{N+1}:0 \leq\sum_{m=1}^M\tau_m\leq 1,
\end{array}
\label{LP}
\end{equation}
%=============================================
where $\mathcal{N}=\{1,2,\ldots,N\}$ is the set of all users. Note that \eqref{LP} is the epigraph representation of the max-min (initial) problem and the constraints $\mathrm{C}_n$, $\forall{n\in\mathcal{N}},$ represent the hypograph of the initial optimization problem \cite{Boyd1}. For brevity, the initial max-min problem has been omitted.

The optimization problem in (\ref{LP}) is a \textit{linear programming} one and can be efficiently solved by well-known methods in the literature, such as simplex or interior-point method \cite{Boyd1}. In general, the complexity of these methods is a function of the dimensions of the problem, i.e. the number of constraints and the number of variables. In the case of \eqref{LP}, the corresponding problem dimensions are $\left((N+1)\times(M+1)\right)$. Note that, depending on the linear optimization method, the complexity can range from polynomial to exponential in the problem dimensions.

The full-space search, i.e. $M=N!$, is generally optimal, but can prove to be inefficient when the number of users is large. A large number of permutations, $M$, means that the decoding order of the users' messages has to change many times, which could also create synchronization issues. To this end, a more efficient method will be discussed in the next subsection, while its effectiveness will be verified in the simulation results, where it will be compared with the full-space search. 
%\vspace{-0.15in}
\subsubsection{A Greedy Algorithm for Efficient TS}\label{greedy}
The complexity of the solution of the problem in (\ref{LP}) increases with the number of permutations, i.e., the inserted variables, which, in turn, increases considerably with the number of users. For a relatively small number of users, e.g., when $N=5$, $120$ permutations have to be taken into account. For the practical implementation of the TS technique, considering such a number of permutations may be prohibitive. On the other hand, a priori exclusion of some permutations might cause severe degradation to the system performance in terms of minimum rate and fairness. In order to efficiently set the TS configuration, an iterative method is proposed below. 

With the proposed method, instead of a priori considering all permutations, a subset of permutations is dynamically constructed, while the corresponding TS variables, $\tau_m$, are also optimized. The main advantage of Algorithm 1 is the joint calculation of the optimal $\mathcal{R}_{\min}$ and the corresponding TS arguments, using a minimum subset of all possible permutations for the greedy search, without the need of a priori knowledge of the point of operation within the achievable rate region. Note that the proposed algorithm is also applicable to conventional NOMA schemes, in the case that the objective is to optimize the performance of the weakest user, when the sum rate is maximized. The steps are discussed in detail below:
\begin{enumerate}
\item \textit{Initialization}: The users' indices are assigned in descending order with respect to $g_n$, in order to construct the first permutation, i.e., $\mathbf{A}(1,j_{1,n})$. The achievable throughput of each user is calculated using (\ref{user rate}) and (\ref{N user rate}). 
\item \textit{Main loop (iteratively)}:
\begin{enumerate}[label=\roman*)]
\item The users' decoding order is rearranged in descending order with respect to the throughput they achieve so far, forming the new candidate permutation to be inserted in $\mathbf{A}$. This step can only improve the minimum rate which is achieved so far.
\item If the new permutation is not already in $\mathbf{A}$, it is added, while a new variable is inserted in $\mathbold{\tau}$. Adding new permutations in this way gives the opportunity to the users that achieve small throughput to improve their rates, while at the same time, the minimum achievable throughput is never reduced. 
\item The linear optimization problem in (\ref{LP}) is solved for the updated $\mathbold{\tau}$.
\item  The new users' rates are calculated using (\ref{with sharing}).
\end{enumerate}
\item \textit{Convergence evaluation}: The main loop of the algorithm is repeated until the maximum number of iterations $K$ is reached, or a permutation is already included in $\mathbf{A}$. Please note that only new permutations are inserted in $\mathbf{A}$, because, otherwise, there would be two variables in $\mathbold{\tau}$ with exactly the same physical meaning.
\end{enumerate}
The above procedure can be summarized in Algorithm 1. 
Note that Algorithm 1 substitutes an optimization problem of large dimensions with multiple smaller problems. More specifically, the number of optimization variables is drastically reduced, by assuming only a subset of all possible permutations. As the number of users increases, solving a number of optimization problems of small dimensions offers great complexity savings compared to the original problem, since its number of optimization variables is factorial in the number of users, as it will be illustrated in the results section.

\begin{algorithm}\label{Alg}
\caption{: Greedy Algorithm for Efficient TS Configuration}
\begin{algorithmic}[1]
\State \underline{\textbf{Initialization}} 
\State For the first permutation is $\mathbf{A}(1,j_{1,n}),\,\forall n\in \mathcal{N} $, assign $j_{1,n}$ so that $g_1\geq g_2\geq...\geq g_n\geq...\geq g_N$.
\State Calculate $R^\mathrm{u}_n$ using (\ref{user rate}) and (\ref{N user rate}).
\State Set $k=0$, $\tilde{R}^\mathrm{u}_n[0]=R^\mathrm{u}_n,\, \forall n\in \mathcal{N}$.
\State \underline{\textbf{Main loop}} \Repeat 
\State Set $k=k+1$.
\State Assign $j_{k+1,n}$ so that the values of $\tilde{R}^\mathrm{u}_n[k-1]$ are in descending order. Thus, $\forall n,l \in \mathcal {N}$: \If {$\tilde{R}^\mathrm{u}_n[k-1]\leq \tilde{R}^\mathrm{u}_l[k-1]$}
\State Select $j_{k+1,n},j_{k+1,l}:j_{k+1,n}\geq j_{k+1,l}$. \EndIf
\State Update $\mathbf{A}$.
\If {$ \mathbf{A}(k+1,:)\neq \mathbf{A}(m,:),\, \forall m\leq k$}
\State Solve (\ref{LP}), setting $M=k+1$.
\State Update the individual data rates $\tilde{R}^\mathrm{u}_n[k]$ using (\ref{with sharing}).
\EndIf
\Until $k=K$ or $\exists l\leq k: \mathbf{A}(k+1,:)= \mathbf{A}(l,:)$.
\end{algorithmic}
\end{algorithm}

%\vspace{-0.35in}
\subsection{Equal Individual Data Rate Optimization}
In this section, the aim is to maximize the equal individual data rate, i.e.  the minimum user throughput in the case where all users aim to transmit with an equal rate, $\mathcal{R}_\mathrm{eq}$, and, thus, $T$ can be adjusted accordingly. First,  the corresponding optimization problem for fixed descending decoding order is presented and efficiently solved.
%\vspace{-0.2in}
\subsubsection{Fixed Descending Decoding Order}
The problem of equal individual data rate maximization, when the message of the users with the best channel conditions is decoded first, can be written as:
%===========================================================
\begin{equation}
\begin{array}{ll}
\underset{T,R_\mathrm{eq}}{\text{\textbf{max}}}& \mathcal{R}_\mathrm{eq} \\
\,\,\,\text{\textbf{s.t.}}&\mathrm{C}_n: R^\mathrm{u}_n\geq  \mathcal{R}_\mathrm{eq},\,\forall{n\in\mathcal{N}}, \quad \mathrm{C}_{N+1}:0<T<1.
\end{array}
\label{opt_dec}
\end{equation}
%=============================================
The first $N$ constraints of the optimization problem in (\ref{opt_dec}) are strictly concave since
\begin{equation}
\frac{d^2R_{n}}{dT^2}=-
\frac{\tilde{a}_n\left( (\tilde{a}_n+2\tilde{b}_n)(1-T)+2\tilde{a}_n\tilde{b}_nT+2\tilde{b}_n^2T\right)}{\ln(2)\left(1-T+\tilde{b}_nT\right)^2\left(1-T+\tilde{a}_nT+\tilde{b}_nT\right)^2}<0,
\label{concavity}
\end{equation}
where $\tilde{a}_n=\eta\rho_0 g_n$, $\tilde{b}_n=\eta\rho_0\sum_{j=n+1}^Ng_j$, and $\tilde{b}_N=0$. Also, the objective function, as well as the $(N+1)$-th constraint, are linear, and therefore  (\ref{opt_dec}) is a convex optimization problem, which can be solved by standard numerical methods such as interior point and bisection method.
%\footnote{The bisection method is used in \cite{Ju}, in a problem of similar structure in order to handle the linear objective function. In this paper, we do not use the bisection method, in order to avoid the corresponding extra complexity.}. 
However, Lagrange dual decomposition is used, which proves to be extremely efficient, since, given the LMs, the optimal $T$ and $R_\mathrm{eq}$ can be directly calculated, without the utilization of matrices which may increase memory requirements. More importantly, using the adopted method, it is guaranteed that the optimal solution can be obtained in polynomial time \cite{Boyd1}. Also, note that by using dual-decomposition our work is directly comparable to \cite{Ju}, in terms of complexity of the provided solution for the equal individual data rate maximization problem, among others.

\paragraph{Dual Problem Formulation and Solution of (\ref{opt_dec})}
In order to handle the linear objective function in (\ref{opt_dec}), it is replaced with $\ln(R_\mathrm{eq})$, without affecting the convexity. Since the primal problem is convex and satisfies the Slater's condition qualifications, strong duality holds, i.e., solving the dual is equivalent to solving the primal problem \cite{Boyd1}. In order to formulate the dual problem, the Lagrangian is needed, which is given by

\begin{equation}
\begin{split}
\mathcal{L}(\mathbold{\lambda},T,\mathcal{R}_\mathrm{eq})&=\ln(\mathcal{R}_\mathrm{eq})+\sum_{n=1}^N\lambda_n\left((1-T)\log_2\left(1+\frac{\tilde{a}_n}{\tilde{b}_n+\frac{1-T}{T}}\right)-\mathcal{R}_\mathrm{eq}\right),
\label{lagrangian1}
\end{split}
\end{equation}
where $\lambda_n\geq 0$ is the LM, which corresponds to the constraint $\mathrm{C}_n$ and $\mathbold{\lambda}$ is the Lagrange multiplier vector with elements $\lambda_n$. The constraint $\mathrm{C}_{N+1}$ is absorbed into the KKT conditions, and is presented in detail in the next subsection.

The dual problem is now given by \begin{equation}
\underset{\mathbold{\lambda}}{\text{\text{min}}}\,\,\,\,\underset{T,\mathcal{R}_\mathrm{eq}}{\text{\text{max}}}\,\,\mathcal{L}(\mathbold{\lambda},T,\mathcal{R}_\mathrm{eq}).
\label{dual problem 1}
\end{equation}

Note that the Lagrangian in (\ref{lagrangian1}) is strictly concave in $(0,1)$ with respect to $T$ and $R_\mathrm{eq}$ for a specific $\mathbold{\lambda}$, and thus it has a unique maximization point in $(0,1)$, so its first derivative must have a unique zero point.  According to the KKT conditions, the optimal values of $\mathcal{R}_{\mathrm{eq}}$ and $T$ are given by 
%\begin{equation}
%\frac{\partial \tilde{\mathcal{L}}(\mathbold{\lambda},T)}{\partial T}=0,
%\end{equation}
\begin{equation}
\mathcal{R}_{\mathrm{eq}}^*=\frac{1}{\sum_{n=1}^N\lambda_n}
\label{optimal_rate}
\end{equation}
and
\begin{equation}
T^*\!=\!T\in(0,1)\!:\!\sum_{n=1}^N\lambda_n\left(\ln\left(1+\frac{\tilde{a}_n}{\tilde{b}_n+\frac{1-T}{T}}\right)-\frac{\tilde{a}_n(1-T)}{(1-T+\tilde{b}_nT)(1-T+\tilde{a}_nT+\tilde{b}_nT)}\right)\!=\!0.
\label{optimalT2}
\end{equation}
The above equation can be solved numerically (e.g., using Newton-Raphson).

The dual problem in (\ref{dual problem 1}) can be solved iteratively. In each iteration, the optimal $\mathcal{R}_{\mathrm{eq}}$ and $T$ are calculated for a fixed LM vector, $\mathbold{\lambda}$, using (\ref{optimal_rate}) and (\ref{optimalT2}), while $\mathbold{\lambda}$ is then updated using the sub-gradient method as follows
\begin{equation}
\lambda_n[j+1]=\left[\lambda_n[j]-\hat{\lambda}_n[j]\times
\left((1-T)\log_2\left(1+\frac{\tilde{a}_n}{\tilde{b}_n+\frac{1-T}{T}}\right)-\mathcal{R}_\mathrm{eq}\right)\right]^+,\,\forall n \in \mathcal{N},
\end{equation}
where $j$ is the iteration index, $\hat{\lambda}_n,\,n\in\mathcal{N}$ are positive step sizes, and $[\cdot]^+=\min(\cdot,0)$. Since Problem 1 is concave, it is guaranteed that the iterations between the two layers converge to the optimal solution if the size of the chosen step satisfies the infinite travel condition \cite{Boyd2}. As it can be observed from the solution in (\ref{optimal_rate}), the equal individual data rate is inversely proportional to the sum of the LMs. This result is consistent with the physical interpretation of the  LMs, which are indicative of how active the corresponding constraints are, depicting the impact of the weakest users, via the violated constraints, on the optimal value \cite{Boyd1}. 
%If $\lambda^*_n$ is  small it means that the effect of the $n$-th constraint on the determination of $\mathcal{R}_{\mathrm{eq}}^*$ in (\ref{optimal_rate}), as well as of $x^*$ in (\ref{optimalT2}), is not significant. On the other hand, if $\lambda_n^*$ is large it means that if the constraint is loosened or tightened a bit, the effect on $\mathcal{R}_{\mathrm{eq}}^*$ will be great \cite{Boyd1}. In this case, the throughput of the $n$-th user is in high priority when optimizing the time that is dedicated to energy transfer.
%\vspace{-0.15in}
\subsubsection{Time-Sharing}
In contrast to the previous subsection, where fixed descending decoding order was assumed, here the aim is to maximize the equal individual data rate, while utilizing the TS technique. For this purpose, $T$ as well as the TS configuration need to be optimized. Please note that in contrast to the TS configuration discussed in section \ref{maxthrdes}, the solution provided in this subsection does not necessarily maximize the system throughput. 
 
Next, the indices of the users are ordered according to $g_1\geq g_2\geq\ldots\geq g_N$, however, the order of decoding depends on the TS. 
Taking into account the above considerations, the problem of equal individual data rate maximization can be formulated as
%===========================================================
\begin{equation}
\begin{array}{ll}
\underset{\mathbold{\tau},T,\mathcal{R}_\mathrm{eq}}{\text{\textbf{max}}}& \mathcal{R}_\mathrm{eq} \\
\,\,\,\,\text{\textbf{s.t.}}&\mathrm{C}_n: \tilde{R}^\mathrm{u}_n\geq  \mathcal{R}_\mathrm{eq},\,\forall{n\in\mathcal{N}},\\
&\mathrm{C}_{N+1}:0<T<1,\\
&\mathrm{C}_{N+2}:0\leq\sum_{m=1}^M\tau_m\leq 1.
\end{array}
\label{opt}
\end{equation}
%=============================================

Apparently, the optimization problem in (\ref{opt}) is non-convex, due to the coupling of the variables $T$ and $\mathbold{\tau}$. It is noted that there is no standard approach for solving non-convex optimization problems in general. In order to derive an efficient and optimal time allocation method for the considered problem, the following observations  are taken into account.

\begin{remark}\label{problem seperation}
A selection for $T$ corresponds to a specific capacity region for the set of users $\mathcal{N}$, where the TS technique can also be used. On the other hand, each of the points of this capacity region corresponds to a different selection of $\mathbold{\tau}$. As it has already been mentioned, with proper selection of the TS variables, any point of the capacity region can be achieved.
\end{remark}

Taking into account Remark \ref{problem seperation}, for a given time $T$, the achievable rate region is defined by the inequalities
\begin{equation}
\begin{split}
\tilde{R}^\mathrm{u}_n(T)&\leq (1-)\log_2\left(1+\tfrac{\eta\rho_0 Tg_n}{1-T}\right),\,\forall n\in\mathcal{N}\\
\sum_{n\in\mathcal{M}_k} \tilde{R}^\mathrm{u}_n(T)&\leq (1-T)\log_2\left(1+\tfrac{\eta\rho_0 T\sum g_n}{1-T}\right),\,\forall i: \mathcal{M}_k\subseteq\mathcal{N},
\end{split}
\label{time_sharing_rates_constraint}
\end{equation}
where the second inequality holds for any sum set, $\mathcal{M}_k\subseteq\mathcal{N}$. Now, suppose that the BS cancels all other users' messages, except the user with the weakest link. In this case it is desired that its throughput is at least equal to the final achievable $\mathcal{R}_\mathrm{eq}$, i.e.
\begin{equation}
\underbrace{(1-T)\log_2\left(1+\frac{\eta\rho_0Tg_N}{1-T}\right)}_{\tilde{R}^\mathrm{u}_N}\geq\mathcal{R}_\mathrm{eq}.
\end{equation}
Accordingly, for the two weakest users, that is for $n=N$ and $n=N-1$, their sum-throughput is maximized when the BS cancels out all other users' messages, while one of the two messages is also canceled. Since they can allow TS for the time that each user's message will be canceled, for the sum of the throughput of these two users it must hold that
\begin{equation}
\underbrace{(1-T)\log_2\left(1+\frac{\eta\rho_0T\sum_{n=N-1}^Ng_n}{1-T}\right)}_{\tilde{R}^\mathrm{u}_{N-1}+\tilde{R}^\mathrm{u}_N}\geq2\mathcal{R}_\mathrm{eq}.
\end{equation}

Following the same strategy for all other users, it yields that $\mathcal{R}_{\mathrm{eq}}$ is bounded by the following set of inequalities
\begin{equation}
\mathcal{R}_{\mathrm{eq}} \leq \frac{(1-T)\log_2\left(1+\frac{\eta \rho_0\sum_{i=n}^{N}g_i}{\frac{1-T}{T}}\right)}{(N+1-n)},\,\forall{n\in\mathcal{N}},
\end{equation}
in which $\mathbold{\tau}$ does not appear. In this way, $\mathcal{R}_{\mathrm{eq}}$ is bounded by an alternative representation of the capacity region.
Consequently, the optimization in \eqref{opt} can be optimally solved by reducing it into two disjoint problems, after minimizing the initial
search space. These optimization problems are:

%===========================================================
\emph{Problem 1: Optimization of $T$}
\begin{equation}
\begin{array}{ll}
\underset{T,R_\mathrm{eq}}{\text{\textbf{max}}}& \mathcal{R}_\mathrm{eq} \\
\,\,\,\text{\textbf{s.t.}}&\textit{C}_n: (1-T)\log_2\left(1+\frac{\eta \rho_0\sum_{i=n}^{N}g_i}{\frac{1-T}{T}}\right)\geq (N+1-n)\mathcal{R}_\mathrm{eq},\,\forall{n\in\mathcal{N}},\\
&\mathrm{C}_{N+1}:0<T<1,
\end{array}
\label{opt2}
\end{equation}
%=============================================

%================================
\emph{Problem 2: Calculation of the TS vector $\mathbold{\tau}$}
\begin{equation}
\begin{array}{ll}
{\text{\textbf{find}}}& \mathbold{\tau} \\
\,\,\text{\textbf{s.t.}}&\mathrm{C}_n: \tilde{R}_n(T^*)\geq  \mathcal{R}^*_\mathrm{eq},\,\forall{n\in\mathcal{N}},\\
&\mathrm{C}_{N+1}:0 \leq\sum_{m=1}^M\tau_m\leq 1.
\end{array}
\label{LP2}
\end{equation}
%===============================

In the above, $\mathcal{R}^*_\mathrm{eq}$, denotes the optimal solution for $\mathcal{R}_\mathrm{eq}$, which is calculated by solving Problem 1. The geometrical interpretation of \eqref{opt2} is that $\mathcal{R}_\mathrm{eq}$ is maximized up to a value, such that there is a point within the capacity region, whose coordinates are all equal to $\mathcal{R}_\mathrm{eq}$. The solution of Problem 2 is calculated after the solution of Problem 1. More specifically, Problem 2 calculates the value of the TS vector, $\boldsymbol{\tau}$, such that $\mathcal{R}^*_\mathrm{eq}$ by each user individually, i.e. the capacity of each user is at least $\mathcal{R}^*_\mathrm{eq}$, which is the main motivation for using TS. Please note that, when solving Problem 2, since $T^*$ and $\mathcal{R}^*_\mathrm{eq}$ have already been fixed, this is a linear optimization problem, with similar structure to (\ref{LP}). Thus, it can be solved by utilizing the same linear programming methods or  by using Algorithm 1. On the other hand, Problem 1 is jointly concave with respect to $T$ and $\mathcal{R}_\mathrm{eq}$, and satisfies Slater's constraint qualification. Thus, it is a convex optimization problem, which can be solved by following similar steps as in the solution of (\ref{opt_dec}).

\paragraph{Solution of Problem 1}
In this subsection, the optimization problem (\ref{opt2}), i.e. Problem 1, is solved by \textit{Lagrange dual decomposition}. The Lagrangian of Problem 1, after replacing the initial objective function with $\ln(\mathcal{R}_{\mathrm{eq}})$, is given by
\begin{equation}
\mathcal{L}(\mathbold{\mu},T,\mathcal{R}_\mathrm{eq})=\ln(\mathcal{R}_\mathrm{eq})+\sum_{n=1}^N\lambda_n
\left(\frac{(1-T)\log_2\left(1+\frac{\tilde{c}_n}{\frac{1-T}{T}}\right)}{\tilde{d}_n}-\mathcal{R}_\mathrm{eq}\right),
\label{lagrangian}
\end{equation}
where $\tilde{c}_n=\eta \rho\sum_{i=n}^{N}g_i$ and $\tilde{d}_n=N+1-n$. 
%$\mu_n\geq 0$ is the Lagrange multiplier, which corresponds to the constraint $\mathrm{C}_n$, and $\mathbold{\mu}$ is the Lagrange multiplier vector with elements $\mu_n$. 

The dual problem is now given by 
\begin{equation}
\underset{\mathbold{\lambda}}{\text{\text{min}}}\,\,\,\,\underset{T,\mathcal{R}_\mathrm{eq}}{\text{\text{max}}}\,\,\mathcal{L}(\mathbold{\lambda},T,\mathcal{R}_\mathrm{eq}).
\label{dual problem}
\end{equation}

The dual problem in (\ref{dual problem}) can be iteratively solved, similarly to the solution of problem (\ref{dual problem 1}). In each iteration, the optimal values of $\mathcal{R}_{\mathrm{eq}}$ and $T$ are given by 
\begin{equation}
\mathcal{R}_{\mathrm{eq}}^*=\frac{1}{\sum_{n=1}^N\lambda_n}
\end{equation}
and
\begin{equation}
T^*=T\in(0,1):\sum_{n=1}^{N}\frac{\lambda_n}{\tilde{d}_n}\left(\ln\left(1+\frac{\tilde{c}_nT}{1-T}\right)\frac{\tilde{c}_n}{T(\tilde{c}_n-1)+1}\right)=0.
\label{optimalT3}
\end{equation}

Furthermore, the LMs can be updated as follows
\begin{equation}
\lambda_n[j+1]=\left[\lambda_n[t]-\hat{\lambda}_n[j]\left(\frac{(1-T)\log_2\left(1+\frac{\tilde{c}_n}{\frac{1-T}{T}}\right)}{\tilde{d}_n}-\mathcal{R}_{\mathrm{eq}}\right)\right]^+,\,\forall n\in\mathcal{N}.
\end{equation}

%\vspace{-0.15 in}
\subsection{Illustrative Examples}\label{examples}
In this section, two examples for the capacity region of a simple two-user system are presented, with the aim to highlight the differences between schemes (a)-(d). The following cases are considered: i) similar channel conditions and, ii) asymmetric channel conditions. In both examples, the EH efficiency of each user is assumed to be $\eta_1=0.5$, and the amplifier's efficiency is $\eta_2=0.38$. Without loss of generality, the focus is on the effect of distance, denoted by $d_n$, thus all directional antenna gains are assumed to be equal to 0 dB and fading parameters are set to $h_1=h_2=1$. Also, $P_0=30$ dBm, $N_0W=-114$ dBm and a carrier center frequency of $470$ MHz is considered, which will be used in the standard Institute of Electrical and Electronics Engineers (IEEE) 802.11 for the next generation of wireless fidelity (Wi-Fi) systems \cite{wifi, Kwan}. The TGn path loss model for indoor communication is adopted \cite{tgn,Kwan}, with the breakpoint distance being 5 m. The free space loss up to and after the breakpoint distance, are assumed to be slope of 2 and 3.5, respective;y. 

\begin{figure}
\centering
\includegraphics[width=0.8\linewidth]{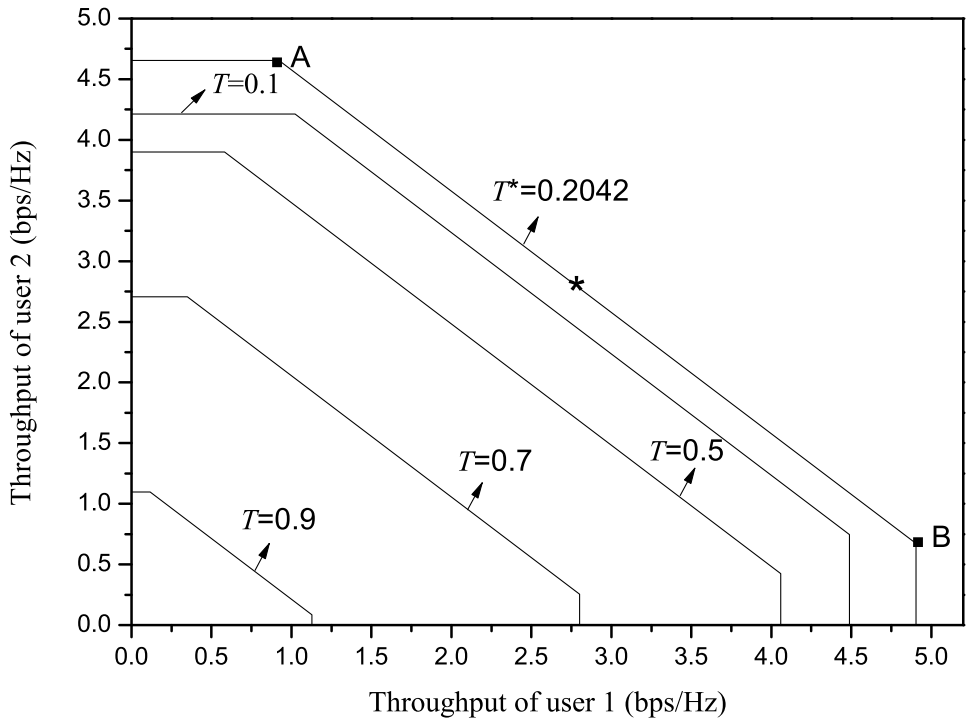}
%\vspace{-0.2in}
\caption{Example 1: Achievable throughput region.}
\label{Fig2section2}
\end{figure}

\textit{Example 1 (Similar distance)}:
For the first example, it is assumed that $d_1=9.9$ m and $d_2=10.1$ m, which corresponds to ${L}_{1}=2.4067 \cdot 10^{-6}$ and ${L}_{2}=2.156 \cdot 10^{-6}$ . This example is representative of the case of two users located at similar distance from the BS. Fig. \ref{Fig2section2} depicts the capacity region for different choices of $T$, as well as, for the optimal value of $T$, which is $T^*=0.2042$. There are two fixed decoding orders available, thus matrix $\mathbf{A}$ is given by
\begin{equation}
\mathbf{A}=\left[\begin{array}{cc}\mathbf{A}(1,1)&\mathbf{A}(1,2)\\\mathbf{A}(2,1)&\mathbf{A}(2,2)\end{array}\right]=\left[\begin{array}{cc}1&2\\2&1\end{array}\right].
\end{equation}
Descending decoding order with respect to the channel values (first row of $\mathbf{A}$) is in favor of the user with the worst channel conditions (user 2) whose message is decoded free of interference, i.e. $R^\mathrm{u}_1=0.92253$ bps/Hz and $R^\mathrm{u}_2=4.65538$ bps/Hz. This point corresponds to point A of the capacity region. Comparing to point B (ascending decoding order - second row of $\mathbf{A}$), where $R^\mathrm{u}_1=4.90526$ bps/Hz and $R^\mathrm{u}_2=0.6726$ bps/Hz, it is observed that descending order results in higher minimum individual rate, therefore increasing user fairness. The selection of point A corresponds to scheme (a).

Interestingly, the capacity region which is formed when $T=T^*=0.2042$, includes a set of solutions that dominates, in terms of both achievable system throughput and minimum user throughput, any other set of solutions, imposed by the capacity region formed by any other value of $T$. This is an important observation, taking into account that any point of the capacity region can be achieved with proper TS configuration. In this example, by choosing $\tau_1=0.4688$ and $\tau_2=0.5312$, the users' achievable throughput becomes $\tilde{R}^\mathrm{u}_1=\tilde{R}^\mathrm{u}_2=2.7891$ bps/Hz. This configuration corresponds to the point that is marked with asterisk, which simultaneously maximizes the system throughput and the minimum individual data rate. Note that the point with asterisk is the solution for both schemes (b) and (d).

\begin{figure}
\centering
\includegraphics[width=0.8\linewidth]{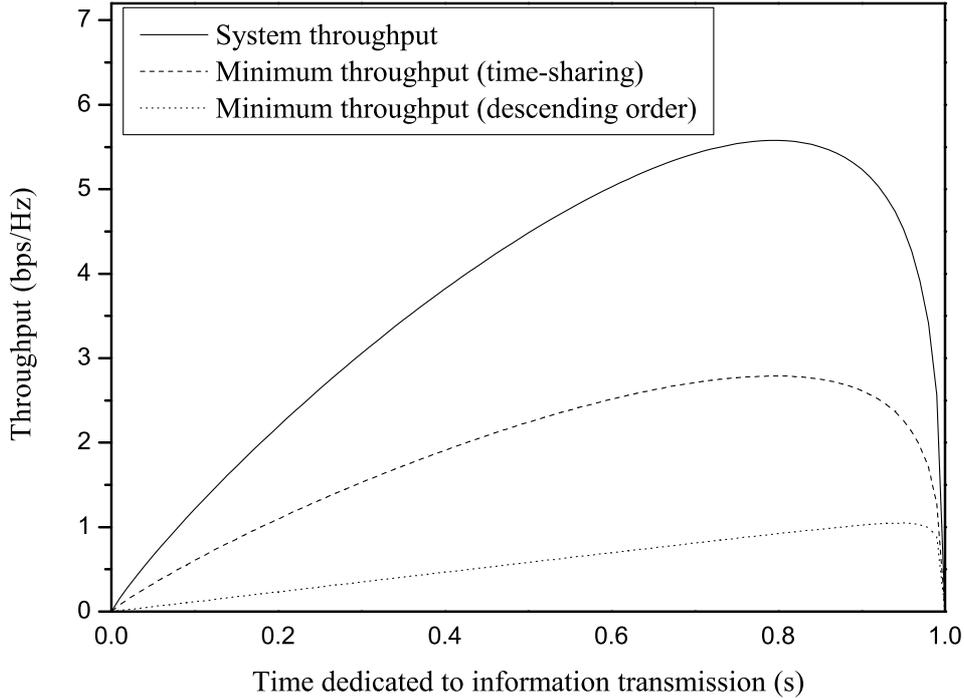}
%\vspace{-0.2in}
\caption{Example 1: System throughput and minimum-throughput.}
\label{Fig3section2}
\end{figure}

In Fig. \ref{Fig3section2}, the minimum throughput between the two users, with respect to the value of T is depicted, while the achievable system throughput of the users is also depicted as a reference. It is shown that by using and optimizing the TS, the value of $T$ that maximizes the system throughput also maximizes the minimum user throughput. This is because $\mathcal{R}^*_\mathrm{min}=\frac{\mathcal{R}^*_{\mathrm{tot}}}{2}$. On the other hand, it is illustrated that when the fixed descending decoding order is chosen, then the value of $T$ that maximizes the minimum user throughput is higher than $T^*$, which corresponds to a lower value of system throughput. The latter solution corresponds to scheme (c), however it is observed that either the minimum rate or system throughput or both can be further increased by choosing one of schemes (a), (b), or (d).
%However, as it will be shown in the next example, the solution that maximizes the system throughput does not always maximize the minimum-throughput, even with proper TS configuration. 

\begin{figure}
\centering
\includegraphics[width=0.8\linewidth]{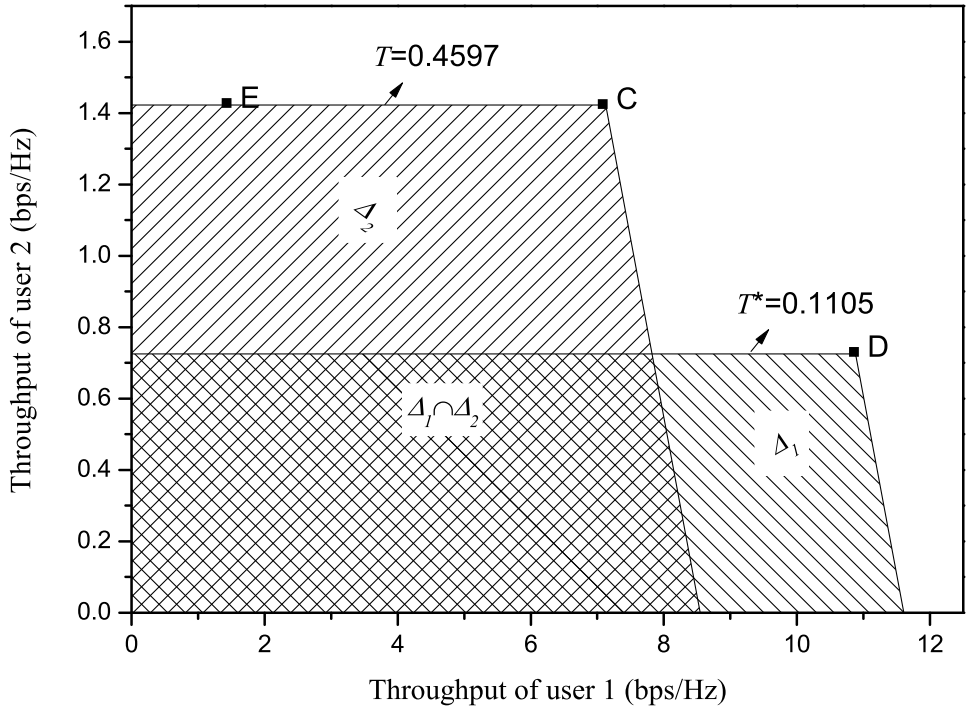}
%\vspace{-0.2in}
\caption{Example 2: Achievable throughput region.}
\label{Fig4section2}
\end{figure}

\textit{Example 2 (The doubly near-far problem)}:
For this example, the distances between the users and the BS are chosen in a way that $\mathcal{L}_{1}\gg \mathcal{L}_{2}$, i.e. $d_1=6$ m and $d_2=14$ m, such as that ${L}_{1}=3.7808 \cdot 10^{-5}$ and ${L}_{2}=2.5786 \cdot 10^{-7}$. This configuration is a representative of the doubly near-far phenomenon \cite{Ju}, which has been discussed in the previous section. When NOMA is used, this phenomenon directly affects the capacity region, as it is evident from Fig. \ref{Fig4section2}. As it can be observed, the value of $T$ that maximizes the system throughput is denoted by $T^*=0.1105$. 
When descending decoding order is utilized, then the achievable throughput values are $R^\mathrm{u}_1=10.8823$ bps/Hz and $R^\mathrm{u}_2=0.7251$ bps/Hz. This point corresponds to point D of the region $\Delta_1$ (scheme (a)). It is remarkable that the set of solutions included in the capacity region that is formed when $T=T^*$ does not dominate any other set of solutions both in terms of system throughput and minimum user throughput, e.g., $\Delta_2$ is not a subset of $\Delta_1$. Furthermore, TS technique cannot improve the minimum throughput, i.e. point D is the solution for both schemes (a) and (b). However, minimum data rate can be improved by a different selection of $T$. For example, when $T=0.46$ the point of the capacity region which maximizes the minimum rate is point C, where the users' capacity values become $R^\mathrm{u}_1=7.1242$ bps/Hz and $R^\mathrm{u}_2=1.4223$ bps/Hz. However, this selection does not maximize the system throughput. Moreover, since schemes (c) and (d) assume equal individual data rates, the solution point for both schemes will be point E, where $\mathcal{R}_\mathrm{eq}=R^\mathrm{u}_1=R^\mathrm{u}_2=1.4223$ bps/Hz. Note that, although the sum rate in this case is low, both users' rate is double compared to point D, showing that schemes (c)-(d) substantially improve the minimum rate in case of asymmetric channel conditions.

%\vspace{-0.15in}
\subsection{Simulations and Discussion}

For the simulations, it is assumed that the users are uniformly distributed in a ring-shaped surface, with $r_{c1}=5$ m and $r_{c2}=20$ m being the radii of the inner and the outer circle, respectively, while the BS is located at the center of the ring. This setup corresponds to the distribution of users within their serving cell in practical systems.  The path loss model, as well as the EH and the amplifier's efficiency are set according to section \ref{examples}. Besides, it is assumed that the small scale fading coefficient is given by the complex random variable $H_{n}\sim\mathcal{CN}(0,1)$. All statistical results are averaged over $10^5$ random channel realizations.  The receiver of the BS is assumed to have a white power spectral density of $N_0=-174$ dBm/Hz, while all directional antenna gains are assumed to be 7.5 dB, and the available bandwidth is assumed as 1 MHz. Finally, all permutations are considered when optimizing TS, i.e., $M=N!$, unless stated otherwise.

Next, the performance metrics of the proposed optimization schemes, i.e. (a)-(d), are presented against the corresponding results of the baseline orthogonal (TDMA) scheme, which is considered in \cite{Ju}. For the readers' convenience, the following notation regarding the comparison with the TDMA approach \cite{Ju} is used: [A] - System throughput maximization, [B] - Equal individual data rate maximization. Note that both TDMA schemes have been discussed in the previous section, where, as in \cite{Ju}, case [B] is referred to as ``common-throughput''.
%\vspace{-0.15in}
\subsubsection{Throughput Comparison}
\begin{figure}
\centering
\includegraphics[width=0.8\linewidth]{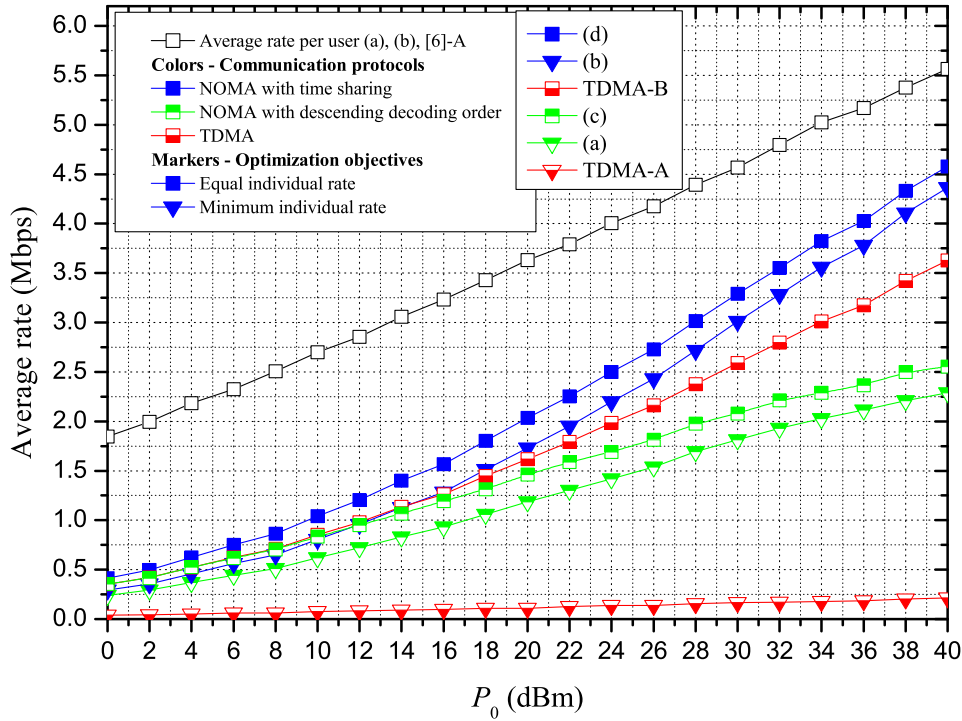}
%\vspace{-0.2in}
\caption{Comparison of average user rates, when $N=3$.}
\label{Fig5}
\end{figure}

In Fig. \ref{Fig5}, the average individual data rate that is achieved by all schemes (a)-(d), is illustrated and compared for the case of $N=3$. 
For the schemes with symmetric rates, i.e., for schemes (c), (d), and [B]-TDMA, the common rate of all users is illustrated. On the other hand, for schemes with asymmetric rates (schemes (a), (b), and [A]-TDMA), both the average rate per user, i.e., normalized sum rate, and the rate of the weakest user are depicted, in order to fully describe these schemes' performance. Note that the normalized sum rate for all three schemes with asymmetric rates is the same, since they all maximize the achievable system throughput.
%Different communication protocols and communication objectives are illustrated by curves of different color and markers, respectively. Furthermore, the average rate per user, i.e., normalized sum rate,  for the schemes with asymmetric rates (a), (b), and [A]-TDMA, is also included, because since the minimum rate does not fully describe their performance. From this curve it is evident that the average rate per user for these three schemes is the same, since they are directly comparable due to common objective, i.e., they maximize the achievable system throughput. For the schemes which achieve asymmetric data rates, i.e., for schemes (a), (b), and [A]-TDMA, the minimum individual data rate is depicted. 
It is also notable that both NOMA schemes (a) and (b) greatly outperform [A]-TDMA in terms of minimum throughput, and especially scheme (b), which employs TS. This is due to the fact that, when fixed descending decoding order is used (scheme (a)), only the corner points of the capacity region can be achieved. However, this is not the case when TS is applied (scheme (b)), since the degrees of freedom increase and any point of the capacity region can be achieved. On the other hand, when the equal individual data rates that are achieved by schemes (c), (d), and [B]-TDMA are explored, it is observed that scheme (c) does not outperform [B]-TDMA. However, comparing the minimum rate of scheme (a) and the common rate of scheme (c), it is evident that, although a fixed decoding order is still used, scheme (c) improves the rate achieved by the weakest user, since this is indeed its optimization objective. When TS is applied, scheme (d) substantially outperforms [B]-TDMA, notably improving the equal rate at which the users transmit, especially for the medium/high $P_0$ region. Since fixed decoding order (scheme (c)) cannot improve the equal rate when compared to TDMA, it is concluded that TS (scheme (d)) should be preferred when the objective is to maximize the equal individual rate. Therefore, in the next figures, schemes (a), (b), and (d) are mainly considered.

\begin{figure}
\centering
\includegraphics[width=0.8\linewidth]{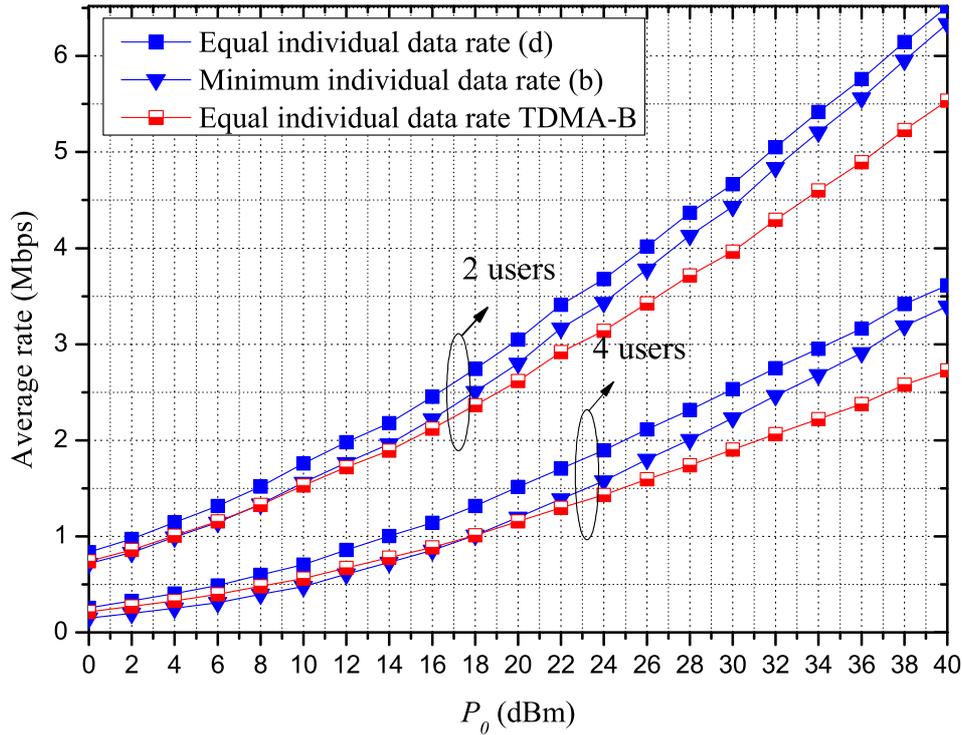}
%\vspace{-0.2in}
\caption{Comparison of (b), (d) and [6]-B in terms of average throughput of the weakest user.}
\label{Fig6}
\end{figure}

Fig. \ref{Fig6} focuses on the comparison of scheme (d) and [B]-TDMA, i.e., the schemes that maximize the minimum individual data rate, for NOMA and TDMA, respectively. Scheme (d) always outperforms TDMA, for the whole range of $P_0$ values, and for any number of users, by increasing margin as the SNR increases. Apart from the aforementioned schemes, the minimum rate of scheme (b) is also depicted in this figure, due to a remarkable observation: although the optimization objective of scheme (b) is the maximization the total system throughput and not of the minimum rate, scheme (b) still outperforms [B]-TDMA in terms of minimum rate, for medium and high $P_0$ values. This is evident in the figure, mainly for small number of users (e.g., $N=2$) where schemes (b) and (d) have very similar performance. Consequently, the combination of NOMA with TS, regardless of the optimization objective, leads to results which greatly improve the minimum rate, and consequently the user fairness, in contrast to the benchmark orthogonal technique of TDMA.

\begin{figure}
\centering
\includegraphics[width=0.8\linewidth]{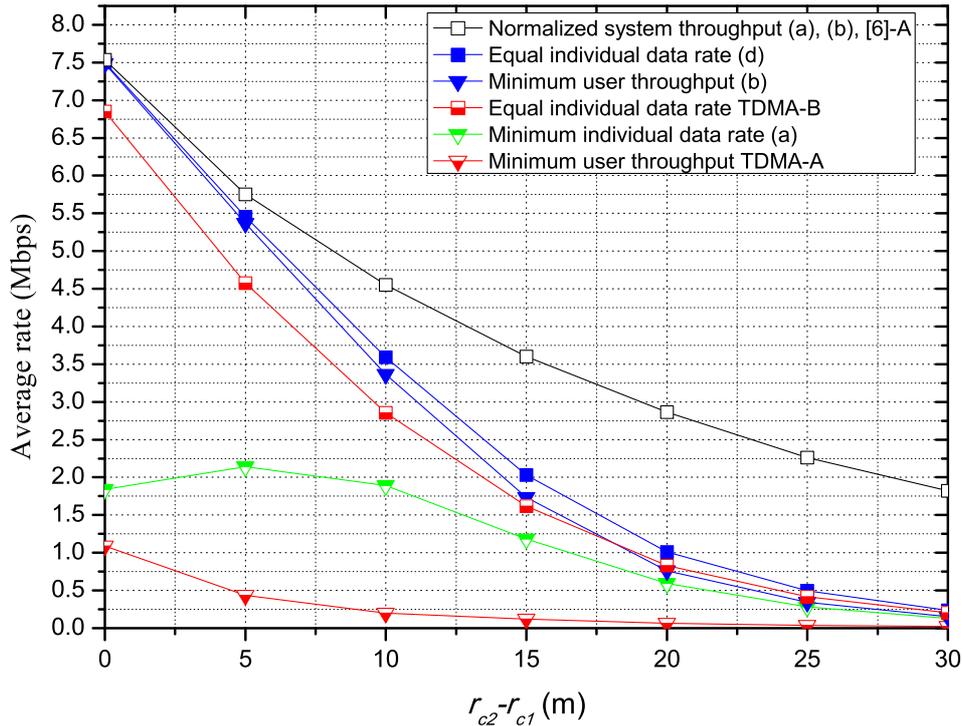}
%\vspace{-0.2in}
\caption{Comparison of average throughput, when $N=3$.}
\label{distance}
\end{figure}

In Fig. \ref{distance}, the impact of distance on performance is illustrated, by altering the radius of the outer circle $r_{c2}$, such as $0\leq r_{c2}-r_{c1}\leq 30 $. As it can be observed, similarly to TDMA, when NOMA with TS is used, both average and minimum throughput reduce as $r_{c2}$ increases, since the users are situated further from the BS. An interesting observation can be drawn from the curve corresponding to the minimum rate achieved by scheme (a). One can notice that, for similar $r_{c1}$ and $r_{c2}$, the minimum rate is lower than for greater distances, in contrast to all other schemes where the rate values decrease with distance. This is because when $r_{c1}\approx r_{c2}$ the users tend to have very similar channel conditions. Thus, the user with slightly better conditions sees strong interference from all the other users for the whole period of time, $1-T$, which leads to a severe degradation of its rate, i.e. the minimum rate among users. This does not happen for greater distance, since the users present variety regarding their channel conditions, so a descending decoding order can improve the minimum rate. Similarly, when TS is used for small distances, it improves the minimum rate, since it interchanges the users that suffer from interference, offering fairness. Comparing the minimum rate (triangle markers) of schemes (a), (b), and [A]-TDMA which maximize the system throughput, both NOMA schemes outperform TDMA, but it is also evident that scheme (b) which employs TS can offer substantial gains, especially for small distances. Finally, scheme (d) outperforms scheme [B]-TDMA in terms of achieved equal data rate, regardless of the distance.

%\vspace{-0.15in}
\subsubsection{Trade-off Between EH and Information Transmission}
\begin{figure}[t!]
\centering
\includegraphics[width=0.8\linewidth]{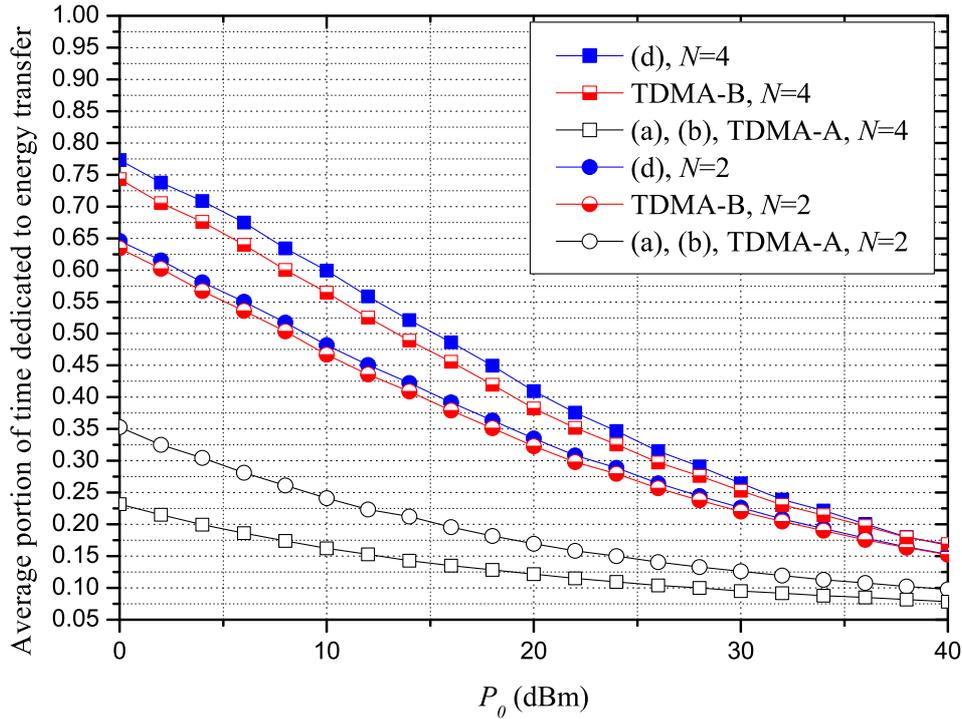}
%\vspace{-0.2in}
\caption{Comparison among (a), (b), (d), [6]-A, and [6]-B in terms of portion of time dedicated to energy transfer.}
\label{Fig7}
\end{figure}

In Fig. \ref{Fig7}, the charging time is depicted when the system throughput and the equal individual data rate are maximized, for $N=2$ and $N=4$. As it can be observed, when the aim is to maximize the system throughput and the number of users increases, the portion of time dedicated to energy transfer is reduced. This happens because the system throughput is mainly affected by the individual data rates of the users with good channel conditions, the average number of which increases with $N$, since uniform distribution has been assumed for the users' locations. Moreover, the users with good channel conditions tend to prefer lower values of $T^*$, compared to those with worse channel conditions, in order to improve their individual data rates. In other words, they have enough energy to transmit and, as a result, their sensitivity to the resource of time dedicated to information transmission increases. 

On the other hand, when the equal individual data rate is maximized, the weakest user, i.e. the one with the worst channel conditions, must have enough energy supply to achieve the equal individual data rate. In this case, as the number of users increases, the portion of time dedicated to energy transfer also increases. Moreover, NOMA dedicates slightly more time to EH compared to TDMA. The reason for this is that NOMA exploits more efficiently the time dedicated to information transmission, requiring less time for achieving the same equal data rate with TDMA. However, when the time dedicated to energy transfer increases, the energy consumption to the BS's side also increases. The last observation motivates the comparison of the two schemes, i.e. NOMA and TDMA, in terms of energy efficiency.
%\vspace{-0.15in}
\subsubsection{Energy Efficiency}
\begin{figure}
\centering
\includegraphics[width=0.8\linewidth]{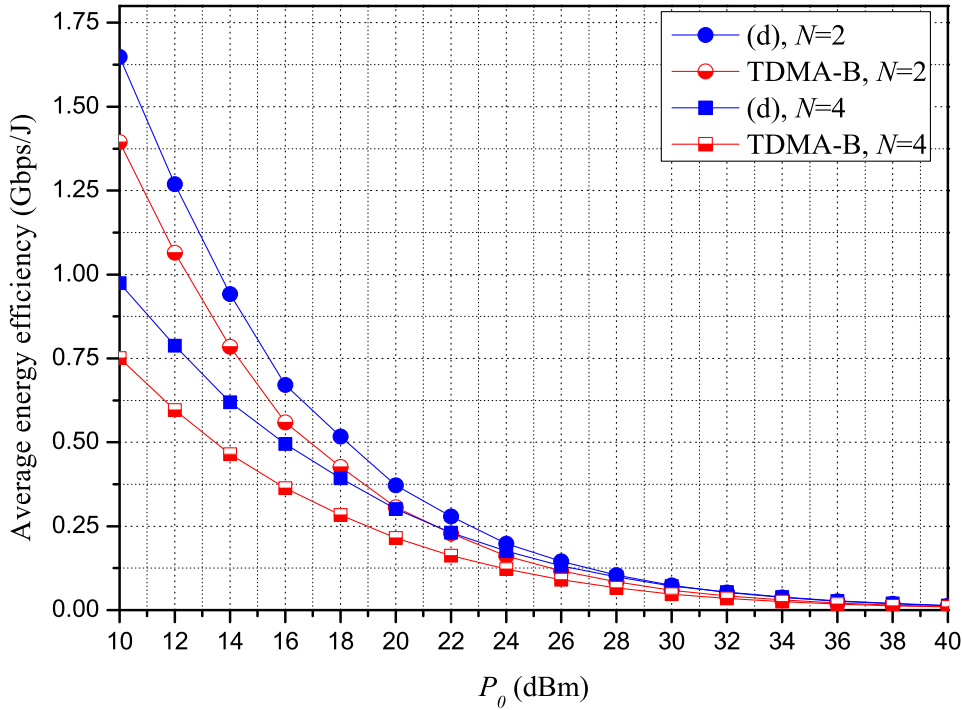}
%\vspace{-0.2in}
\caption{Comparison of average energy efficiency between (d) and [6]-B.}
\label{Fig8}
\end{figure}

When the optimization objective is the maximization of the total system throughput, schemes (a), (b), and [A]-TDMA achieve the exact same system throughput, as seen by Fig. \ref{Fig5}, and harvest the exact same amount of energy, as implied by Fig. \ref{Fig7}. Thus, their energy efficiency is the same. Consequently, this subsection compares the energy efficiency of the schemes which aim to maximize the common rate, i.e., schemes (d) and [B]-TDMA.
The efficiency of the energy transmitted by the BS, denoted by $\mathcal{E}$, when equal transmission rate is required among users, is defined as
\begin{equation}
\mathcal{E}=\frac{N\mathcal{R}_{\mathrm{eq}}}{P_0T}.
\end{equation}
In Fig. \ref{Fig8}, NOMA and TDMA are compared in terms of energy efficiency. It is remarkable that, although NOMA dedicates more time to EH when compared to TDMA, i.e. more energy is transmitted, it achieves higher energy efficiency for the whole range of $P_0$. This is because NOMA achieves much higher individual rates compared to TDMA.
%Another important observation is that the energy efficiency is decreased for both schemes, when the value of $P_0$ increases. This indicates that, increasing the power transmitted by the BS, cannot proportionally improve the performance, introducing an interesting trade-off between desired energy efficiency and achieved data rates.
%\vspace{-0.15in}
\subsubsection{Fairness Comparison}
\begin{figure}
\centering
\includegraphics[width=0.8\linewidth]{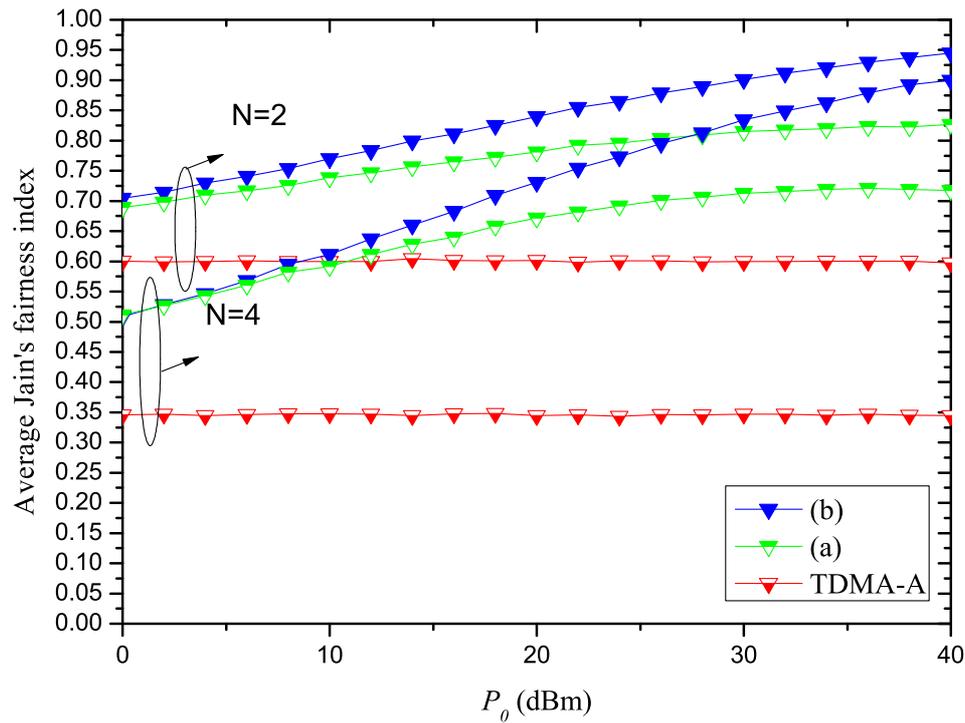}
%\vspace{-0.2in}
\caption{Jain's fairness index comparison.}
\label{Fig9}
\end{figure}

In order to fairly compare the two schemes (NOMA and TDMA) when asymmetric user rates are allowed, in Fig. \ref{Fig9} the Jain's fairness index, $\mathcal{J}$, is used,  which is given by \cite{Tafazoli}
\begin{equation}
\mathcal{J}=\frac{(\sum_{n=1}^NR^\mathrm{u}_n)^2}{N\sum_{n=1}^N(R^\mathrm{u}_n)^2}.
\end{equation}
Note that Jain's fairness index is bounded between 0 and 1, with unitary value indicating equal users' rates. It is seen in Fig. \ref{Fig4section2} that NOMA provides more fairness compared to TDMA, for the whole range of $P_0$, without affecting the total system throughput. This observation is valid also for different number of users. More specifically, scheme (b) is fairer than scheme (a), due to the fact that TS and the corresponding optimization offer a better balance between different users' rates. When $P_0$ is low, schemes (a) and (b) present similar results, because the difference between achievable throughput of the users is greater, and in these cases TS might not offer substantial improvement, i.e. the weakest user might need to decode always free from interference, and still achieve very low rates, as it was also noted in the doubly near-far case. However, as the transmitted power grows, so do the achievable throughput values of the users, making the decoding order permutations performed by TS meaningful. Another observation is that, when the number of users increases the Jain's fairness index decreases. This is because there is greater diversity of user channel conditions, therefore making fairness improvement a more difficult objective.
%\vspace{-0.15in}

%\begin{figure}
%\centering
%\subfigure[Scheme (b)]{
%\includegraphics[width=0.8\linewidth]{figures/chapter3a/fig11.eps}
%\label{Fig10}}
%\subfigure[Scheme (d)]{
%\includegraphics[width=0.8\linewidth]{figures/chapter3a/fig12.eps}
%\label{Fig12}}
%%\vspace{-0.1in}
%\caption{Evaluation of the convergence speed of the greedy algorithm.}
%\end{figure}

\begin{figure}
\centering
\includegraphics[width=0.8\linewidth]{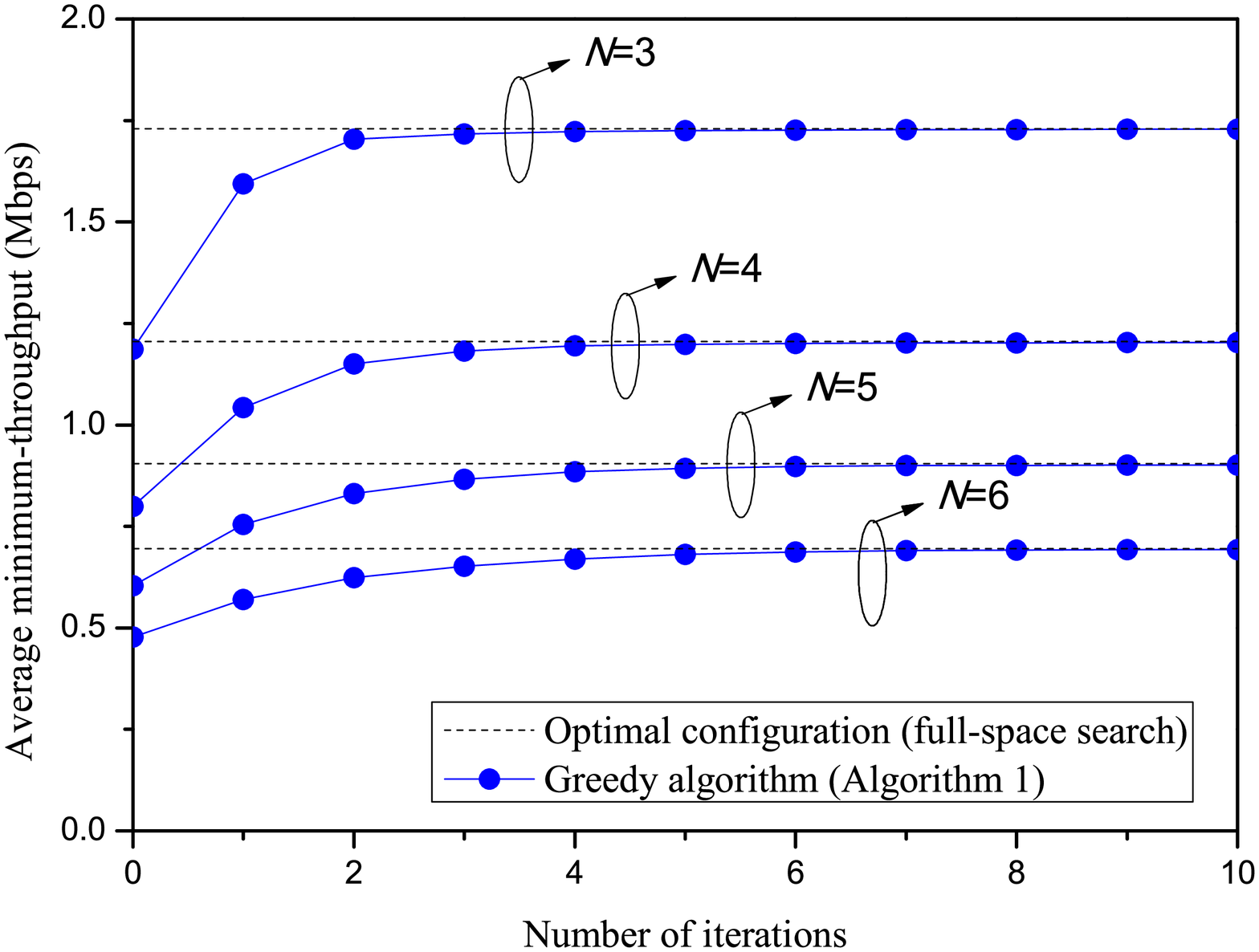}
%\vspace{-0.2in}
\caption{Evaluation of the convergence speed of the greedy algorithm for Scheme (b).}
\label{Fig10}
\end{figure}

\begin{figure}
\centering
\includegraphics[width=0.8\linewidth]{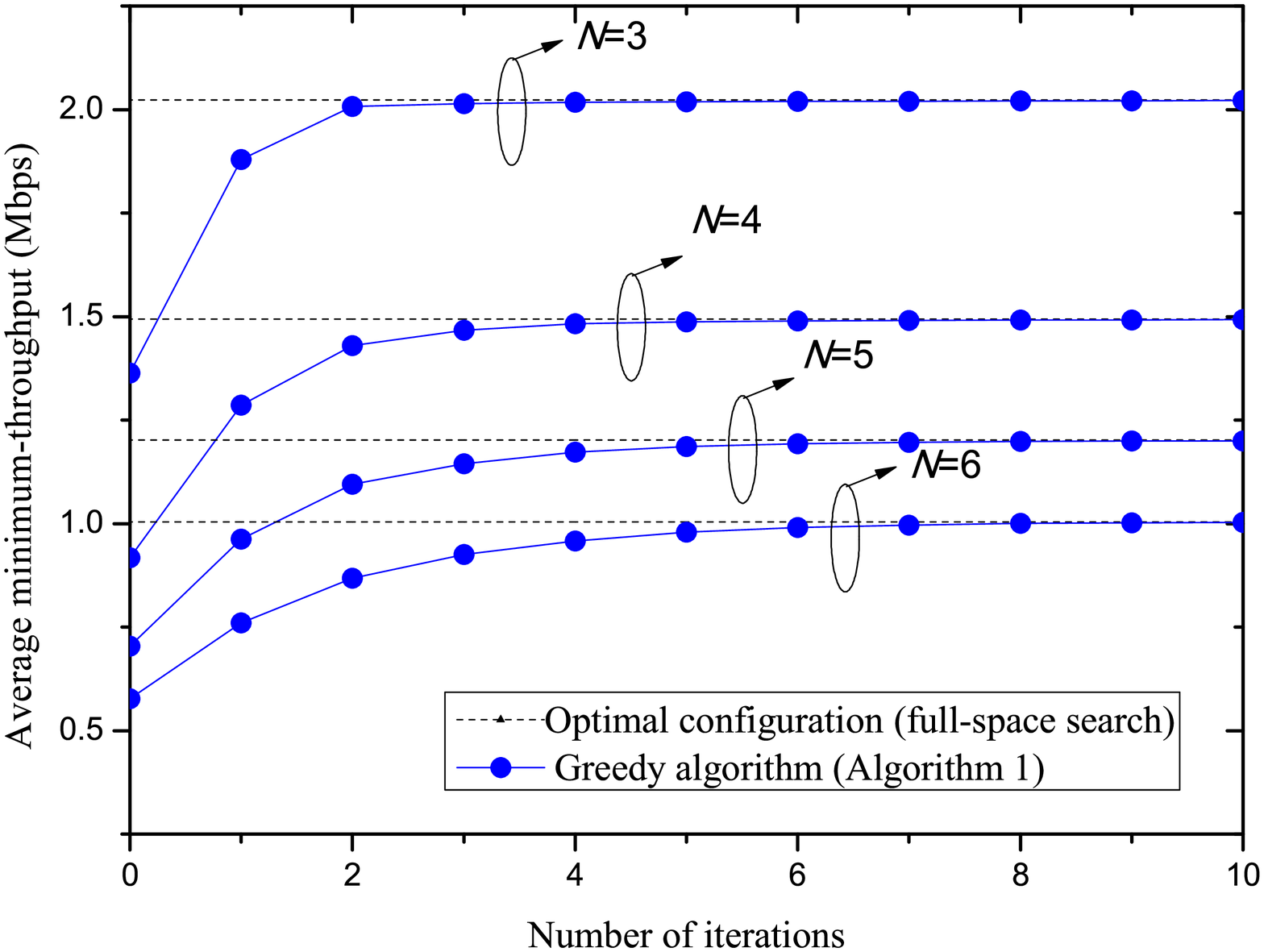}
%\vspace{-0.2in}
\caption{Evaluation of the convergence speed of the greedy algorithm for Scheme (d).}
\label{Fig11}
\end{figure}

\begin{table*}
\centering \caption{Algorithm complexity}\label{dimensions}
\begin{tabular}{lll}
\hline 
\textbf{Number of users} &\textbf{Optimal Scheme}&\textbf{Algorithm 1}\\
&$(N+1)\times (N!+1)$&$(N+1)\times (k+2),\,\forall k=\{1,...,N+1\}$\\
\hline
$N=3$& $4\times 7$& $4\times (3,4,5,6)$\\
$N=4$& $5\times25$& $5\times(3,4,5,6,7)$\\
$N=5$ & $6\times 121$& $5\times (3,4,5,6,7,8)$\\
$N=6$ &$7\times 721$& $7\times(3,4,5,6,7,8,9)$\\
$N=7$&$ 8\times 5041$& $8\times (3,4,5,6,7,8,9,10)$\\
$N=8$& $9\times 40321$& $9\times (3,4,5,6,7,8,9,10,11)$\\
\hline
\end{tabular}
\end{table*}
\subsubsection{Convergence of the Greedy Algorithm}

Fig. \ref{Fig10} illustrates the evolution of the average minimum user throughput when the proposed greedy algorithm is used for the TS configuration of scheme (b). Similarly, Fig. \ref{Fig11} presents the corresponding results for scheme (d). In particular, the convergence speed of the proposed algorithm for $P_0=20$ dBm and $N=3,4,5,6$ is investigated. The dashed lines denote the minimum user throughput for each case study.

The number of iterations that are required for Algorithm 1 to converge increases with the number of users. This is because, as the number of users increases, the possible permutations also increase, slowing the convergence of Algorithm 1. However, it is observed that the proposed iterative algorithm converges to the optimal value within $N+1$ iterations. Thus, the proposed technique reduces the maximum number of permutations that have to be considered by the full-space search from $N!$ (when all permutations are considered) to $N+2$. This is also evident in Table \ref{dimensions}, where the problem dimensions of the full-space search and of Algorithm 1 are shown, for different number of users, with $k$ being the index of the iteration. Although Algorithm 1 needs to solve more optimization problems, their dimensions are orders of magnitude smaller, especially as the number of users increases, enabling the utilization of the TS technique in practical implementations.

\section{Maximizing Proportional Fairness in Wireless Powered Communications}\label{Maximizing Proportional Fairness in Wireless Powered Communications}
In this section, the proportional fairness is maximized, as a means to balance the trade-off between the sum-rate and fairness, which has been presented in the previous sections. Two communication protocols are, i.e., TDMA and NOMA with TS (NOMA-TS), when asymmetrical (unequal) rates are allowed. The formulated problems are optimally and efficiently solved, while simulate results illustrate the effectiveness of the proposed methods. The results of the research presented in this section are included in \cite{Proportional_fairness}.

\subsection{Related Work and Motivation}
As it has been shown in the previous section the harvest-then-transmit protocol induces a fundamental trade-off between 
sum rate and fairness. More speicifically, when the sum rate is maximized, fairness is considerably reduced, due to the 
doubly near-far problem. On the other hand, equal rate maximization considerably 
reduces the sum-rate, which, in case of NOMA, also happens because the achievable capacity of some users might exceed the 
maximum equal rate.

In order to balance the trade-off between performance and fairness, an alternative metric, such as the proportional fairness, can be used and maximized. This metric is widely applied in wireless networks in order to balance user fairness and network sum-rate. This scenario has been partially studied in \cite{benchmark}, under the assumption of NOMA with fixed decoding order and adaptive power allocation, which is considered as benchmark in the present section. As defined in the previous section, the fixed decoding order implies that the BS decodes the users messages during the SIC process with a constant order. This assumption reduces the degrees of freedom and degrades the system's performance, since some users always experience more interference and their messages are always decoded last. Moreover, NOMA with adaptive power allocation induces higher complexity than NOMA with fixed power allocation and especially TDMA.

\subsection{Contribution}
In this work, the proportional fairness is maximized by proper time and rates allocation, considering two communication protocols, i.e., TDMA and NOMA-TS, when asymmetrical (unequal) rates are allowed.  Note that NOMA-TS is a generalization of NOMA with fixed decoding order, so that a user, whose message suffers from strong interference for a specific decoding order, can experience a better reception reliability for another decoding order, during the implementation of SIC. The corresponding optimization problems are efficiently solved in closed form, by using convex optimization, and more specifically Lagrange dual decomposition. The evaluation of the proposed strategies through extensive simulations reveals that NOMA-TS maximizes proportional fairness and outperforms the benchmark, while TDMA proves to be an appropriate choice, when all users are located in similar distances from the BS.

\subsection{System model}
%\vspace{-0.1 in}
The uplink of a wireless network is considered, consisting of $N$ EH users and one BS. The harvest-then-transmit is utilized, with similar assumptions as in the previous section. As mentioned above, two different communication protocols are considered during uplink (Phase $2$), namely TDMA and NOMA-TS. The achievable rates that correspond to each of the considered protocols are described in the following subsections. According to \eqref{time constraint} and \eqref{userthroughputtdma} when TDMA is assumed the achievable rates are constrained by 
\begin{equation}
R^\mathrm{u}_n\leq t_n\log_2\left(1+\frac{\eta \rho_0 g_nT}{t_n}\right).
\label{TDMA capacity constraint}
\end{equation}
and
\begin{equation}
\sum_{n=1}^Nt_n\leq 1-T.
\label{TDMA time constraint}
\end{equation}

Also, when NOMA is used the achievable rates, denoted by $\tilde{R}^\mathrm{u}_n$, are bounded by \eqref{time_sharing_rates_constraint}.
In contrast to the previous section, the aim of the proposed analysis is to maximize the system's performance while achieving a balance between the sum-rate and fairness. To this end, the \textit{proportional fairness} is used, which is defined as the sum of logarithms of the individual rates. Next, the proportional fairness maximization problem is defined and solved, for TDMA and NOMA-TS.

\subsection{Problem Formulation and Solution}
%In general, for the solution of the corresponding optimization problems, $\textit{Lagrange dual decompostion}$ is used, which proves to be extremely efficient, since, given the Lagrange multipliers (LMs), the optimal rates and time-allocation can be directly calculated in parallel.
\subsubsection{TDMA}
When TDMA is used the proportional fairness maximization problem can be expressed as
%===========================================================
\begin{equation}
\begin{array}{ll}
\underset{R^\mathrm{u}_n,t_n,\forall n\in \mathcal{N},T}{\text{\textbf{max}}}& \sum_{n=1}^N \ln(R^\mathrm{u}_n)\\
\quad\,\,\,\,\text{\textbf{s.t.}}& \mathrm{C}_1: R^\mathrm{u}_n\leq t_n\log_2\left(1+\frac{\eta \rho_0 g_nT}{t_n}\right),\\
&\mathrm{C}_2: \sum_{n=1}^Nt_n\leq 1-T,\\
&\mathrm{C}_3:t_n,T\geq 0.   \\
\end{array}
\label{opt TDMA}
\end{equation}
%=============================================

%===========================================================
\begin{equation}
\begin{array}{ll}
\underset{R^\mathrm{u}_n,t_n,\forall n\in \mathcal{N},T}{\text{\textbf{max}}}& \sum_{n=1}^N \ln(R^\mathrm{u}_n)\\
\quad\,\,\,\,\text{\textbf{s.t.}}& \mathrm{C}_1: \text{Eq.\,}\eqref{TDMA capacity constraint}, \mathrm{C}_2: \text{Eq.\,} \eqref{TDMA time constraint}, \mathrm{C}_3:t_n,T\geq 0.   \\
\end{array}
\label{opt TDMA}
\end{equation}
%=============================================

The optimization problem in \eqref{opt TDMA} is concave and can be solved by standard numerical methods, such
as interior point or bisection method. Here, the \textit{Lagrange dual decomposition} is used, which proves to be very efficient, since, given the LMs, closed form expressions for $R_n,t_n,\forall n\in \mathcal{N},T$ are derived. Thus, it is guaranteed that the optimal solution can be obtained in a polynomial time \cite{Boyd1}.

\begin{theorem}\label{theorem1}
Considering the optimization problem in \eqref{opt TDMA}, let $\mu\geq 0$ and $\lambda_n\geq 0$ denote the LMs, which corresponds to the constraints $\mathrm{C}_1$ and $\mathrm{C}_2$, respectively. Then, the optimal rate allocation policy is given by
\begin{equation}
{R^\mathrm{u}_n}^*=\frac{1}{\lambda_n}.
\label{optimal rate TDMA}
\end{equation}
The optimal $t_n$ is given by
\begin{equation}
t_n^*=\frac{\eta \rho_0 g_nT^*}{\tilde{A}_n^*-1},
\label{optimal tn TDMA}
\end{equation}
where
\begin{equation}
\tilde{A}_n^*=-\frac{1}{\mathcal{W}\left(-\exp\left(-\frac{1+\mu\ln(2)}{\lambda_n}\right)\right)},
\label{optimal A TDMA}
\end{equation}
with $W_0(\cdot)$ being the principal branch of the Lambert W function \cite{lambert,harvesting} and $\mu$ can be found after solving (\ref{optimal m})
\begin{equation}
\sum_{n=1}^N\frac{\eta \rho_0 g_n\lambda_n}{\tilde{A}_n^*\ln(2)}-\mu=0.
\label{optimal m}
\end{equation}
Finally, the optimal time allocation between the two phases is
\begin{equation}
T^*=\frac{1}{\sum_{n=1}^N\frac{\eta \rho_0 g_n}{\tilde{A}_n^*-1}+1}.
\label{optimal T TDMA}
\end{equation}
\end{theorem}

\begin{proof}
Since \eqref{opt TDMA} is a concave optimization problem, it can be optimally solved by \textit{Lagrange dual decomposition} \cite{Boyd1}. To this end, the Lagrangian is needed, which is given by
\begin{equation}
\mathcal{L}=\sum_{n=1}^N \ln(R^\mathrm{u}_n)-\mu \left(\sum_{n=1}^Nt_n+T-1\right)
-\lambda_n\left(R^\mathrm{u}_n-t_n\log_2\left(1+\frac{\eta \rho_0 g_nT}{t_n}\right)\right).
\end{equation}
%\begin{equation}
%\mathcal{L}=\sum_{n=1}^N \ln(R_n)-\mu \left(\sum_{n=1}^Nt_n+T-1\right)-\lambda_n\left(R_n-t_n\log_2\left(1+\frac{a_nT}{t_n}\right)\right).
%\end{equation}
According to the KKT conditions, it must hold that $\frac{\partial \mathcal{L}}{\partial R^\mathrm{u}_n}=0$, $\frac{\partial \mathcal{L}}{\partial T}=0$, and $\frac{\partial \mathcal{L}}{\partial t_n}=0,\forall n\in \mathcal{N}$, which yield \eqref{optimal rate TDMA}, \eqref{optimal m}, and
\begin{equation}
\frac{(\tilde{A}_n-1)\lambda_n}{\tilde{A}_n}+\mu\ln(2)-\lambda_n \ln(\tilde{A}_n)=0,
\label{optimal tn}
\end{equation}
respectively, where $\tilde{A}_n=\frac{\eta \rho_0 g_nT}{t_n}+1$.
Using \eqref{optimal tn} and after some manipulations \eqref{optimal tn TDMA} is derived. Finally, it is easy to prove that the constraint $\mathrm{C}_1$ should hold with equality, thus the optimal time allocation between the two phases is given by \eqref{optimal T TDMA}. Note that in (\ref{optimal rate TDMA}) and (\ref{optimal A TDMA}), $\lambda_n$ is calculated iteratively by using the subgradient method.
\end{proof}

\subsubsection{NOMA-TS}
When NOMA-TS is used the proportional fairness maximization problem can be expressed as
%===========================================================
\begin{equation}
\begin{array}{ll}
\underset{\tilde{R}^\mathrm{u}_n\forall n\in \mathcal{N},T}{\text{\textbf{max}}}& \sum_{n=1}^N \ln(\tilde{R}^\mathrm{u}_n)\\
\quad\,\text{\textbf{s.t.}}& \text{Eq.\,}\eqref{time_sharing_rates_constraint}, 0\leq T\leq1.   \\
\end{array}
\label{opt1}
\end{equation}
%=============================================

For a specific value of $T$, in which case all constraints are linear, the form of the objective function of \eqref{opt1} infers that,  the achievable rates of the users with the worst channel conditions will be increased as much as possible, with the aim to become equal with the ones with better channel conditions, as long as they do not exceed the sum achievable rate of the belonging subset. Taking this into account, the proportional fairness is only constrained by the achievable rate of the subsets consisting of the users with the weaker channel conditions, among the ones with the same cardinality. The above can easily be verified by considering the Lagrangian and the KKT conditions that correspond to \eqref{opt1}, and by noticing that the (non-negative) LMs that correspond to the constraints that limit the sum achievable rate of subsets with the same cardinality cannot be simultaneously non-zero. Thus, the sole active constraints are:
\begin{equation}
\sum_{n=N+1-i}^N\tilde{R}^\mathrm{u}_n\leq (1-T)\log_2\left(1+\frac{\eta\rho_0 \sum_{n=N+1-i}^N{g}_nT}{1-T}\right),\forall i=1,...,N,
\label{NOMA capacity constraint}
\end{equation}
 and \eqref{opt1} is simplified to
%===========================================================
\begin{equation}
\begin{array}{ll}
\underset{\tilde{R}_n\forall n\in \mathcal{N},T}{\text{\textbf{max}}}& \sum_{n=1}^N \ln(\tilde{R}^\mathrm{u}_n)\\
\quad\,\text{\textbf{s.t.}}& \text{Eq.\,}\eqref{NOMA capacity constraint}, 0\leq T\leq1.   \\
\end{array}
\label{opt1_simplified}
\end{equation}
%=============================================

Regarding \eqref{opt1_simplified}, it is hard to directly solve for $T$, since it appears in all the capacity equations in \eqref{NOMA capacity constraint}. Thus, in order  to avoid the utilization of Newton-Raphson method \cite{NOMA_WPT_joutnal}, and decrease the corresponding complexity, the auxiliary variables $T_i$ are introduced. Thus, \eqref{opt1} can be reformulated as
%===========================================================
\begin{equation}
\begin{array}{ll}
\underset{R_n\forall n\in \mathcal{N},T_m}{\text{\textbf{max}}}& \sum_{n=1}^N \ln(\tilde{R}^\mathrm{u}_n)\\
\text{\quad\,\,\,\,\textbf{s.t.}}&\mathrm{C}_1: \sum_{n=N+1-i}^N\tilde{R}^\mathrm{u}_n\leq (1-T_i)\log_2\left(1+\frac{\eta\rho_0 \sum_{n=N+1-i}^N{g}_nT_i}{1-T_i}\right),\\
&\quad\quad\quad\quad\quad\quad\quad\quad\quad\quad\quad\forall i=1,...,N,   \\
&\mathrm{C}_2: T_i=T_N,\, \forall i=1,...,N-1,\\
&\mathrm{C}_3: 0\leq T_i\leq 1,\, \forall i=1,...,N.\\
\end{array}
\label{opt2section3}
\end{equation}
%=============================================

\begin{theorem}\label{theorem2}
Let $\lambda_i\geq 0$ and $\mu_i$ denote LMs, which correspond to the constraints $\mathrm{C}_1$ and $\mathrm{C}_2$, respectively. Then, the optimal rate allocation policy is given by
\begin{equation}
{\tilde{R}}^{\mathrm{u}*}_n=\frac{1}{\sum_{i=N+1-n}^N \lambda_i},
\label{optimal rate NOMA}
\end{equation} 
and the optimal time allocation policy, $T^*$, is  
\begin{equation}
T^*=T^*_i=\frac{\tilde{B}_i^*-1}{\eta\rho_0 \sum_{n=N+1-i}^N{g}_i+\tilde{B}_i^*},
\label{optimal T NOMA}
\end{equation}
where
\begin{equation}
\tilde{B}^*_i=\frac{\eta\rho_0 \sum_{n=N+1-i}^N{g}_i-1}{W\left((\eta\rho_0 \sum_{n=N+1-i}^N{g}_i-1)\exp(\frac{\tilde{\mu}_i}{\lambda_i}-1)\right)},
\label{optimal B NOMA}
\end{equation}
and
\begin{equation}
\tilde{\mu}_i=\begin{cases}&\mu_i\ln(2),\,i=1,...,N-1,\\&-\sum_{j=1}^{N-1}\mu_j\ln(2),\,i=N.\end{cases}
\end{equation}
\end{theorem}

\begin{proof}
The optimization problem in \eqref{opt2section3} is concave and it can be also solved by using {Lagrange dual decomposition} \cite{Boyd1}.
The Lagrangian is given by 
\begin{equation}
\begin{split}
\mathcal{L}=&\sum_{n=1}^N \ln(\tilde{R}^\mathrm{u}_n)-\sum_{i=1}^{N-1}\mu_i\left(T_i-T_N\right)\\
&-\sum_{i=1}^N\lambda_i\left(\sum_{n=N+1-i}^N\tilde{R}^\mathrm{u}_n- (1-T_i)\log_2\left(1+\frac{\eta\rho_0 \sum_{n=N+1-i}^N{g}_nT_i}{1-T_i}\right)\right).\\
\end{split}
\end{equation}
According to KKT conditions, $\frac{\partial \mathcal{L}}{\partial \tilde{R}^\mathrm{u}_n}=0,\,\forall{n}$, from which \eqref{optimal rate NOMA} is derived, and  $\frac{\partial \mathcal{L}}{\partial T_i}=0,\forall i=1,..,N$, which yields
\begin{equation}
\lambda_i\frac{\eta\rho_0 \sum_{n=N+1-i}^N{g}_n+\tilde{B}_i-1}{\tilde{B}_i}-\lambda_i\ln(\tilde{B}_i)-\tilde{\mu}_i=0,\forall i,
\label{KKT NOMA 1}
\end{equation}
where
\begin{equation}
\tilde{B}_i=1+\frac{\eta\rho_0 \sum_{n=N+1-i}^N{g}_nT_i}{1-T_i}.
\label{KKT NOMA 3}
\end{equation}
Solving \eqref{KKT NOMA 1} and \eqref{KKT NOMA 3} for $T_i$, $T^*$, is given by \eqref{optimal T NOMA}. Also, $\lambda_i$ is calculated iteratively by the subgradient method.
\end{proof}

%\vspace{-0.1 in}

\subsection{Simulations and Discussion}
%\vspace{-0.05 in}

\begin{figure}[t!]
\centering
\includegraphics[width=0.8\columnwidth]{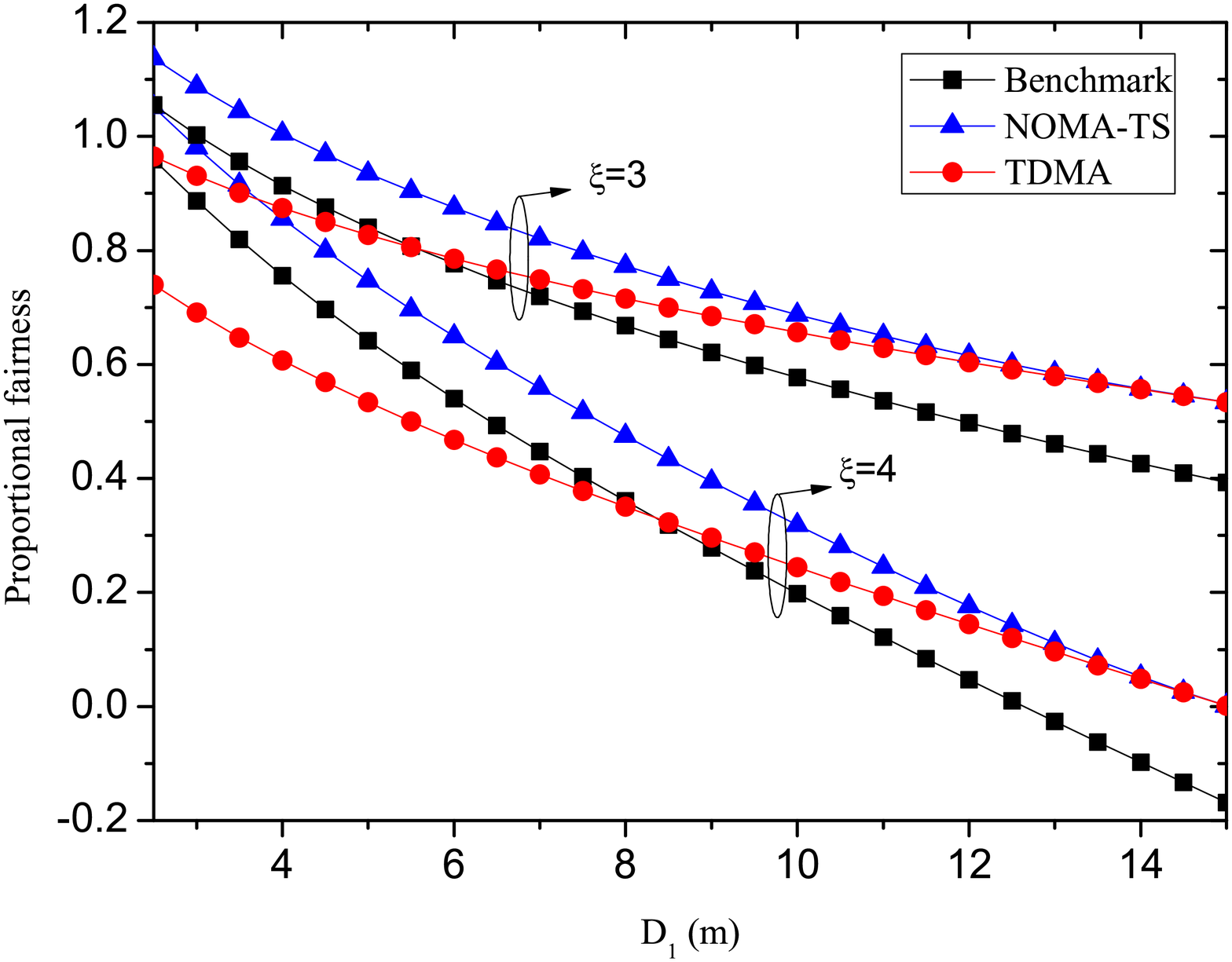}
\caption{Proportional fairness versus $D_1$ for $N_1=7$ and $N_2=3$.}
\label{Fig1a}
\end{figure}

\begin{figure}[t!]
\centering
\includegraphics[width=0.8\columnwidth]{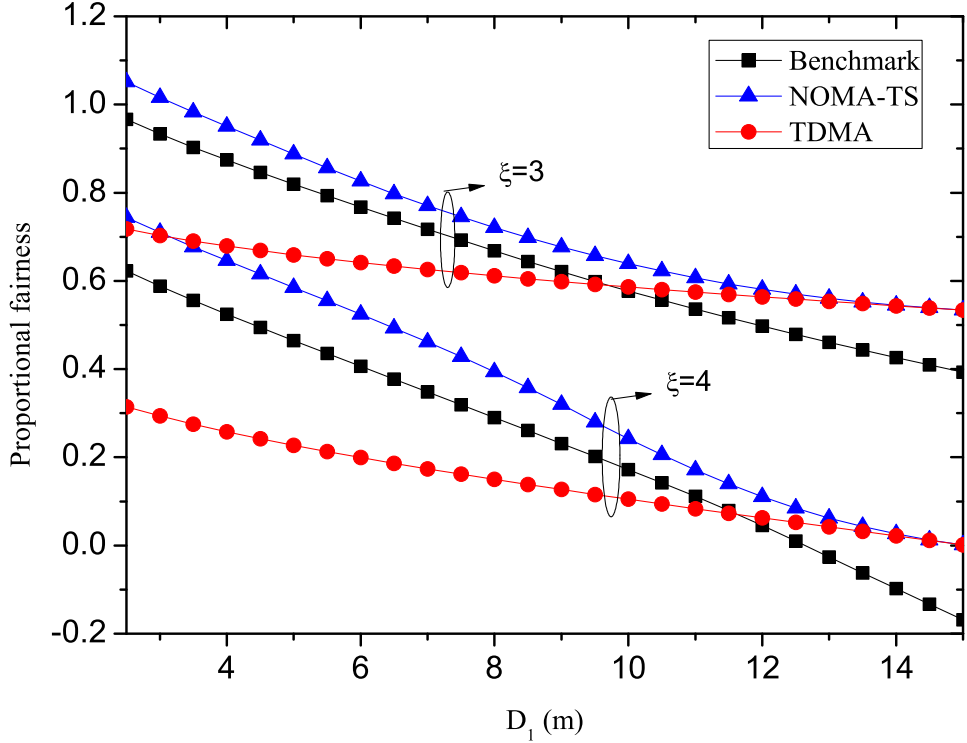}
\caption{Proportional fairness versus $D_1$ for $N_1=3$ and $N_2=7$.}
\label{Fig1b}
\end{figure}

Two users' groups are considered, each of which is located in different distance, $D_{i\in\{1,2\}}$, from the BS. Let $N_{i}$ denote the number of users of the $i$-th group. It is assumed that the $1$-st group is located closer to the BS.  It is further assumed that $\eta=0.5$, $\gamma_n=10^{-3}d_n^\xi$, where $d_n$ is the distance of the $n$-th user from the BS, $\xi$ is the pathloss exponent, $N_0W=-160$ dBm, and $P_0=1$ W \cite{benchmark}. The provided results focus on the impact of the number of users, pathloss value, and distances $D_1$ and $D_2$ on performance, while the proposed schemes are compared to the benchmark one in \cite{benchmark}, i.e., NOMA with fixed decoding order and adaptive power allocation. 

\begin{figure}[t!]
\centering
\includegraphics[width=0.8\columnwidth]{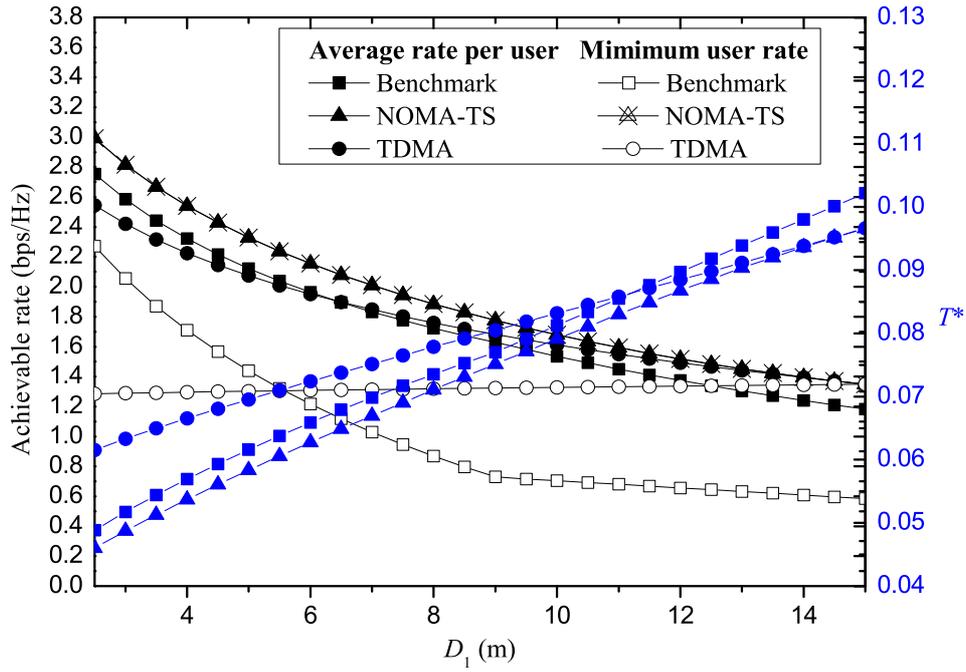}
%\vspace{-0.15in}
\caption{Achievable rate and $T^*$ for $N1=7$ and $N_2=3$.}
\label{Fig2}
%\vspace{-0.15in}
\end{figure}

\begin{figure}[t!]
\centering
\includegraphics[width=0.8\columnwidth]{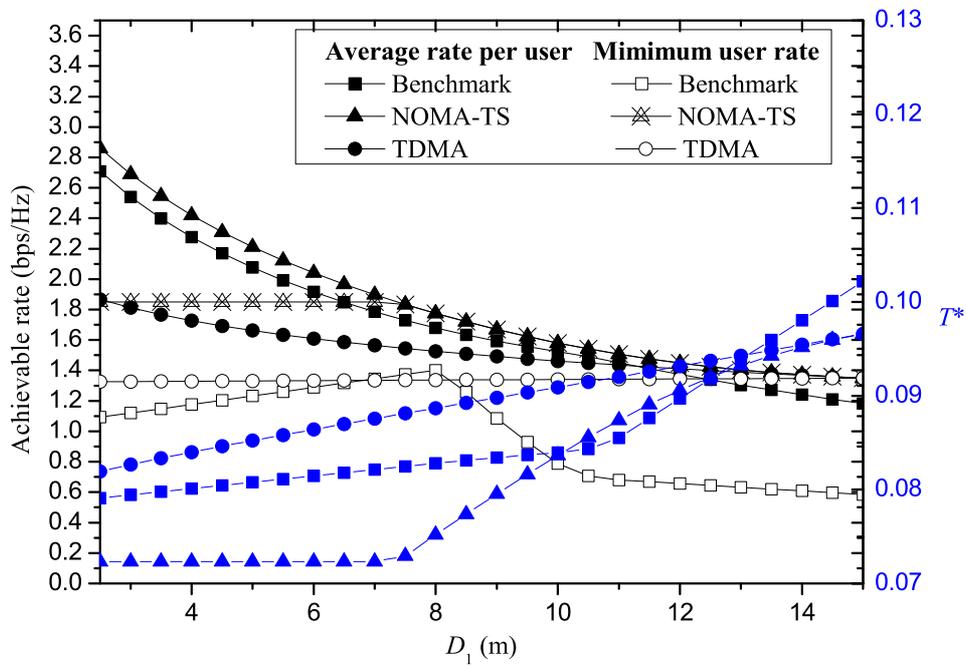}
%\vspace{-0.15in}
\caption{Achievable rate and $T^*$ for $N1=3$ and $N_2=7$.}
\label{Fig3}
%\vspace{-0.15in}
\end{figure}

Figs. \ref{Fig1a}, \ref{Fig1b}, \ref{Fig2}, and \ref{Fig3} focus on the effect of $D_1$ on performance, while assuming $D_2=15$ m and groups with unequal number of users, i.e., $N_1\neq N_2$. More specifically, In Figs. \ref{Fig1a} and \ref{Fig1b}, the maximum proportional fairness achieved by each scheme is illustrated, for two values of $\xi$, i.e., $\xi=3$ and $\xi=4$. When $N_1>N_2$, higher proportional fairness is achieved for all the considered protocols, especially when $D_1$ is small. Another notable observation from Figs. \ref{Fig1a} and \ref{Fig1b} is that NOMA-TS outperforms the other two schemes for the whole range of $D_1$ despite the values of $\alpha,N_1,N_2$, while it performs much better than TDMA, when $N_2>N_1$ and $\alpha$ increases, indicating the resilience of NOMA to higher pathloss values. Also, TDMA outperforms the benchmark NOMA scheme for the higher values of $D_1$ and especially when $N_1>N_2$. Finally, when $D_1=D_2$ the two proposed schemes achieve exactly the same performance. In this case, TDMA should be preferred, since in general, NOMA increases the decoding complexity at the BS.

In Figs. \ref{Fig2} and \ref{Fig3} the average rate per user, i.e. the normalized sum rate, the minimum user rate, and the time allocated to EH, are illustrated, versus $D_1$, assuming $\xi=3.5$. It is seen that NOMA-TS outperforms the other two protocols, both in terms of normalized sum rate and minimum rate. This is because TS increases the degrees of freedom and any point of the capacity region can be achieved.  Moreover, Figs. \ref{Fig2} and \ref{Fig3} illustrate that, when $N_1>N_2$, the minimum rate achieved by NOMA-TS is equal to the average rate per user, which however does not happen when $N_2>N_1$ and $D_1$ is small. In contrast to NOMA, the minimum rate achieved by TDMA is not influenced by $D_1$. Furthermore, Figs. \ref{Fig2} and \ref{Fig3} reveal that the number of users in the two groups also affects the optimal value of $T$. More specifically, when $N_2>N_1$ and the $1$-st user group is located close to the BS, the time allocated to the EH phase increases. Finally, it is remarkable that NOMA-TS allocates less time to EH compared to TDMA, for the whole range of $D_1$, increasing the energy efficiency.

%It is evident that when the users have similar rate requirements, TDMA is a better option compared to the benchmark scheme, for medium and large values of $D_1$. Interestingly, for high values of $D_1$, TDMA outperforms the benchmark scheme both in terms of sum rate and minimum rate. 

%\vspace{-0.05 in}
%\begin{figure}[t!]
%\centering
%\includegraphics[width=0.85\columnwidth]{fairness1_case1.eps}
%\vspace{-0.32in}
%\caption{Proportional fairness for $N_1=7$ and $N_2=10$.}
%\label{Fig1}
%\end{figure}
%
%\begin{figure}[t!]
%\centering
%\includegraphics[width=0.85\columnwidth]{fairness1_case2.eps}
%\vspace{-0.32in}
%\caption{Proportional fairness for $N_1=3$ and $N_2=7$.}
%\label{Fig1}
%\end{figure}

%\vspace{-0.05 in}

\begin{figure}[t!]
\centering
\includegraphics[width=0.8\columnwidth]{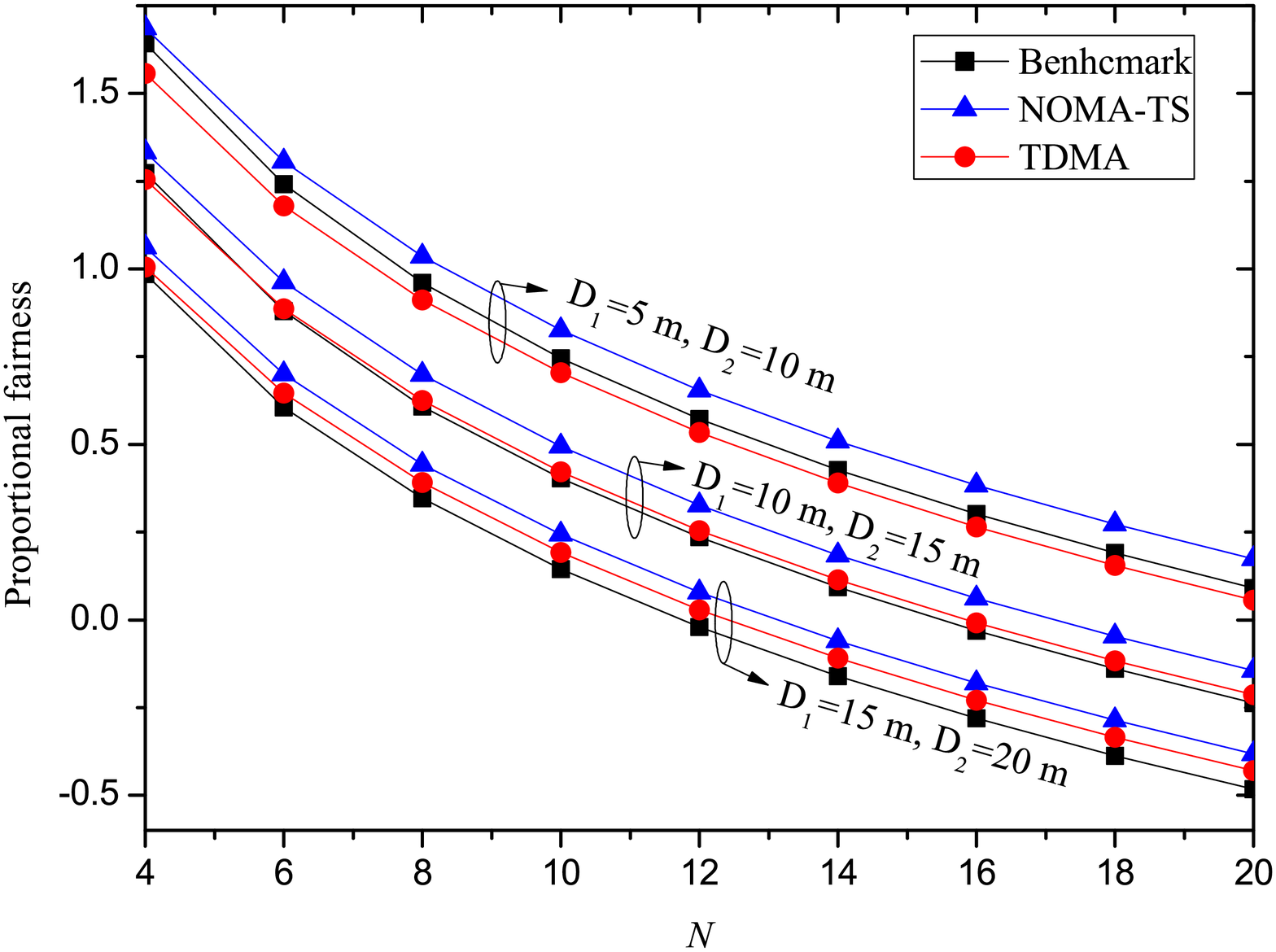}
%\vspace{-0.15in}
\caption{Proportional fairness versus $N$.}
\label{Fig4}
%\vspace{-0.2 in}
\end{figure}

In Fig. \ref{Fig4} the considered schemes are evaluated with respect to the number of users, for
three different distances realizations, i.e., i) $D_1=5$ m and $D_2=10$ m, ii) $D_1=10$ m and $D_2=15$ m, and iii) $D_1=15$ m and $D_2=20$ m, assuming that $N_1=N_2$ and $\xi=3.5$. It is observed that NOMA-TS achieves the highest proportional fairness, which implies that when it is used, instead of the other two protocols, more users can enter to the system. When both groups of user are closely located to the BS (e.g $D_1=5$ m and $D_2=10$ m) NOMA with fixed decoding order and adaptive power allocation outperforms TDMA. The opposite happens when both groups are located far from the BS, e.g., when $D_1=10$ m and $D_2=15$ m and especially when $D_1=15$ m and $D_2=20$ m, in which case the benchmark scheme becomes more prone to interference.

%\vspace{-0.05 in}
%\vspace{-0.05 in}
%\begin{figure}[t!]
%\centering
%\includegraphics[width=0.85\columnwidth]{rates2.eps}
%\vspace{-0.32in}
%\caption{Achievable rates for $D1=10$ m.}
%\label{Fig4}
%\vspace{-0.2 in}
%\end{figure}

%In Fig. \ref{Fig1} the average prices $p_1^*$ and $p_2^*$ are illustrated. It is seen that the optimal prices increase with the number of users. This implies that the demand is increasing, while the available resources and energy are limited. On the other hand, when the harvested energy increases, the price per unit of energy reduces (e.g. $p_1^*=0.00109$ for $N=5$ and $\frac{E}{QN_0}=30$ dB), since the BS has more energy to cover the demand. In contrast to $p_1^*$, increasing the harvested energy leads to an increase in the optimum price $p_2^*$. This is a notable observation, because it indicates that the more the transmitted power increases, the smaller the increase of the users' utilities is, and thus, instead of demanding energy, the demand for the other resources increases.

%%CHAPTER 5 (CONCLUSIONS) %%%%%%%%%%%%%%%%%%%%%%%%%%%%%%%%%%%%%%%%%%%%%%%%%%%%%%%
%%\else\fi
%
%%\ifchapteroneflag
%%CHAPTER 1%%%%%%%%%%%%%%%%%%%%%%%%%%%%%%%%%%%%%%%%%%%%%%%%%%%%%%%%%%%%%%%%%%%%%%
\chapter[Simultaneous Wireless Information and Power  Transfer (SWIPT)][Simultaneous Wireless Information and Power  Transfer (SWIPT)]{Simultaneous Wireless Information and Power  Transfer (SWIPT)}\label{ch:chapter4}

This chapter focuses on SWIPT, which enables  the simultaneous transfer of data using the same RF signals, that are used for WPT. SWIPT can be used, among others, in the downlink of wireless powered communications, as it has already been mentioned in the previous chapter, as well as to support the first hop of wireless powered relaying systems. Both scenarios will be investigated in the following sections, where the corresponding contributions will be presented. More specifically, the general aim in both scenarios is to maximize the performance in terms of achievable rate, taking into account the energy constraints imposed by the SWIPT techniques and their impact on the considered communication protocols.

\section{Introduction to SWIPT} \label{ch:chapter4_1}
SWIPT presupposes the efficient design of the communication system that receives information and energy simultaneously \cite{Grover, Varshney}, which also depends on the specific system implementation \cite{book, RuiChallenges}. The idea of SWIPT has been reported in various scenarios, such as one source-destination pair \cite{KwanWPT, Kwan2}, MIMO communications systems \cite{RuiMIMO, Xiaoming, krik2, Suraweera1, Suraweera2, EE_MISO, EE_MIMO, Xiang}, orthogonal frequency division multiple access (OFDMA) \cite{KwanWPT, OFDMA, EE_uplink_downlink}, cooperative networks \cite{Esnaola, krik1, Ding_path_loss, Matthaiou, OWPR, Gzhu, Dingnew, Dingcooperative, Nasir}, communication systems with security \cite{Hong_Xing, RuiSec1,RuiSec2}, and cognitive radio \cite{Nallanathan1, LiuWang, Mohjazi, SLee, 7122349}.

\subsection{SWIPT Strategies}

In accordance to all previous prominent works, practical circuits for harvesting energy from radio signals are not yet able to decode the carried information directly. In other words, the signal that is used for harvesting
energy cannot be reused for decoding information. In order to overcome this problem in single antenna EH nodes, two strategies have been proposed, i.e., \textit{time-switching (TSw)} and \textit{power-splitting (PS)}. In TSw, during a portion of time, the received signal is used solely for EH, instead of decoding, while PS is based on the division of the signal's power into two streams \cite{architecture, krik3, Caijun1, Liu}. On the other hand, when MIMO configurations are used, expect for TSw and PS, SWIPT can also be achieved by using \textit{antenna switching (AS)} or \textit{spatial switching (SS)}.  These techniques are discussed in more detail in the following subsection. However, it is noted that MIMO configurations are out of the scope of this thesis.

\subsubsection{Time-Switching}

\begin{figure*}
\centering
\includegraphics[width=0.8\linewidth]{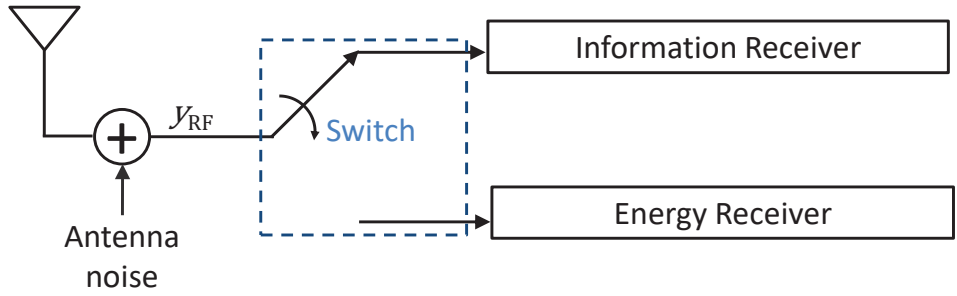}
\caption{Time-switching architecture.}
\label{Time-switching architecture}
\end{figure*}
TSw enables SWIPT by switching the receiver between information decoding and energy power transfer, as it is shown in Fig. \ref{Time-switching architecture}. Signal splitting is performed in the time domain, and, thus, the entire received signal received is
utilized either for information decoding or EH (Fig. 3a). Although the hardware implementation of TS at the receiver is simple, this technique requires accurate time synchronization and scheduling.

\subsubsection{Power-Splitting}

\begin{figure*}
\centering
\includegraphics[width=0.8\linewidth]{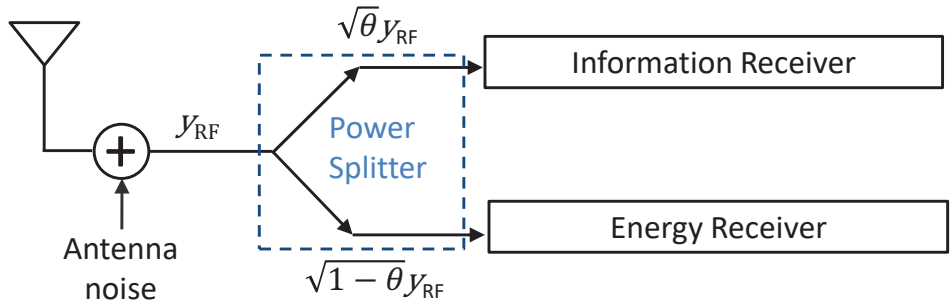}
\caption{Power-splitting architecture.}
\label{Power-splitting architecture}
\end{figure*}

PS enables the receiver to decode information and harvest energy from the same received by splitting the received signal into two streams of different power levels using a PS component, with the power ratio $\frac{\theta}{1-\theta}$, where $\theta$ is the power fraction used for information processing. Thus, as it is shown in Fig. \ref{Power-splitting architecture}, one signal stream is sent to the rectenna circuit for EH, and the other is converted to baseband for information decoding.   The PS technique, apart from being closer to the information theoretical optimum, compared to TSw, it achieves continuous SWIPT, as the signal
received in one time slot is used for both information decoding and power transfer, however, at the expense of increased receiver complexity compared to TSw. Therefore, it is more suitable for applications with critical information/energy or delay constraints.

\subsubsection{Antenna-Switching}
In AS, different antennas are used for information and energy reception, which requires the division of the receiving antennas in two groups. The number of antennas assigned to each group is subject to optimization.

\subsubsection{Spatial-Switching}
SS is based on the singular value decomposition (SVD) of the MIMO channel, which enables the transformation of the communication link into parallel eigenchannels. Thus, the main idea behind SWIPT with SS is that the eigenchannels, instead of the receiving antennas, can be divided in two groups, the output of which is driven to either the decoding circuit or the rectification one.

\subsection{Trade-off between the Harvested Energy and the Achievable Rate}
Due to the aforementioned strategies, SWIPT creates an interesting trade-off between the harvested energy and the achievable rate \cite{architecture}. More specifically, assuming a point-to-point communication system, when the harvested energy is maximized the achievable rate becomes zero, since all the signal's power or duration of the block is utilized for energy transfer. The opposite happens when the aim is to maximize the information rate. Similarly, in a two-hop relaying system, with an EH relay that employs SWIPT, if too much of the received power is directed to the energy harvesting or the detection circuit then the first or the second hop might be a burden for the end-to-end quality of communication, respectively. For example, assuming a self-powered relay that performs decode-and-forward (DF) relaying with PS, as in \cite{Esnaola}, if $\theta$ is set too low, too much power is directed to the energy harvesting circuit and there is not sufficient signal power for decoding, which leads to a decoding failure. On the other hand, if $\theta$ is selected too high, too much signal power is directed to the detection circuit, at the expense of the relaying transmission power.

%\newpage

\section{Joint Downlink/Uplink Design for Wireless Powered Networks with Interference}\label{Joint Downlink/Uplink Design for Wireless Powered Networks with Interference}

In this section, the downlink and the uplink of a wireless powered communication network are jointly optimized, while their rates are improved, by utilizing corresponding priority weights. In contrast to the system model investigated in Chapter \ref{ch:chapter3}, it is assumed that information is also transferred during downlink, except for energy, by using SWIPT.  Two communication protocols are taken into account, namely TDMA and NOMA. Also, the role of interference is also investigated. The considered multidimensional non-convex optimization problems are transformed into the equivalent convex ones and can be solved with low complexity. The results of the research presented in this section are included in \cite{joint_d_u_access, joint_d_u_ICT}.

\subsection{Related Work and Motivation}

As it has already been shwon in  the previous section, which investigated  the joint design of downlink energy  transfer and uplink information  tranmission in multiuser communications systems,   the rate and fairness can be substantially improved, when uplink NOMA- TS is utilized. Power allocation in uplink NOMA scenarios, with fixed decoding order of the users' messages, has been investigated in \cite{benchmark}, \cite{Dingnew}, and \cite{massive_mimo}, considering a single-antenna, multiantenna, and massive MIMO BS, respectively.

Downlink NOMA with SWIPT has been proposed in \cite{Dingcooperative}, which provides closed form expressions for the outage probability of the users, assuming a cooperative communication system with multiple wireless powered relays. Moreover, in \cite{BNBF}  the outage performance of cooperative relaying for two-user downlink NOMA systems is investigated, while a best near best far user selection scheme is proposed. Also, SIC in the downlink with SWIPT has been investigated in \cite{Krikidis_SIC}, which focuses on the coverage probability of a random user in bipolar ad hoc networks. It should be highlighted that the concept of downlink is different from that of the uplink NOMA, since in the downlink all users receive the interfering messages from the same source, i.e., via the same link \cite{VLC, Krikidis_NOMA, wei2016survey}.  For example, TS is a technique that cannot be applied in downlink NOMA. Interestingly, it has not be shown yet if and under which circumstances NOMA outperforms orthogonal schemes, e.g. TDMA, when used for the downlink of WPNs. Regarding this issue, it should also be considered that the utilization of downlink NOMA, in contrast to uplink NOMA, implies that SIC takes place at the EH users, and, thus, the corresponding complexity is increased. 

On the other hand, the joint optimization of downlink and uplink information transmission in WPNs has been studied in \cite{EE_uplink_downlink}, when the aim is to maximize the energy efficiency, while utilizing OFDMA. Interestingly, a user far from the BS of a WPN receives less power than a nearer user, therefore its uplink rate is negatively affected. A cascade effect of this phenomenon appears when information is also transmitted during the EH phase, using SWIPT, since the downlink rate of the far user is also affected. Moreover, the distance of a user from the BS also affects the level of the received interference, since, usually, users near the BS receive less interference compared to the cell-edge users, the performance of which is more severely impaired, despite the potential gain due to EH from interference signals. This effect, which hereafter will be called \textit{cascaded near-far problem (CnfP)}, has not been investigated in the existing literature. 

\subsection{Contribution}
In this work, a WPN is considered in the presence of interference. The communication is performed in two phases; during the first phase, the BS transmits information to the users, while the users also harvest energy, and during the second phase, the users utilize the harvested energy in order to transmit their messages towards the BS. In this network setup, the CnfP is caused by: i) the difference in achievable user rates during downlink, due to their asymmetric positioning, ii) the difference in achievable user rates during uplink, due to different harvested energy during downlink, iii) the asymmetric impact of interference on the users, both for the information reception and the EH.

The presented analysis focuses on the optimal system design, in order to reduce the impact of CnfP in WPNs with interference, considering a sole communication channel and nodes with single antennas. More specifically, the following aspects are considered and optimized:
\begin{itemize}
\item Two well-known multiple access schemes are considered for the downlink, i.e., NOMA and TDMA, in order to investigate their performance in WPNs with interference. For the uplink, NOMA-TS is assumed, based on the results of the previous chapter.
\item The minimum downlink and uplink rate is joinly maximized, while achieving a balance between them, by adding a desirable weight for each rate in the optimization formulation. It is shown that the resulting high dimensional non-convex optimization problems can be transformed to convex ones and, thus, be optimally solved by well-known methods with low complexity.
\item Based on the above optimization solutions, the CnfP and its impact on the performance of WPNs are investigated, for both communication protocols. The implementation of NOMA in the downlink is proved to offer gain over the TDMA protocol, especially in the case that the users are located at different distances from the BS, i.e., in the case that the CnfP is strong. 
\item Extensive comparison between the two considered protocols
for the downlink also verifies that NOMA is a more energy efficient solution than TDMA for usage in the downlink of WPNs, both in the presence or the absence of interfering sources.
\end{itemize}

\subsection{System model}\label{S:System}
\begin{figure*}[t!]
\centering
\includegraphics[width=0.9\linewidth]{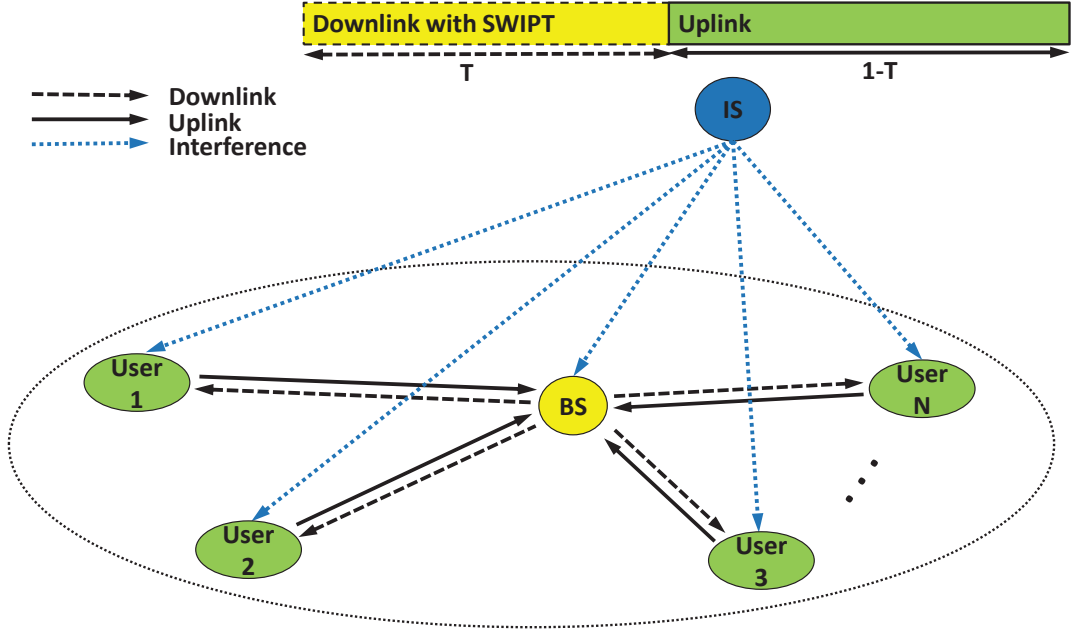}
\caption{Sequential downlink (with simultaneous energy transfer) and uplink in wireless powered networks with multiple users and interference.}
\label{Fig1}
\end{figure*}

Both the downlink and the uplink of a wireless network are considered, assuming $N$ users, with $n\in\mathcal{N}=\{1,...,N\}$ and one BS. It is assumed that all users share the same bandwidth resources and all nodes are equipped with a single antenna.  Assuming channel reciprocity, the channel between the BS and the $n$-th user, and the corresponding reciprocal, are denoted by $h_n$ and $\bar{h}_n$, respectively, where $\bar{(\cdot)}$ denotes the conjugate of $(\cdot)$, while the channel power gain is $\gamma_n=|h_n|^2=|\bar{h}_n|^2$. It is further assumed that all nodes consume energy only for information transmission. Moreover, an interfering source (IS) is assumed. In line with Fig. \ref{Fig1}, where the considered system model is presented, the communication is divided into time frames of unitary duration, each of which consists of two distinct phases:
 
\underline{Phase $1$ (downlink with SWIPT)}: The BS transmits with power $P_0$, which is used by the users in order to decode the BS's messages, as well as to charge their batteries. The duration of this phase is denoted by $0\leq  T\leq1$. Two different protocols are considered, namely NOMA and TDMA.

\underline{Phase $2$ (uplink)}: The remaining amount of time, i.e., $1- T$ is assigned to the users, in order to transmit their messages. It is considered that NOMA-TS is used, since it was proven in \cite{NOMA_WPT_joutnal} that it maximizes the rates and fairness among users. 

\subsubsection{Downlink with NOMA}
In this section, the downlink phase is described, when downlink NOMA and simultaneous power transfer towards the users is applied. NOMA allows the BS to simultaneously serve all users by using the entire bandwidth
to transmit data, through a superposition coding technique at the transmitter side. According to the NOMA protocol, the BS transmits the sum of the users' messages with the corresponding power, that is, $\sum_{n=1}^N\sqrt{P^\mathrm{d}_n}s^\mathrm{d}_n$, where $P^\mathrm{d}_n$ and $s^\mathrm{d}_n$, with $\mathbb{E}[|s^\mathrm{d}_n|^2]=1$, are the allocated power and the message for the $n$-th user, respectively, while the superscript $(\cdot)^\mathrm{d}$ denotes a value for the downlink phase. Moreover, the transmitting power is subject to
\begin{equation}
\sum_{n=1}^NP^\mathrm{d}_n\leq P_0.
\label{power constraint}
\end{equation}

It is assumed assume that the signal received by each user is
split into two streams, and the power fraction, $0\leq\theta_n\leq 1$, is used for information processing, while the fraction $1-\theta_n$ is devoted to EH. The observation at the $n$-th user which is used for information decoding, considering unitary antenna gains, is given by
\begin{equation}\label{received_signal_with_noise}
y^\mathrm{d}_n=h_n\sqrt{\theta_n}\sum_{i=1}^N\sqrt{P^\mathrm{d}_i}s^\mathrm{d}_i+\sqrt{\theta_n}I_n+\nu_n,
\end{equation}
where $\nu_n$ denotes the additive noise at the $n$-th user and $I_n$ is the interfering signal. In fact, noise is added in two parts of the receiver, i.e. the receive antenna noise and the circuit noise \cite{architecture, krik2}. However, the power of the antenna noise is too small and can be neglected, in line with \cite{Ju, Dingcooperative}. Thus, in \eqref{received_signal_with_noise}, only one additive noise parameter is included.

The $j$-th user carries out SIC, by detecting and removing the $n$-th user's message, for all $n<j$, from its observation \cite{Ding3,Krikidis_NOMA}.  Thus, the achievable rate of the $n$-the user, $n\in\{1, 2, \cdots, N\}$, is bounded by
\begin{equation}
R^\mathrm{d}_n=\min(R^\mathrm{d}_{n\rightarrow n},R^\mathrm{d}_{n\rightarrow n+1},...,R^\mathrm{d}_{n\rightarrow N})
\label{ordering_general}
\end{equation} 
%\begin{equation}
%R^\mathrm{d}_n\leq R^\mathrm{d}_{n\rightarrow n},
%\end{equation} 
%under the constraint that 
%\begin{equation}
%R^\mathrm{d}_{n\rightarrow j}\geq R^\mathrm{d}_{n\rightarrow n},\,\forall j>n
%\end{equation}
where  $R^\mathrm{d}_{n\rightarrow j}$ denotes the rate at which user $U_j$ detects the message intended for the $n$-th user. In the above, 
%\begin{equation}
%R^\mathrm{d}_{n\rightarrow j}=\!
%\begin{cases}\! T\log_2\left(1+\frac{p^\mathrm{d}_n\theta_j{\gamma_j}}{\theta_j{\gamma_j}\sum_{i=n+1}^N p^\mathrm{d}_i+\theta_jp_{I,j}+1}\right),\\
%\quad\quad\quad\quad\quad\quad\quad\quad\quad n=\{1,...,N-1\},\\
%\! T\log_2\!\left(1+\frac{p^\mathrm{d}_N\theta_j{\gamma_j}}{\theta_jp_{I,j}+1}\right),\,n=N,
%\end{cases}
%\end{equation}
\begin{equation}
R^\mathrm{d}_{n\rightarrow j}= T\log_2\left(1+\frac{p^\mathrm{d}_n\theta_j\gamma_j}{\theta_j\gamma_j\sum_{i=n+1}^N p^\mathrm{d}_i+\theta_jp_{\mathrm{I},j}+1}\right),
\label{rate downlink NOMA}
\end{equation}
where $p^\mathrm{d}_n=\frac{P^\mathrm{d}_n}{N_0W}$ and $p_{\mathrm{I},j}=\frac{P_{\mathrm{I},j}}{N_0W}$, in which $P_{\mathrm{I},j}$ is the power of the received interference by the $j$-th user, $N_0$ is the the white power spectral density of the AWGN and $W$ is the channel bandwidth.
It is assumed that $P_{\mathrm{I},j}$ is perfectly sensed by the $j$-th user and reported to the BS in order to properly allocate the available resources.  Note than when $n=N$, \eqref{rate downlink NOMA} is written as
\begin{equation}
R^\mathrm{d}_{N\rightarrow N}=\log_2\!\left(1+\frac{p^\mathrm{d}_N\theta_N\gamma_N}{\theta_Np_{\mathrm{I},N}+1}\right).
\end{equation}
Hereafter, $\boldsymbol{p}=\{p^\mathrm{d}_1,\ldots,p^\mathrm{d}_N\}$ denotes the set of values of transmit power among users and, $\boldsymbol{\theta}=\{\theta_1,\ldots,\theta_N\}$, the set of PS factors among users.

The harvested energy by each user is given by
\begin{equation}
E_n=\eta_1  T(1-\theta_n)\left(\gamma_n\sum_{i=1}^NP^\mathrm{d}_i+P_{\mathrm{I},n}\right),
\label{energy}
\end{equation}
where $0<\eta_1<1$ is the efficiency of the energy harvester.

\paragraph{Special Case: Interference-free Downlink}
In the case of absence of interfering sources and without loss of generality, the values $\theta_ng_n$ enforced to be sorted according to the users' ordering, i.e., 
\begin{equation}
\theta_1\gamma_1 \leq \theta_2\gamma_2 \leq \cdots \leq \theta_N\gamma_N.
\label{ordering}
\end{equation} 
Thus, the achievable data rate of the $n$-th user, $n\in\{1, 2, \cdots, N\}$, can be obtained as 
%\begin{figure*}
%\begin{equation}\
%R^\mathrm{d}_n\!=\!
%\begin{cases}\! T\log_2\!\left(\!1+\frac{p^\mathrm{d}_n\theta_ng_n}{\theta_ng_n\sum_{i=n+1}^N p^\mathrm{d}_i+1}\!\right)\!,\,n=\{1,...,N-1\},\\
%\! T\log_2\!\left(1+p^\mathrm{d}_N\theta_Ng_N\right)\!,\,n=N.
%\end{cases}
%\label{user rate}
%\end{equation}
\begin{equation}
R^\mathrm{d}_n= T\log_2\!\left(\!1+\frac{p^\mathrm{d}_n\theta_n\gamma_n}{\theta_n\gamma_n\sum_{i=n+1}^N p^\mathrm{d}_i+1}\!\right),
\label{user downlink rate}
\end{equation}
which for $n=N$ is written as
\begin{equation}
R^\mathrm{d}_N= T\log_2\!\left(1+p^\mathrm{d}_N\theta_Ng_N\right).
\end{equation}

Note that (\ref{user downlink rate}) is conditioned on $R^\mathrm{d}_{n\rightarrow j}\geq R^\mathrm{d}_{\mathrm{t},n}$, $\forall n<j$, where $R^\mathrm{d}_{\mathrm{t},n}$ denotes the targeted rate of the $n$-th user. When $R^\mathrm{d}_{\mathrm{t},n}$ is determined opportunistically through the user's channel condition, i.e., $R^\mathrm{d}_{\mathrm{t},n}\leq R^\mathrm{d}_n$, it can be easily verified that the condition $R^\mathrm{d}_{n\rightarrow j}\geq R^\mathrm{d}_{\mathrm{t},n}$ always holds since $\theta_j{\gamma_j}\geq \theta_ng_n$ for $j>n$. Consequently, the users' data rates can be given directly by (\ref{user downlink rate}). 
%\hrulefill
%\end{figure*}

\subsubsection{Downlink with TDMA}
When TDMA is used in the downlink, the BS serves by sequentially sending the non-interfering signals, $s^\mathrm{d}_n,n\in\mathcal{N}$, with transmit power $P_0$. In this case,
\begin{equation} 
\sum_{n=1}^Nt_n\leq  T,
\label{time-constraint}
\end{equation}
where $t_n\geq 0$ denotes the amount of time that is allocated to each user. Hereinafter, $\boldsymbol{t}=\{t_1,\ldots,t_N\}$, will be used to denote the set of values of allocated time among users.

Thus, during the time allocated for the $j$-th user, considering unitary antenna gains, the $n$-th user receives 
\begin{equation}
y^\mathrm{d}_n=h_n\sqrt{P_0}s^\mathrm{d}_j+I_n+\nu_n,\quad j\neq n.
\end{equation}
It is assumed that when the BS transmits the message of the $j$-th user, the $n$-th user utilizes all the received power for harvesting. On the other hand, when $j=n$, its own message is transmitted by the BS. Then, it is assumed that the $n$-th user splits the received power in two streams, i.e., the power fraction $\theta_n$ is used for information processing, while the fraction $1-\theta_n$ is used for harvesting. In that case, the received signal is given by
\begin{equation}
y_n=h_n\sqrt{\theta_nP_0}s^\mathrm{d}_n+\sqrt{\theta_n}I_n+\nu_n,
\end{equation}
and the corresponding rate is 
\begin{equation}
R^\mathrm{d}_n=t_n\log_2(1+\frac{\theta_n\rho_0\gamma_n}{\theta_np_{\mathrm{I},n}+1}),
\label{rate TDMA}
\end{equation}
with $\rho_0=\frac{P_0}{N_0W}$. Finally, the total harvested energy is given by
\begin{equation}
\begin{split}
E_n=\eta_1 ({\gamma_n}P_0+P_{\mathrm{I},n})\sum_{i\in\mathcal{N}}t_i -\eta_1 \theta_n t_n({\gamma_n}P_0 +P_{\mathrm{I},n}).
\end{split}
\label{energy TDMA}
\end{equation}

\subsubsection{Uplink}
It is highlighted that Phase $2$, i.e., the uplink phase, is common for both methods assumed for the downlink. TS can be combined with NOMA for the uplink, since the decoding of all messages takes place at the BS, in contrast to downlink NOMA. Therefore, NOMA-TS has been selected for the uplink, according to which all users simultaneously send their messages, $s^\mathrm{u}_n$, where $\mathbb{E}[|s^\mathrm{u}_n|^2]=1$, with transmit power $P^\mathrm{u}_n$ for the $n$-th user, while the superscript $(\cdot)^\mathrm{u}$ denotes a value for the uplink phase. Thus, the observation at the BS, assuming unitary antenna gains and efficiency of users' amplifiers, is given by
\begin{equation}
y^\mathrm{u}=\sum_{n=1}^N\bar{h}_n\sqrt{P^\mathrm{u}_n}s^\mathrm{u}_n+I+\nu_0,
\end{equation}
where $I$ denotes the interfering signal and $\nu_0$ denotes the additive noise at the BS. As mentioned in Section \ref{chapter1section2}, by using SIC and TS, the capacity region is bounded by \cite{NOMA_WPT_joutnal}
\begin{equation}
\sum_{n\in\mathcal{M}_k} \tilde{R}^\mathrm{u}_n\leq (1- T)\log_2\left(1+\frac{\sum_{n\in\mathcal{M}_k} p^\mathrm{u}_n{\gamma_n}}{p_{\mathrm{I},0}+1}\right),\,\forall \mathcal{M}_k: \mathcal{M}_k\subseteq\mathcal{N},
\label{uplink rate general limitation}
\end{equation}
with $\tilde{R}^\mathrm{u}_n$ being the uplink rate achieved by the $n$-th user, $p^\mathrm{u}_n=\frac{P^\mathrm{u}_n}{N_0W}$, $p_{\mathrm{I},0}=\frac{P_{\mathrm{I},0}}{N_0W}$,  and $P_{\mathrm{I},0}$ is the power of the interference received by the BS. It is assumed that $P_{\mathrm{I},0}$ is perfectly sensed by the BS. Finally, $\mathcal{M}_k$ denotes any possible subset of the users. 

It is assumed that the energy required to receive/process information
is negligible compared to the energy required for
information transmission \cite{Nasir, LiuWang, Dingcooperative}. Thus, when users utilize solely the energy that they harvest during the $1$-st phase, denoted by $E_n$, to transmit their information, and unitary efficiency of the user's amplifier is assumed, then $P^\mathrm{u}_n$ can be calculated as
\begin{equation}
P^\mathrm{u}_n=\frac{E_n}{1- T}.
\label{power}
\end{equation}
Note that the harvested energy, $E_n$, depends on the selected protocol for the downlink, i.e.  NOMA or TDMA.

\subsection{Resource Allocation}\label{S:Problem}
In this section, both the downlink and the uplink rate are maximized, while achieving: i) fairness among users, by ensuring that the maximized rate can be achieved by each of them, and ii) a balance between the downlink and the uplink rate. To this end, an auxiliary variable $\mathcal{R}$ is used, which denotes the lower bound of the weighted downlink/uplink rates, i.e. $\frac{R^\mathrm{d}_n}{\alpha}$ and $\frac{\tilde{R}^\mathrm{u}_n}{\beta}$, where $\alpha,\beta\geq0$, with $\alpha+\beta=1$, correspond to the weights used for the downlink and uplink, respectively. 
Thus, according to the above, it must hold that 
\begin{equation}
R^\mathrm{d}_n\geq \alpha \mathcal{R},
\label{min rate downlink}
\end{equation} and 
\begin{equation}
\tilde{R}^\mathrm{u}_n\geq \beta \mathcal{R},
\label{min rate uplink}
\end{equation} 
For example, when $\alpha=1$ or $\alpha=0$, only the downlink or uplink is optimized, respectively. By setting $\alpha=0.5$, the aim is to achieve the same rate for both the downlink and the uplink. Moreover, in the problem formulation, regarding the downlink, the specific formulation according to both protocols that are presented in Section \ref{S:System} is taken into account.

\subsubsection{Downlink with NOMA}
Taking into account \eqref{power} and \eqref{energy}, the constraint in \eqref{uplink rate general limitation} can be rewritten as
\begin{equation}
\sum_{n\in\mathcal{M}_k} \tilde{R}^\mathrm{u}_n\leq (1- T)\log_2\left(1+\frac{\eta_1 T\! \sum\limits_{n\in\mathcal{M}_k} (1-\theta_n){\gamma_n}\left({\gamma_n}\sum\limits_{i=1}^N\!p_i^\mathrm{d}+p_{\mathrm{I},n}\right)}{(1- T)(p_{\mathrm{I},0}+1)}\right),\,\forall \mathcal{M}_k: \mathcal{M}_k\subseteq\mathcal{N}.
\label{uplink NOMA rates general}
\end{equation}
The minimum rate maximization problem can be written as
%===========================================================
\begin{equation}
\begin{array}{ll}
\underset{\mathcal{R},  T, \boldsymbol{p}, {\boldsymbol{\theta}}}{\text{\textbf{max}}}& \mathcal{R} \\
\,\,\,\text{\textbf{s.t.}}&\mathrm{C}_1: \min(R^\mathrm{d}_{n\rightarrow n},R^\mathrm{d}_{n\rightarrow n+1},...,R^\mathrm{d}_{n\rightarrow N})\geq \alpha \mathcal{R},\,\forall n\in\mathcal{N},\\
&\mathrm{C}_2: (1- T)\log_2\!\!\left(1\!\!+\!\!\frac{\eta_1 T\!\!\!\! \sum\limits_{n\in\mathcal{M}_k}\!\!\!\! (1-\theta_n){\gamma_n}\left({\gamma_n}\sum\limits_{i=1}^N\!p_i^\mathrm{d}+p_{\mathrm{I},n}\right)}{(1- T)(p_{\mathrm{I},0}+1)}\right)\geq\beta||\mathcal{M}_k||\mathcal{R},\,\forall \mathcal{M}_k\subseteq\mathcal{N},\\
&\mathrm{C}_3:\sum_{n=1}^Np_n^\mathrm{d}\leq \rho_0,\\
&\mathrm{C}_4:0\leq \theta_n\leq 1,\,\forall{n\in\mathcal{N}},\\
&\mathrm{C}_5:p_n^\mathrm{d}\geq 0,\,\forall{n\in\mathcal{N}},\\
&\mathrm{C}_6:0\leq  T\leq 1,\\
\end{array}
\label{opt NOMA general}
\end{equation}
%=============================================
where $||.||$ denotes cardinality and $\mathrm{C}_1$, $\mathrm{C}_2$, $\mathrm{C}_3$  correspond to \eqref{ordering_general} and \eqref{min rate downlink}, \eqref{min rate uplink} and \eqref{uplink NOMA rates general}, and \eqref{power constraint}, respectively, while the remaining constraints (i.e., $\mathrm{C}_4$-$\mathrm{C}_6$) force the optimized variables not to exceed their maximum/minimum value.

Using the epigraph form of \eqref{opt NOMA general}, it can be rewritten as
%===========================================================
\begin{equation}
\begin{array}{ll}
\underset{\mathcal{R},  T, \boldsymbol{p}, {\boldsymbol{\theta}}}{\text{\textbf{max}}}& \mathcal{R} \\
\,\,\,\text{\textbf{s.t.}}&\mathrm{C}_1:  T\log_2\left(1+\frac{p^\mathrm{d}_n\theta_j{\gamma_j}}{\theta_j{\gamma_j}\sum_{i=n+1}^N p^\mathrm{d}_i+\theta_j p_{\mathrm{I},j}+1}\right)\geq \alpha \mathcal{R},\,\forall{n\in\mathcal{N}},j\in\{n,...,N\},\\
&\mathrm{C}_2: (1- T)\log_2\left(1+\frac{\eta_1 T\sum\limits_{n\in\mathcal{M}_k}(1-\theta_n){\gamma_n}\left({\gamma_n}\sum\limits_{i=1}^N\!p_i^\mathrm{d}+p_{\mathrm{I},n}\right)}{(1- T)(p_{\mathrm{I},0}+1)}\right)\geq\beta||\mathcal{M}_k||\mathcal{R},\,\forall \mathcal{M}_k\subseteq\mathcal{N},\\
&\mathrm{C}_3:\sum_{n=1}^Np_n^\mathrm{d}\leq \rho_0,\\
&\mathrm{C}_4:0\leq \theta_n\leq 1,\,\forall{n\in\mathcal{N}},\\
&\mathrm{C}_5:p_n^\mathrm{d}\geq 0,\,\forall{n\in\mathcal{N}},\\
&\mathrm{C}_6:0\leq  T\leq 1.\\
\end{array}
\label{opt NOMA general_2}
\end{equation}
%=============================================
Note that the epigraph form is a useful tool from optimization theory. It represents
a set of points (i.e., a graph) above or below the considered function \cite{Boyd1}.
\begin{proposition}\label{Proposition 1}
The inequality in $\mathrm{C}_3$ can be replaced by equality, without excluding the optimal from the set of all solutions.
\end{proposition}
\begin{proof}
Let's assume that the optimal $\mathcal{R}$, denoted by $\mathcal{R}^*$ is achieved when $\boldsymbol{p}^*=\{p^*_1,...,p^*_N\}$, for which $\sum_{n=1}^Np^*_n<\rho_0$. Let $\boldsymbol{p}'$ be another set, for which $\boldsymbol{p}'=\{\rho_0-\sum_{n=2}^Np^*_n,p^*_2,...,p^*_N\}$, i.e., it is the same vector as $\boldsymbol{p}^*$, apart from the power allocated to the first user. Since $\sum_{n=1}^Np'_n=\rho_0$, it is $p'_1>p^*_1$ and thus the rates at which all users (including $U_1$) detect the message of the $U_1$ is improved. Thus, since $R_1^\mathrm{d}=\min(R^\mathrm{d}_{1\rightarrow 1},R^\mathrm{d}_{1\rightarrow 2},...,R^\mathrm{d}_{1\rightarrow N})$, $R_1^\mathrm{d}$ is increased. At the same time, the values for the rest users' rates are retained, since the message of the first user is canceled by the rest of the users. Therefore, all users' rates remain the same, while $R_1^\mathrm{d}$ is increased. In this way,
 all inequalities regarding $\mathcal{R}^*$ are still satisfied. Thus, at least $\mathcal{R}^*$ can be achieved by $\boldsymbol{p}'$, contradicting the sole optimality of $\boldsymbol{p}^*$.  
\end{proof}

Proposition \ref{Proposition 1} is critical for the replacement of the constraint in $\mathrm{C}_2$ with 
\begin{equation}
(1- T)\log_2\left(1+\frac{\eta_1 T \sum\limits_{n\in\mathcal{M}_k} (1-\theta_n){\gamma_n}\left({\gamma_n}\rho_0+p_{\mathrm{I},n}\right)}{(1- T)(p_{\mathrm{I},0}+1)}\right)\geq\beta||\mathcal{M}_k||\mathcal{R},\,\forall \mathcal{M}_k\subseteq\mathcal{N}.
\label{uplink NOMA rates simplified}
\end{equation}

The time splitting parameter, $ T$, which appears in the capacity formula in both the downlink and uplink, couples the power allocation variables $\boldsymbol{p}$ and ${\boldsymbol{\theta}}$ and results in a non-convex problem. It is noted that there is no standard approach for
solving non-convex optimization problems in general. 

In order to overcome this issue and provide a tractable solution, a full search with respect to $ T$ is performed. Particularly, for a given value of $ T$, the variables $\boldsymbol{p}$ and ${\boldsymbol{\theta}}$ are optimized with the aim to maximize the corresponding minimum rate. The procedure is repeated for all possible values of $ T$ and record the corresponding achieved values of $\mathcal{R}$.

However, even with this simplification the problem remains non-convex, with respect to $\boldsymbol{p}$ and ${\boldsymbol{\theta}}$, which are coupled. To this end, the initial variables are replaced by $p_n^\mathrm{d}\triangleq \exp(\tilde{p}_n)$, $\theta_n\triangleq\exp(\tilde{\theta}_n)$, and $\mathcal{R}\triangleq \exp(\tilde{\mathcal{R}})$, and the optimization problem in \eqref{opt NOMA general_2}, after some mathematic manipulations, can now be written as
%===========================================================
\begin{equation}
\begin{array}{ll}
\underset{\tilde{\mathcal{R}}, \boldsymbol{\tilde{p}}, {\boldsymbol{\tilde{\theta}}}}{\text{\textbf{max}}}& \tilde{\mathcal{R}} \\
\,\,\,\text{\textbf{s.t.}}&\mathrm{C}_1:  T\log_2\left(1+\frac{\exp(\tilde{p}_n)\exp(\tilde{\theta}_j){\gamma_j}}{\exp(\tilde{\theta}_j){\gamma_j}\sum_{i=n+1}^N \exp(\tilde{p}_i)+\exp(\tilde{\theta}_j)p_{\mathrm{I},j}+1}\right)\\
&\quad\quad\quad\quad \geq\alpha\exp(\tilde{\mathcal{R}}),\,\forall{n\in\mathcal{N}},j\in\{n,...,N\},\\
&\mathrm{C}_2: (1- T)\log_2\left(1+\frac{\eta_1 T \sum\limits_{n\in\mathcal{M}_k} \left(1-\exp(\tilde{\theta}_n)\right){\gamma_n}\left({\gamma_n}\rho_0+p_{\mathrm{I},n}\right)}{(1- T)(p_{\mathrm{I},0}+1)}\right)\\&\quad\quad\quad\geq\beta||\mathcal{M}_k||\exp(\tilde{\mathcal{R}}),\,\forall \mathcal{M}_k\subseteq\mathcal{N},\\
&\mathrm{C}_3:\sum_{n=1}^N\exp(\tilde{p}_n)= \rho_0,\\
&\mathrm{C}_4:0\leq\exp(\tilde{\theta}_n)\leq 1,\\
&\mathrm{C}_5:\exp(\tilde{p}_n)\geq 0,\\
\end{array}
\label{opt_simplified_NOMA_1 general}
\end{equation}
%=============================================
which is still non-convex. However, after some mathematic manipulations and by relaxing the equality in $\mathrm{C}_3$ with inequality, the optimization problem in \eqref{opt_simplified_NOMA_1 general} can be rewritten as
%===========================================================
\begin{equation}
\begin{array}{ll}
\underset{\tilde{\mathcal{R}}, \tilde{\boldsymbol{p}}, \tilde{{\boldsymbol{\theta}}}}{\text{\textbf{max}}}& \tilde{\mathcal{R}} \\
\,\,\,\text{\textbf{s.t.}}&\mathrm{C}_1:\ln\left(\frac{p_{\mathrm{I},j}\exp(-\tilde{p}_n)+\exp(-\tilde{p}_n-\tilde{\theta}_j)}{{\gamma_j}}+\sum\limits_{i=n+1}^{N}\exp(\tilde{p}_i-\tilde{p}_n)\right)\\
&\quad+\ln(2^\frac{\alpha \exp(\tilde{\mathcal{R}})}{ T}-1)\leq 0,\,\forall{n\in\mathcal{N}},j\in\{n,...,N\},\\
&\mathrm{C}_2:\sum\limits_{n\in\mathcal{M}_k}\exp(\tilde{\theta}_n){\gamma_n}({\gamma_n}\rho_0+p_{\mathrm{I},n})+
\frac{(1- T)(P_{\mathrm{I},0}+1)}{\eta_1  T}2^\frac{\beta||\mathcal{M}_k||\exp(\tilde{\mathcal{R}})}{1- T}\\
&\quad\leq\sum\limits_{n\in\mathcal{M}_k}{\gamma_n}({\gamma_n}\rho_0+p_{\mathrm{I},n})\quad+\frac{(1- T)(p_{\mathrm{I},0}+1)}{\eta_1  T} ,\,\forall\mathcal{M}_k\subseteq\mathcal{N},\\
 &\mathrm{C}_3:\sum_{n=1}^N\exp(\tilde{p}_n)\leq \rho_0,\\
&\mathrm{C}_4:\tilde{\theta}_n\leq 0,\,\forall{n\in\mathcal{N}}.\\
\end{array}
\label{opt_simplified_NOMA_1 general_2}
\end{equation}
%=============================================
Note that the left inequality of $\mathrm{C}_4$ and $\mathrm{C}_5$ of the optimization problem in \eqref{opt_simplified_NOMA_1 general} are always valid, thus, they vanish from \eqref{opt_simplified_NOMA_1 general_2}.
\begin{proposition}\label{proposition2}
The optimization problem in \eqref{opt_simplified_NOMA_1 general_2} is convex.
\label{convexity of NOMA}
\end{proposition}

\begin{proof}
The objective function of \eqref{opt_simplified_NOMA_1 general_2} and $\mathrm{C}_4$ are linear. Regarding $\mathrm{C}_1$, the first term is a convex log-sum-exp function \cite{Boyd1}, while the second term, i.e., 
\begin{equation}
f=\ln(2^\frac{\alpha \exp(\tilde{\mathcal{R}})}{T}-1)
\end{equation}
is also convex, considering that $\frac{\partial^2 f}{\partial \mathcal{R}^2}\geq 0$ \cite{Boyd1}. This can be easily proved, since
\begin{equation}
\frac{\partial^2 f}{\partial \mathcal{R}^2}=\frac{2^{\tilde{z}_1}{\tilde{z}_1}\ln(2)\left(2^{\tilde{z}_1}-{\tilde{z}_1}\ln(2)-1\right)}{(2^{\tilde{z}_1}-1)^2},
\end{equation}
with ${\tilde{z}_1}=\frac{\alpha \exp(\tilde{\mathcal{R}})}{ T}$. Note that $w=2^{\tilde{z}_1}-{\tilde{z}_1}\ln(2)-1$ is an increasing function with respect to ${\tilde{z}_1}$ and when ${\tilde{z}_1}\rightarrow 0$, $w\rightarrow 0$. Finally, the left side of the constraints $\mathrm{C}_2$ and $\mathrm{C}_3$ are sum-exp functions and, thus, convex. 
\end{proof}

Proposition \ref{proposition2} is also critical, since it proves that \eqref{opt_simplified_NOMA_1 general_2} can be optimally solved in polynomial time, by well-known algorithms, such as the interior-point method \cite{Boyd1}.

\paragraph{Special Case: Interference-free Communication}
This subsection focuses on the absence of interference, and, thus, mainly on the parts of \eqref{opt NOMA general} that change.
First, the constraint $\mathrm{C}_1$ can be replaced by two simpler constraints i.e. \eqref{ordering} and \eqref{user rate}.
Moreover, when interference is zero, the inequality in \eqref{uplink NOMA rates general}  can be rewritten as
\begin{equation}
\sum_{n\in\mathcal{M}_k} \tilde{R}^\mathrm{u}_n\leq (1- T)\log_2\left(1+\frac{(\eta_1 T \sum\limits_{i=1}^N\!p_i^\mathrm{d})\sum\limits_{n\in\mathcal{M}_k}\! (1-\theta_n){\gamma_n^2}}{1- T}\right),\,\forall \mathcal{M}_k: \mathcal{M}_k\subseteq\mathcal{N}.
\label{uplink NOMA rates}
\end{equation}
Consequently, using the epigraph form, the minimum rate maximization problem can be expressed as
%===========================================================
\begin{equation}
\begin{array}{ll}
\underset{\mathcal{R},  T, \boldsymbol{p}, {\boldsymbol{\theta}}}{\text{\textbf{max}}}& \mathcal{R} \\
\,\,\,\text{\textbf{s.t.}}&\mathrm{C}_{1a}: \theta_{n}\gamma_{n} \leq \theta_{n+1}\gamma_{n+1},\,\forall{n\in\{1,...,N-1\}},\\
&\mathrm{C}_{1b}:  T\log_2\left(1+\frac{p_n^\mathrm{d}\theta_n{\gamma_n}}{\theta_n{\gamma_n}\sum\limits_{i=n+1}^N p_i^\mathrm{d}+1}\right)\geq \alpha \mathcal{R},\,\forall{n\in\mathcal{N}},\\
&\mathrm{C}_2: (1- T)\log_2\left(1+\frac{(\eta_1 T \sum\limits_{i=1}^Np_i^\mathrm{d})\!\!\sum\limits_{n\in\mathcal{M}_k} (1-\theta_n){\gamma_n^2}}{1- T}\right)\geq\beta||\mathcal{M}_k||\mathcal{R},\,\forall \mathcal{M}_k\subseteq\mathcal{N},\\
&\mathrm{C}_3:\sum_{n=1}^Np_n^\mathrm{d}\leq \rho_0,\\
&\mathrm{C}_4:0\leq \theta_n\leq 1,\,\forall{n\in\mathcal{N}},\\
&\mathrm{C}_5:p_n^\mathrm{d}\geq 0,\,\forall{n\in\mathcal{N}},\\
&\mathrm{C}_6:0\leq  T\leq 1.\\
\end{array}
\label{opt NOMA}
\end{equation}
%=============================================
Subsequently, using one-dimensional search for the optimization of $ T$ and similar steps as the ones in the previous subsection, \eqref{opt NOMA} can be rewritten as
%===========================================================
\begin{equation}
\begin{array}{ll}
\underset{\tilde{\mathcal{R}}, \tilde{\boldsymbol{p}}, \tilde{{\boldsymbol{\theta}}}}{\text{\textbf{max}}}& \tilde{\mathcal{R}} \\
\,\,\,\text{\textbf{s.t.}}&\mathrm{C}_{1a}:\tilde{\theta}_n-\tilde{\theta}_{n+1}\leq \ln\left(\frac{\gamma_{n+1}}{{\gamma_n}}\right),\,\forall n\in \{1,...,N-1\},\\
&\mathrm{C}_{1b}:\ln\left(\frac{\exp(-\tilde{p}_n-\tilde{\theta}_n)}{{\gamma_n}}+\sum\limits_{i=n+1}^{N}\exp(\tilde{p}_i-\tilde{p}_n)\right)+\ln(2^\frac{\alpha \exp(\tilde{\mathcal{R}})}{ T}-1)\leq 0,\,\forall{n\in\mathcal{N}},\\
&\mathrm{C}_2:\sum\limits_{n\in\mathcal{M}_k}\!\!\!\exp(\tilde{\theta}_n){\gamma_n^2}+\frac{1- T}{\eta_1 \rho_0  T}2^\frac{\beta||\mathcal{M}_k||\exp(\tilde{\mathcal{R}})}{1- T}\leq\sum\limits_{n\in\mathcal{M}_k}\!\!\!\!\!{\gamma_n^2}+\frac{1- T}{\eta_1 \rho_0  T} ,\,\forall\mathcal{M}_k\subseteq\mathcal{N},\\
 &\mathrm{C}_3:\sum_{n=1}^N\exp(\tilde{p}_n)\leq \rho_0,\\
&\mathrm{C}_4:\tilde{\theta}_n\leq 0,\,\forall{n\in\mathcal{N}}.\\
\end{array}
\label{opt_simplified_NOMA_1}
\end{equation}
%=============================================

Considering Proposition \ref{convexity of NOMA} and the linearity of  $\mathrm{C}_{1a}$, it can be easily proved that the optimization problem in \eqref{opt_simplified_NOMA_1} is a convex one. Taking into account the replacement of $\mathrm{C}_1$ with $\mathrm{C}_{1a}$ and $\mathrm{C}_{1b}$, it is observed that \eqref{opt_simplified_NOMA_1} is simpler than \eqref{opt_simplified_NOMA_1 general_2} since the $\sum_{i=0}^{N-1}(N-i)$ nonlinear constraints are replaced by $N-1$ linear constraints and solely $N$ nonlinear ones.

\subsubsection{Downlink with TDMA}
The minimum rate maximization problem, taking into acount \eqref{power}, \eqref{energy TDMA} and the constraint in \eqref{uplink rate general limitation}, can be rewritten as
\begin{equation}
\begin{split}
\sum_{n\in\mathcal{M}_k} \tilde{R}^\mathrm{u}_n\leq (1- T)&\log_2\left(1+\frac{\eta_1  \sum\limits_{n\in\mathcal{M}_k}{\gamma_n}\left(({\gamma_n}\rho_0+p_{\mathrm{I},n})\sum\limits_{i\in\mathcal{N}}t_i -\theta_n({\gamma_n}\rho_0 t_n+p_{\mathrm{I},n})\right)}{(1- T)(1+p_{\mathrm{I},0})}\right),\\
&\forall \mathcal{M}_k: \mathcal{M}_k\subseteq\mathcal{N}.
\end{split}
\label{uplink rate limitation}
\end{equation}
The minimum rate maximization problem, using the epigraph form, as in \eqref{opt NOMA general}, can be written as
%===========================================================
\begin{equation}
\begin{array}{ll}
\underset{\mathcal{R},  T, \boldsymbol{t}, {\boldsymbol{\theta}}}{\text{\textbf{max}}}& \mathcal{R} \\
\,\,\,\text{\textbf{s.t.}}&\mathrm{C}_1: t_n\log_2(1+\frac{\theta_np{\gamma_n}}{\theta_np_{\mathrm{I},n}+1})\geq \alpha \mathcal{R},\,\forall{n\in\mathcal{N}},\\
&\mathrm{C}_2: (1- T)\log_2\left(\!1+\frac{\eta_1 \!\!\!\!\! \sum\limits_{n\in\mathcal{M}_k}\!\!\!\!\!{\gamma_n}\left(({\gamma_n}\rho_0+p_{\mathrm{I},n})\sum\limits_{i\in\mathcal{N}}t_i -\theta_nt_n({\gamma_n}\rho_0 +p_{\mathrm{I},n})\right)}{(1- T)(1+p_{\mathrm{I},0})}\!\right)\\ &\quad\quad\quad\geq\beta||\mathcal{M}_k||\mathcal{R},\,\forall \mathcal{M}_k\subseteq\mathcal{N},\\
&\mathrm{C}_3: \sum_{n=1}^Nt_n\leq  T,\\
&\mathrm{C}_4:0\leq \theta_n\leq 1,\,\forall{n\in\mathcal{N}},\\
&\mathrm{C}_5: t_n\geq 0,\,\forall{n\in\mathcal{N}},\\
&\mathrm{C}_6:0\leq  T\leq 1,
\end{array}
\label{opt_TDMA}
\end{equation}
%=============================================
where $\mathrm{C}_1$, $\mathrm{C}_2$, and $\mathrm{C}_3$ correspond to \eqref{rate TDMA} and \eqref{min rate downlink}, \eqref{min rate uplink} and \eqref{uplink rate limitation}, and \eqref{time-constraint}, respectively, while the rest of the  constraints (i.e., $\mathrm{C}_4$-$\mathrm{C}_6$) limit the optimized variables not to exceed their maximum/minimum value.

\begin{proposition}\label{Proposition 3}
The inequality in $\mathrm{C}_3$ can be replaced by equality without excluding the optimal from the set of all solutions.
\end{proposition}
\begin{proof}
The proof is similar to that of Proposition \ref{Proposition 1}.
\end{proof}
Considering  Proposition \ref{Proposition 3}, $\mathrm{C}_2$ can be replaced by 
\begin{equation}
\begin{split}
\sum_{n\in\mathcal{M}_k} \tilde{R}^\mathrm{u}_n\leq (1- T)&\log_2\left(1+\frac{\eta_1 \sum\limits_{n\in\mathcal{M}_k}{\gamma_n}\left(({\gamma_n}\rho_0+p_{\mathrm{I},n}) T -\theta_nt_n({\gamma_n}\rho_0 +p_{\mathrm{I},n})\right)}{(1- T)(1+p_{\mathrm{I},0})}\right),\\&\forall \mathcal{M}_k: \mathcal{M}_k\subseteq\mathcal{N}.
\end{split}
\label{uplink rate limitation_2}
\end{equation}
Moreover, one-dimensional search is assumed for the optimization of $ T$. However, even with these simplifications, the optimization problem in \eqref{opt_TDMA}, remains non-convex due to the coupling of the variables $\boldsymbol{\theta}$ and $\boldsymbol{t}$.

Next, by setting $t_n\triangleq \exp(\tilde{t}_n)$, $\theta_n\triangleq\exp(\tilde{\theta}_n)$, and $\mathcal{R}\triangleq \exp(\tilde{\mathcal{R}})$, the optimization problem in \eqref{opt_TDMA} can be rewritten as
 %===========================================================
\begin{equation}
\begin{array}{ll}
\underset{\tilde{\mathcal{R}}, \boldsymbol{\tilde{t}}, {\boldsymbol{\tilde{\theta}}}}{\text{\textbf{max}}}& \tilde{\mathcal{R}} \\
\,\,\,\text{\textbf{s.t.}}
&\mathrm{C}_1: \exp(\tilde{t}_n)\log_2\left(1+\frac{\exp(\tilde{\theta}_n)\rho_0{\gamma_n}}{{\exp(\tilde{\theta}_n)p_{\mathrm{I},n}+1}}\right)\geq \alpha \exp(\tilde{\mathcal{R}}),\,\forall{n\in\mathcal{N}},\\
&\mathrm{C}_2: (1- T)\log_2\left(\!1+\frac{\eta_1  \sum\limits_{n\in\mathcal{M}_k}{\gamma_n}\left(({\gamma_n}\rho_0+p_{\mathrm{I},n}) T -\exp(\tilde{\theta}_n+\tilde{t}_n)({\gamma_n}\rho_0+p_{\mathrm{I},n})\right)}{(1- T)(1+p_{\mathrm{I},0})}\right)\\
&\quad\geq\beta||\mathcal{M}_k||\exp(\tilde{\mathcal{R}}),\,\forall \mathcal{M}_k\subseteq\mathcal{N},\\
&\mathrm{C}_3: \sum_{n=1}^N\exp(\tilde{t}_n)= T,\\
&\mathrm{C}_4:0\leq\exp(\tilde{\theta}_n)\leq 1,\\
&\mathrm{C}_4:\exp(\tilde{t}_n)\geq 0,\\
\end{array}
\label{opt_TDMA_simplified1}
\end{equation}
%=============================================
which, after some mathematic manipulations and by relaxing the equality in $\mathrm{C}_3$ with inequality, can be expressed as
 %===========================================================
\begin{equation}
\begin{array}{ll}
\underset{\tilde{\mathcal{R}}, \boldsymbol{\tilde{t}}, {\boldsymbol{\tilde{\theta}}}}{\text{\textbf{max}}}& \tilde{\mathcal{R}} \\
\,\,\,\text{\textbf{s.t.}}&\mathrm{C}_1: \ln\left(2^{\alpha \exp(\tilde{\mathcal{R}}-\tilde{t}_n)}-1\right)+\ln\left(p_{\mathrm{I},n}+\exp(-\tilde{\theta}_n)\right)\leq\ln(\rho_0{\gamma_n}),\,\forall{n\in\mathcal{N}},\\
&\mathrm{C}_2: \sum\limits_{n\in\mathcal{M}_k}\exp(\tilde{\theta}_n+\tilde{t}_n){\gamma_n}({\gamma_n}\rho_0+p_{\mathrm{I},n})+\frac{(1- T)(1+p_{\mathrm{I},0})}{\eta_1}2^\frac{\beta||\mathcal{M}_k||\exp(\tilde{\mathcal{R}})}{1- T}\\
&\quad\leq  T\sum\limits_{n\in\mathcal{M}_k}{\gamma_n}({\gamma_n}\rho_0+p_{\mathrm{I},n})+\frac{(1- T)(1+p_{\mathrm{I},0})}{\eta_1},\,\forall \mathcal{M}_k\subseteq\mathcal{N},\\
 &\mathrm{C}_3: \sum_{n=1}^N\exp(\tilde{t}_n)\leq  T,\\
 &\mathrm{C}_4:\tilde{\theta}_n\leq 0,\,\forall{n\in\mathcal{N}}.\\
\end{array}
\label{opt_TDMA_simplified2}
\end{equation}
%=============================================
Note that the left inequality of $\mathrm{C}_4$ and $\mathrm{C}_5$ in the optimization problem \eqref{opt_TDMA_simplified1} are always valid, thus, they vanish from \eqref{opt_TDMA_simplified2}.
\begin{proposition}
The optimization problem in \eqref{opt_TDMA_simplified2} is convex.
\end{proposition}
\begin{proof}
The proof is similar to that of Proposition \ref{proposition2}. Note that the first term of the left side of $\mathrm{C}_1$, i.e., 
\begin{equation}
f=\ln\left(2^{\alpha \exp(\tilde{\mathcal{R}}-\tilde{t}_n)}-1\right)
\end{equation}
is a function of the variables $\tilde{\mathcal{R}}$ and $\tilde{t}_n$, thus its convexity must be proved considering its Hessian matrix rather than its second derivatives. More specifically, its Hessian matrix has non-negative eigenvalue, which is
\begin{equation}
\phi=\frac{2^{{\tilde{z}_2}+1}{\tilde{z}_2}\ln(2)\left(2^{\tilde{z}_2}-{\tilde{z}_2}\ln(2)-1\right)}{(2^{\tilde{z}_2}-1)^2},
\end{equation}
where ${\tilde{z}_2}=\alpha \exp(\tilde{\mathcal{R}}-\tilde{t}_n)$.
\end{proof}

It needs to be mentioned that the optimization problem in \eqref{opt_TDMA_simplified2} is simpler than \eqref{opt_simplified_NOMA_1 general_2}, since it has a lower number of non-linear constraints due to $\mathrm{C}_1$.

\subsection{Simulations and Discussion}

In this section, simulation results are presented for a system with $N=2$ or $N=3$ users, for $\eta_1=0.5$. When $N=2$, the distances of the users from the BS are $d_1=5$ m and $d_2=1$ m, while for $N=3$, it is $d_1=5$ m, $d_2=3$ m, and $d_3=1$ m, respectively. A bounded path loss model is adopted
\begin{equation}
{\gamma_n}=\frac{1}{1+d_n^\xi},
\end{equation}
as in \cite{Ding_path_loss}, where $\xi$ is the path-loss exponent, with $\xi=2$, while fast fading is neglected, in order to focus on the asymmetry of the system due to different user distances from the BS. The indexing of the users is in ascending order with respect to their channel gains, ${\gamma_n}$. Finally, one-dimensional search is performed for the optimization of $ T$, with a step of $0.01$.

\subsubsection{Communication in the Presence of Interference}
Regarding the source of interference, for the sake of convenience for the illustration, a sole interfering source (IS) is considered, the distance of which from the BS is denoted by $D_0$. It is assumed that the BS, the users and the IS are located on a single line, connecting the BS and the IS. Then, the received interference by each user (normalized by the noise power) is given by
\begin{equation}
p_{\mathrm{I},n}=\frac{p_\mathrm{IS}}{1+(D_0-d_n)^\xi},
\end{equation}
where $p_\mathrm{IS}=\frac{P_{IS}}{N_0W}$, with $P_\mathrm{IS}$ being the transmit power of the IS. Also, the normalized interference received by the BS is calculated as
\begin{equation}
p_{\mathrm{I},0}=\frac{p_\mathrm{IS}}{1+D_0^\xi}.
\end{equation}
Hereinafter, it is assumed that $p_\mathrm{IS}=40$ dB.

\begin{figure}[t!]
\centering
\includegraphics[width=0.8\columnwidth]{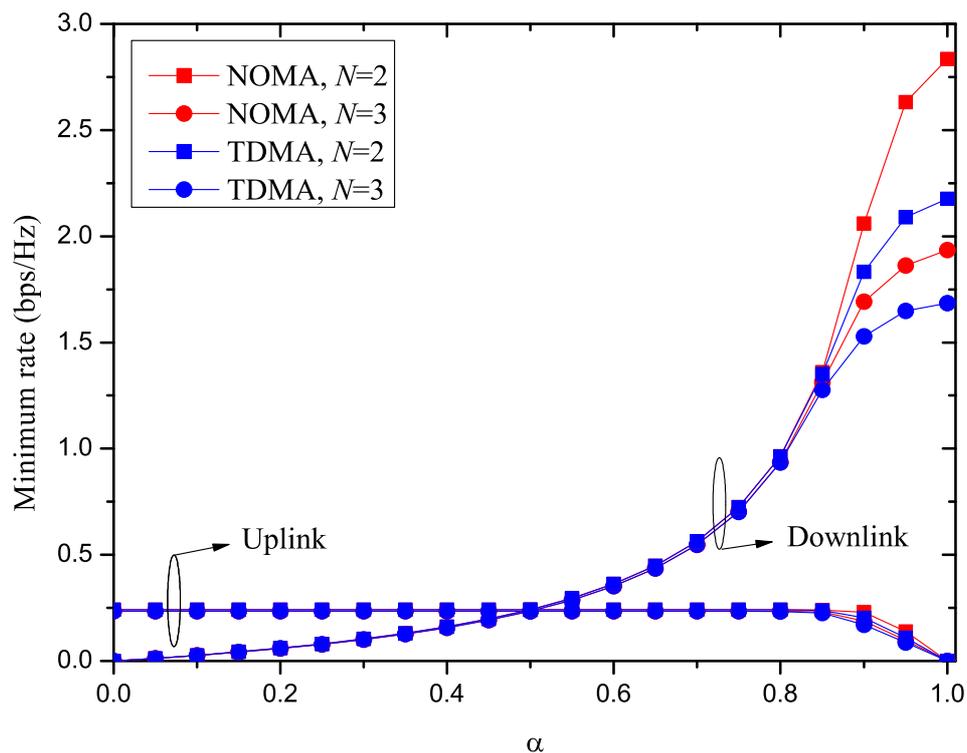}
\caption{Impact of $\alpha$ on the minimum rate, for $\rho_0=40$ dB, $p_\mathrm{IS}=40$ dB, and $D_0=20$ m.}
\label{Fig:rates_vs_a_IS}
\end{figure}

In Fig. \ref{Fig:rates_vs_a_IS}, the rate achieved in the uplink and in the downlink, for $N=2,3$, is depicted with respect to the value of $\alpha$, in the presence of the IS. It is obvious that in the case of $\alpha<0.5$, the uplink rate cannot be substantially increased, by either of the two protocols used during the downlink, mainly because of the power that can be harvested and then reused during uplink. However, when priority is given to the downlink rate, i.e., for $\alpha>0.5$, the downlink rate is substantially improved. Furthermore, for values $\alpha>0.85$, the use of NOMA during downlink offers a considerable gain in the achieved rate, for both values of the number of users, compared to TDMA. Therefore, it is concluded that the NOMA protocol in the downlink can provide more fair performance to the users than TDMA, even in the presence of interference. 

\begin{figure}[t!]
\centering
\includegraphics[width=0.8\columnwidth]{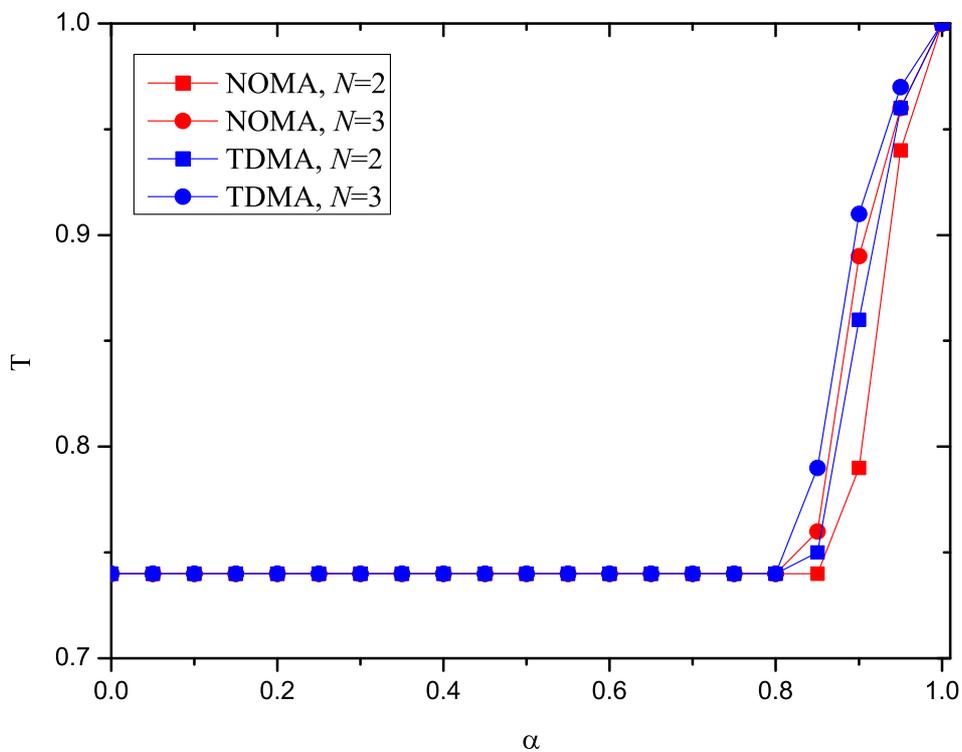}
\caption{Impact of $\alpha$ on $ T$, for $\rho_0=40$ dB, $p_\mathrm{IS}=40$ dB, and $D_0=20$ m.}
\label{Fig:time_vs_a_IS}
\end{figure}

In Fig. \ref{Fig:time_vs_a_IS}, the optimized $ T$ that is dedicated to the downlink is depicted with respect to the value of $\alpha$, for the same setup as in Fig. \ref{Fig:rates_vs_a_IS}. It is easily observed that, for $\alpha<0.8$, the time allocated for the downlink is practically unaltered. Thus, comparing to Fig. \ref{Fig:rates_vs_a_IS}, one can conclude that the achieved minimum uplink rates and the optimal time allocation do not change considerably for $\alpha<0.8$, while the increase in the value of the minimum downlink rates is mainly due to the different power allocation and PS, and not due to a different optimal value for the time allocation factor $ T$. However, for $\alpha>0.8$, when priority is given mainly to the downlink, the time allocated for downlink (and thus for EH as well) substantially increases, which leads to a considerable increase in the downlink rates. It is further observed by Fig. \ref{Fig:time_vs_a_IS} that the time allocated for downlink is higher in the case of TDMA, rather than for NOMA. This indicates that more harvested energy is needed for TDMA. Taking into account that NOMA achieves better rates with less harvested power, it is induced that NOMA is more energy efficient than TDMA for the downlink.

\begin{figure}[t!]
\centering
\includegraphics[width=0.8\columnwidth]{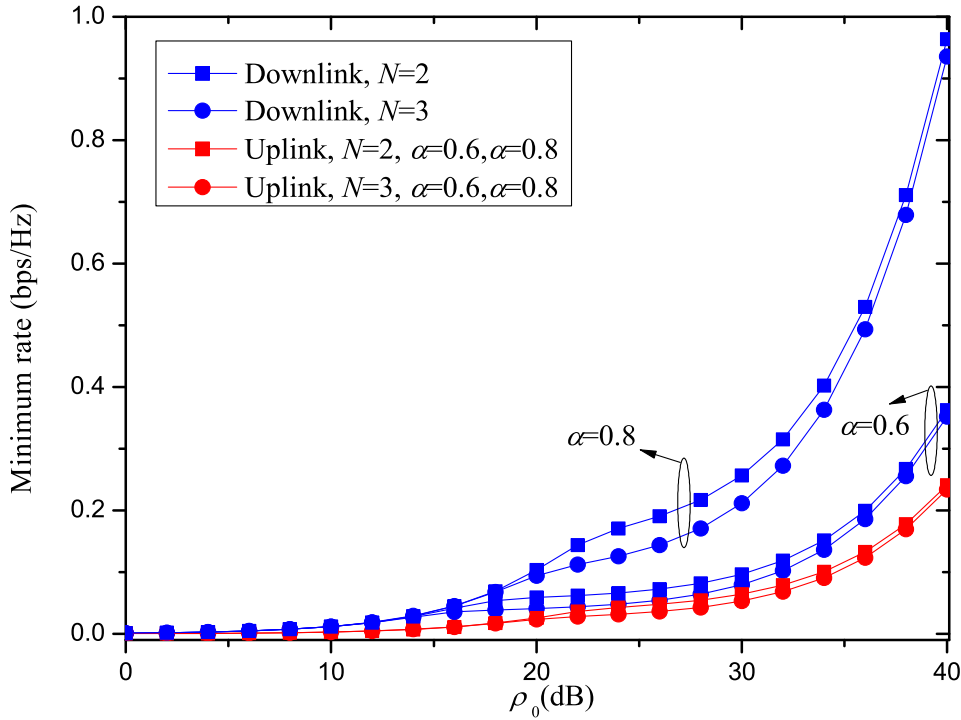}
\caption{Impact of $\rho_0$ on the minimum rate for $p_\mathrm{IS}=40$ dB, $D_0=20$ m, and different values of $\alpha$.}
\label{Fig:rate_vs_p_IS}
\end{figure}

In Fig. \ref{Fig:rate_vs_p_IS}, the achieved minimum rate in the uplink and the downlink is investigated, when $\alpha=0.6$ and $\alpha=0.8$, with respect to the total transmit power of the BS, $\rho_0$. The IS is located again at distance $D_0=20$ m, with transmit power $p_\mathrm{IS}=40$ dB. From Fig. \ref{Fig:rates_vs_a_IS}, one can observe that, when $\rho_0=40$ dB, the uplink rate is practically the same, for both values of $\alpha$. Furthermore, both NOMA and TDMA achieve the same uplink rate for these values of $\alpha$. This is observed for other values of $\rho_0$ as well, thus the uplink rate is plotted only once in Fig. \ref{Fig:rate_vs_p_IS} for each number of users. However, the downlink rate, although it is practically the same for both protocols, it differs according to the choice of $\alpha$, since $\alpha=0.8$ leads to higher rate, i.e., when priority is given to the downlink. For both  values of $\alpha$, it is easily seen that, for transmit power $\rho_0>30$ dB, the rate increases faster, compared to transmit power values between 20 and 30 dB. This indicates that, when $\rho_0=30$ dB, the interference imposed by the IS can now be mitigated easier, due to the available transmit power at the BS, achieving increasing data rates. This is more obvious for higher values of $\alpha$. 

%Regarding Fig \ref{Fig9} higher $\alpha$ leads to lower allocated power to the users with the worst channel conditions.
%This is reasonable, because the selection of $\theta_n$ is different (higher). For example (NOMA) in $40$ dB, when $a=0.8$ then $ T^*=0.74$, $p_1^{d}=0.9190$, $p_2^{d}=0.0682$, $p_3^{d}=0.0127$, $\theta_1=0.052$, $\theta_2=0.7079$, $\theta_3=0.8772$.
%When $\alpha=0.6$, then $ T^*=0.74$, $p_1^{d}=0.9792$, $p_2^{d}=0.0164$, $p_3^{d}=0.0041$, $\theta_1=0.0011$, $\theta_2=0.3884$, $\theta_3=0.1999$.

\begin{figure}[t!]
\centering
\includegraphics[width=0.8\columnwidth]{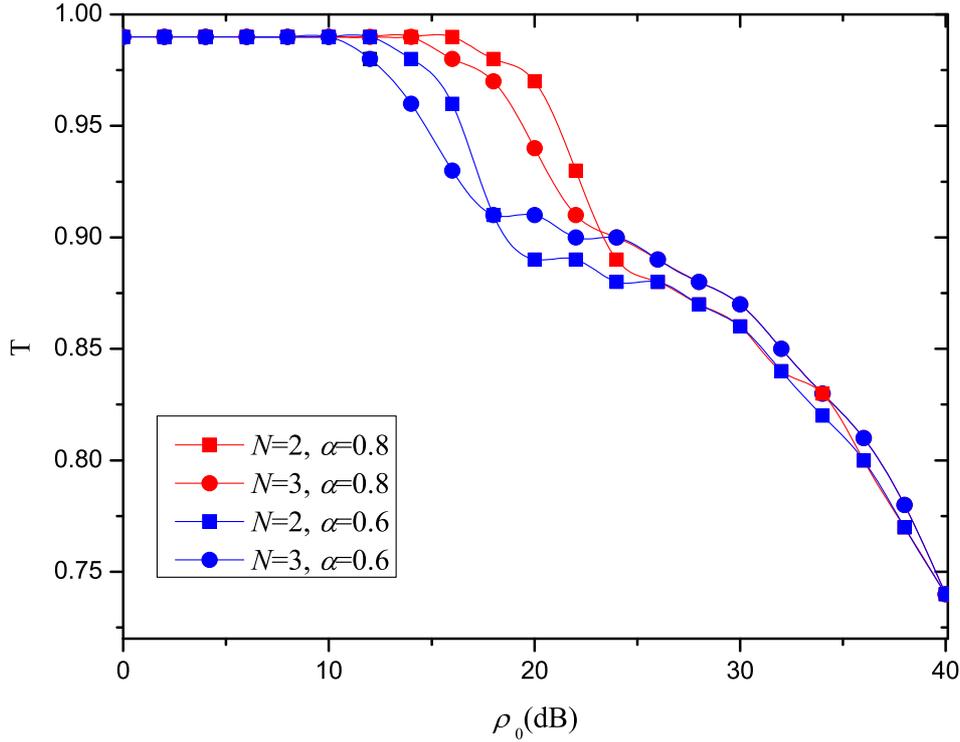}
\caption{Impact of $\rho_0$ on $ T$ for $p_\mathrm{IS}=40$ dB, $D_0=20$ m, and different values of $\alpha$.}
\label{Fig:time_vs_p_IS}
\end{figure}

Accordingly, in Fig. \ref{Fig:time_vs_p_IS}, the impact of $\rho_0$ on the allocated time $ T$ to the downlink, is illustrated for $\alpha=0.6$ and $\alpha=0.8$. Again, the results of both NOMA and TDMA are the same, so they are plotted only once. It is easily seen that, for both numbers of users, $N=2,3$, when $\alpha=0.8$, more time is allocated to the downlink and, consequently, to the EH, which is expected, since the downlink is given higher priority.

\begin{figure}[t!]
\centering
\includegraphics[width=0.8\columnwidth]{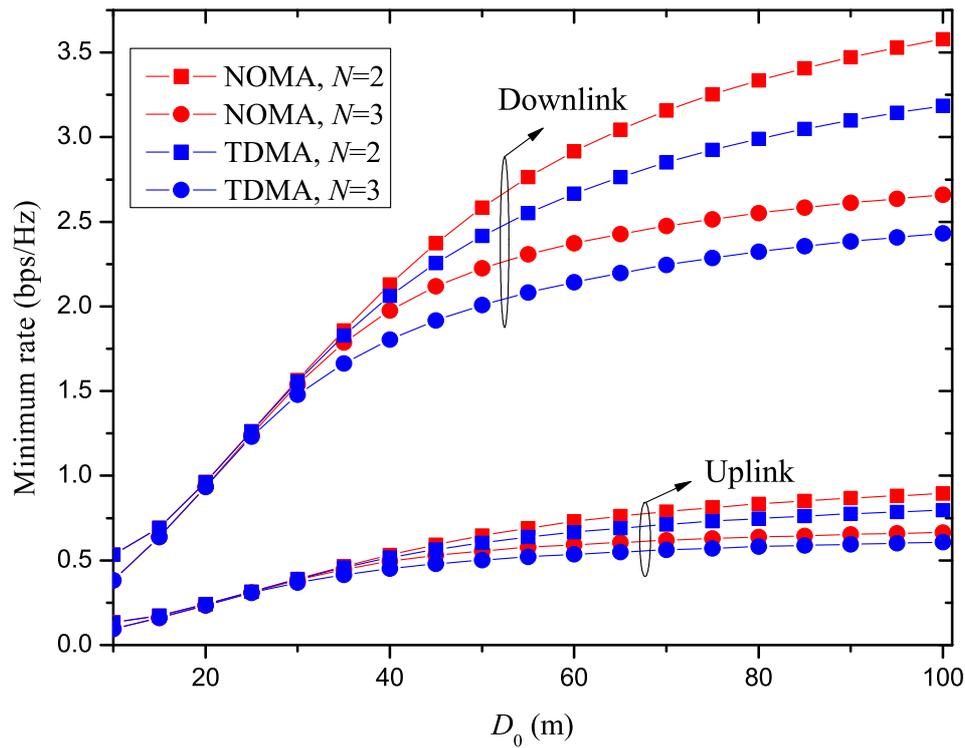}
\caption{Impact of $D_0$ on the minimum rate, for $\rho_0=40$ dB, $p_\mathrm{IS}=40$ dB, $\alpha=0.8$.}
\label{Fig:rate_vs_dist_IS}
\end{figure}

\begin{figure}[t!]
\centering
\includegraphics[width=0.8\columnwidth]{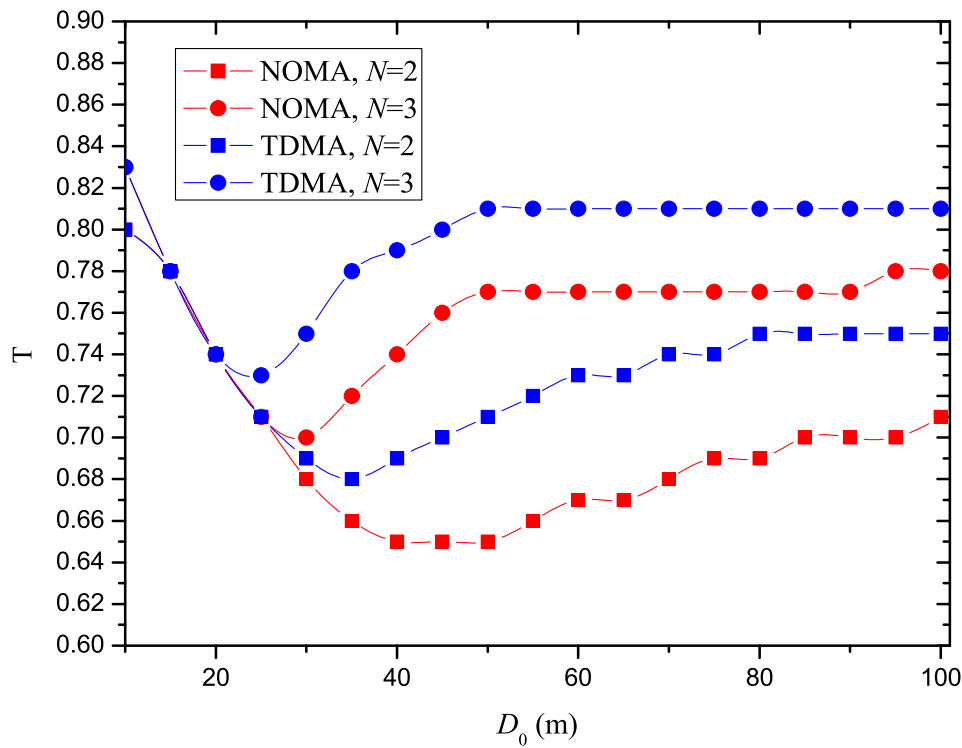}
\caption{Impact of $D_0$ on $ T$, for $\rho_0=40$ dB, $p_\mathrm{IS}=40$ dB, $\alpha=0.8$.}
\label{Fig:time_vs_dist_IS}
\end{figure}

In Figs. \ref{Fig:rate_vs_dist_IS} and \ref{Fig:time_vs_dist_IS}, a system with $N=2,3$ users as in the previous cases is considered, while the impact of the distance $D_0$ is examined, at which the IS is located, on the achieved uplink/downlink rate and the optimized allocated time $ T$, for both NOMA and TDMA protocols. More specifically, the distance $D_0$ varies between $D_0=10$ and $D_0=100$ m. From Fig. \ref{Fig:rate_vs_dist_IS}, it is easily observed that, when the IS is located further from the users and the BS, i.e., when the power of the interference is low, NOMA achieves substantial gains, both for the uplink and the downlink rates, compared to TDMA. This is mostly evident for $D_0>40$ m. Therefore, NOMA seems to be less prone to interference than TDMA, when the received unwanted power is low. Furthermore, from Fig. \ref{Fig:time_vs_dist_IS}, the TDMA protocol requires more time $T$ allocated to the downlink and therefore, to EH, especially when the IS is located further from the BS and the users. This indicates that the NOMA protocol is more energy efficient from TDMA, since it achieves better performance, with less harvested energy, for varying power levels of interference. 

Next, motivated by the energy efficiency and the resilience towards low levels of interference that NOMA presents compared to TDMA,  numerical results for the case of interference-free communication are presented, in order to investigate the performance gains offered by NOMA in the downlink, compared to TDMA, in absence of interfering sources.

\subsubsection{Interference-free Communication}

\begin{figure}[t!]
\centering
\includegraphics[width=0.8\linewidth]{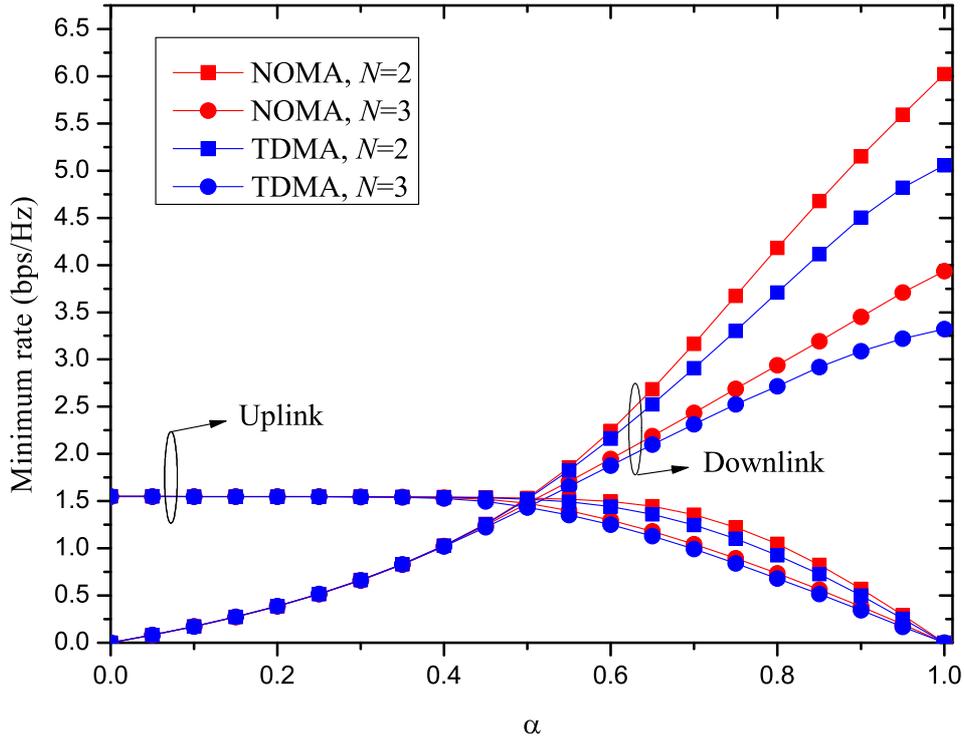}
\caption{Impact of $\alpha$ on the minimum rate, for $\rho_0=40$ dB.}
\label{Fig:rate_vs_a_free}
\end{figure}

This subsection presents numerical results for the special case when no interference is considered. More specifically, in Fig. \ref{Fig:rate_vs_a_free}, the rate achieved in the uplink and in the downlink, for $N=2,3$, is depicted with respect to the value of $\alpha$. As expected, when $\alpha>0.5$, since the downlink is prioritized over the uplink, the achieved rate for the downlink is higher. However, in the absence of interference, the impact of the value of  $\alpha$ is more evident on the achieved rates, since for $\alpha>0.5$, the uplink rate decreases, while the downlink rate is substantially increased. Regarding the comparison between NOMA and TDMA for the downlink, the two protocols seem to perform similarly, when priority is given for the uplink rate, i.e., when $\alpha<0.5$. However, for $\alpha>0.5$, NOMA outperforms TDMA in the end-to-end optimization, achieving higher rates for the uplink and downlink, when compared to TDMA, in contrast to the case of interference, when NOMA outperformed TDMA only for values of $\alpha>0.8$. In Fig. \ref{Fig:rate_vs_a_free}, it can be seen that NOMA can achieve the same downlink rate with TDMA but for a lower value of $\alpha$, which translates in higher uplink rate. For example, the highest downlink rate achieved by TDMA, which is for $\alpha=1$ when the uplink rate is zero, is achieved by NOMA for $\alpha\approx0.85$, where the uplink rate is non-zero. When $N$ increases, the achieved rate is reduced, however it also depends on the choice of $\alpha$, thus revealing a tradeoff between the desired rate and the prioritization between the downlink and the uplink.

\begin{figure}[t!]
\centering
\includegraphics[width=0.8\linewidth]{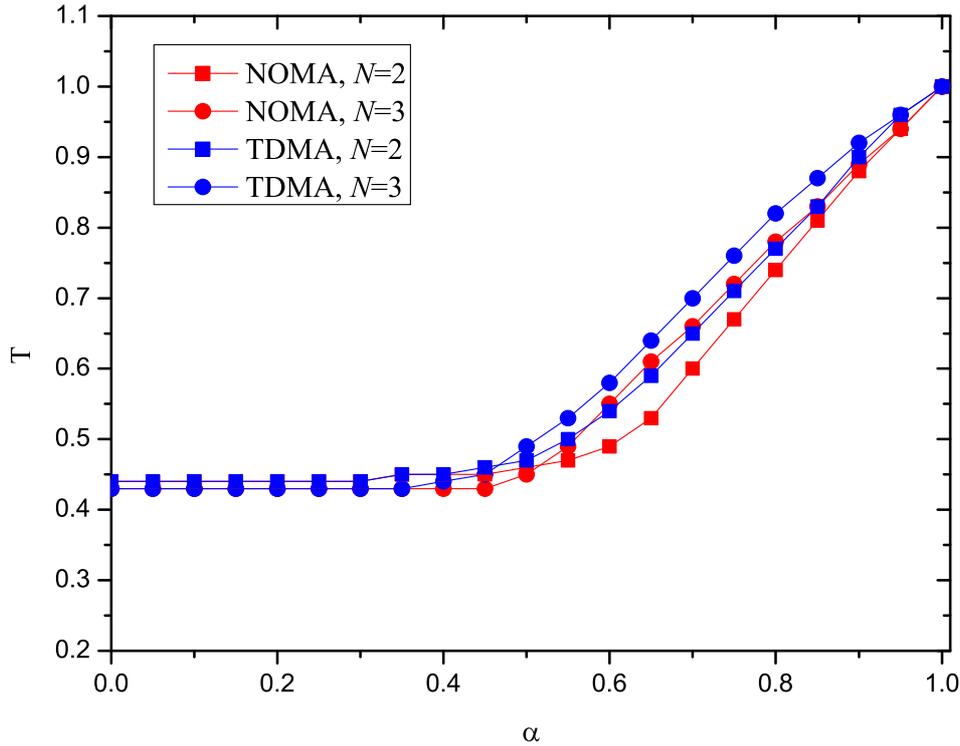}
\caption{Impact of $\alpha$ on $ T$, for $\rho_0=40$ dB.}
\label{Fig:time_vs_a_free}
\end{figure}
In Fig. \ref{Fig:time_vs_a_free}, the same setup as Fig. \ref{Fig:rate_vs_a_free} is examined, but the optimized time fraction dedicated to the downlink phase when the users harvest energy is depicted with respect to the value of $\alpha$, for both protocols used in the downlink. Comparing with Fig. \ref{Fig:time_vs_a_IS} where interference is present, it is observed that the time allocated for the downlink increases for values of $\alpha>0.5$, instead of $\alpha>0.8$. In the case of interference-free communication, similarly to Fig. \ref{Fig:time_vs_a_IS}, TDMA requires more time dedicated to the downlink and thus for EH, indicating once more that NOMA is a more energy-efficient solution than TDMA.

\begin{figure}[t!]
\centering
\includegraphics[width=0.8\linewidth]{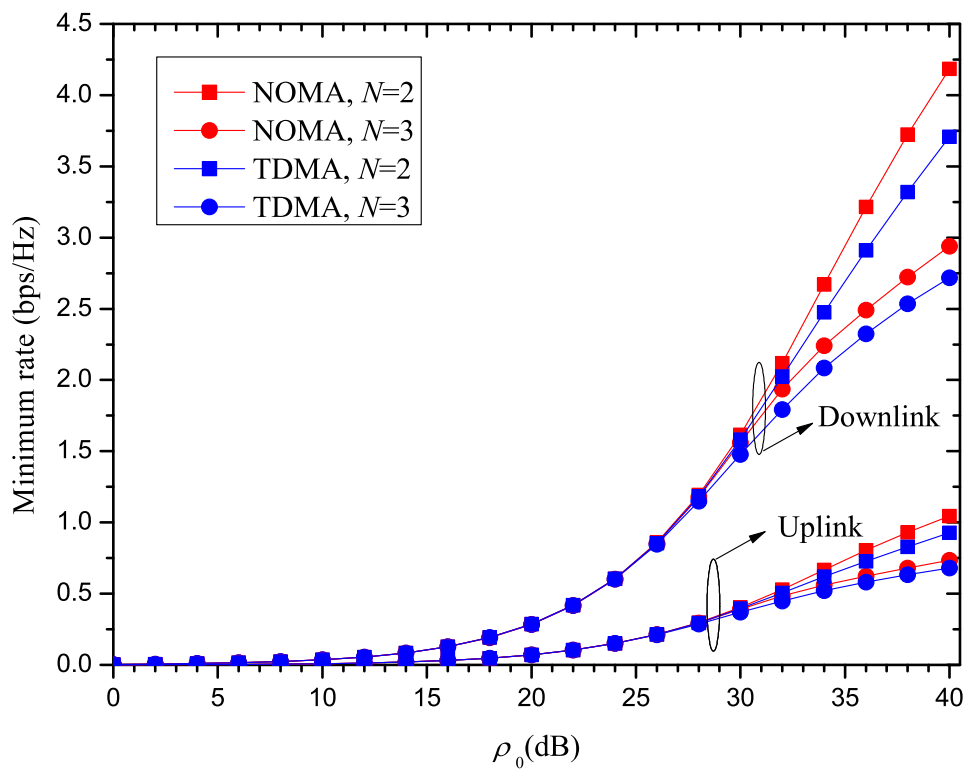}
\caption{Impact of $\rho_0$ on the minimum rate, for $\alpha=0.8$.}
\label{Fig:rate_vs_p_free}
\end{figure}

In Fig. \ref{Fig:rate_vs_p_free}, the achieved rate for the downlink and the uplink is presented, with respect to the transmit signal-to-noise ratio, $\rho_0$, when $\alpha=0.8$. One can observe that NOMA performs better than TDMA, as $\rho_0$ increases, in contrast to the case of interference, when both protocols achieved the same performance, for $\alpha=0.8$. Another useful observation from this figure, but also from Fig. \ref{Fig:rate_vs_p_IS}, is the fact that, for $N=3$, the rate increases with a smaller slope as $\rho_0$ increases, which is expected since it reflects the congestion of the multiple access schemes in use, as the number of users increases.

\begin{figure}[t!]
\centering
\includegraphics[width=0.8\linewidth]{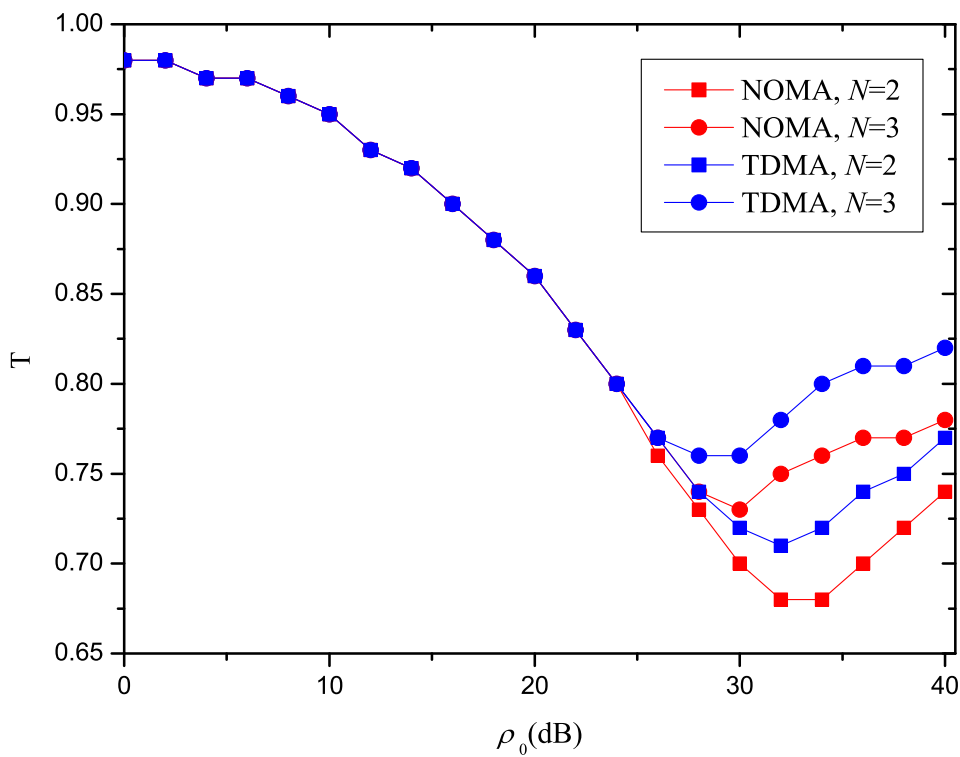}
\caption{Impact of $\rho_0$ on $ T$, for $\alpha=0.8$.}
\label{Fig:time_vs_p_free}
\end{figure}

In Fig. \ref{Fig:time_vs_p_free}, the optimal value of $ T$ is plotted against the value of $\rho_0$, when $\alpha=0.8$, in the absence of interference. A very interesting observation is that, although the time dedicated to the downlink - and, consequently, to the EH - decreases as $\rho_0$ increases, this is reversed after a value for $\rho_0$, for both $N=2$ and $N=3$, implying that higher availability of power at the BS will require more time dedicated to the downlink, after that value of $\rho_0$. This can be explained as follows: as observed in Fig. \ref{Fig:rate_vs_p_free} for $N=3$, the slope of the rate increase is smaller for large $\rho_0$. Thus, increasing only the available power at the BS leads to saturation regarding the achievable rate, and therefore, further optimization can be achieved mainly by increasing the time dedicated to downlink, and not the transmit power.

\begin{figure}[t!]
\centering
\includegraphics[width=0.8\linewidth]{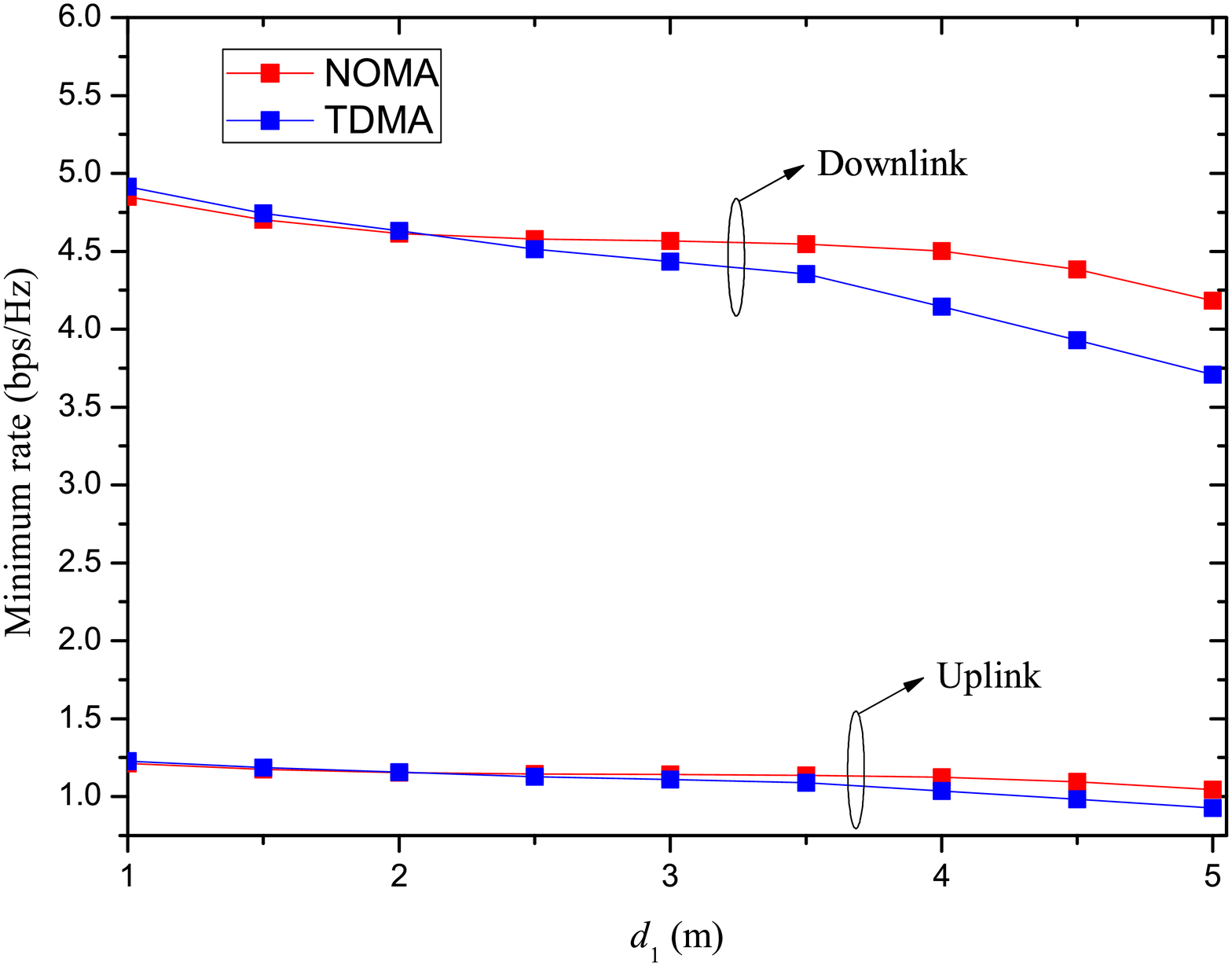}
\caption{The impact of distance on the minimum rate, for $\alpha=0.8$, $\rho_0=40$ dB, and $d_2=1$ m.}
\label{Fig:rate_vs_dist_free}
\end{figure}

Finally, in Fig. \ref{Fig:rate_vs_dist_free}, where only two users are assumed in the absence of interfering source, the impact of asymmetric distances between the users and the BS is investigated. More specifically, the achieved rate is illustrated with respect to the distance of the first user, when the distance of the second user is fixed to $d_2=1$ m, while $\alpha=0.8$ and $\rho_0=40$ dB. The gains in terms of achieved rate that NOMA can offer compared to TDMA in the downlink are greater, as the CnfP becomes more intense. Thus, it is clear that the NOMA scheme can offer more fairness than TDMA, when users are asymmetrically located with respect to the BS.

\section{Throughput Maximization in Multicarrier Wireless Powered Relaying Networks}\label{Throughput Maximization in Multicarrier Wireless Powered Relaying Networks}

Motivated by the implementation of WPT in cooperative networks, the application of SWIPT in multicarrier relaying network is investigated, with the aim to to optimize the QoS and the utilization of the available
resources. More specifically, the section is devoted on the joint optimization of the achievable rate, the allocated power among the different carriers, and the PS ratio, when the amplify-and-forward (AF) relaying protocol is considered. The results of the research presented in this section are included in \cite{Panos}.

%Wireless power transfer is an upcoming EH technique, aiming to overcome the constraint of fixed energy supplies and provide self-sustainability to network nodes \cite{Varshnev, Grover}. Interestingly, wireless signals can be used for the simultaneous wireless information and power transfer (SWIPT). However, in practice, the node cannot harvest power and process the information from the received signal at the same time. Power splitting provides an efficient solution to this problem, as the received signal can be divided into two streams, one for information processing and one for EH \cite{Liu}.  SWIPT has also been investigated in the context of an orthogonal frequency division multiplexing (OFDM) point-to-point communication system\cite{Kwan1, Kwan2}. 

\subsection{Related Work and Motivation}

The employment of relays is known to enhance the QoS and increase the network coverage, especially
when there is no line of sight (LoS) between the source and the destination. With simultaneous employment of energy
harvesting and WPT, the relay nodes can be self-powered and independent \cite{Michalopoulos, Esnaola, Ding_path_loss}. Thus, the utilization of wireless powered relay nodes proves to be extremely efficient, when the relay nodes have to be placed in remote positions and supplying them with the required power is difficult and costly. 

It is noted that in \cite{Michalopoulos, Esnaola, Ding2}, each source was assumed to have access to a sole communication channel, while only the DF relaying prtocol has been considered. Thus, although an expression for the optimal PS parameter has been derived in \cite{Esnaola, Ding_path_loss}, the presented analysis cannot be directly extended to the case of multiple channels, where the joint optimization of dynamic PS and power allocation becomes challenging.

\subsection{Contribution}
In this section, a source and a destination are considered, which communicate through a wireless-powered relay over multiple channels. The AF protocol is considered, for low-complexity relay nodes, since it does not require decoding of the received signal. Note that AF relaying is included in the LTE-A standard \cite{3GPP_relays}. It is further assumed that the relay is able to harvest energy from other sources -if available- such as solar power, wind etc., presenting a general optimization framework, which can accommodate various EH techniques, combined with WPT. Finally, for the problem of dynamic PS and power allocation on each of the available channels at the source and the relay, the maximization of the total throughput and the limitations of WPT process are taken into account. In order to solve this problem, two solutions are proposed: An optimal one and a fast-converging low-complexity iterative one. Simulation results reveal
the importance of the proposed dynamic optimization problem,
while they verify the effectiveness of the fast-converging proposed
iterative solution.

\subsection{System Model}
\begin{figure}[h!]
\centering%
\includegraphics[keepaspectratio,width=0.7\linewidth]{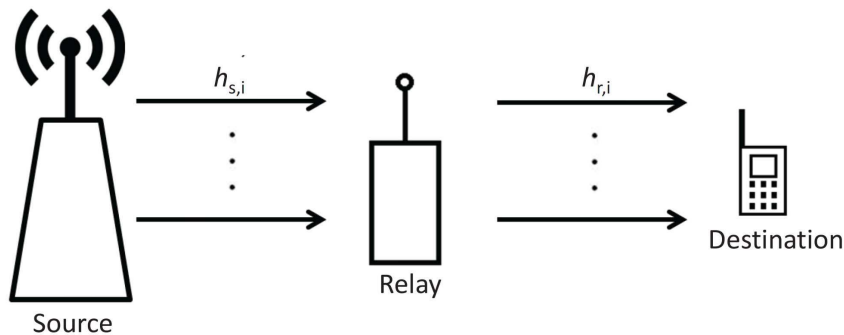}
\caption{System model overview.}
\label{systemmodel}
\end{figure}

A source (S), which communicates with a destination (D) via a relay (R) over $N_\mathrm{c}$ independent channels, is assumed, as shown in Fig.~\ref{systemmodel}. The relay is wireless powered and amplifies and forwards the received signals.  Moreover, no LoS between the source and the destination exists, which makes the relay's utilization of significant importance.

The S-R and R-D channel coefficients over the $i$-th channel are denoted by $h_{\mathrm{s},i}$ and $h_{\mathrm{r},i}$ respectively, and are considered independent complex Gaussian random variables with zero mean, i.e. $h_{\mathrm{s},i}\sim\mathcal{CN}(0,\sigma_{s,i}^2)$ and $h_{\mathrm{r},i}\sim \mathcal{CN}(0,\sigma_{r,i}^2)$, where $\sigma_{s,i}^2$ and $\sigma_{r,i}^2$ represent both path loss and shadowing. Antenna gains and the eficciency of relay's amplifier are not taken into account, i.e., they are assumed to be unitary.  Furthermore, it is assumed that the relay performs EH by PS. The signal is split into two streams, and the power fraction, $\theta$, is used for harvesting, while the fraction $1-\theta$ is devoted to information processing. The received signal over the $i$-th channel is \cite{Esnaola}
\begin{equation}\label{received_signal}
y_{\mathrm{r},i}=\sqrt{(1-\theta)P_{\mathrm{s},i}}h_{\mathrm{s},i}s_{s,i}+\nu_{\mathrm{r},i}
\end{equation}
where $P_{\mathrm{s},i}$ is the transmitted power by the source on the $i$-th channel, $s_{s,i}$ is the transmitted signal on the $i$-th channel and $\nu_{r,i}$ is the AWGN at R, i.e. $\nu_{r,i}\sim\mathcal{CN}(0,N_0W_i)$, with $W_i$ being the $i$-th channel bandwidth. The total harvested power at R is
\begin{equation}\label{harvested_energy}
P_\mathrm{rt}=\eta_1\theta\sum_{i=1}^{N_\mathrm{c}}P_{\mathrm{s},i}|h_{\mathrm{s},i}|^2+P_{\mathrm{r}0}
\end{equation}
where  $0<\eta_1<1$ is the efficiency of the energy harvester and $P_{\mathrm{r}0}$ denotes the available power from other sources (apart from SWIPT) and/or other fixed power needs of R (except for retransmission). A case of special interest is when $P_{\mathrm{r}0}=0$, in which the relay utilizes all the harvested energy by SWIPT for retransmission.

%\subsection{Amplify-and-Forward (AF) relaying}

The relay multiplies the received signal in (\ref{received_signal}) by
\begin{equation}\label{aplify_factor}
G_{\mathrm{r},i}=\sqrt{\frac{P_{\mathrm{r},i}}{(1-\theta)P_{\mathrm{s},i}|h_{\mathrm{s},i}|^2+N_0W_i}}
\end{equation}
where, $P_{\mathrm{r},i}$, is the power transmitted by the relay over the $i$-th channel. Since R harvests energy from the first hop, which is then redistributed over the available channels, the transmission power values $P_{\mathrm{r},i}$ of R depend on the values $P_{\mathrm{s},i}$ of S, the ratio $\theta$, the efficiency $\eta_1$ and the channel coefficients $h_{\mathrm{s},i}$. More specifically, it holds that 
\begin{equation}
\sum_{i=1}^{N_\mathrm{c}} P_{\mathrm{r},i}\leq P_\mathrm{rt}, 
\end{equation}
where $P_\mathrm{rt}$ is given in \eqref{harvested_energy}. Thus, each $P_{\mathrm{r},i}$ needs to be optimally determined.

The signal which is received at the destination over the $i$-th channel is given by
\begin{equation}\label{received_d}
y_{\mathrm{d},i}=h_{\mathrm{r},i}G_{\mathrm{r},i}\,y_{r,i}+\nu_{\mathrm{d},i}
\end{equation}
where $\nu_{d,i}$ is the AWGN added at the destination with $\nu_{d,i}\sim\mathcal{CN}(0,N_0W_i)$. If the SNR of the S-R and R-D communication links on the $i$-th channel are
\begin{equation}\label{g_sr}
\gamma_{\mathrm{s},i}=\frac{(1-\theta)P_{\mathrm{s},i}|h_{\mathrm{s},i}|^2 }{N_0W_i}
\end{equation}
and
\begin{equation}
\gamma_{r,i}=\frac{P_{\mathrm{r},i}|h_{\mathrm{r},i}|^2}{N_0W_i},
\end{equation}
respectively, then the total achievable rate is
\begin{equation}\label{eq.11}
\mathcal{R}_\mathrm{tot}=\sum_{i=1}^{N_\mathrm{c}} \frac{1}{2}W_i\log_2(1+\gamma_{\mathrm{tot},i}),
\end{equation} where the factor $\frac{1}{2}$ is due to the half duplex operation of the relay and $\gamma_i$ is the end-to-end SNR, given by
\begin{equation}\label{g_af}
\gamma_{\mathrm{tot},i} = \frac{\gamma_{\mathrm{s},i} \gamma_{\mathrm{r},i}}{\gamma_{\mathrm{s},i} + \gamma_{\mathrm{r},i}+1}.
\end{equation}
\subsection{Power Allocation and Splitting Optimization}
In this section, the joint dynamic power allocation and splitting optimization problem is solved. The optimization is performed by a node with full channel state information (CSI). During estimation, R acquires CSI regarding the S-R link, while D acquires CSI regarding the R-D link, from pilot symbols sent by S and R respectively. The corresponding CSI is sent via feedback to the node which performs the optimization. The optimization problem can be defined as
\begin{equation}
\begin{array}{ll}
\underset{\mathbf{P_\mathrm{s}}, \theta, \mathbf{P_\mathrm{r}}}{\text{\textbf{max}}}
&\mathcal{R}_\mathrm{tot} \\
\,\,\,\text{\textbf{s.t.}}
&\mathrm{C}_1:\ \sum_{i=1}^{N_\mathrm{c}}P_{\mathrm{r},i}\leq P_\mathrm{rt},\\
&\mathrm{C}_2:\ \sum_{i=1}^{N_\mathrm{c}}P_{\mathrm{r},i}\leq P_\mathrm{rm},\\
&\mathrm{C}_3:\ \sum_{i=1}^{N_\mathrm{c}}P_{\mathrm{s},i}\leq P_\mathrm{sm},\\
&\mathrm{C}_4:\  0\leq\theta\leq 1,\\
&\mathrm{C}_5:\ P_{\mathrm{s},i}\geq 0,\, \forall i,\\
&\mathrm{C}_6:\  P_{\mathrm{r},i}\geq 0,\, \forall i,
\end{array}
\label{opt14b}
\end{equation}
%=============================================
where $\mathbf{P_\mathrm{s}}$ and $\mathbf{P_\mathrm{r}}$ are the sets of the allocated power, $P_{\mathrm{s},i}$ and $P_{\mathrm{r},i}$, respectively. Constraint $C_1$ represents the limited harvested power which is available for retransmission. Constraints $C_2$ and $C_3$ include the hardware and regulations limitations $P_\mathrm{rm}$ and $P_\mathrm{sm}$ on the total transmitted power by R and S, respectively.
%and $C5$ and $C6$ are necessary for the natural limitations of the physical system.
\subsubsection{Optimization with One-dimensional Search for $\theta$}
The optimization problem (\ref{opt14b}) is non-convex, therefore the complexity to solve it is high, mainly due to the existence of the PS ratio $\theta$, which couples the power allocation variables and results in a non-convex function. In order to overcome this limitation and derive a tractable power allocation algorithm,  a full search with respect to $\theta$ is performed, as in \cite{KwanWPT}. In practice, in order to solve (\ref{opt14b}), the range of $\theta\in[0,1]$ in $K+1$ equally spaced intervals is discretized, i.e. $\theta\in\{0,\theta_1,\theta_2,\ldots,\theta_K,1\}$. For the interval width, $\Delta\theta=\theta_n-\theta_{n-1}$, it holds that $0<\Delta\theta<<1$, while for each $\theta_n$, $n=1,\ldots,K$, (\ref{opt14b}) has to be solved (values $0$ and $1$ are excluded from the search because they lead to $R_\mathrm{tot}=0$). Therefore, the complexity of one-dimensional search is proportional to the number of value intervals for $\theta$, while it is optimal only for infinitely small value intervals. However, even with the aid of the above, the optimization problem in (\ref{opt14b}) is still non-convex, so the following well-known approximation for the end-to-end SNR is utilized, which has been shown to be tight, especially in the medium and high SNR region \cite{Alouini}. Thus,
\begin{equation}\label{approx_snr}
\gamma_{\mathrm{tot},i}=\frac{\gamma_{\mathrm{s},i}\gamma_{\mathrm{r},i}}{\gamma_{\mathrm{s},i}+\gamma_{\mathrm{r},i}+1}\simeq \frac{\gamma_{\mathrm{s},i}\gamma_{r,i}}{\gamma_{\mathrm{s},i}+\gamma_{\mathrm{r},i}}=\tilde{\gamma}_{\mathrm{tot},i},
\end{equation}
\begin{equation}
\mathcal{R}_\mathrm{tot}\simeq\tilde{\mathcal{R}}_\mathrm{tot}=\sum_{i=1}^{N_\mathrm{c}}\frac{1}{2}W_i\log_2\left(1+\tilde{\gamma}_{\mathrm{tot},i}\right).
\end{equation}
Now, for a specific value of $\theta$, problem (\ref{opt14b}) simplifies to
%===========================================================
\begin{equation}
\begin{array}{ll}
\underset{\mathbf{P_\mathrm{s}}, \mathbf{P_\mathrm{r}}}{\text{\textbf{max}}}&
 \tilde{\mathcal{R}}_\mathrm{tot}\\
\,\,\,\text{\textbf{s.t.}}
&\mathrm{C}_1,\mathrm{C}_2,\mathrm{C}_3,\mathrm{C}_5,\mathrm{C}_6
\label{opt2_1}
\end{array}
\end{equation}
%=============================================
%===========================================================
%\begin{equation}
%\begin{array}{ll}
%\underset{\mathcal{P}_s, \mathcal{P}_r}{\text{\textbf{max}}}
%& \tilde{\mathcal{R}} \\
%\text{\textbf{s.t.}}
%&C_1,C_2,C_3,C_5,C_6
%\end{array}
%\label{opt2_1}
%\end{equation}
%=============================================
which is jointly concave with respect to the optimization variables, since the Hessian matrix of its objective function is negative semi-definite. Moreover, it satisfies Slater's constraint qualification, and, thus, it can now be optimally and efficiently solved with dual decomposition, since the duality gap between the dual and the primal solution is zero \cite{Boyd1}. More importantly, it is guaranteed that its global optimum solution can now be obtained in polynomial time.

The Lagrangian of the primal problem (\ref{opt2_1}) is given by
\begin{equation}
\mathcal{L}=\tilde{\mathcal{R}}_\mathrm{tot}-\lambda_1\left(\sum_{i=1}^{N_\mathrm{c}}P_{\mathrm{r},i}-P_\mathrm{rt}\right)-\lambda_2\left(\sum_{i=1}^{N_\mathrm{c}}P_{\mathrm{r},i}-P_\mathrm{rm}\right)-\lambda_3\left(\sum_{i=1}^{N_\mathrm{c}}P_{\mathrm{s},i}-P_\mathrm{sm}\right),
\end{equation}
where $\lambda_1, \lambda_2,\lambda_3 \geq 0$ are the LMs of the constraints $\mathrm{C}_1, \mathrm{C}_2, \mathrm{C}_3$, correspondingly. The constraints $\mathrm{C}_5$ and $\mathrm{C}_6$ will be absorbed into the KKT conditions and, thus, the dual problem is given by
\begin{equation}
\underset{\lambda_1,\lambda_2,\lambda_3}{\text{\textbf{min}}}\,\,\,\,\underset{\mathbf{P_\mathrm{s}}, \mathbf{P_\mathrm{r}}}{\text{\textbf{max}}}\,\mathcal{L}.
\label{dual}
\end{equation}

The dual problem in (\ref{dual}) can be recursively solved in two consecutive layers, namely \textit{Layer 1} and \textit{Layer 2}. In each recursion, the subproblem of power allocation at S and R is solved in Layer 1 by using the KKT conditions for a fixed set of LMs, which are then updated in Layer 2. For this purpose, the subgradient method is used, which enables the parallelized solution of $N$ identically structured problems, corresponding to the optimization of $P_{\mathrm{s},i}$ and $P_{\mathrm{r},i}$ (Layer 1) and requiring only knowledge of the updated values of the LMs. This two-layer approach, which converges after a reasonable number of recursions, reduces considerably the required computational and memory resources. The two layers are explained in detail below.

\textit{Layer 1}: Using the KKT conditions, the optimal power allocation on the $i$-th channel is given by
\begin{subequations}
\begin{equation}
\begin{array}{l}
P_{\mathrm{s},i}=\left[P_{\mathrm{s},i}\in\mathbb{R}:\frac{\partial \mathcal{L}}{\partial P_{\mathrm{s},i}}=0\cap \frac{\partial \mathcal{L}}{\partial P_{\mathrm{r},i}}=0\right]^+,
\end{array}
\end{equation}
\begin{equation}
\begin{array}{l}
P_{\mathrm{r},i}=\left[P_{\mathrm{r},i}\in\mathbb{R}:\frac{\partial \mathcal{L}}{\partial P_{\mathrm{s},i}}=0\cap \frac{\partial \mathcal{L}}{\partial P_{\mathrm{r},i}}=0\right]^+,
\end{array}
\end{equation}
\end{subequations}
%\begin{equation}
%\begin{array}{ll}
%&P_{\mathrm{s},i}=\left[P_{\mathrm{s},i}\in\mathbb{R}:\frac{\partial \mathcal{L}}{\partial P_{\mathrm{s},i}}=0\cap \frac{\partial \mathcal{L}}{\partial P_{\mathrm{r},i}}=0\right]^+\\
%&P_{\mathrm{r},i}=\left[P_{\mathrm{r},i}\in\mathbb{R}:\frac{\partial \mathcal{L}}{\partial P_{\mathrm{s},i}}=0\cap \frac{\partial \mathcal{L}}{\partial P_{\mathrm{r},i}}=0\right]^+
%\end{array}
%\end{equation}
where $[\cdot]^+=\max\left(\cdot,0\right)$.
%The analytical expressions of the optimal power allocation can be derived with simple mathematical calculations, which are omitted due to space limitations. 
It is remarkable that the allocated power in each channel can be calculated in parallel, which further reduces the complexity of the proposed solution.
%
%\begin{equation}
%\{P_{\mathrm{s},i},P_{\mathrm{r},i}\}=\left[\{P_{\mathrm{s},i},P_{\mathrm{r},i}\}\in\mathbb{R}^2:\frac{dL}{dP_{\mathrm{s},i}}=0\cap \frac{dL}{dP_{\mathrm{r},i}}=0\right]^+
%\end{equation}
%where
%\begin{equation}
%[\{w_1,\ldots,w_N\}]^+=\left\{\begin{array}{ll}\{w_1,\ldots,w_N\},&w_1,\ldots,w_N>0,\\\{0,\ldots,0\},&\textit{elsewhere}\end{array}\right..
%\end{equation}

\textit{Layer 2}: Since the dual function is differentiable, the subgradient method can be used to update the LMs as follows
\begin{subequations}
\begin{equation}
\begin{array}{l}
\lambda_1(j+1)=\left[\lambda_1(j)-\hat{\lambda}_1(j)\left(P_\mathrm{rt}-\sum_{i=1}^{N_\mathrm{c}}P_{\mathrm{r},i}\right)\right]^+,
\end{array}
\end{equation}
\begin{equation}
\begin{array}{l}
\lambda_2(j+1)=\left[\lambda_2(j)-\hat{\lambda}_2(j)\left(P_\mathrm{rm}-\sum_{i=1}^{N_\mathrm{c}}P_{\mathrm{r},i}\right)\right]^+,
\end{array}
\end{equation} 
\begin{equation}
\begin{array}{l}
\lambda_3(j+1)=\left[\lambda_3(j)-\hat{\lambda}_3(j)\left(P_\mathrm{sm}-\sum_{i=1}^{N_\mathrm{c}}P_{\mathrm{s},i}\right)\right]^+,
\end{array}
\end{equation}
\end{subequations}
where index $j > 0$ is the recursion index and $\hat{\lambda}_1(t),\hat{\lambda}_2(t),\hat{\lambda}_3(t)$ are positive step sizes, chosen
in order to satisfy the \textit{diminishing step size rules} \cite{Boyd2}.
Since the transformed problem is concave, it is guaranteed that the iteration between the two layers converges to the primal optimal solution \cite{Boyd1,Boyd2}.
\subsubsection{Iterative Solution}
In this subsection, inspired by the alternating optimization \cite{alternating}, a suboptimal iterative solution is introduced, in order to solve (\ref{opt14b}). In each iteration, first, problem (\ref{opt2_1}) is solved with respect to $\mathbf{P_\mathrm{s}}$ and $\mathbf{P_\mathrm{r}}$, for a specific $\theta$ (a given initialization value). Then, the derived solution for $\mathbf{P_\mathrm{s}}$ is used as an input in order to solve the following problem
%===========================================================
\begin{equation}
\begin{array}{ll}
\underset{\theta,\mathbf{P_\mathrm{r}}}{\text{\textbf{max}}}&
\tilde{R}_\mathrm{tot}\\
\,\,\,\text{\textbf{s.t.}}
&\mathrm{C}_1,\mathrm{C}_2,\mathrm{C}_4,\mathrm{C}_6.
\label{opt3}
\end{array}
\end{equation}
%=============================================

%===========================================================
%\begin{equation}
%\begin{aligned}
%& \underset{\theta,\mathcal{P}_r}{\text{\textbf{max}}}
%& &\tilde{R} \\
%& \text{\textbf{s.t.}}
%& &C_1,C_2,C_4,C_6.
%\end{aligned}
%\label{opt3}
%\end{equation}
%=============================================
Note that, $\mathbf{P_\mathrm{r}}$ is also optimized in \eqref{opt3} and is not considered as given, using the solution of \eqref{opt2_1}. $\mathbf{P_\mathrm{r}}$ is a dependent variable on $\mathbf{P_\mathrm{s}}$ and $\theta$ and, thus, it is jointly optimized with each of these two variables, in each step. The new value of $\theta$, which is given by the solution of (\ref{opt3}), is used in the next iteration in order to find the new values for $\mathbf{P_\mathrm{s}}$ and $\mathbf{P_\mathrm{r}}$. Each iteration is a sequential solution of problems \eqref{opt2_1} and \eqref{opt3}. Thus, the complexity of the iterative method is proportional to the number of iterations.

The optimization problem in (\ref{opt3}) can be solved similarly to that in (\ref{opt2_1}), since it is also concave with respect to the optimization variables.
Specifically, the Lagrangian of the primal solution is given by
\begin{equation}
\begin{array}{ll}
\mathcal{L}=\tilde{\mathcal{R}}_\mathrm{tot}-\lambda_1\left(\sum_{i=1}^{N_\mathrm{c}}P_{\mathrm{r},i}-P_\mathrm{rt}\right)-\lambda_2\left(\sum_{i=1}^{N_\mathrm{c}}P_{\mathrm{r},i}-P_\mathrm{rm}\right),
\end{array}
\end{equation}
where $\lambda_1,\lambda_2 \geq 0$ are the LMs.

The dual problem in this case is defined as
\begin{equation}
\underset{\lambda_1,\lambda_2}{\text{\textbf{min}}}\,\,\,\,\underset{\theta, \mathbf{P_\mathrm{r}}}{\text{\textbf{max}}}\,\mathcal{L}.
\end{equation}
Thus, an approach of two layers can be also used in order to find the optimum solution.

\textit{Layer 1}: Using again  the KKT conditions, the optimal values of $\theta$ and $\mathcal{P}_r$ are given by
\begin{subequations}
\begin{equation}
\begin{array}{l}
\theta=\left[\theta\in\mathbb{R}:\frac{\partial \mathcal{L}}{\partial \theta}=0\cap\frac{\partial \mathcal{L}}{\partial P_{\mathrm{r},i}}=0,\,\forall i\right]_0^1,
\end{array}
\end{equation}
\begin{equation}
\begin{array}{l}
P_{\mathrm{r},i}=\left[P_{\mathrm{r},i}\in\mathbb{R}:\frac{\partial \mathcal{L}}{\partial \theta}=0\cap\frac{\partial \mathcal{L}}{\partial P_{\mathrm{r},i}}=0,\,\forall i\right]^+,
\end{array}
\end{equation}
\end{subequations}
where $[\cdot]_0^1=\max\left(\min\left(\cdot,1\right),0\right)$.

\textit{Layer 2}: The LMs are updated by
\begin{subequations}
\begin{equation}
\begin{array}{l}
\,\lambda_1(j+1)=\left[\lambda_1(j)-\hat{\lambda}_1(j)\left(P_\mathrm{rt}-\sum_{i=1}^{N_\mathrm{c}}P_{\mathrm{r},i}\right)\right]^+,
\end{array}
\end{equation}
\begin{equation}
\begin{array}{l}
\lambda_2(j+1)=\left[\lambda_2(j)-\hat{\lambda}_2(t)\left(P_\mathrm{rm}-\sum_{i=1}^{N_\mathrm{c}}P_{\mathrm{r},i}\right)\right]^+,
\end{array}
\end{equation}
\end{subequations}
where $\hat{\lambda}_1(j),\hat{\lambda}_1(j)$ are again positive step sizes. Note that, since $P_\mathrm{rt}$ is dependent on $\theta$ as shown in \eqref{harvested_energy}, it is updated in each recursion.
\begin{algorithm}\label{Alg:suboptimal}
\caption{}
\begin{algorithmic}[1]
\State Initialize $\theta \in (0,1)$.
\Repeat
\State Solve (\ref{opt2_1}) with given $\theta$.
\State Return $\mathbf{P_\mathrm{s}}$.
\State Solve (\ref{opt3}) with given $\mathbf{P_\mathrm{s}}$.
\State Return $\mathbf{P_\mathrm{r}},\theta$.
\Until Number of iterations or another criterion is reached.
\end{algorithmic}
\end{algorithm}
%%%%%%%%%%%%%%%%%%%%%%%%%%%%%%%%%%%%%%%%%%%%%%%%%%%%%%%%%%

The exact steps in order to find the solution of the initial problem with the proposed iterative method are described in Algorithm 1, which converges very close to the optimal solution in a few iterations, as shown by simulation in the next section. 

%Apart from the number of iterations, another general criterion to stop the algorithm is when $\left(\Delta{\mathbf{P_\mathrm{s}}}<\epsilon_1\right)\bigcap\left(\Delta{\mathbf{P_\mathrm{r}}}<\epsilon_2\right)\bigcap\left(\Delta{\theta}<\epsilon_3\right)$ where $\epsilon_1,\epsilon_2,\epsilon_3 \rightarrow 0$.

\subsection{Simulations and Discussion}

\begin{figure}[h]
\centering%
\includegraphics[keepaspectratio,width=0.8\columnwidth]{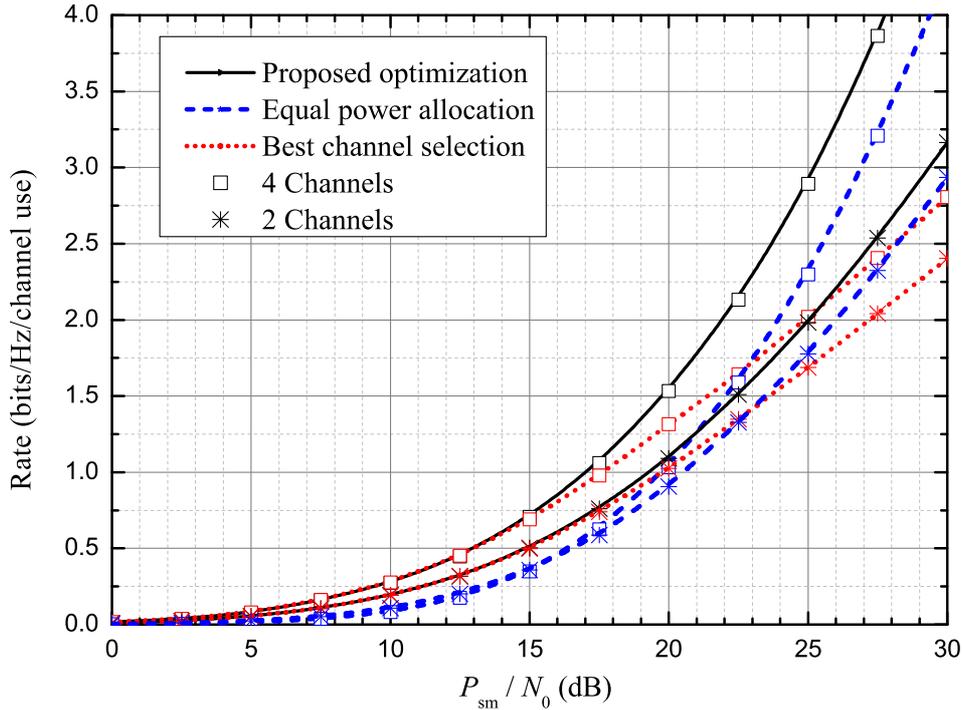}
\caption{Rate against SNR for different number of channels.}
\label{Fig:snrfig}
\end{figure}
\begin{figure}
\centering%
\includegraphics[keepaspectratio,width=0.8\columnwidth]{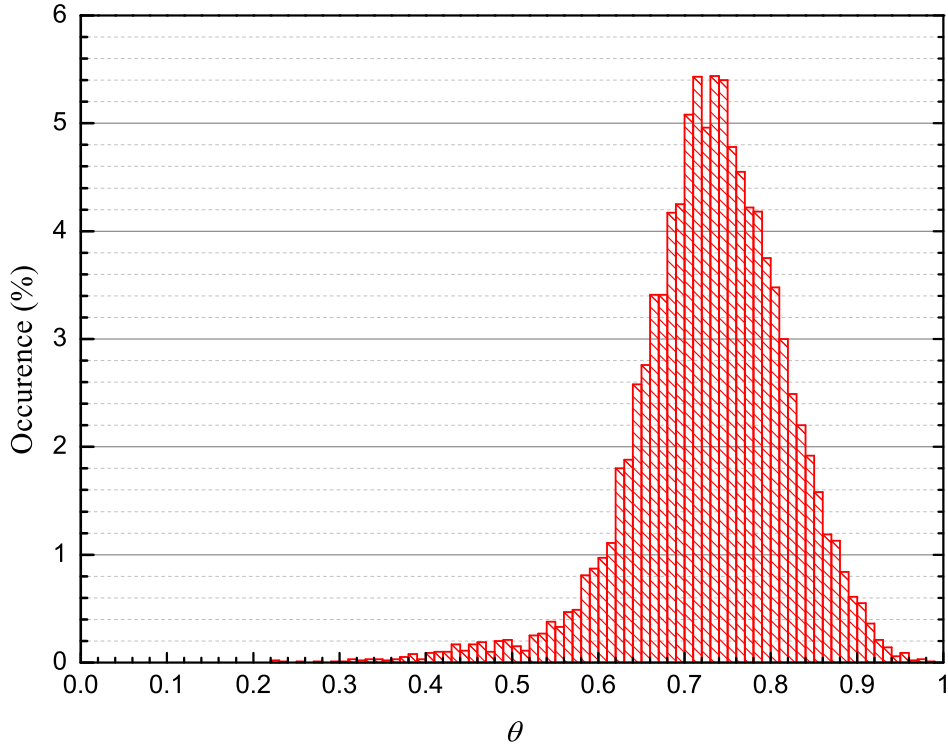}
\caption{Occurrence of values for the optimal $\theta$ in 100 intervals between $\theta=0$ and $\theta=1$.}
\label{Fig:theta_stat}
\end{figure}

This section presents results from simulations for a system which utilizes two or four channels of unitary bandwidth, i.e., $W_i=1$ Hz, $\forall i$. Furthermore, the focus is on the case where the relay harvests energy only via SWIPT and utilizes that for retransmission, i.e., $P_{\mathrm{r}0}=0$. It is also assumed that $P_\mathrm{rm}=P_\mathrm{sm}$. The coordinates of the source and destination are $(-1,0)$ and $(1,0)$, respectively, while the relay is placed at $(-0.25,0.5)$. A bounded path loss model is assumed, for which $\sigma_{\mathrm{s},i}^2=\frac{1}{1+d_\mathrm{sr}^{a_i}}$ and $\sigma_{\mathrm{r},i}^2=\frac{1}{1+d_\mathrm{rd}^{a_i}}$, where $d_\mathrm{sr}$ and $d_\mathrm{rd}$ are the ditances from S to R and R to D, rectively, and $\xi_i$ is the path loss exponent of the $i$-th channel, as in \cite{Ding_path_loss}. This model ensures that the path loss model is valid even for distance values lower than $1$ m \cite{Haenggi}. Specifically, it is considered that $\xi_1=2$ and $\xi_2=2.5$, in order to capture the different propagation characteristics of each channel. When four channels are used, it is further assumed that $\xi_3=3$ and $\xi_4=3.5$. Finally, the efficiency of the energy harvester is set to $\eta_1=0.3$, as a worst case, capturing the effects of low-cost hardware.

\begin{figure}[h!]
\centering%
    \subfigure[Rate]{\includegraphics[keepaspectratio,width=0.7\linewidth]{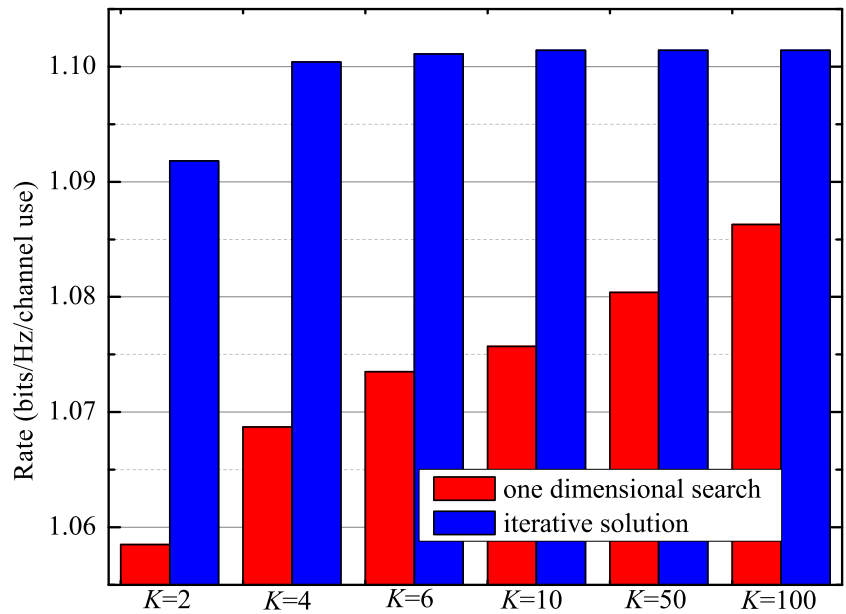}\label{Fig:conva}}
    %\hspace{-10px}
    \subfigure[$\theta$]{\includegraphics[keepaspectratio,width=0.7\linewidth]{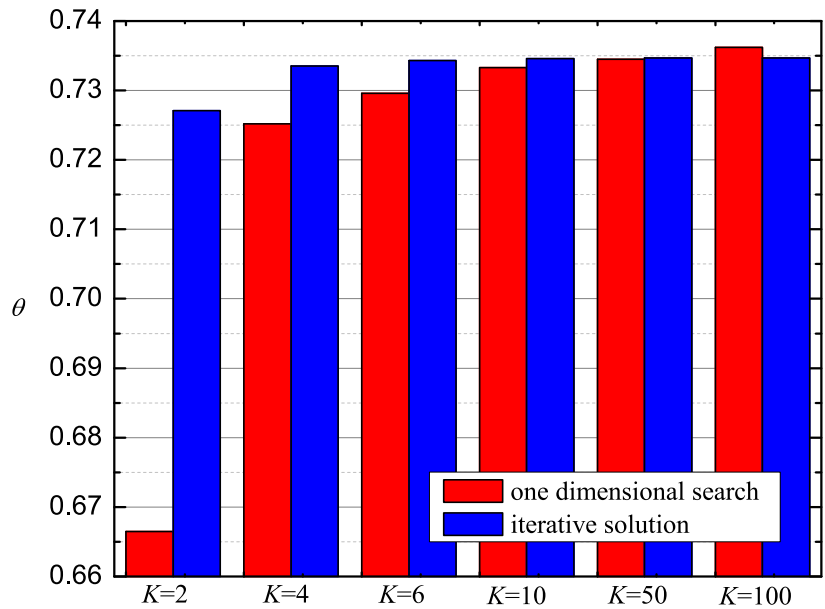}\label{Fig:convb}}
     \caption{Comparison of the two optimization methods for $K$ optimization runs.}
    \label{Fig:conv}
\end{figure}

In Fig. \ref{Fig:snrfig}, the average achievable rate of the proposed optimization is illustrated, for different values of the power ratio, $\frac{P_\mathrm{sm}}{N_0}$. In conjunction with the proposed method, two other power allocation strategies are illustrated for comparison: a) equal PS between the channels, i.e., $P_{\mathrm{s},i}=\frac{P_\mathrm{sm}}{N_\mathrm{c}}\, \forall i$ and $P_{\mathrm{r},i}=\frac{\min(P_\mathrm{rt},P_\mathrm{rm})}{N_\mathrm{c}} \,\forall i$, and b) selection of only the best channel $j$ (the one which maximizes the achievable rate if only one is selected) for each realization, i.e., $P_{s,j}=P_\mathrm{sm}$, $P_{r,j}=\min(P_\mathrm{rt},P_\mathrm{rm})$, and $P_{\mathrm{s},i}=P_{\mathrm{r},i}=0\, \forall i\neq j$. For all methods, the optimal PS ratio is dynamically computed. The proposed power allocation clearly outperforms both schemes, especially when the number of channels increases. For low values of the power ratio, the results for the proposed method and the best channel selection are very close, which implies that the dynamic power allocation preferably allocates more power over the best channel. However, for power ratio values over $15$ dB, the proposed method clearly outperforms the best channel selection, while it displays a performance gain gap over the equal power allocation method of about $1$ dB or $2$ dB when 2 or 4 channels are used, respectively.

In Fig. \ref{Fig:theta_stat}, the distribution of the optimal value of $\theta$ is illustrated when 2 channels are used, for $10,000$ channel realizations. The transmitted power to noise ratio is set to $\frac{P_\mathrm{sm}}{N_0}=20$ dB and the values of $\theta$ are grouped in intervals of width equal to $0.01$. One can observe that the values of optimal $\theta$ which occur more often are within $0.71$ and $0.75$, while more than $90\%$ of the channel realizations lead to an optimal value for $\theta$ in the interval between $0.6$ and $0.9$. These values, which correspond to harvesting more than $60\%$ of the received power, are expected due to the relative position of the network nodes, since more power is needed for transmission over the second hop which is of longer distance.

In Fig. \ref{Fig:conv}, the two proposed optimization methods, i.e., the one-dimensional search for $\theta$ and the iterative solution, are directly compared in terms of complexity and convergence, for $\frac{P_\mathrm{sm}}{N_0}=20$ dB, for two utilized channels and for the same number of optimization operations.
As it can be observed, the iterative method is suboptimal, but it converges fast, while it outperforms the one-dimensional search for low number of operations. More specifically, the solution of the one-dimensional search reaches closer to the optimum when more values of $\theta$ are searched, i.e., for high values of $K$. Each value of $\theta$ which is searched corresponds to solving one optimization problem in \eqref{opt2_1}. Similarly, the iterative method converges to the optimal solution as more iterations are performed. Each iteration comprises two steps, i.e., one solution of \eqref{opt2_1} and one solution of \eqref{opt3}. Thus, for the sake of a fair complexity comparison, $K$ values of $\theta$ in the one-dimensional search are compared to $\frac{K}{2}$ iterations of the iterative method, corresponding to solving $K$ optimizations. Specifically, in Fig. \ref{Fig:conva}, the rate which is achieved after $K=2,4,6,10,50,100$ optimization runs is compared. It is easily observed that the iterative solution converges faster than the one-dimensional search, since in the latter, $\theta$ is restricted to specific quantized values. The resulting rate for the iterative method reaches its optimal value for 5 iterations ($K=10$) and remains unchanged afterwards. Note, however, that both methods achieve solutions for the rate that are less than $4\%$ lower than the optimal value, even for $K=2$. Similar observations can be made in Fig. \ref{Fig:convb}, where the optimal values of $\theta$ are compared.

%\begin{figure}
%\centering%
%\includegraphics[keepaspectratio,width=1\linewidth, trim=1.8cm 1.3cm 3cm 2cm, clip=true]{R_d_1000.eps}
%\caption{Rate against source-relay distance.}
%\label{distfig}
%\end{figure}

%%CHAPTER 1%%%%%%%%%%%%%%%%%%%%%%%%%%%%%%%%%%%%%%%%%%%%%%%%%%%%%%%%%%%%%%%%%%%%%%
%%\else\fi
%
%\part{Physical Layer Key Exchange}\label{Part:PhyLayerKeyExchange}
%
%%\ifchaptereightflag
%%CHAPTER 5 (CONCLUSIONS) %%%%%%%%%%%%%%%%%%%%%%%%%%%%%%%%%%%%%%%%%%%%%%%%%%%%%%%
\chapter[Simultaneous Lightwave Information and Power
Transfer (SLIPT)]{Simultaneous Lightwave Information and Power
Transfer (SLIPT)}\label{ch:chapter5}

This chapter extends the concept of SWIPT in the direction of wireless optical technology, which can be used instead of RF, to simultaneously transfer energy and data. It is called from now on as Simultaneous Lightwave Information and Power Transfer (SLIPT). Note that SLIPT is fundamentally different to SWIPT and creates many new challenges, especially regarding the optimization of the related parameters.

\section{ A Brief Introduction to Optical Wireless Communication (OWC) }\label{Ch5:Introduction}
\begin{figure}
\centering
\includegraphics[width=\linewidth]{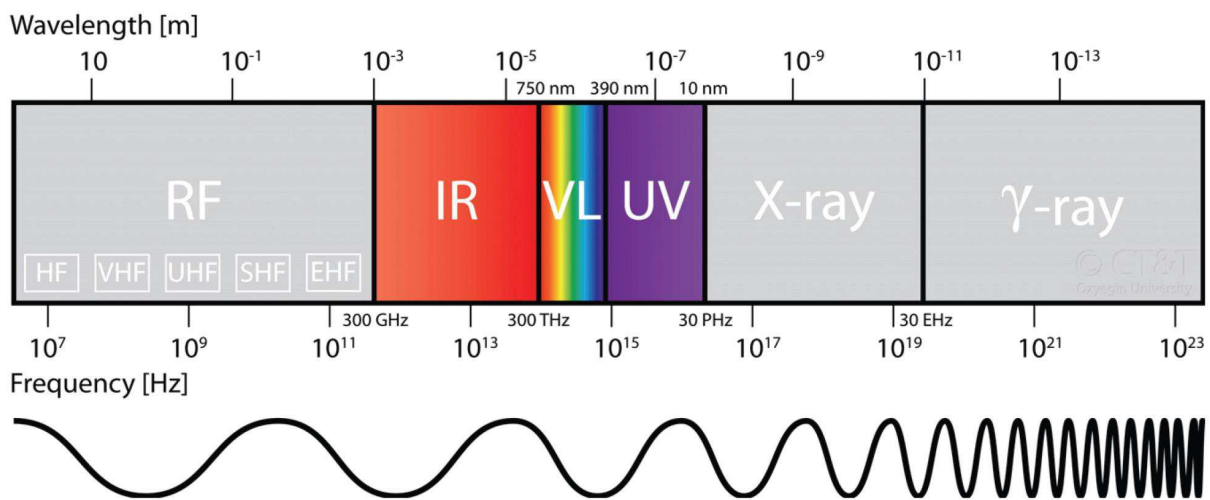}
\caption{The EM spectrum \cite{MUysal}}
\label{figsspectrumchapter5}
\end{figure}

\begin{figure}
\centering
\includegraphics[width=\linewidth]{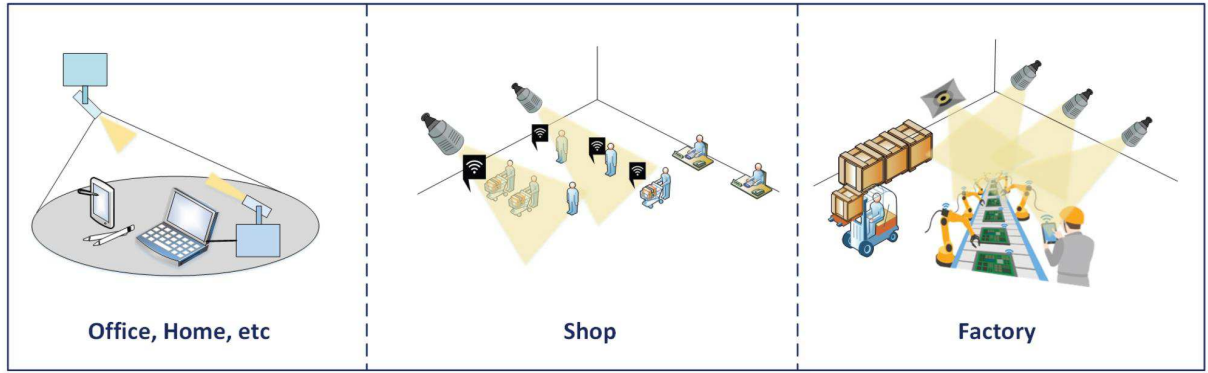}
\caption{Scenarios for indoor OWC applications}
\label{figscenarioschapter5}
\end{figure}

OWC systems utilize optical carrier to wirelessly transfer information from one site to another in uguided propagation media \cite{Kaverhard}.  This technology offers practically an unlimited bandwidth (400 THz) and includes IR, visible (VL) and ultraviolet (UV), as depicted in Fig. \ref{figsspectrumchapter5}. Typically, OWC are based on laser or LED transmitters, due to their ability to be switched on and off at a very high speed. On the receiver side, positive-intrinsic-negative (PIN) and avalanche PDs are commonly used, due to their ability to provide highspeed linear photodetection.

A widely used variation of OWC is free space optics (FSO), which refers to point-to-point information transmission at the near IR frequencies \cite{Haas2, K9}. FSO communications include both short and long range applications. It is noted that short range FSO systems were widely utilized before their RF counterparts became prevalent in the market \cite{Kaverhard}. These were for indoor use and appeared in a number of applications, e.g., remote control units. On the other hand, long range applications include terrestrial systems, which provide a promising solution for the backhaul bottleneck, since they can achieve very high data rates \cite{uysal2016optical, MUysal, K10, Vahid1, Vahid2}. Other uses of long range FSO include aeronautical and space communications. The performance of outdoor FSO systems depends on the atmospheric conditions, as well as precise pointing between the transmitter and receiver\cite{K1, K2, K3, K4, K5, K7,K8, GK7}, which calls for very small irradiance angles.  

However, different variations of OWC, e.g. visible light communications (VLC), do not require pointing and wider irradiance angles can be used \cite{Komine,Hranilovic2, Kahn, Wangrate}. Also, the communication links can be either direct or diffuse \cite{Kaverhard}. This enables the development of a diverse field of communication applications, including wireless access points, wireless local area networks, body area networks, and vehicular networks among others.

Interestingly, the era of IoT opens up the opportunity for a number of promising applications in smart buildings, health monitoring, and predictive maintenance. In the context of wireless access to IoT devices,  RF technology is the main enabler. Furthermore, the exponential growth in the data traffic puts tremendous pressure on the existing global telecommunication networks and the expectations from the 5G wireless networks. Also,  it is remarkable that most of the data consumption/generation, which are related to IoT applications, occurs in  indoor environments \cite{Volker}. Motivated by this, OWC,  such as VLC or  IR, have been recognized as  promising  alternative/complementary technologies to RF, in order to give access to IoT devices in indoor applications \cite{Volker}.  Consequently, OWC are envisioned to be used in a vast number of scenarios, such as in offices, commercial centers, airports, hospitals, industrial environments, etc., as depicted in Fig. \ref{figscenarioschapter5}.

The data rates reported for indoor VLC/IR networking are much higher than those achieved by WiFi, 
especially when client and server are closely located. Apart from the very high data rates \cite{Hranilovic1}, the advantages of VLC/IR include:  i) easy bandwidth reuse, ii) increase of energy efficiency  and considerable energy savings, iii) no RF contamination, iv) free from RF interference, v) and inherent security, since light can be confined within a room. Moreover, nowadays, LEDs and photodetectors (PDs) tend to be considerably cheaper than their RF counterparts, while the cost-efficiency is further improved due to the potential to use the existing lighting infrastructure \cite{Kaverhard, Arnon1, Arnon2}.

Recent research results on OWC open doors of new opportunities and application fields, such as the development of an alternative VLC-based indoor localization technology \cite{uysal2016optical} or the simultaneous information and energy transfer using lightwave technology, with the latter being investigated in the next section.

\section{SLIPT for Indoor Optical Wireless Systems}\label{Ch5:SLIPT}

This section, proposes novel SLIPT strategies for indoor IoT applications, which can be implemented through VL or IR communication systems, with solar panel-based receivers. The proposed strategies are performed at the transmitter or at the receiver, or at both sides, named \textit{Adjusting transmission}, \textit{Adjusting reception}, and \textit{Coordinated adjustment of transmission and reception}, correspondingly. For each strategy, different policies are also proposed, which correspond to different degrees of freedom (DoF). In order to balance the trade-off between harvested energy and  QoS, two optimization problems are formulated and optimally solved. The derived optimization framework can be used to increase the feasibility and efficiency of SLIPT, as it is verified with simulation results. The results of the research presented in this section are included in \cite{SLIPT}.

\subsection{Related Work and Motivation}
Although RF based WPT is a well investigated topic in the last five years, optical wireless power transfer (OWPT)  is a new topic and only a few works have been reported so far in the open literature.  In the pioneering work of Fakidis et. al. \cite{Fakidis}, the visible and  IR parts of the EM spectrum was  used for OWPT, through laser or LEDs at the transmitter and solar cells at the receiver side. Also, in \cite{carvalho} and \cite{Nasiri} EH was performed by using the existing lighting fixtures for indoor IoT applications. Regarding the simultaneous optical wireless information and power transfer, in \cite{Li} the sum rate maximization problem has been optimized in a downlink VLC system with SWIPT.  However, in this paper the utilized EH model does not correspond to that of the solar panel, where only the DC component of the modulated light can be used for EH, in contrast to the alternating current (AC) component, which carries the information. The separation of the DC and AC  components was efficiently achieved by the self-powered solar panel receiver proposed in \cite{Haas1, Haas2}, where it was proved that the use of the solar panel  for communication purposes does not limit its EH capabilities. Thus, the utilization of the power-splitting in the useful recent work \cite{sandalidis}, where the received photocurrent is split in two parts with each of them to include  both a  DC and a AC part, reduces the EH efficiency. Moreover, in \cite{sandalidis} an oversimplified EH model was used, assuming that the harvested energy is linearly proportional to the received optical power, while an optimization of the splitting technique was not presented. Furthermore, in the significant research works \cite{Alouini_optical1, Alouini_optical2}, a dual-hop hybrid VLC/RF communication system is considered, in order to extend the coverage. In these  papers, besides detecting the information over the VLC link, the relay is also able to harvest energy from the first-hop VLC link, by extracting the DC component of the received optical signal. This energy can be used to re-transmit the data to a mobile terminal over the second-hop RF link. Also,  in \cite{Alouini_optical1} the proposed hybrid system was optimized, in terms of data rate maximization, while in \cite{Alouini_optical2} the packet loss probability was evaluated.

\subsection{Contribution}
This section presents for first time a framework for simultaneous optical wireless information and power transfer, called from now on as  \textit{Simultaneous Lightwave Information and Power Transfer (SLIPT)}, which 
can be efficiently used for indoor IoT applications through VLC or IR systems. More specifically, novel 
and fundamental strategies are proposed, in order to increase the feasibility and efficiency of SLIPT, when a solar panel-based receiver is used. These strategies are performed at the transmitter or at the receiver, or at both sides, named \textit{Adjusting transmission}, \textit{Adjusting reception}, and \textit{Coordinated adjustment of transmission and reception}. Regarding adjusting transmission 
two policies are proposed: 
\begin{enumerate}[label=\roman*.]
\item \textit{Time-splitting (TSp)}, according to which the time frame is separated in two 
distinct phases, where in each of them the main focus is either on communication  or energy transfer and,
\item \textit{Time-splitting with DC bias optimization}, which is a generalization of TSp. 
\end{enumerate}
In contrast to RF-based wireless 
powered networks, where the TSw strategy and adjustment of the related parameters takes place at the receiver's side, 
TSp in SLIPT refers to the adaptation of specific parameters of the transmitted signal. Regarding adjusting reception, the 
\textit{Field-of-view (FoV) adjustment} policy is proposed, while according to the coordinated adjustment of transmission and reception strategy, the simultaneous optimization of the former policies at both transmitter and receiver is proposed, in order to maximize the harvested energy, while achieving the required QoS (e.g.  data rate and 
SINR). Finally, the resulting two optimization problems are formulated and optimally solved.

\subsection{System and Channel Model}\label{S:Intro}
\begin{figure}[h!]
\centering
\includegraphics[width=0.8\columnwidth]{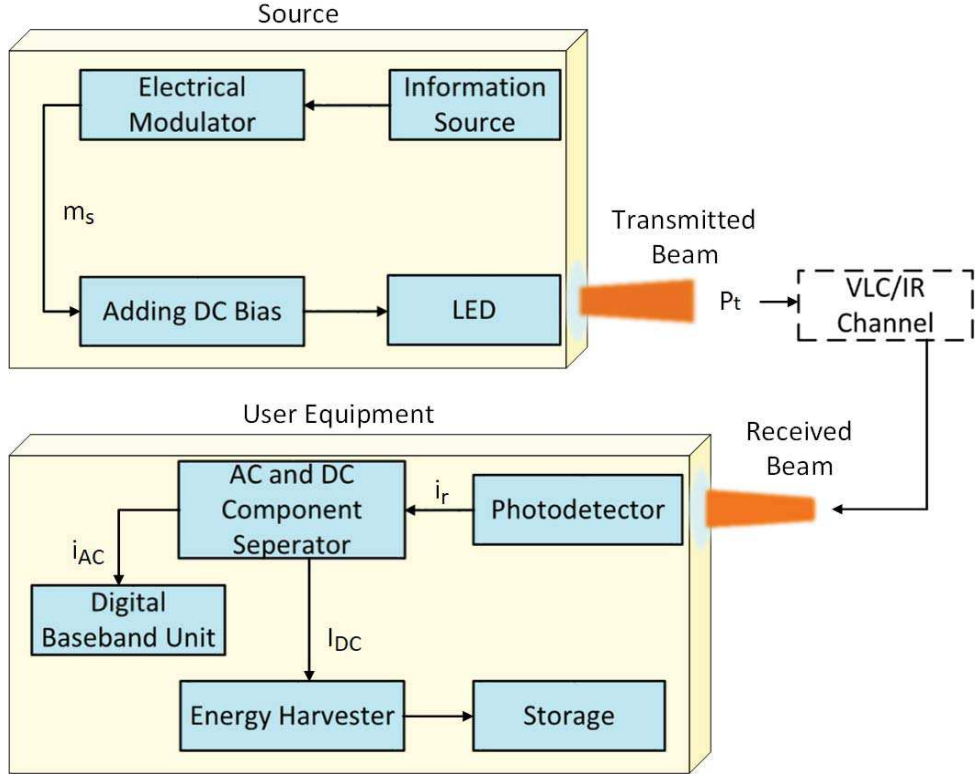}
\vspace{-0.05 in}
\caption{SLIPT transceiver design}
\label{Fig1chapter5}
\end{figure}

The downlink transmission of an OWC system is considered, which consists of one LED and a single user. It is also assumed that the user is equipped with the functionality of EH. The VLC/IR transmitter/receiver design is shown in Fig. \ref{Fig1chapter5}, while the VLC/IR downlink communication is depicted in Fig. \ref{Fig2chapter5}.

\subsubsection{Optical Wireless Transmission}
Let $m_\mathrm{s}(t_\mathrm{in})$ denote the modulated electrical signal that corresponds to the bit stream from the information source. A DC 
bias $B$ is added to $m_\mathrm{s}(t_\mathrm{in})$ to ensure that the resulting signal is non-negative,  before being used to 
modulate the optical intensity of the LED and regulate the LED in the proper operation mode.
The transmitted optical signal from the LED is \cite{Alouini_optical1}
\begin{equation}
P_\mathrm{t}(t_\mathrm{in})=P_\mathrm{LED}[B+m_\mathrm{s}(t_\mathrm{in})],
\end{equation}
where $P_\mathrm{LED}$ is the LED power.
The electrical signal varies around the DC bias $B\in[I_L,I_H]$ with peak amplitude  $A$, where $I_\mathrm{L}$ is the minimum and $I_\mathrm{H}$ is the maximum input bias currents, correspondingly.  In order to avoid clipping distortion by the nonlinearity of the LED, by restraining the input electrical signal to the LED within the linear region of the LED operation, the following limitation is induced
\begin{equation}
A\leq\min(B-I_\mathrm{L},I_\mathrm{H}-B),
\label{constraint}
\end{equation}
where $\min(\cdot)$ denotes the minimum between the elemts.

\subsubsection{Channel Model}
The channel gain is given by \cite{Komine,Hranilovic2, Kahn}
\begin{equation}\label{channelpowergain}
h=\frac{L_\mathrm{r}}{d^2}R_0(\varphi)g_\mathrm{f}(\psi)g_\mathrm{c}(\psi)\cos(\psi),
\end{equation}
where $L_r$ is the physical area of
the photo-detector, $d$ is the transmission distance from the LED to the illuminated
surface of the photo-detector, $g_\mathrm{f}(\psi)$ is the gain of the optical filter and $g_\mathrm{c}(\psi)$ represents the gain of the optical concentrator, given by \cite{Komine,Kahn}
\begin{equation}
g_\mathrm{c}(\psi)=\begin{cases}
\frac{\rho^2}{\sin^2 (\Psi_\mathrm{FoV})},\,0\leq \psi\leq \Psi_\mathrm{FoV},\\
0,\, \psi> \Psi_\mathrm{FoV}.
\end{cases}
\end{equation}
with $\rho$ and $\Psi_\mathrm{FoV}$ being the refractive index and FoV, respectively.
Also in (\ref{channelpowergain}),  $R_0(\varphi)$ is the Lambertian radiant intensity of the LED, given by
\begin{equation}
R_0(\varphi)=\frac{\xi_0+1}{2\pi}\cos^{\xi_0} \varphi,
\end{equation}
where $\varphi$ is the irradiance angle, $\psi$ is the incidence angle, and
\begin{equation}
\xi_0=-\frac{1}{\log_2\cos(\Phi_{1/2})},
\end{equation}
with $\Phi_{1/2}$ being the semi-angle at half luminance.

\begin{figure}
\centering
\includegraphics[width=0.8\columnwidth]{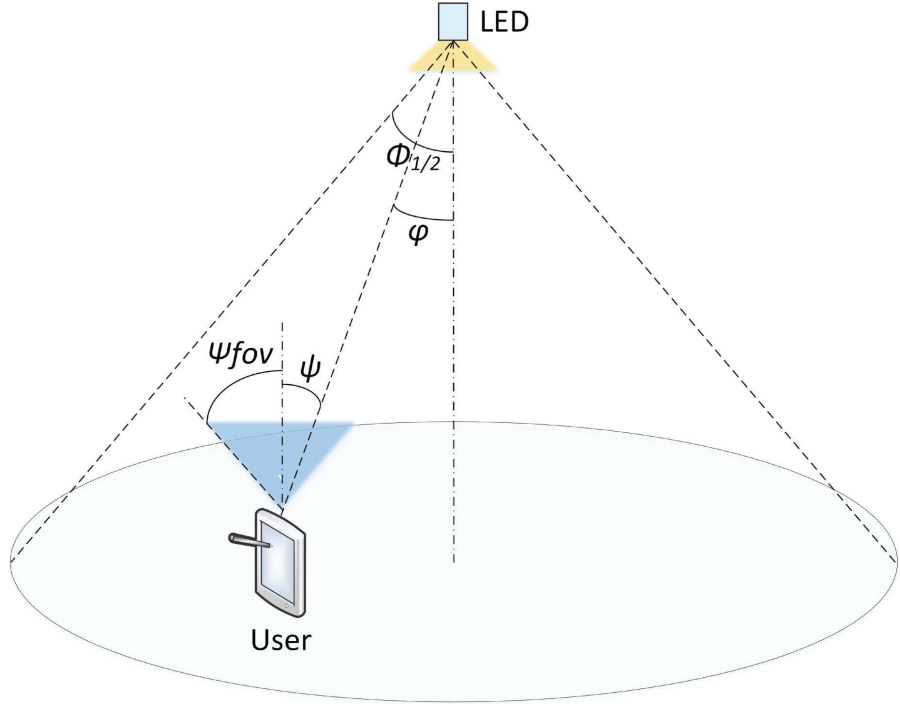}
\vspace{-0.05 in}
\caption{VLC/IR downlink communication}
\label{Fig2chapter5}
\end{figure}

\subsubsection{Received Electrical SINR}
The electrical current $i_r(t_\mathrm{in})$ at the output of the PD can be written as
\begin{equation}
i_\mathrm{r}=\eta_0 (h P_\mathrm{t}(t_\mathrm{in})+P_\mathrm{ro})+\nu(t_\mathrm{in})=I_{\mathrm{DC}}(t_\mathrm{in})+i_\mathrm{AC}(t_\mathrm{in})+\nu(t_\mathrm{in}),
\end{equation}
where $\eta_0$ is the photo-detector responsivity in $\mathrm{A/W}$, $P_\mathrm{ro}$ is the received optical signal from other sources, e.g. other neighboring LEDs, $I_{\mathrm{DC}}$ is the DC component, $i_\mathrm{AC}(t)$ is the AC component, and $\nu(t_\mathrm{in})$ is the AWGN, which is created from background shot noise and thermal noise.

The AC component $i_\mathrm{AC}(t)$ is composed of two terms, i.e. 
\begin{equation}
i_\mathrm{AC}(t_\mathrm{in})=i_{\mathrm{AC},1}(t_\mathrm{in})+i_{\mathrm{AC},2}(t_\mathrm{in}),
\end{equation}
where
\begin{equation}
i_{\mathrm{AC},1}(t_\mathrm{in})=\eta_0 h P_\mathrm{LED}m_\mathrm{s}(t_\mathrm{in})
\end{equation}
is due to the dedicated LED, and $i_{\mathrm{AC},2}(t_\mathrm{in})$ is due to other interfering sources.
Thus, the received SINR can be written as
\begin{equation}
\gamma_0=\frac{(\eta_0 h P_\mathrm{LED}A)^2}{P_I+\sigma^2},
\label{SINR}
\end{equation}
where $\sigma^2$ is the noise power and $P_I$ is the electrical power of the received interference.

\subsubsection{EH Model}
As it has already been mentioned the photocurrent consists of both the DC and AC signals. In order to perform EH, the DC component is blocked by a capacitor and passes through the EH branch \cite{Haas1}. The harvested energy is given by \cite{li2011solar}
\begin{equation}
E=F_0I_\mathrm{DC}V_\mathrm{oc},
\label{EH}
\end{equation}
with $F_0$ being the fill factor \cite{li2011solar} and $I_\mathrm{DC}=I_{\mathrm{DC},1}+I_{\mathrm{DC},2}$
being the DC component of the output current, where
\begin{equation}
I_{\mathrm{DC},1}=\eta hP_\mathrm{LED}B
\end{equation}
is due to the dedicated LED, while $I_{\mathrm{DC},2}$ is due to different light sources, e.g. neighboring LEDs. Also, $V_{\mathrm{oc}}$ is the open circuit voltage and is given by
\begin{equation}
V_{\mathrm{oc}}=V_\mathrm{t}\ln\left(1+\frac{I_{\mathrm{DC}}}{I_0}\right),
\end{equation}
where $V_\mathrm{t}$ is the thermal voltage and $I_0$ is the dark saturation current of the
PD. 
%Moreover, $F_0$ is the fill factor, defined as the ratio of the maximum power from the solar cell to the product of the open-circuit voltage $V_\mathrm{oc}$ and the short circuit current $I_\mathrm{sc}$.

\subsection{SLIPT Strategies}
In this section fundamental SLIPT strategies are proposed, for use in VLC/IR communication systems. These strategies are  performed either at the transmitter  or at the receiver, or at both sides: \textit{Adjusting transmission}, \textit{Adjusting reception}, and \textit{Coordinated adjustment of transmission and reception}.

\subsubsection{Adjusting Transmission}
Next, two policies for the adjusting transmission strategy are introduced, named \textit{Time-splitting} and  \textit{Time-splitting with DC Bias Optimization}.
\paragraph{Time-splitting}
\label{strategy 1}
According to the Time-splitting policy the received optical signal is  used for a portion of time solely for  EH, instead of decoding. During this period of time the LED  transmits by using the maximum DC bias, in order to maximize the harvested energy by the receiver. Thus, assuming time frames of unitary duration, there are the following two distinct phases during a time frame:

\underline{Phase $1$}: The AC component of the received signal is used for information decoding and the DC component for EH. Let  $A_1$ and $B_1\in[I_L,I_H]$ denote the peak amplitude of $m_\mathrm{s}(t_\mathrm{in})$ and DC bias, respectively. During Phase 1, the aim is to maximize the received SINR.  Since SINR is an increasing function with respect to $A_1$, then $A_1$ takes its maximum value, which, considering \eqref{constraint} is given by $A_1=\frac{I_H-I_L}{2}$ and similarly, $B_1=\frac{I_H+I_L}{2}$. The duration of this phase is denoted by $0\leq T\leq 1$, which can be optimized according to the QoS requirements. For a specific value of $T$, the amount of harvested energy is given by
\begin{equation}
E_{TS}^{[1]}=F_0T\left(\eta_0 hP_\mathrm{LED}\frac{I_H+I_L}{2}+I_{\mathrm{DC},2}\right)V_\mathrm{t} \ln\left(1+\frac{\eta_0 hP_\mathrm{LED}\frac{I_H+I_L}{2}+I_{\mathrm{DC},2}}{I_0}\right).\\
\end{equation}

\underline{Phase $2$}: In the time period $1-T$, the aim is to maximize the harvested energy, which is an increasing function with respect to $B$. Thus, during Phase 2 the transmitter eliminates the AC part and maximizes the DC bias, i.e., $A=0$ and $B=I_H$, where  $A_2$ and $B_2\in[I_L,I_H]$ denote the values of $A$ and $B$, respectively. 
%Thus, \textbf{the received SINR is zero }and the amount of harvested energy during this phase, is given by
Thus, the amount of harvested energy during this phase, is given by
\begin{equation}
E_{TS}^{[2]}=F_0(1-T)(\eta_0 hP_\mathrm{LED}I_H+I_{\mathrm{DC},2})V_\mathrm{t}\ln\left(1+\frac{\eta_0 hP_\mathrm{LED}I_H+I_{\mathrm{DC},2}}{I_0}\right).
\end{equation}

Considering both phases, the total harvested energy is given by
\begin{equation}
E_{TS}=E_{TS}^{[1]}+E_{TS}^{[2]}.
\label{harvested energy TS}
\end{equation}

\paragraph{Time-Splitting with DC Bias Optimization}
\label{strategy 2}
This policy is a generalization of Time-splitting. During Phase $1$, the DC bias is optimized in order to increase the harvested energy, while it simultaneously enables information transfer, i.e., $A_1> 0$. In this case, the total harvested energy is given by
\begin{equation}
E_\mathrm{TSBO}=F_0T\left(\eta_0 hP_\mathrm{LED}B_1+I_{\mathrm{DC},2}\right)V_\mathrm{t}\ln\left(1+\frac{\eta_0 hP_\mathrm{LED}B_1+I_{\mathrm{DC},2}}{I_0}\right)+E_{TS}^{[2]},\\
\label{harvested energy TSBO}
\end{equation}
where $B_1$ is the DC bias during Phase $1$.

\subsubsection{Adjusting Reception}
\label{strategy 3}
We propose the \textit{Adjustment of the FoV} policy for the adjusting reception strategy, in order  to balance the trade-off between  harvested energy and SINR. Controlling FoV is particularly important especially when, except for the used VLC/IR LED, there are extra light sources in the serving area \cite{Ghamdi}, e.g. neighboring LEDs that serve other users. For the practical and efficient implementation of this policy, electrically controllable liquid crystal (LC) lenses is a promising technology \cite{Tsou}.

When the aim is to maximize the SINR, the FoV is tuned up to receive the beam of the dedicated LED only (if possible), in order to reduce the beam overlapping. This is achieved by tuning the FoV to the narrowest setting, that allows reception only from that LED. On the other hand, when the aim is to achieve a balance between SINR and harvested energy, a wider FoV setting could be selected. 

For the sake of practicality, it is assumed that the VLC/IR receiver has discrete FoV settings, i.e. $\Psi_\mathrm{FoV}\in\{\Psi_\mathrm{FoV}^{[1]},..,\Psi_\mathrm{FoV}^{[M]}\}$. Also, note that except for $h$, both $P_I$ and $I_{\mathrm{DC},2}$ are also discrete functions of $\Psi_\mathrm{FoV}$, i.e., $P_I=P_I(\Psi_\mathrm{FoV})$ and $I_{\mathrm{DC},2}=I_{\mathrm{DC},2}(\Psi_\mathrm{FoV})$.

\subsubsection{Coordinated Transmission and Reception Adjustment}
\label{strategy 4}
Considering \eqref{SINR}, \eqref{harvested energy TS}, and \eqref{harvested energy TSBO}, it is revealed that both   SINR and  harvested energy -apart from $A_1$, $B_1$ and $T$-  also depend on the selection of $\Psi_\mathrm{FoV}$, despite the utilized adjusting transmission technique. This dependence motivates the coordinated transmission and reception adjustment, i.e. the coordination between the strategy \ref{strategy 1} or \ref{strategy 2} and \ref{strategy 3}, which results in the following two policies, i.e.
\begin{itemize}
\item Policy 1: Time-splitting with tunable FoV (\ref{strategy 1} and \ref{strategy 3})
\item Policy 2: Time-splitting with DC bias optimization and tunable FoV (\ref{strategy 2} and \ref{strategy 3})
\end{itemize}
Note that in both policies, during Phase 2, where the aim is to maximize the harvested energy, the FoV setting that maximizes $E_{TS}^{[2]}$ should be used. This is not necessarily the widest setting, because although it increases the received beams (if there are neighboring LEDs), it reduces $g_\mathrm{c}(\psi)$. On the other hand, the preferable FoV setting during phase 1, denoted by $\Psi_\mathrm{FoV,1}$, cannot be straightforwardly determined, since it also depends  on the required QoS.

\subsection{SLIPT Optimization}

SLIPT induces an interesting trade-off between harvested energy and  QoS. This subsection aims to balance this trade-off by maximizing the harvested energy, while achieving the required user QoS. The focus is on the coordinated adjustment of transmission and reception strategy, which can be considered as a generalization of the other SLIPT strategies. The following optimization problems can be formulated, based on the two techniques presented in subsection \ref{strategy 4}.

Regarding the QoS, two different criteria are taken into account, namely SINR and information rate. Note that these two criteria are not equivalent to each other, when either of the two techniques is used, due to the time-splitting.  More specifically,  since only Phase $1$ is used for information transmission (the duration of which is $T$), the lower bound of the capacity is given by \cite{Wangrate}
\begin{equation}
R^\mathrm{d}=T\log_2\left(1+\frac{e}{2\pi}\gamma_0\right).
\label{rate}
\end{equation}

\subsubsection{Time-Splitting with Tunable FoV}
The corresponding optimization problem can be expressed as

%===========================================================
\begin{equation}
\begin{array}{ll}
\underset{T, \Psi_\mathrm{FoV,1}}{\text{\textbf{max}}}&  E_\mathrm{TS} \\
\quad\text{\textbf{s.t.}}&\mathrm{C}_1: R^\mathrm{d}\geq R_{\mathrm{th}},\\
& \mathrm{C}_2: \gamma_0\geq \gamma_{\mathrm{th}},\\
&\mathrm{C}_3:0 \leq T\leq 1 ,\\
&\mathrm{C}_4:\Psi_\mathrm{FoV,1}\in\{\Psi_\mathrm{FoV}^{[1]},..,\Psi_\mathrm{FoV}^{[M]}\},\\
\end{array}
\label{opt TS}
\end{equation}
%=============================================
where $R_{\mathrm{th}}$ and $\gamma_{\mathrm{th}}$ denote the information rate SINR and threshold, respectively. 
\begin{theorem}
The optimal value of $T$ in (\ref{opt TS}) is given by
\begin{equation}
T^*=\frac{R_{\mathrm{th}}}{\log_2\left(1+\frac{e\left(\eta_0 h P_\mathrm{LED}(I_H-I_L)\right)^2}{8\pi(P_I(\Psi^*_\mathrm{FoV,1})+\sigma^2)}\right)},
\label{optimal T TS}
\end{equation}
where $(\cdot)^*$ denotes optimality.
\end{theorem}

\begin{proof}
The optimization problem \eqref{opt TS} is a combinatorial one. In order to find the optimal solution, all possible values of $\Psi_\mathrm{FoV,1}$  have to be checked before selecting the value that maximizes the harvested energy, $E_h^\mathrm{TS}$, while satisfying the constraints $\mathrm{C}_1$, $\mathrm{C}_2$, and $\mathrm{C}_3$. For a specific value of $\Psi_\mathrm{FoV,1}$, if
\begin{equation}
\frac{(\eta_0 h P_\mathrm{LED}\frac{I_H-I_L}{2})^2}{P_I(\Psi_\mathrm{FoV,1})+\sigma^2}<\gamma_{\mathrm{th}},
\end{equation}
then the optimization problem is infeasible, since $\mathrm{C}_2$ is not satisfied.
Also, due to constraint $\mathrm{C}_1$, the following limitation is induced for $T$, 
\begin{equation}
T \geq\frac{R_{\mathrm{th}}}{\log_2\left(1+\frac{e(\eta_0 h P_\mathrm{LED}\frac{I_H-I_L}{2})^2}{2\pi(P_I(\Psi_\mathrm{FoV,1})+\sigma^2)}\right)}.
\end{equation}
Moreover, the harvested energy is decreasing with respect to $T$. Thus, the optimal value of $T$ is given by
\eqref{optimal T TS} and the proof is completed.
\end{proof}
Note that if $T^*>1$, the optimization problem in \eqref{opt TS} is infeasible, due to $\mathrm{C}_3$.

\subsubsection{Time-Splitting with DC Bias Optimization and Tunable FoV}

The corresponding optimization problem can be formulated as

%===========================================================
\begin{equation}
\begin{array}{ll}
\underset{B_1, A_1, T, \Psi_\mathrm{FoV,1}}{\text{\textbf{max}}}&  E_\mathrm{TSBO} \\
\,\,\,\text{\textbf{s.t.}}&\mathrm{C}_1: R^\mathrm{d}\geq R_{\mathrm{th}},\\
&\mathrm{C}_2: \gamma_0\geq \gamma_{\mathrm{th}},\\
&\mathrm{C}_3:A_1\leq \min(B_1-I_L,I_h-B_1),\\
&\mathrm{C}_4:0 \leq T\leq 1 ,\\
&\mathrm{C}_5:A_1\geq 0,\\
&\mathrm{C}_6:I_L\leq B_1\leq I_H,\\
&\mathrm{C}_7:\Psi_\mathrm{FoV,1}\in\{\Psi_\mathrm{FoV}^{[1]},..,\Psi_\mathrm{FoV}^{[M]}\}.\\
\end{array}
\label{opt TS DC}
\end{equation}
%=============================================
\begin{proposition}
The optimal value of $B$ in (\ref{opt TS DC}) belongs in the range $\left[\frac{I_H+I_L}{2},I_H\right]$.
\label{Proposition1}
\end{proposition}

\begin{proof}
The constraint $\mathrm{C}_3$ can be rewritten as
\begin{equation}
\mathrm{C}_{3\mathrm{a}}:A_1\leq B_1-I_L, \mathrm{C}_{3\mathrm{b}}:A_1\leq I_H-B_1.
\end{equation}
For a specific value of $B_1$, only one of the constraints $\mathrm{C}_{3\mathrm{a}}$ and $\mathrm{C}_{3\mathrm{b}}$ is activated. Now, let assume that the optimal solution is  $B_1^*<\frac{I_H+I_L}{2}$, for which all the constraints are satisfied. In this case, $\mathrm{C}_{3\mathrm{a}}$ is activated.  However, by setting $B_1=\frac{I_H+I_L}{2}$ the objective function is increased, while the constraints are still satisfied. Thus, it becomes evident that $B_1^*$ is not optimal. Consequently, Proposition \ref{Proposition1} has been proved by contradiction.
\end{proof}

The optimal value $\Psi_\mathrm{FoV,1}$ is calculated similarly to the solution  of \eqref{opt TS}. Regarding the rest optimization variables of \eqref{opt TS DC} they are optimized according to the following theorem:
\begin{theorem}
\label{Theorem2}
For a specific value of $\Psi_\mathrm{FoV,1}$, the optimal value of $T$ is given by
\begin{equation}
T^*=\underset{K_1\leq T\leq K_2}{\mathrm{argmax}} \tilde{E}_\mathrm{TSBO}
\label{equivalent}
\end{equation}
with $\tilde{E}_\mathrm{TSBO}$ being solely a function of $T$ and given by \eqref{harvested energy TSBO}, by replacing $A_1$ and $B_1$ by
\begin{equation}
A_1=\frac{1}{\eta_0hP_\mathrm{LED}}\sqrt{\frac{2\pi(P_I(\Psi_\mathrm{FoV,1})+\sigma^2)(2^{\frac{R_\mathrm{th}}{T}}-1)}{e}},
\label{A1}
\end{equation}
and
\begin{equation}
B_1=I_H-A_1,
\label{B1}
\end{equation}
respectively.
Also,
\begin{equation}
K_1=\frac{R_\mathrm{th}}{\log_2\left(1+\frac{e\left(\eta_0 hP_\mathrm{LED}(I_H-I_L)\right)^2}{8\pi\left(P_I(\Psi^*_\mathrm{FoV,1})+\sigma^2\right)}\right)}
\end{equation}
\begin{equation}
K_2=\min\left(\frac{R_\mathrm{th}}{\log_2\left(1+\frac{e\gamma_\mathrm{th}}{2\pi}\right)},1\right).
\end{equation}
Finally, the optimal values of $A_1$ and $B_1$ are given by \eqref{A1} and \eqref{B1}, by replacing $\Psi_\mathrm{FoV,1}$ and $T$ by $\Psi^*_\mathrm{FoV,1}$ and $T^*$, respectively.
\end{theorem}

\begin{proof}
Considering Proposition \ref{Proposition1} and for a specific value of $\Psi_\mathrm{FoV,1}$ the optimization problem in \eqref{opt TS DC} can be re-formulated as
%===========================================================
\begin{equation}
\begin{array}{ll}
\underset{B_1, A_1, T}{\text{\textbf{max}}}&  E_\mathrm{TSBO} \\
\,\,\,\,\text{\textbf{s.t.}}&\mathrm{C}_1: R^\mathrm{d}\geq R_{\mathrm{th}},\\
&\mathrm{C}_2: \gamma_0\geq \gamma_{\mathrm{th}},\\
&\mathrm{C}_3:A_1+B_1\leq I_H,\\
&\mathrm{C}_4:0 \leq T\leq 1,\\
&\mathrm{C}_5:A_1\geq 0,\\
&\mathrm{C}_6:B_1\geq \frac{I_H+I_L}{2}.\\
\end{array}
\label{reformulated}
\end{equation}
%=============================================

%\begin{figure*}[h!]
%\begin{equation}
%\begin{split}
%&E_\mathrm{TSBO}=\\
%&0.75T(\eta h_nP_\mathrm{LED}B+I_2)V_t\ln(1+\frac{\eta h_nP_\mathrm{LED}B_1+I_2}{I_0})+\\
%&0.75(1-T)(\eta h_nP_\mathrm{LED}I_HV_t+I_2)\ln(1+\frac{\eta h_nP_\mathrm{LED}I_H+I_2}{I_0}),
%\end{split}
%\label{harvesting energy single user}
%\end{equation}
%\hrulefill
%\end{figure*}
The optimization problem in \eqref{reformulated} still cannot be easily solved in its current form, since the objective function as well as the constraints $\mathrm{C}_1$ and $\mathrm{C}_2$ are not concave. However, it can be solved with low complexity by using the following reformulation.

First, the inequalities in $\mathrm{C}_1$ and  $\mathrm{C}_3$ are replaced by equalities. Then, $A_1$ and $B_1$ are given by \eqref{A1} and \eqref{B1}, respectively. By substituting $A_1$ and $B_1$ by \eqref{A1} and \eqref{B1}, $\mathrm{C}_1$,  $\mathrm{C}_3$, and $\mathrm{C}_3$ of \eqref{reformulated} vanish, and the optimization problem is rewritten as

%===========================================================
\begin{equation}
\begin{array}{ll}
\underset{B_1, A, T_1\forall n}{\text{\textbf{max}}}&  \tilde{E}_h^\mathrm{TSBO} \\
\,\,\,\text{\textbf{s.t.}}& \mathrm{C}_2: T\leq \frac{R_\mathrm{th}}{\log_2\left(1+\frac{e\gamma_\mathrm{th}}{2\pi}\right)},\\
&\mathrm{C}_4:0\leq T\leq 1 ,\\
&\mathrm{C}_6:T\geq \frac{R_\mathrm{th}}{\log_2\left(1+\frac{e\left(\eta_0 hP_\mathrm{LED}(I_H-I_L)\right)^2}{8\pi\left(P_I+\sigma^2\right)}\right)},\\
\end{array}
\end{equation}
%=============================================
which is equivalent to \eqref{equivalent}, and, thus, the proof is completed.

\end{proof}

\vspace{-0.1 in}
\subsection{Simulations and Discussion}

The downlink VLC/IR system of Fig. \ref{Fig2chapter5}, where  the user is located in a distance $d=1.5$ m from the LED, $\psi=0$, and the transmitter plane is parallel to the receiver one, i.e., $\varphi=\psi$. In the same room there are $N$ other LEDs, which simultaneously use the same frequency band. The distance between each of them and from the dedicated LED is $1.5$ m. It is also assumed that $F_0=0.75$, $P_{LED}=20$ W/A, $\Phi_{1/2}=60$ deg, $\sigma^2=10^{-15}$ A$^2$, $L_r=0.04$ m$^2$, $\eta_0=0.4$ A/W, $I_0=10^{-9}$ A, $I_L=0$ A, $I_H=12$ mA \cite{Alouini_optical1}, $g_\mathrm{f}=1$, $\rho=1.5$, $\gamma_{th}=10$ dB, and  two settings for the FoV, i.e., $\Psi_\mathrm{FoV}\in\{30,50\}$ deg, are considered. 

Regarding the neighboring LEDs, it is assumed that the DC bias and the peak amplitude are given by $A_n'=B_n'=6$ mA, $\forall n\in\{1,...,N\}$, while the rest parameters are equal to those of the dedicated LED. Furthermore, the channel between them and the user's receiver, denoted by $h'_n$ is modeled according to \eqref{channelpowergain}, using the corresponding parameters. Thus, when the widest FoV setting is selected, $P_I$ and $I_{\mathrm{DC},2}$ are given by
\begin{equation}
P_I=\sum_{n=1}^N(\eta_0 h'_nP_{LED}A_n')^2
\end{equation}
and
\begin{equation}
I_{\mathrm{DC},2}=\sum_{n=1}^N\eta_0 h'_nP_{LED}B_n',
\end{equation}
otherwise their values are zero.

\begin{figure}[t!]
\centering
\includegraphics[width=0.8\columnwidth]{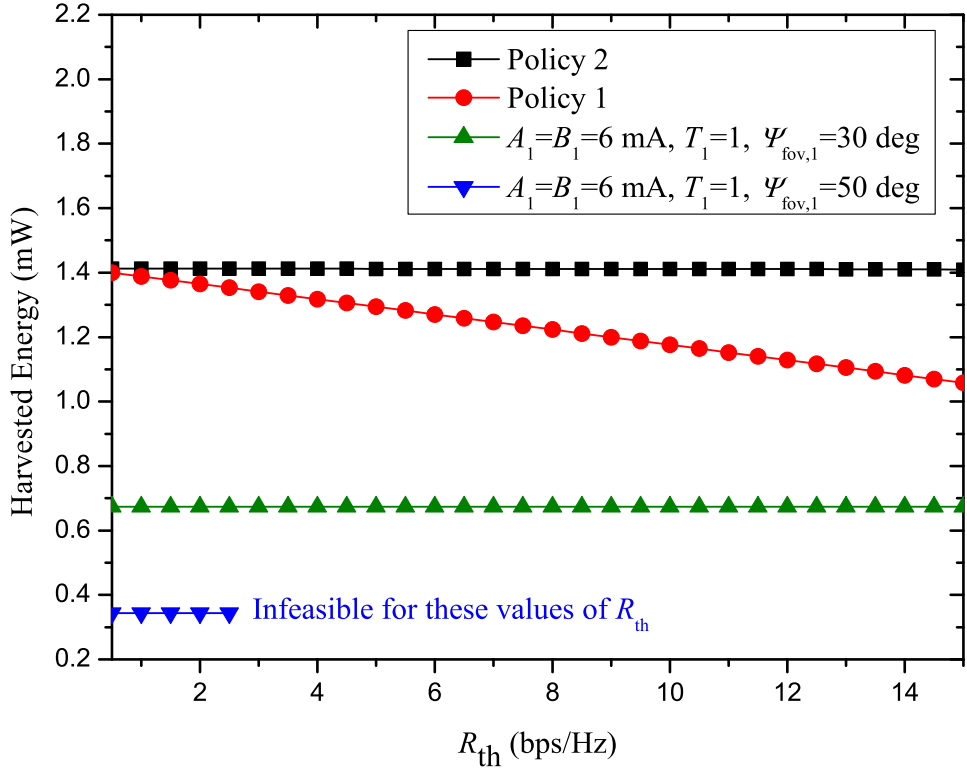}
\caption{Harvested energy vs $R_{th}$ for $N=1$.}
\label{Fig3chapter5}
\end{figure}

\begin{figure}[t!]
\centering
\includegraphics[width=0.8\columnwidth]{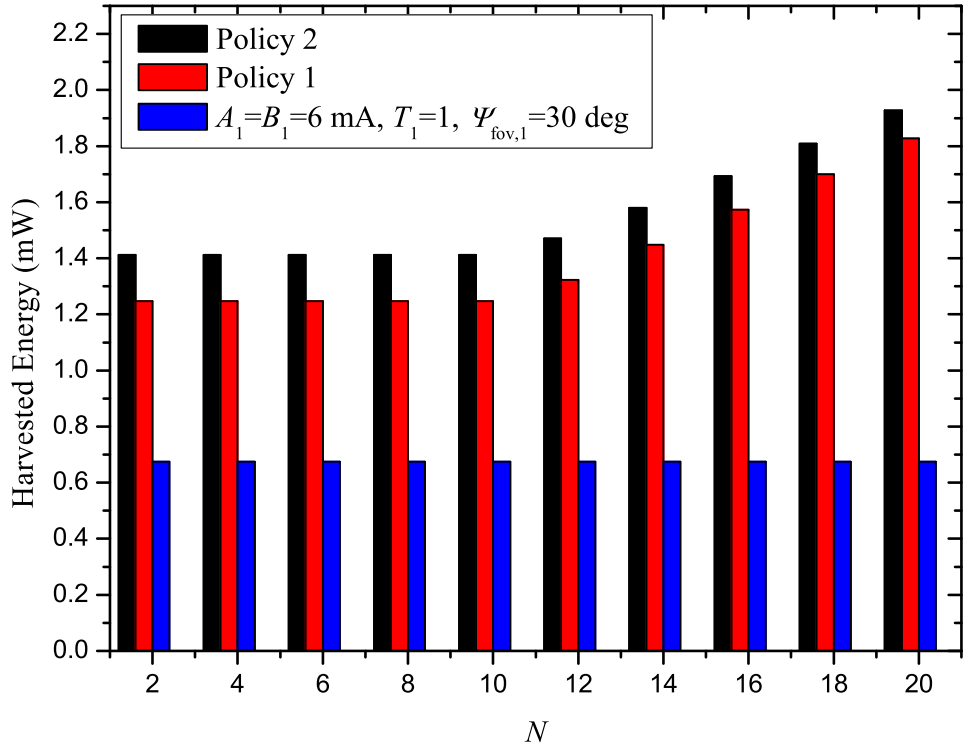}
\caption{Harvested energy vs $N$ for $R_{th}=7$ bps/Hz.}
\label{Fig4chapter5}
\end{figure}

The performance of both optimized policies of Section \ref{strategy 4} are compared for $N=1$, while they are also presented against the case of fixed $A_1$, $B_1$, $T$, and $\Psi_{\mathrm{FoV},1}$, which is considered as the baseline policy. More specifically, in Fig. \ref{Fig3chapter5} the harvested energy is plotted against the rate threshold. As it is observed, both policies significantly outperform the baseline for both values of $\Psi_{\mathrm{FoV},1}$. Regarding the baseline, the value $\Psi_{\mathrm{FoV},1}=50$ deg reduces the harvested energy compared to $\Psi_{\mathrm{FoV},1}=30$ deg, because $g_\mathrm{c}(\psi)$ decreases and thus,  cancels the benefit of receiving the beam of the neighboring LED. Also, the baseline policy with $\Psi_{\mathrm{FoV},1}=50$  deg is infeasible  for medium and high values of $R_{th}$, because the rate threshold cannot be reached, due to the received interference. Interestingly, Policy 2 outperforms Policy 1, especially for the high region of $R_{th}$, which is due to the extra degrees of freedom. Similar conclusions can be obtained by Fig. \ref{Fig4chapter5}, where the harvested energy is plotted against the number of neighboring LEDS. For this specific setup, the baseline with $\Psi_{\mathrm{FoV},1}=50 $ deg is not feasible, and, thus, it is omitted. It is observed here that for a small number of neighboring LEDs, the harvested energy remains constant, since the receiver prefers the smallest FoV setting. However, as the number of neighboring LEDs increases, the receiver prefers the widest FoV setting and the harvested energy increases with the increase of LEDs.
%%CHAPTER 5 (CONCLUSIONS) %%%%%%%%%%%%%%%%%%%%%%%%%%%%%%%%%%%%%%%%%%%%%%%%%%%%%%%
%%\else\fi
%
%%\ifchaptersevenflag
%%CHAPTER 5 (CONCLUSIONS) %%%%%%%%%%%%%%%%%%%%%%%%%%%%%%%%%%%%%%%%%%%%%%%%%%%%%%%
\chapter[Conclusions and Future Work]{Conclusions and Future Work}\label{ch:chapter6}

This chapter sums up the conclusions drawn from each part of this research, while it also presents some possible future extensions.

\section{Conclusions and Future Work}\label{Ch6:S:Conclusions}
Chapter \ref{ch:chapter2} has investigated efficient resource allocation in EHBSs, which has been introduced in Section \ref{ch:chapter2_1}. To this end, in Section \ref{Autonomous Energy Harvesting Base Stations}, a generalized Stackelberg game has been used to investigate the optimal energy and resource allocation problem of an autonomous energy harvesting base station. Existence and uniqueness of the variational equilibrium has been proved. Also, an efficient distributed algorithm has been proposed, which can be adopted by the players in order to maximize their utility functions. The simulation results have revealed that by applying the proposed method, the users maximize their level of satisfaction while the BS maximizes its revenue for all values of harvested energy and for an arbitrary number of users. The investigated system model can be extended to the direction of multiple competing BSs or time evolution of the users' requirements \cite{PD7}.

Chapter \ref{ch:chapter3} has focused on wireless powered communication networks, when the harvest then transmit protocol is adopted.

More specifically, in Section \ref{Introduction to Trade-offs in Wireless Powered Networks}, the dependence between harvested energy, achieved throughputs, fairness,  and energy efficiency has been introduced. Extensive simulations have shown, among others, that the increase of the energy arrival rate reduces the portion of time that is allocated to energy harvesting, and have revealed and interesting trade-off between sum-throughput and fairness.

Section \ref{Wireless Powered Communications with Non-Orthogonal Multiple Access} has investigated the performance and user fairness improvement of wireless powered networks, which apply the harvest-then-transmit protocol, when NOMA is applied for the uplink. Four different schemes were studied, which aimed at the maximization of the total system throughput, or the maximization of equal individual data rates for the users. The analysis and extensive results showed that:
\begin{itemize}
\item NOMA can offer substantial improvement of the rate of the weakest user, in comparison to TDMA, without affecting the total throughput of the network, when maximization of the system throughput is required. 
\item The user fairness was further increased, when NOMA was combined with time-sharing. On the other hand, when symmetric rates are required for the users, the combination of NOMA and time-sharing also outperforms TDMA, when maximizing the equal individual user rate. 
\item There is still an interesting dependence among system throughput, minimum individual data rate, and harvested energy. However, compared to orthogonal conventional schemes,the implementation of a non-orthogonal technique can offer substantial increase in user fairness,  when the users' transmit power is constrained by the harvested energy. 
\end{itemize}
The system model which was adopted in this Section can be considered as a benchmark for more advanced configuration. Apart from its combination with decoding techniques such as time-sharing, further improvement in throughput, energy efficiency and user fairness is expected, when more complex configuration at the BS is assumed, such as multiple antennas, beamforming or scheduling, but also when more advanced hardware capabilities are assumed for the users, such as energy storage units. Of course, this will introduce optimization problems, where the challenge will be to solve them with acceptable complexity.

Section \ref{Maximizing Proportional Fairness in Wireless Powered Communications} has investigated the maximization of proportional fairness, as a means to balance the trade-off between the sum-rate and fairness. The simulation results have revealed that the proposed techniques offer an efficient solution to time and rates allocation problem in wireless-powered communication systems with asymmetrical rates.

Chapter \ref{ch:chapter4} has been dedicated on SWIPT. The available techniques to enable SWIPT have been introduced in Section \ref{ch:chapter4_1}. SWIPT can be used duirng the downlink of a wireless powered network or during the first hop of a relaying system, to enable the uplink of energy harvesting nodes or the second hop of energy harvesting relays. These scenarios have been investigated in Sections \ref{Joint Downlink/Uplink Design for Wireless Powered Networks with Interference} and \ref{Throughput Maximization in Multicarrier Wireless Powered Relaying Networks}, respectively.

In Section \ref{Joint Downlink/Uplink Design for Wireless Powered Networks with Interference}, both the downlink and the uplink of a wireless powered network, in the presence of interference, were optimized. Two different protocols were utilized for the downlink, i.e., NOMA and TDMA, while NOMA with time sharing was used for the uplink. The formulated optimization problems maximize the minimum rate among users, which is achieved both in the downlink and the uplink, by introducing corresponding priority weights. Furthermore, all the parameters regarding the energy harvesting of the users were optimized during the downlink, both for NOMA and TDMA. For this reason, the structure of the formulated non-convex multidimensional optimization problems was investigated, while the initial optimization problems were successfully transformed into the equivalent convex ones, which can be solved with polynomial complexity. The results revealed an interesting dependence among the harvested energy, the achieved minimum downlink/uplink rate, the interference which is imposed on the communication network, and energy efficiency achieved by the implemented protocols. More specifically, the results showed that:
\begin{itemize} 
\item A relatively high downlink rate can be achieved, while the required energy is simultaneously harvested by the users for the uplink, even at the presence of interference.
\item When NOMA is utilized in the downlink, it can offer substantial gains, compared to TDMA, especially in the cases when the downlink is prioritized, and when the users are asymmetrically positioned, i.e., when the cascaded near-far problem appears. This gain offered by the NOMA protocol is especially achieved when the interference power level is low, or in the absence of interference. 
\item The performance of the network, when NOMA is utilized, is achieved requiring less energy transmission by the BS, revealing the energy efficiency of the NOMA protocol, compared to TDMA, when applied to wireless powered networks.
\end{itemize}
The considered system model can be extended to a scenario of heterogeneous users that need access to different applications, and, thus, they do not acquire the same quality of service, where different priority must be given to each user. Also, because interference influences dynamically both the information rate and the amount of energy to be harvested, finding effective interference management techniques is of crucial importance for future networks which employ SWIPT.

In Section \ref{Throughput Maximization in Multicarrier Wireless Powered Relaying Networks}, the achievable rate of a two-hop multicarrier link was maximized, with respect to the power allocation at the source and the relay, and the power splitting ratio for energy harvesting at the relay. Two different solution methods were proposed, while they were compared in terms of complexity and convergence. It was shown that the proposed strategy with any of the two methods leads to a notable increase of the total achievable rate, compared to equal power allocation or best channel selection. The ideas developed in this section can be extended in several directions. First, the presented results can be considered as a benchmark for all other practical cases, such as imperfect CSI or imperfect feedback \cite{GKK4}. Moreover, different communication protocols can be considered, such as two-way, semi-blind AF \cite{GK5}, multihop, and compute-and-forward relaying \cite{GKK3, GK5, GK9, PD3, PD4}. Rather than this, the investigation of the proposed in methods under the assumption of OFDM and/or carrier aggregations is also challenging \cite{PD1,PD2}. Furthermore, since the hardware of the low-cost relay nodes is most likely to be of poor quality, relays are more prone to hardware imperfections \cite{PD5, Alex1, Alex2, Alex3, Alex4}. Also, when SWIPT is applied, energy harvesting is performed at a cost on the signal quality used for detection. Thus, hardware imperfections have a more profound impact on the throughput performance of SWIPT systems and need to be taken into account when analysing the performance of relaying topologies.

Chapter \ref{ch:chapter5} has investigated the simultaneous information and power transfer in OWC systems, termed as SLIPT. Section \ref{Ch5:Introduction} has served as an overview of basic characteristics, applications, advantages, and opportunities of OWC. Section \ref{Ch5:SLIPT} has proposed and optimized new strategies and policies in order to balance the trade-off between the harvested energy and the QoS, when SLIPT  with a solar panel based receiver is utilized. Considering that only the DC component can be used for energy harvesting, in contrast to AC component, which carries the information, the proposed optimization framework has focused on the appropriate selection of the DC bias, FoV, as well as the time dedicated solely to energy harvesting. The presented simulation results have verified that the proposed strategies considerably increase the harvested energy, compared to SLIPT with fixed policies. It is worth-noting that SLIPT creates a vast number of challenges and future research directions. First, it is highlighted that solar panels provide lower speed than PIN and Avalanche PDs, which creates the need for using separate receivers or new generation solar cells (e.g., organic). Also, there is a trade-off between the size of the mobile devices versus the PD's light-collecting area. Moreover, the efficiency of SLIPT can be further improved by the exploitation of new bulbs, which utilize both VL and IR, as well as the multiuser and multiLED coordination.

\ifbibliographyflag
%BIBLIOGRAPHY%%%%%%%%%%%%%%%%%%%%%%%%%%%%%%%%%%%%%%%%%%%%%%%%%%%%%%%%%%%%%%%%%%%
%IEEEtranN BibTeX style is compatible with natbib citation style
\bibliographystyle{IEEEtranN} %other BibTeX style: unsrt
%\nocite{*} %Lists all entries of the bib files
%Add more databases with commas. The abbreviation files first.
%\bibliography{\arenachapterbibpath}
%\bibliography{\mythesisbibpath}

%BIBLIOGRAPHY%%%%%%%%%%%%%%%%%%%%%%%%%%%%%%%%%%%%%%%%%%%%%%%%%%%%%%%%%%%%%%%%%%%
\else\fi

\newpage

\ifpublicationsflag
%LIST OF PUBLICATIONS%%%%%%%%%%%%%%%%%%%%%%%%%%%%%%%%%%%%%%%%%%%%%%%%%%%%%%%%%%%
\chapter{Publications}\label{ch:publications}
%\thispagestyle{empty} %Uncomment in the case of no chapter title
%\null\vfill %Uncomment in the case of no chapter title

%Other interesting choices are \textasteriskcentered, \textborn, \textleaf
\begin{itemize}[\textreferencemark]
\item K. N. Pappi, V. M. Kapinas, and G. K. Karagiannidis, \textquotedblleft A theoretical limit for the ML 
performance of MIMO systems based on lattices,\textquotedblright\ in \emph{Proc. IEEE International Conference 
on Communications (IEEE ICC)}, Budapest, Hungary, June 2013, pp.~1791--1796.
\item V. M. Kapinas and G. K. Karagiannidis, \textquotedblleft A universal MIMO approach for 3GPP wireless 
standards,\textquotedblright\ in \emph{Proc. IEEE Wireless Communications and Networking Conference (IEEE 
WCNC)}, Paris, France, April 2012, pp.~1936--1941.
\item V. M. Kapinas, V. K. Paraforou, C. A. Liondas, and G. K. Karagiannidis, \textquotedblleft Power allocation 
for quasi-orthogonal space-time block codes with $1$ or $2$ bits feedback,\textquotedblright\ in \emph{Proc. 
IEEE International Conference on Communications (IEEE ICC)}, Cape Town, South Africa, May 2010, pp.~1--6.
\item V. M. Kapinas, S. K. Mihos, and G. K. Karagiannidis, \textquotedblleft On the monotonicity of the generalized 
Marcum and Nuttall $Q$-functions,\textquotedblright\ \emph{IEEE Transactions on Information Theory}, vol.~55, 
no.~8, pp.~3701--3710, August 2009.
\item S. K. Mihos, V. M. Kapinas, and G. K. Karagiannidis, \textquotedblleft Lower and upper bounds for the 
generalized Marcum and Nuttall $Q$-functions,\textquotedblright\ in \emph{Proc. 3rd International Symposium on 
Wireless Pervasive Computing (ISWPC)}, Santorini, Greece, May 2008, pp.~736--739.
\item V. M. Kapinas, M. Ili\'{c}, G. K. Karagiannidis, and M. Pejanovi\'{c}-{\DJ}uri\v{s}i\'{c}, \textquotedblleft 
Aspects on space and polarization diversity in wireless communication systems,\textquotedblright\ in 
\emph{Proc. 15th Telecommunications Forum (TELFOR)}, Belgrade, Serbia, November 2007, pp.~183--186.
\item V. M. Kapinas, P. Horv\'{a}th, G. K. Karagiannidis, and I. Frigyes, \textquotedblleft Time synchronization 
issues for quasi-orthogonal space-time block codes,\textquotedblright\ in \emph{Proc. International Workshop on 
Satellite and Space Communications (IWSSC)}, Salzburg, Austria, September 2007, pp.~66--70.
\item V. M. Kapinas and G. K. Karagiannidis, \textquotedblleft On decoupling of quasi-orthogonal space-time block 
codes based on inherent structure,\textquotedblright\ in \emph{Proc. 16th IST Mobile and Wireless 
Communications Summit (IST Summit)}, Budapest, Hungary, July 2007, pp.~1--5.
\end{itemize}

\newpage\thispagestyle{empty}

\phantom{for blank page} %This command creates a blank page (it assists the colophon to be placed on the recto) 

%LIST OF PUBLICATIONS%%%%%%%%%%%%%%%%%%%%%%%%%%%%%%%%%%%%%%%%%%%%%%%%%%%%%%%%%%%
\else\fi

\newpage

\ifcolophonpageflag
%COLOPHON PAGE%%%%%%%%%%%%%%%%%%%%%%%%%%%%%%%%%%%%%%%%%%%%%%%%%%%%%%%%%%%%%%%%%%
%\chapter*{}\label{ch:colophon}
\thispagestyle{empty} %Uncomment for no pagination
\null\vfil %Uncomment in the case of no chapter title
\begin{center}
\LARGE{\textsf{Colophon}}
\end{center}
This thesis was typeset with \LaTeXe\ in A4 (no trimmed) stock paper, using the \textsf{memoir} class created by Peter
Wilson. For the body text, the standard \textsf{Computer Modern} and \AmS\ fonts in Type~1 (PostScript) format and
\nprounddigits{2}\lptnum{\fonttypesize}pt size have been used with \nprounddigits{2}\lptnum{\baselineskip}pt leading
(line spacing). The Greek text, where available, was typeset with the Type~1 \textsf{CB} Greek fonts developed by Claudio Beccari and facilitated by the \textsf{babel} (English, Greek support), \textsf{teubner} and \textsf{inputenc} (\verb?iso-8859-7? option) packages. The bibliography items and associated citations have been generated with the \textsf{IEEEtranN} \BibTeX\ and \textsf{natbib} (\verb?numbers?, \verb?sort&compress? options) styles. The \textsf{acronym} (\verb?printonlyused?, \verb?smaller? options) and \textsf{nomencl} (\verb?intoc?, \verb?noprefix? options) packages have been used for the abbreviation and nomenclature lists, respectively, while the \textsf{bigstrut}, \textsf{multirow} and \textsf{blkarray} have assisted the design of matrices. The \textsf{psfrag} and \textsf{subfigure} (\verb?small?, \verb?bf?, \verb?tight? options) have been found useful for some figures. The rest of packages called include the \textsf{numprint}, \textsf{calc}, \textsf{dsfont}, \textsf{textcomp}, and \textsf{dtklogos}, apart from the most common \textsf{amsmath}, \textsf{amssymb} (superset of \textsf{amsfonts}), and \textsf{amsthm}. Finally, the \textsf{mathrsfs} has been invoked for enabling the use of the Ralph Smith's Formal Script font (in math mode). Below, the values of the most important page and class layout parameters are given in terms of common length units.
\begin{table}[h]\centering
\nprounddigits{5}%number of decimal points
\npdigits{3}{5}%defines the alignment in the tabular environment
\legend{\textsc{Page\slash Class Layout Parameters}}
\begin{tabular}{>{\bfseries}l c c c}
\toprule
 & \textbf{points} & \textbf{inches} & \textbf{centimeters}\\
 & \textbf{(pt)}   & \textbf{(in)}   & \textbf{(cm)}\\
\midrule[\heavyrulewidth]
\textbackslash stockheight    &  \lptnum{\stockheight}    &  \linnum{\stockheight}    &  \lcmnum{\stockheight}\\
\textbackslash paperheight    &  \lptnum{\paperheight}    &  \linnum{\paperheight}    &  \lcmnum{\paperheight}\\
\textbackslash textheight     &  \lptnum{\textheight}     &  \linnum{\textheight}     &  \lcmnum{\textheight}\\
\textbackslash stockwidth     &  \lptnum{\stockwidth}     &  \linnum{\stockwidth}     &  \lcmnum{\stockwidth}\\
\textbackslash paperwidth     &  \lptnum{\paperwidth}     &  \linnum{\paperwidth}     &  \lcmnum{\paperwidth}\\
\textbackslash textwidth      &  \lptnum{\textwidth}      &  \linnum{\textwidth}      &  \lcmnum{\textwidth}\\
\textbackslash uppermargin    &  \lptnum{\uppermargin}    &  \linnum{\uppermargin}    &  \lcmnum{\uppermargin}\\
\textbackslash topskip        &  \lptnum{\topskip}        &  \linnum{\topskip}        &  \lcmnum{\topskip}\\
\textbackslash lowermargin    &  \lptnum{\lowermargin}    &  \linnum{\lowermargin}    &  \lcmnum{\lowermargin}\\
\textbackslash footskip       &  \lptnum{\footskip}       &  \linnum{\footskip}       &  \lcmnum{\footskip}\\
\textbackslash spinemargin    &  \lptnum{\spinemargin}    &  \linnum{\spinemargin}    &  \lcmnum{\spinemargin}\\
\textbackslash foremargin     &  \lptnum{\foremargin}     &  \linnum{\foremargin}     &  \lcmnum{\foremargin}\\
\textbackslash oddsidemargin  &  \lptnum{\oddsidemargin}  &  \linnum{\oddsidemargin}  &  \lcmnum{\oddsidemargin}\\
\textbackslash evensidemargin &  \lptnum{\evensidemargin} &  \linnum{\evensidemargin} &  \lcmnum{\evensidemargin}\\
\textbackslash beforechapskip &  \lptnum{\beforechapskip} &  \linnum{\beforechapskip} &  \lcmnum{\beforechapskip}\\
\textbackslash midchapskip    &  \lptnum{\midchapskip}    &  \linnum{\midchapskip}    &  \lcmnum{\midchapskip}\\
\textbackslash afterchapskip  &  \lptnum{\afterchapskip}  &  \linnum{\afterchapskip}  &  \lcmnum{\afterchapskip}\\
\bottomrule
\end{tabular}
\end{table}
\vfil
\begin{leftbar}
\noindent Digital version is recommended to be printed in black \& white on both sides of A4 (8.3x11.7in) paper.
\end{leftbar}

%\newpage\thispagestyle{empty}
%\phantom{for blank page} %This command creates a blank page for the back cover

%COLOPHON PAGE%%%%%%%%%%%%%%%%%%%%%%%%%%%%%%%%%%%%%%%%%%%%%%%%%%%%%%%%%%%%%%%%%%
\else\fi

\newpage\thispagestyle{empty}\phantom{nothing}

\end{document}